\pdfoutput=1
\documentclass[runningheads,openany]{svmult}

\usepackage{palatino}

\usepackage{euler}

\usepackage{helvet}

\usepackage{graphicx}  % standard LaTeX graphics tool

\usepackage{physprbb}  

\usepackage{mathrsfs}

\usepackage{proc}  

\usepackage{epsfig}

\usepackage{amssymb}

\makeindex             % used for the subject index

\usepackage{amsmath}   % useful for coding complex math
\usepackage{multicol}
\usepackage{bm}
\usepackage[dvips,cam]{crop}

\let\tocauthor\authorrunning

\begin{document}

\frontmatter

%%\input{cover} % this is now generated separately (bcover)

                % in order not to confuse page numbering

%%
\thispagestyle{empty}
\parindent=0pt

{\Large\sc Blejske delavnice iz fizike \hfill Letnik~10, \v{s}t. 2}

\smallskip

{\large\sc Bled Workshops in Physics \hfill Vol.~10, No.~2}

\smallskip

\hrule

\hrule

\hrule

\vspace{0.5mm}

\hrule

\medskip
{\sc ISSN 1580-4992}

\vfill

\bigskip\bigskip
\begin{center}

{\bfseries 
{\Large  Proceedings to the $12^\textrm{th}$ Workshop}\\
{\Huge What Comes Beyond the Standard Models\\}
\bigskip
{\Large Bled, July 14--24, 2009}\\
\bigskip
%% space for arXiV reference {\Large [ hep-ph/ \,---\, July 2007]}
}

\vspace{5mm}

\vfill

{\bfseries\large
Edited by

\vspace{5mm}
Norma Susana Manko\v c Bor\v stnik%\rlap{$^{1}$}

\smallskip

Holger Bech Nielsen%\rlap{$^{2}$}

%%\smallskip
%%
%%Colin D. Froggatt%\rlap{$^{3}$}
%%
\smallskip

Dragan Lukman%\rlap{$^1$}

\bigskip

%{\em\normalsize $^1$University of Ljubljana, $^2$ Niels Bohr Institute, %
%$^3$ Glasgow University}

\vspace{12pt}

\vspace{3mm}

\vrule height 1pt depth 0pt width 54 mm}

\vspace*{3cm}

{\large {\sc  DMFA -- zalo\v{z}ni\v{s}tvo} \\[6pt]
{\sc Ljubljana, december 2009}}
\end{center}
\newpage

\thispagestyle{empty}
\parindent=0pt
\begin{flushright}
{\parskip 6pt
{\bfseries\large
                  The 12th Workshop \textit{What Comes Beyond  
                  the Standard Models}, 14.-- 24. July 2009, Bled}

\bigskip\bigskip

{\bfseries\large was organized by}

{\parindent8pt
\textit{Department of Physics, Faculty of Mathematics and Physics,
University of Ljubljana}

%\textit{Primorska Institute of Natural Sciences and Technology, Koper}
}

\bigskip

{\bfseries\large and sponsored by}

{\parindent8pt
\textit{Slovenian Research Agency}

\textit{Department of Physics, Faculty of Mathematics and Physics,
University of Ljubljana}

\textit{Society of Mathematicians, Physicists and Astronomers
of Slovenia}}}
\bigskip
\medskip

{\bfseries\large Organizing Committee}

\medskip

{\parindent9pt
\textit{Norma Susana Manko\v c Bor\v stnik}

\textit{Holger Bech Nielsen}

\textit{Maxim Yu. Khlopov}}

\end{flushright}

\setcounter{tocdepth}{0}

\tableofcontents

\cleardoublepage

\chapter*{Preface}
\addcontentsline{toc}{chapter}{Preface}
The series of workshops on "What Comes Beyond the Standard Model?" started
in 1998 with the idea of organizing a real workshop, in which participants
would spend most of the time in discussions, confronting different
approaches and ideas. The picturesque town of Bled by the lake of the
same name, surrounded by beautiful mountains and offering pleasant walks,
was chosen to stimulate the discussions.

The idea was successful and has developed into an annual workshop, which is
taking place every year since 1998. This year the twelfth workshop took place.
Very open-minded and fruitful discussions
have become the trade-mark of our workshop, producing several published works.
It takes place in the house of Plemelj, which belongs to the Society of
Mathematicians, Physicists and Astrono\-mers of Slovenia.

In this twelfth workshop, which took place from 14th to 24th of July
2009, we were discussing several topics, most of them presented in
this Proceedings mainly as talks and partly in the discussion section. 
The main topic was this time the "approach unifying spin and charges", 
proposed by Norma, as the new way beyond the "standard model of the 
electroweak and colour interactions", accompanied by the critical 
discussions about the chance which this theory has 
to answer the open questions which the "standard model" leaves unanswered. 
Proposing the mechanism for generating families, this "approach" is 
predicting the fourth family to be possibly seen at LHC 
and the stable fifth family which have a chance to form the dark matter. 
The discussions of the questions: Is the "approach unifying spin and charges"  
the right way beyond the standard model? Are the clusters of the fifth family members 
alone what constitute the dark matter? Can the fifth family baryons explain the 
observed properties of the dark matter with the direct measurements included? 
What if such a scenario is not confirmed by the direct measurements? 
What are next steps in evaluating properties of the predicted Yukawa couplings? 
Can we find the way out (besides by a choice of appropriate boundary conditions) 
of the "no go theorem" of Witten, saying that there is 
a little chance  for these kind of theories (to which also the "approach unifying 
spins and charges" belong), since the masses of the fermions, predicted by 
these theories should be too high?

Talks and discussions in our workshop are not at
all talks in the usual way. Each talk or discussions lasted several
hours, divided in two hours blocks, with a lot of questions,
explanations, trials to agree or disagree from the audience or a
speaker side. 
Most of talks are "unusual" in the sense that they are trying to find out new 
ways of understanding and describing the observed  
phenomena. Although we always hope that the discussions will in the very year 
proceedings manifest in the progress published in the corresponding proceedings, it 
happens  many a time that the topics appear in the next or after the next year proceedings. 
This happened also in this year. Therefore neither the discussion section nor the 
talks published in this proceedings, manifest all the discussions and the work 
done in this workshop.  

Several videoconferences were taking place during the Workshop on various
topics. It was organized by the Virtual Institute for Astrophysics
(www.cosmovia.org) of Maxim Khlopov with able support by Didier Rouable. 
We managed to have ample discussions.   The transparent and very 
systematic overview 
of what does the LHC, which is in these days starting again, expect 
to measure in the near future, was presented by John Ellis,  
who stands behind the theoretical understanding of the LHC.
The talks and discussions can be found online at\\ [1ex]
http://viavca.in2p3.fr/bled\_09.html\\[1ex]
The organizers thank all the participants for 
fruitful discussions and talks.

Let us present the starting point of our discussions: 
What science has learned up to now are several effective theories
 which, after making several
starting assumptions, lead to theories (proven or not to be consistent
in a way that they do not run into obvious contradictions), and which,
some of them, are within the accuracy of calculations and experimental
data, (still) in agreement with the observations, the others might be tested 
in future, and might answer at least some of the
open questions, left open by the scientific community accepted
effective theories.
It is a hope that the law of Nature
is "simple" and "elegant", on one or another way, manifesting symmetries or 
complete randomness, whatever the "elegance" and "simplicity" might mean (as few 
assumptions as possible?, very simple starting action?), while the observed 
states are usually not, suggesting that the "effective theories, laws, models" are
usually very complex.

Let us write  in this workshop discussed 
open questions which the two standard models (the electroweak and the
cosmological) leave unanswered:

\begin{itemize}
\item Why has Nature made a choice of four (noticeable) dimensions
    while all the others, if existing, are hidden? And what are the
    properties of space-time in the hidden dimensions?

\item How could "Nature make the decision" about breaking of
    symmetries down to the noticeable ones, if coming from some higher
    dimension d?

\item Why is the metric of space-time Minkowskian and how is the choice
 of metric connected with the evolution of our universe(s)?

\item Why do massless fields exist at the low energy regime at all? 
Where does the weak scale come from?

\item Why do only left-handed fermions carry the weak charge? Why does
    the weak charge break parity?
    
\item Where do families come from?

\item What is the origin of Higgs fields? Where does the Higgs mass
    come from?

\item Can all known elementary particles be understood as different
    states of only one particle, with a unique internal space of
    spins and charges?
    
\item Can one find a loop hole through the Witten's "no-go theorem" and 
      give them back a chance to the Kaluza-Klein-like theories to be the right way 
      beyond the "standard model of the electroweak and colour interaction"? 

\item How can all gauge fields (including gravity) be unified (and
    quantized)?

\item What is our universe made out of besides the (mostly) first family baryonic matter?

\item What is the role of symmetries in Nature?
\end{itemize}

We have discussed these and other questions for ten days.  The reader
can see our progress in some of these questions in this proceedings.
Some of the ideas are treated in a very preliminary way.  Some ideas
still wait to be discussed (maybe in the next workshop) and understood
better before appearing in the next proceedings of the Bled workshops.

The organizers are grateful to all the participants for the lively discussions
and the good working atmosphere. \\[1cm]

\parbox[b]{\textwidth}{%
   \textit{Norma Susana Manko\v c Bor\v stnik, Holger Bech Nielsen,}\\
   \textit{Maxim Yu. Khlopov, Dragan Lukman} 
\hfill\textit{Ljubljana, December 2009}}

\newpage

\cleardoublepage

%%%%%%%%%%%%%%%%%%%%%%%%%%%%%%%%%%%%%%%%%%%%%%%%%%%%%%%%%%%%%

\mainmatter

\parindent=20pt

\setcounter{page}{1}

%%
%% Csaba Balazs, 05.11.2009
%%% Proceedings to the Workshop 
%%% 'What comes beyond the Standard Model'
%%% Bled
%%\documentclass[11pt]{article}
%%\usepackage{graphicx}
\author{C. Bal\'azs\thanks{csaba.balazs@sci.monash.edu.au} and  %
D. Carter\thanks{daniel.carter@sci.monash.edu.au}}
\title{Likelihood Analysis of the Next-to-minimal Supergravity Motivated Model}
\institute{%
School of Physics, Monash University,\\Melbourne Victoria 3800, Australia}

\titlerunning{Likelihood Analysis of the Next-to-minimal Supergravity Motivated Model}
\authorrunning{C. Bal\'azs and D. Carter}
\maketitle

\begin{abstract}
In anticipation of data from the Large Hadron Collider (LHC) and the potential discovery of supersymmetry, we calculate the odds of the next-to-minimal version of the popular supergravity motivated model (NmSuGra) being discovered at the LHC to be 4:3 (57 \%).
We also demonstrate that viable regions of the NmSuGra parameter space outside the LHC reach can be covered by upgraded versions of dark matter direct detection experiments, such as super-CDMS, at 99 \% confidence level.  
Due to the similarities of the models, we expect very similar results for the constrained minimal supersymmetric standard model (CMSSM).
\end{abstract}

\section{Introduction}

% SUSY 

Supersymmetry is one of the most robust theories that can solve outstanding problems of the standard model (SM) of elementary particles.  The theory naturally explains the dynamics of electroweak symmetry breaking while preserving the hierarchy of fundamental energy scales. It also readily accommodates dark matter, the asymmetry between baryons and anti-baryons, the unification of gauge forces, gravity, and more.
%
% SUSY ?
%
But if supersymmetry is the solution to the problems of the standard model, then its natural scale is the electroweak scale, and it is expected to be observed in upcoming experiments, most notably the CERN Large Hadron Collider (LHC).  In this work, we will attempt to determine, quantitatively, what the chances are that this may occur for the simplified case of a constrained supersymmetric model.

% MSSM

The minimal supersymmetric extension of the standard model (MSSM) faces several significant issues, such as the little hierarchy problem \cite{Giudice:2008bi} and the so-called $\mu$ problem \cite{Kim:1983dt}. However extensions of the MSSM by gauge singlet superfields not only resolve the $\mu$ 
problem, but can also ameliorate the little hierarchy problem \cite{Dermisek:2005ar,BasteroGil:2000bw,Gunion:2008kp}.  In the next-to-minimal MSSM (NMSSM), the 
$\mu$ term is dynamically generated and no dimensionful parameters are 
introduced in the superpotential (other than the vacuum expectation values that 
are all naturally weak scale), making the NMSSM a truly natural model (see \cite{Balazs:2008ph} for references).  

% CMSSM

For the sake of simplicity and elegance, we choose to impose minimal super\-gravity-motivated (mSuGra) boundary conditions; specifically, universality of s\-particle masses, gaugino masses, and tri-linear couplings at the grand unification theory (GUT) scale. Thus we define the next-to-minimal supergravity-motivated (NmSuGra) model.

% CNMSSM likelihood analysis

Using a Bayesian likelihood analysis, we identify the regions in the parameter space of the NmSuGra model that are preferred by the present experimental limits from various collider, astrophysical, and low-energy measurements. Thus we show that, given current experimental constraints, the favored parameter space can be detected by a combination of the LHC and an upgraded CDMS at the 95 \% confidence level.

In the next section we define the next-to-minimal version of the supergravity motivated model (NmSuGra).  Then, in Section \ref{sec:Bayes}, we summarize the main concepts of Bayesian inference that we use in this work.  Section \ref{sec:likelihood} contains the numerical results of our likelihood analysis, and Section \ref{sec:detection} gives the outlook for the experimental detection of NmSuGra.

\section{The next-to-minimal supergravity motivated model}
\label{sec:NmSuGra}

% CNMSSM & cNMSSM

The next-to-minimal supersymmetric model (NMSSM) is defined by the superpotential 
\begin{eqnarray}
 W_{NMSSM} = W_{MSSM}|_{\mu = 0} + \lambda \hat{S} \hat{H}_u \cdot \hat{H}_d + \frac{\kappa}{3} \hat{S}^3,
\label{eq:W_NMSSM}
\end{eqnarray}
where $W_{MSSM}|_{\mu = 0}$ is the MSSM superpotential containing only Yukawa terms and having $\mu$ set to zero \cite{Ellwanger:2005dv}, and $\hat{S}$ is a standard gauge singlet with dimensionless couplings $\lambda$ and $\kappa$.  The couplings $\lambda$, $\kappa$, and $y_i$ are dimensionless, and $\hat{X} \cdot \hat{Y} = \epsilon_{\alpha\beta} \hat{X}^\alpha \hat{Y}^\beta$ with the fully antisymmetric tensor normalized as $\epsilon_{11} = 1$.

% CNMSSM

We use supergravity motivated boundary conditions to parametrize the soft masses and tri-linear couplings.  Defining a constrained version of the NMSSM, we assume unification of the gaugino masses to $M_{1/2}$, the sfermion and Higgs masses to $M_0$, and the tri-linear couplings to $A_0$ at the grand unified theory (GUT) scale where the three standard gauge couplings meet $g_1 = g_2 = g_3 = g_{GUT}$.  After electroweak symmetry breaking, our constrained NMSSM model has only five free parameters and a sign. Defining $\tan\beta = \langle H_u \rangle/\langle H_d \rangle$, the parameters of the next-to-minimal supergravity motivated model (NmSuGra) are
\begin{eqnarray}
 P = \{M_0, M_{1/2}, A_0, \tan\beta, \lambda, {\rm sign}(\mu)\} .
\label{eq:5Para}
\end{eqnarray}
Furthermore, from Eq.\ref{eq:W_NMSSM} we see that when the singlet acquires a vev, the MSSM $\mu$ term is dynamically generated as $\mu = \lambda \langle S \rangle$, and thus the NMSSM naturally solves the $\mu$ problem.

% CNMSSM, others !!! update citations !!!
 
Different constrained versions of the NMSSM have been studied in the recent literature \cite{Djouadi:2008yj,Hugonie:2007vd,Belanger:2005kh,Cerdeno:2007sn,Djouadi:2008uw}.  In the spirit of the CMSSM/mSuGra, we adhere to universality and use only $\lambda$ to parametrize the singlet sector.  This way, we keep all the attractive features of the CMSSM/mSuGra while the minimal extension alleviates problems rooted in the MSSM, making the NMSSM a more natural model.

As we have shown in our previous work \cite{Balazs:2008ph}, NmSuGra phenomenology bears a high similarity to the minimal supergravity motivated model.  The most significant departures from a typical mSuGra model are the possibility of a sing\-lino-dominated neutralino and the extended Higgs sector, which may provide new resonance annihilation channels and Higgs decay channels, potentially weakening the mass limit from LEP.

\section{Bayesian inference}
\label{sec:Bayes}

Since several excellent papers have appeared on this subject recently \cite{Feroz:2008wr,Trotta:2008bp,AbdusSalam:2009qd}, in this section, we summarize the concepts of Bayesian inference that we use in our analysis in a compact fashion.  Our starting hypothesis $H$ is the validity of the NmSuGra model.  The conditional probability $\mathcal{P}(P|D;H)$ quantifies the validity of our hypothesis by giving the chance that the NmSuGra model reproduces the available experimental data $D$ with its parameters set to values $P$.  When this probability density is integrated over a region of the parameter space it yields the posterior probability that the parameter values fall into the given region.

% Bayes

Bayes' theorem provides us with a simple way to calculate the posterior probability distribution as
\begin{eqnarray}
 \label{eq:bayes}
 \mathcal{P}(P|D;H) = \mathcal{P}(D|P;H) \frac{\mathcal{P}(P|H)}{\mathcal{P}(D|H)} .
\end{eqnarray}
Here $\mathcal{P}(D|P;H)$ is the likelihood that the data is predicted by NmSuGra with a specified set of parameters.  The a-priori distribution of the parameters within the theory $\mathcal{P}(P|H)$ is fixed by purely theoretical considerations independently from the data.  The evidence $\mathcal{P}(D|H)$ gives the probability of the hypothesis in terms of the data alone, equivalent to integrating out the parameter dependence.

For statistically independent data the likelihood is the product of the likelihoods for each observable. For normally-distributed measurements the likelihood is given by:
\begin{eqnarray}
 {\cal L}_i(D,P;H) = \frac{1}{\sqrt{2 \pi} \sigma_i} \exp(\chi_i^2(D,P;H)/2) ,
\end{eqnarray}
where the exponents $\chi_i^2(D,P;H)/2 = (d_i - t_i(P;H))^2/2 \sigma_i^2$ are defined in terms of the experimental data $D = \{d_i \pm \sigma_{i,e}\}$ and theoretical predictions $T = \{t_i \pm \sigma_{i,t}\}$ for these measurables.  Independent experimental and theoretical uncertainties combine into $\sigma_i^2 = \sigma_{i,e}^2 + \sigma_{i,t}^2$. In cases when the experimental data only specify a lower (or upper) limit, we replace the Gaussian likelihood with a likelihood based on the error function. Often, the profile if the likelihood distribution is used for statistical inference, however this disregards information about the structure of the parameter space itself. In Bayesian statistics we use the so-called marginalized probability, given by the integral of the posterior probability density over all parameter space except the quantity of interest.

\section{Likelihood analysis of NmSuGra}
\label{sec:likelihood}

Our main aim is to calculate the posterior probability distributions for the five continuous parameters of NmSuGra and check the consistency of the model aga\-inst available experimental data.  To this end, we use the publicly available computer code NMSPEC \cite{Ellwanger:2006rn} to calculate the spectrum of the superpartner masses and their physical couplings from the model parameters given in Eq.~(\ref{eq:5Para}).  Then, we use NMSSMTools 2.1.0 and micrOMEGAs 2.2 \cite{Belanger:2006is} to calculate the abundance of neutralinos ($\Omega h^2$) \cite{Komatsu:2008hk}, the spin-independent neutralino-proton elastic scattering cross section ($\sigma_{SI}$) \cite{Ahmed:2008eu}, the NmSuGra contribution to the anomalous magnetic moment of the muon ($\Delta a_{\mu}$) \cite{Jegerlehner:2009ry}, and various b-physics related quantities \cite{Barberio:2008fa,Artuso:2009jw}.  We also impose limits from negative searches for the sparticle masses \cite{AbdusSalam:2009qd}, applying a lower lightest Higgs mass limit where appropriate, as shown in \cite{Barger:2006dh}.  Among the standard input parameters, $m_b(m_b) = 4.214$ GeV and $m_t^{pole}=171.4$ GeV are used.

Using the above specified tools, we generate theoretical predictions for NmSuGra in the following part of its parameter space:
 $0 < M_0 < 5 ~{\rm TeV}, 0 < M_{1/2} < 2 ~{\rm TeV}, -3 ~{\rm TeV} < A_0 < 5 ~{\rm TeV}, 0 < \tan\beta < 60, 10^{-5} < \lambda < 0.6, {\rm sign}(\mu) > 0$.
In this work, we only consider the positive sign of $\mu$ because, similarly to mSuGra \cite{Feroz:2008wr}, the likelihood function is suppressed by $\Delta a_\mu$ and $B(b \to s \gamma)$ in the negative $\mu$ region.
We calculate posterior probabilities using two methods: a uniform random scan, and Markov Chain Monte Carlo (implementing the metropolis algorithm) as described in \cite{Baltz:2006fm}, which is significantly more efficient but marginally less consistent. In general, the two methods are in good agreement. 

\subsection{Posterior probabilities}

\begin{figure}[th]
  \includegraphics[width=0.49\textwidth]{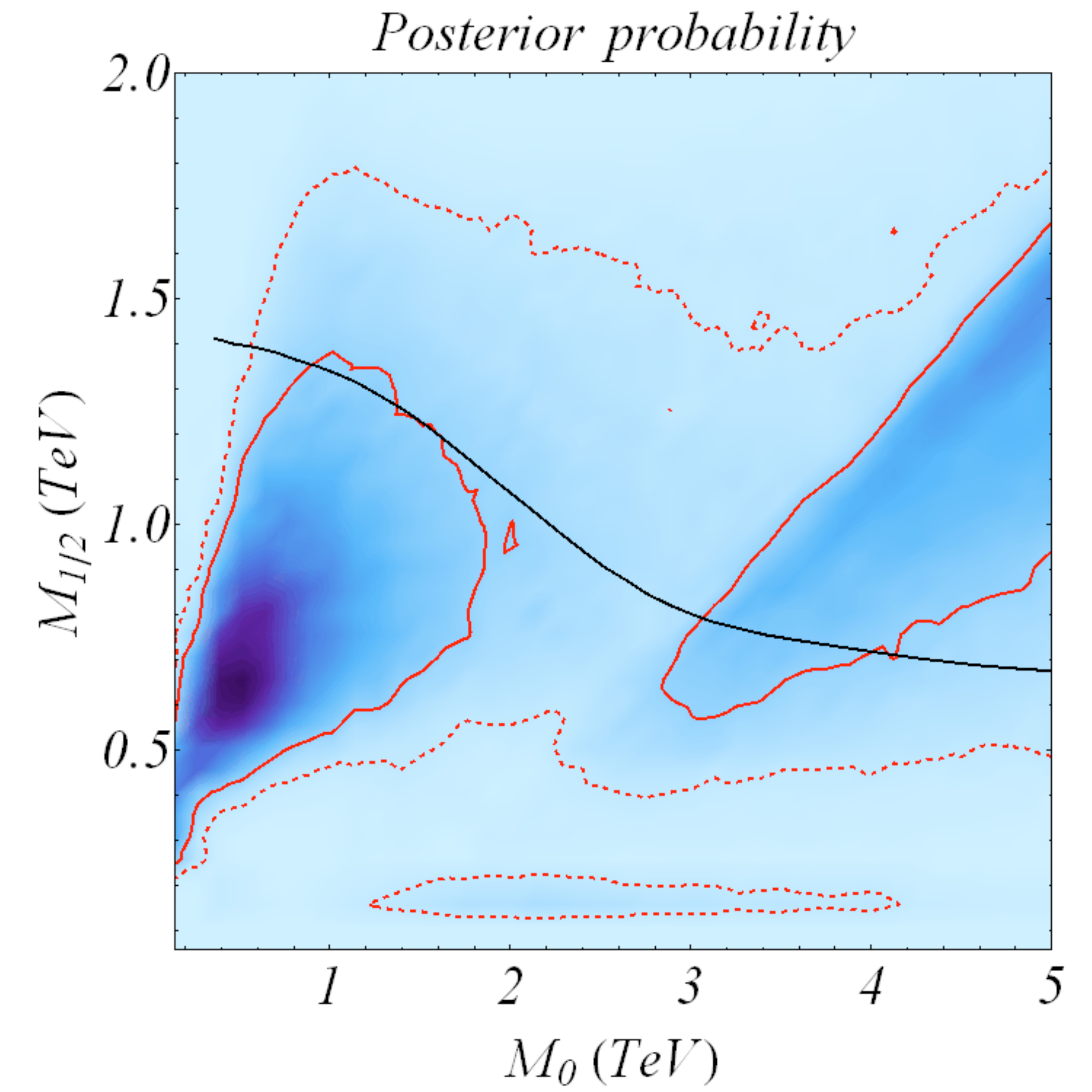}
  \includegraphics[width=0.49\textwidth]{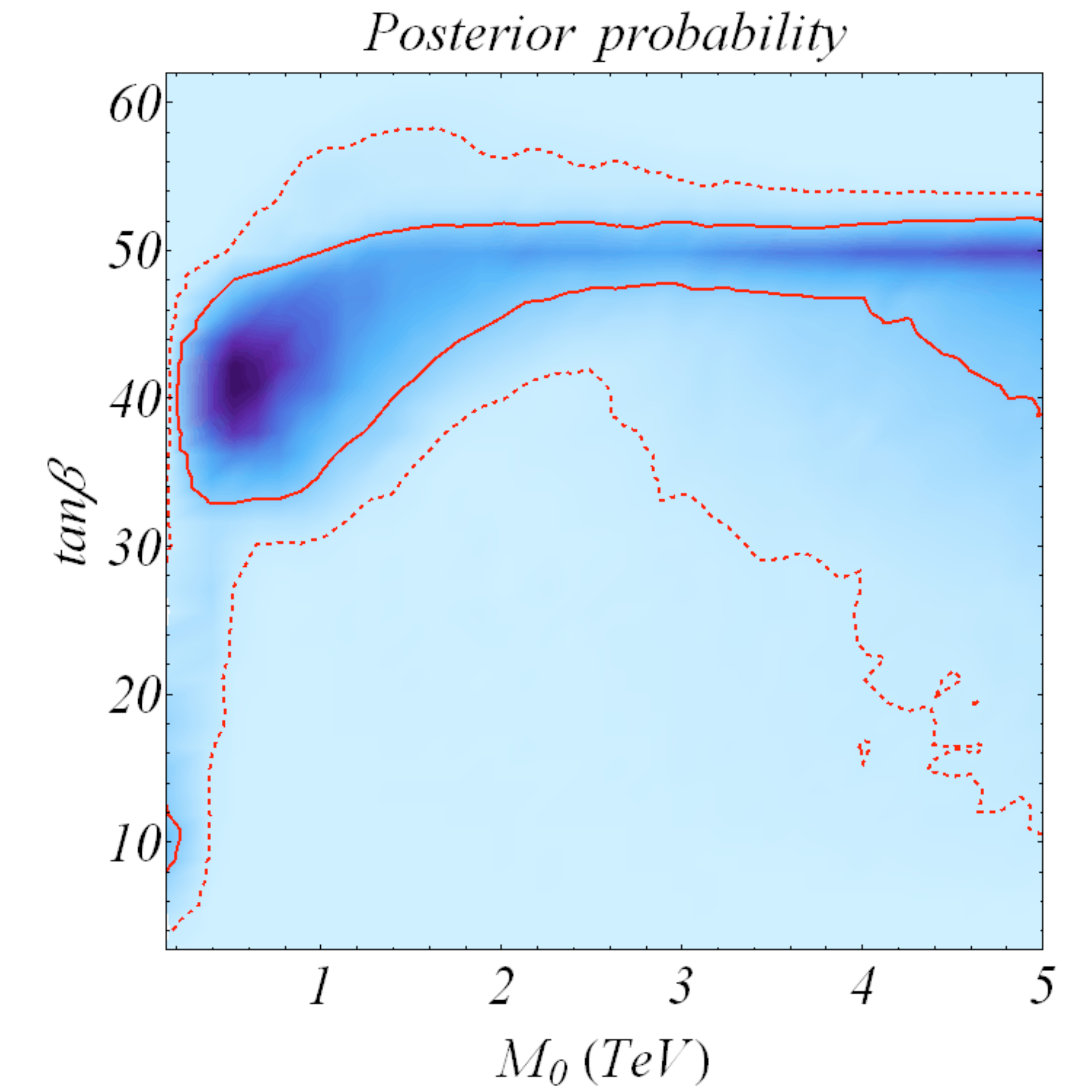}
  \caption{Posterior probabilities marginalized to pairs of NmSuGra input parameters.  The higher probability regions are darker.  Solid (dotted) red lines indicate 68 (95) percent confidence level contours.  On the left frame the black curve shows the estimated reach of the LHC for 100 fb$^{-1}$ luminosity \cite{Baer:2003wx}.}
  \label{fig:MPP2D.inputs}
\end{figure}

We now turn to our numerical results in Figure \ref{fig:MPP2D.inputs}, which shows the posterior probability marginalized to different pairs of NmSuGra input parameters. In the left frame we show the posterior probability marginalized to the plane of the common scalar and gaugino masses, $M_0$ vs. $M_{1/2}$.  The slepton co-annihilation region combined with Higgs resonance corridors, at low $M_0$ and low to moderate $M_{1/2}$ supports most of the probability.  This region is clearly separated from the focus point at high $M_0$ and moderate to high $M_{1/2}$, a large part of which falls in the 68 \% confidence level. While the contribution from $\Delta a_\mu$ strongly suppresses the likelihood at higher values of $M_0$ and $M_{1/2}$, the volume of the focus point region is quite large, contrasted with the highly-sensitive sfermion coannihilation region. This shifts the expectation for $M_0$ much higher than its likelihood distribution might suggest, and implies that it would probably not be reasonable to confine $M_0$ to low values. Most of the focus point happens at high $\tan\beta (\sim 50)$ where the traditional focus point region merges with multiple Higgs resonance corridors creating very wide regions consistent with WMAP.

In $M_{1/2}$, there appears a narrow region close to 150 GeV that corresponds to neutralinos resonantly self-annihilating via the lightest scalar Higgs boson in the s-channel.  This 'sweet spot' emerges as a combined high-likelihood and volume effect. Part of this region is allowed in NmSuGra due to the somewhat relaxed mass limit by LEP on the lightest Higgs.  The narrowness of this strip correlates with the smallness of the lightest Higgs width.

%\FIGURE[th]{
%\includegraphics{file=mpp.m0.m12.eps,width=0.49\textwidth}
%\includegraphics{file=mpp.m0.tanb.eps,width=0.49\textwidth}
%\caption{Posterior probability densities marginalized to pairs of NmSuGra input parameters.  The higher probability regions are darker.  Solid (dotted) red lines indicate 68 (95) percent confidence level contours.  On the left frame the black curve shows the estimated reach of the LHC for 100 fb$^{-1}$ luminosity \cite{Baer:2003wx}.}
%\label{fig:MPP2D.inputs}}

The top right frame of Figure \ref{fig:MPP2D.inputs} shows the distribution of the posterior probability in the $M_0$ vs. $\tan\beta$ frame.  This makes it clear that most of the probable points are carried by Higgs resonant corridors toward higher $\tan\beta$, and the sfermion co-annihilation, due to its narrowness in $M_0$, falls only in the 95 \% confidence, but is outside the 68 \% region.  The exception is a minute corner of the parameter space at very low $M_0$, $M_{1/2}$, and $\tan\beta\sim 10$ where all theoretical results conspire to match experiment, raising the sfermion co-annihilation region into the 68 \% confidence region.  At the opposite, high $M_0$ and $\tan\beta$ corner multiple Higgs resonances combined with neutralino-chargino co-annihilation in the focus point lead to substantial contribution to the total probability. A similar plot shows that positive values of $A_0$ are preferred over negative ones, because Higgs resonance annihilation occurs overwhelmingly at low to moderately positive values of $A_0$, and that $\lambda$ has little impact on the posterior.

\section{Experimental detection of NmSuGra}
\label{sec:detection}

\begin{figure}[th]
  \includegraphics[width=0.49\textwidth]{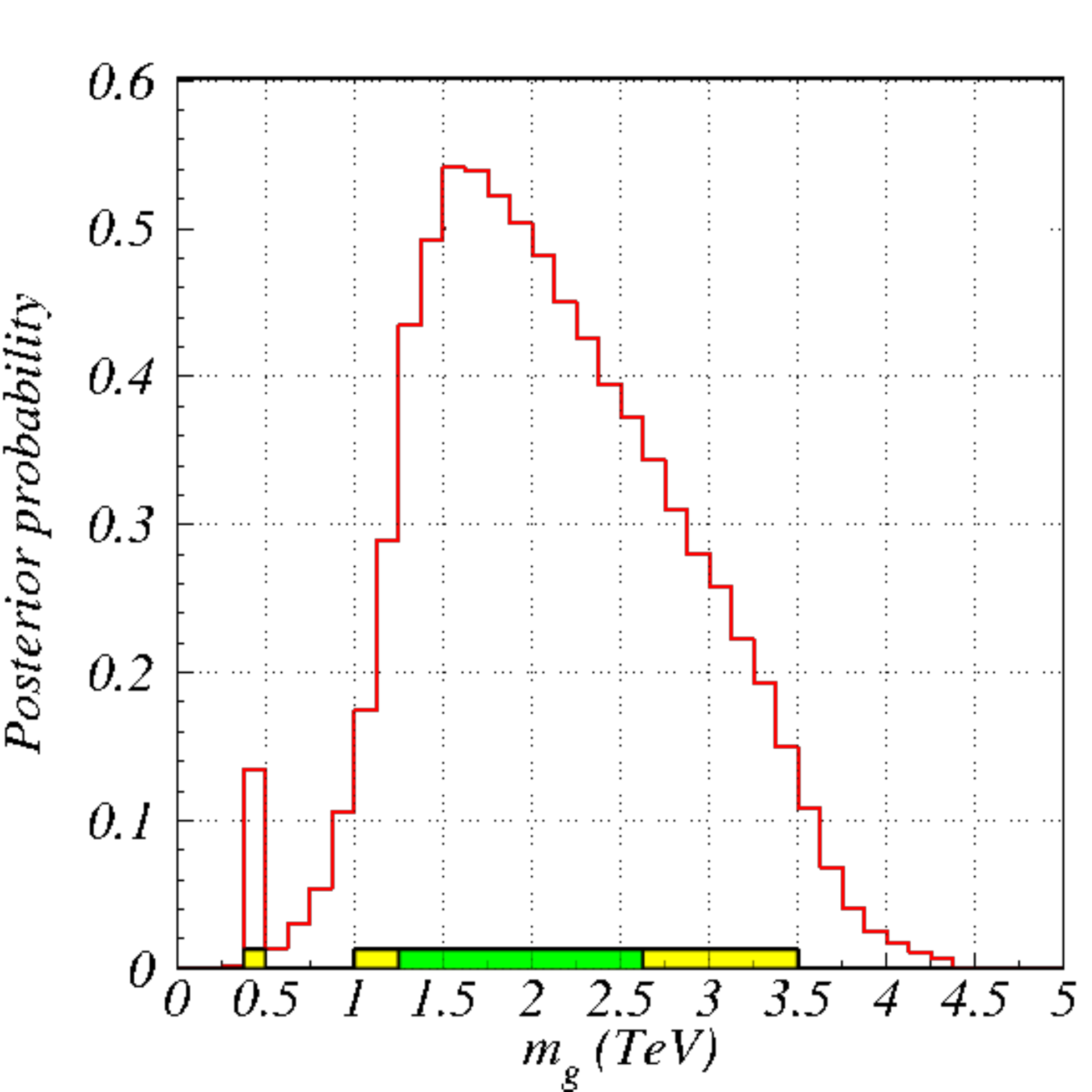}
  \includegraphics[width=0.49\textwidth]{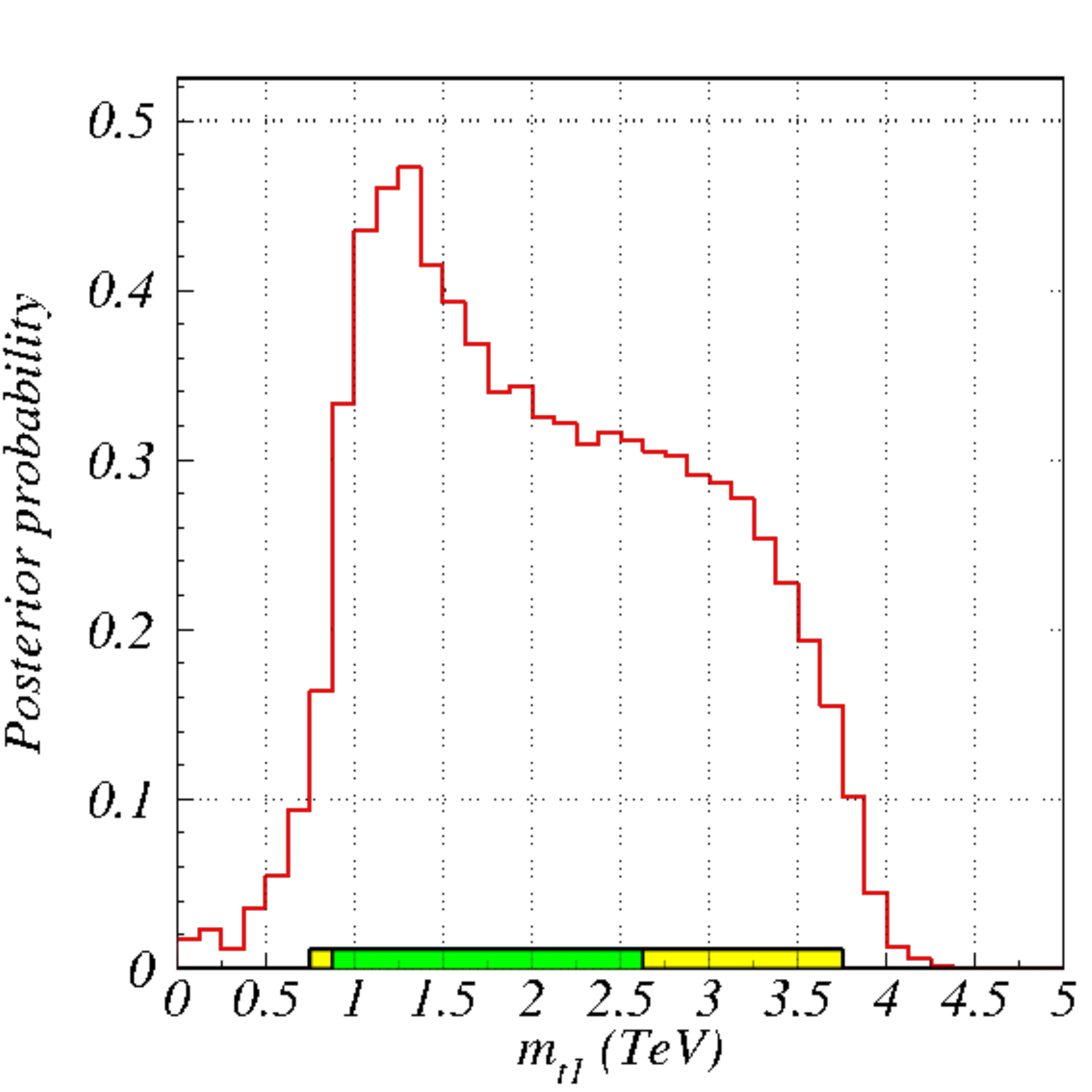}
  \caption{Posterior probability densities marginalized to gluino and stop masses.}
  \label{fig:MPP.masses}
\end{figure}

We examine prospects of NmSuGra being detected at the LHC by plotting the posterior probability marginalized to the masses of relevant sparticles in Figure \ref{fig:MPP.masses}.  Here we see that part of the NmSuGra parameter space, specifically the focus point, is out of the reach of the LHC, as shown by the posterior probability distribution of the gluino mass.  In the mSuGra model the LHC is able to reach about 3 TeV gluinos with 100 fb$^{-1}$ luminosity, provided the model has low $M_0$ \cite{Baer:2003wx}.  In the focus point this reach is reduced to about 1.75 TeV.

%\FIGURE[th]{
%\includegraphics{file=mpp.mgl.eps,width=0.49\textwidth}
%\includegraphics{file=mpp.mt1.eps,width=0.49\textwidth}
%\caption{Posterior probability densities marginalized to gluino and stop masses.}
%\label{fig:MPP.masses}}

In the lower left frame of Figure \ref{fig:MPP.masses} shows that the lighter stop is also expected to be heavier than the likelihood alone would suggest.  Even the sharp peak at low values in the stop likelihood function is overwhelmed due to the minute volume of the parameter space it occupies.

\begin{figure}[th]
\begin{center}
\includegraphics[width=0.49\textwidth]{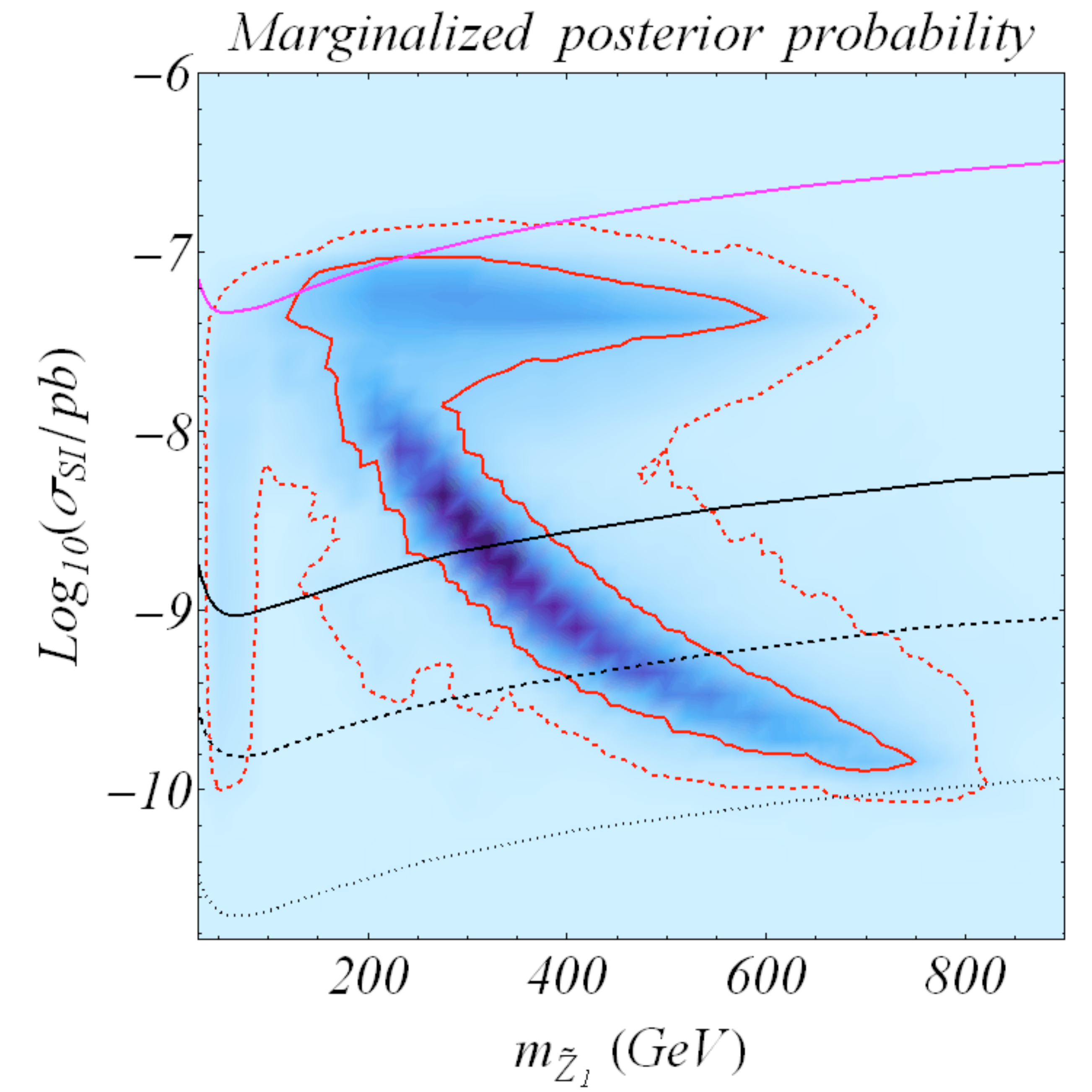}
\end{center}
\caption{Posterior probability density marginalized to the spin-independent neutralino-nucleon elastic recoil cross section and the lightest neutralino mass.  Confidence level contours are shown for 68 (solid red) and 95 (dashed red) \%.  The present (solid magenta) and projected reach of the upgraded CDMS experiment is shown for a 25 (solid black), 100 (dashed black), and a 1000 (dotted black) kg detector.}
\label{fig:MPP.ssi}
\end{figure}

%\FIGURE[th]{
%\includegraphics{file=mpp.mz1.ssi.eps, width=0.49\textwidth}
%\caption{Posterior probability density marginalized to the spin-independent neutralino-nucleon elastic recoil cross section and the lightest neutralino mass.  Confidence level contours are shown for 68 (solid red) and 95 (dashed red) \%.  The present (solid magenta) and projected reach of the upgraded CDMS experiment is shown for a 25 (solid black), 100 (dashed black), and a 1000 (dotted black) kg detector.}
%\label{fig:MPP.ssi}}

While the LHC will not be able to cover the full viable NmSuGra parameter space, fortunately a large part of the remaining region will be accessible to direct detection, measuring the spin-independent neutralino-nucleon elastic recoil cross section, $\sigma_{SI}$.  From several of these experiments, we single out CDMS as the most illustrative example.  Figure \ref{fig:MPP.ssi} shows the posterior probability density marginalized to the plane of $\sigma_{SI}$ and the lightest neutralino mass.

This plot clearly shows that direct detection experiments can play a pivotal role in discovering or ruling out simple constrained supersymmetric scenarios.  Even a 25 kg CDMS will reach a substantial part of the focus point region, complementing the LHC.  

In the possession of the above results, we can quantify the chances for the discovery of NmSuGra at the LHC by calculating the ratio of posterior probabilities inside and outside the reach of the LHC:
\begin{eqnarray}
 \frac{\int_{\rm within~LHC~reach} \mathcal{P}(p_i|D;H) dp_i}
      {\int_{\rm outside~LHC~reach} \mathcal{P}(p_i|D;H) dp_i} = 0.57 .
\end{eqnarray}
{\it According to this the odds of finding NmSuGra at the LHC are 4:3} (assuming, of course, that the model is chosen by Nature).
If we then include the reach of a ton equivalent of CDMS (CDMS1T), the NmSuGra model lies within the combined reach of the LHC and CDMS1T at 99 percent confidence level.  This result strongly underlines the complementarity of collider and direct dark matter searches.

\section{Conclusions}

The next-to-minimal supergravity motivated model is one of the more compelling models for physics beyond the standard model due to its naturalness and simplicity.  In this work we applied a thorough statistical analysis to NmSuGra based on numerical comparisons with present experimental data.  Using Bayesian inference we find that the LHC and future CDMS limits cover the viable NmSuGra parameter region at 99 \% confidence level, underlining the complementarity of these approaches to discovering new physics at the TeV scale.  Thanks to the similarity between our model and the CMSSM, we expect these conclusions to be broadly valid in that model as well.  However, this poses a challenge to the LHC experimentalists to disentangle these models.

\section*{Acknowledgements}
This research was funded by the Australian Research Council under Project ID DP0877916.

%% Don Bennet, 12.11.2009, first contribution
%%% 1 Nov 09  kl. 14:30; 2 Nov 09 00:35; 3 Nov 09 08:07; 03 Nov 09 16:20; 03 Nov 09 22:00
%%%04 Nov 09 05 Nov 09 21:10 06Nov 09; 07 Nov 09; 09 Nov 09 03:33; 9 Nov 09 22:43; 10 Nov 09 21:48: 11 Nov 09 17:07
%%%\documentstyle[11pt]{article}
%%\documentstyle[12pt,epsf]{article}
%%\setlength{\textwidth}{15cm} \setlength{\textheight}{23cm}
%%\setlength{\evensidemargin}{0.15in}%-.2318in}
%%\setlength{\oddsidemargin}{0.15in}%-.2318in}
%%\setlength{\topmargin}{0in} \setlength{\footheight}{0cm}
%%\setlength{\headheight}{0cm}
%%%\setlength{\footskip}{-5.5cm}
%%\setlength{\headsep}{0cm} \setlength{\marginparwidth}{2.5in}
%%%\setlength{\headheight}{-0.7cm}
%%%\newcommand{\marge}{\tiny \marginpar}
\newcommand{\sba}{{\mbox{\scriptsize            \bf             {A}}}}
\newcommand{\sbz}{{\mbox{\scriptsize            \bf             {Z}}}}
\newcommand{\tbz}{{\mbox{\tiny            \bf             {Z}}}}
\newcommand{\sbb}{_{\mbox{\scriptsize            \bf            {B}}}}
\newcommand{\sbc}{_{\mbox{\scriptsize            \bf            {C}}}}
\newcommand{\sbp}{_{\mbox{\scriptsize            \bf            {P}}}}
\newcommand{\sbnul}{_{\mbox{\scriptsize           \bf           {0}}}}
\newcommand{\sbunit}{{\mbox{\scriptsize          \bf           {1}}}}
\newcommand{\bunit}{\mbox{\bf{1}}}
\newcommand{\bpone}{\mbox{\bf{P}}_{1}}
\newcommand{\bptwo}{\mbox{\bf            {P}}_{2}}
\newcommand{\bpprime}{\mbox{\bf  {P'}}}
\newcommand{\baone}{\mbox{\bf{A}}_{1}}
\newcommand{\batwo}{\mbox{\bf            {A}}_{2}}
\newcommand{\sbtwo}{\mbox{\scriptsize            \bf            {2}}}
\newcommand{\sbthree}{\mbox{\scriptsize            \bf            {3}}}
\newcommand{\cartprod}{\mbox{{\bf $\times$}}^{\mbox{{\tiny cart. prod.}}}}
\newcommand{\link}{\mbox{\begin{picture}(4.15,10)
\put(0,3){\circle*{2}}
\put(0,2.75){\line(1,0){7}} \put(7.3,3){\circle*{2}}\mbox{  }
\end{picture}  }  }

\newcommand{\isomorph}{\stackrel{-}{\simeq}}
\newcommand{\fsp}{fundamental spacetime point}
\newcommand{\fsps}{fundamental spacetime points}
\newcommand{\fspx}{fundamental spacetime point }
\newcommand{\fspsx}{fundamental spacetime points }
\newcommand{\site}{\mbox{\begin{picture}(4.15,4)
\put(2,2){\circle*{2}} \end{picture}  }  }
\newcommand{\on}{U_{sign\;U(\Box)}(\Box)}
\newcommand{\sitex}{\site^{\!\!\!x^{\mu}}}
\newcommand{\sitejx}{\site^{\!\!\!x}}
\newcommand{\sitey}{\site^{\!\!\!\!\!y^{\mu}}}
\newcommand{\sitejy}{\site^{\!\!\!\!\!y}}
\newcommand{\sites}{\site^{\!\!\!\!\!s^{\mu}}}
\newcommand{\sitespa}{\site^{\!\!\!\!\!s^{\mu}+a\delta^{\mu}_{\nu}}}
\newcommand{\linklo}{\link^{\!\!\!\!l_0}}
\newcommand{\linkxy}{\;\link^{\!\!\!\!\!\!\!x^{\mu}\;\;y^{\mu}}}
\newcommand{\linkjxy}{\;\link^{\!\!\!\!\!\!\!x\;\;y}}
\newcommand{\linkxaxwss}{\;\link^{\!\!\!\!\!\!\!
x^{\mu}\;\;x^{\mu}+a\delta^{\mu}_{\nu}}}
\newcommand{\linkxax}{\;\link^{\!\!\!\!\!\!\!
x\;\;x+a\delta_{\nu}}}
%%%% FOLLOWING LINE CANNOT BE BROKEN BEFORE 80 CHAR
\newcommand{\linksspa}{\;\link^{\!\!\!\!\!\!\!s^{\mu}\;\;s^{\mu}+
a\delta^{\mu}_{\nu}} }
\newcommand{\linkx}{\;\link^{\!\!\!\!\!\!\!x^{\mu}\;\;}}
\newcommand{\p}{\partial}
\newcommand{\mi}{{\cal M}_i}
\newcommand{\beq}{\begin{equation}}
\newcommand{\eeq}{\end{equation}}
\newcommand{\nin}{\noindent}
\newcommand{\mc}{\multicolumn}
\newcommand{\tla}{\mbox{{\bf{\sf                          T}}}_{L(A)}}
\newcommand{\ba}{\mbox{\bf {A}}}
\newcommand{\dof}{degrees of freedom}
\newcommand{\dofperiod}{degree of freedom.$\:\:$}
\newcommand{\dofx}{degrees of freedom }
\newcommand{\pcp}{partially       confining       phase }
\newcommand{\pcps}{partially        confining         phases }
\newcommand{\ppm}{$A^b_{\mu,\;Peter}, A^b_{\mu,\;Paul}, \cdots ,
A^b_{\mu,\;N_{gen.}}$ }
\newcommand{\ppmd}{$\theta_{Peter}=   \theta_{Paul}=
\cdots =\theta_{N_{gen.}}=\theta_{diag.}$ }

\newcommand{\bb}{\mbox{\bf{B}}}
\newcommand{\bc}{\mbox{\bf    {C}}}
\newcommand{\bp}{\mbox{\bf{P}}}
\newcommand{\bx}{\mbox{\bf    {X}}}
\newcommand{\bz}{\mbox{\bf{Z}}}
\newcommand{\br}{\mbox{\bf{R}}}
\newcommand{\bu}{\mbox{\bf{U}}}
\newcommand{\subnul}{\mbox{\begin{tiny}             \bf{0}\end{tiny}}}
\newcommand{\bnul}{\mbox{\bf{0}}}
\newcommand{\gena}{\frac{\mbox{\boldmath$\lambda^{a}$}}{2}}
\newcommand{\lamb}{\mbox{\boldmath$\lambda^{b}$}}
\newcommand{\genb}{\frac{\mbox{\boldmath$\lambda^{b}$}}{2}}
\newcommand{\genai}{\frac{\mbox{\boldmath$\lambda_{a_i}$}}{2}}
\newcommand{\genbi}{\frac{\mbox{\boldmath$\lambda_{b_i}$}}{2}}
\renewcommand{\arraystretch}{1.5}
\newcommand{\orb}{ORB_{large\;fluc\;in \cartprod_{i\in B(s^{\mu})}
{\cal M}_i}}
%%
%%
%%\hyphenation{Cou-lomb} \hyphenation{cou-pling}
%%\hyphenation{cou-plings} \hyphenation{pro-ducts}
%%\hyphenation{break-downs} \hyphenation{con-fine-ment}
%%\hyphenation{groups} \hyphenation{tends} \hyphenation{pro-duct}
%%\hyphenation{form} \hyphenation{fact} \hyphenation{un-Higgsed}
%%\hyphenation{Higgsed} \hyphenation{re-sults}
%%\hyphenation{ex-act-if-i-ca-tion} \hyphenation{min-i-mal}
%%\hyphenation{so-phis-ti-cat-ed}
%%
\title{The Multiple Point Principle: Characterization of the
Possible Phases for the SMG }
\author{D. L. Bennett\thanks{dlbennett99@gmail.com}}
\institute{%
Brookes Institute for Advanced Studies,\\
 B{\o}gevej 6, 2900 Hellerup,
Denmark} 

\authorrunning{D. L. Bennett}
\titlerunning{The Multiple Point Principle}
\maketitle
%%\begin{document}
%%\bibliographystyle{unsrt}
%%\maketitle
%%
%%\begin{center}
%%
%%
%%
%%
%%\end{center}
%%
%%\vspace{1cm}
%%
%%

\section{Introduction}
 The Multiple Point Principle (MPP)~\cite{abel,nonabel} states that Nature takes on
intensive parameter values (coupling constant values) that
correspond to a maximally degenerate vacuum where these degenerate
vacua all have essentially vanishing cosmological constants. The MPP
was originally applied in the context of lattice gauge theory for
the purpose of predicting the values of the three gauge coupling
constants for the Standard Model Group (SMG). This pursuit entailed
among other things a way in which to characterize the possible
phases of of a non-simple gauge group such as the SMG. Having such a
phase classification scheme, it was subsequently necessary to
parameterize the action in such a way that these various phases
could be provoked. In such an action parameter space our claim is
that Nature takes on parameter values corresponding to the the point
(or surface)  --- the multiple point --- at which a maximum number
phases come together.

The presentation at the 12th International Workshop ``What Comes
Beyond the Standard Model'' in Bled (2009) was an attempt at a
somewhat comprehensive review of the original implementation of the
MPP. I was very happy that my talks were interrupted by so many
questions. Many of these were about the formal way that different
possible phases of the SMG are distinguished. So rather than a
review of MPP I shall in this proceedings contribution address the
questions posed. These were centered around the way in which the
various possible phases for a non-simple gauge group such as the SMG
are characterized in terms of subgroups $K \subseteq SMG$ and
invariant subgroups $H \triangleleft K$. It will be seen that the
subgroups $K$ and $H$ are defined according to the way that they
transform under gauge transformations $\Lambda_{Const}$ and
$\Lambda_{Linear}$ having respectively constant and linear gauge
functions. The quantum fluctuation patterns characteristic of a
given phase are defined in terms of $K \subseteq SMG$ and $H
\triangleleft K$. We are working with a lattice formulation of a
gauge theory. The different phases in such a theory are generally
regarded as lattice artefacts. However we assume that a lattice is
just one implementation of a fundamental really existing Planck
scale regulator. In light of this assumption ``lattice artefact''
phases become ontological. That transitions between such phases are
most often first order plays an important role in the finetunning
mechanism inherent to MPP.

\section{Distinguishing the possible phases of a non-simple
group}

%For illustrative purposes, one can think of
Using a lattice formulation of a gauge theory with gauge group
\footnote{The symbol $G$ denotes a generic gauge group where we
should have the $SMG$ or at least a non-simple gauge group in
mind unless the context indicates otherwise.} $G$, let the
dynamics of the system be described by a Lagrangian ${\cal
L}(A^{\mu},\phi)$ that is invariant under (local) gauge
transformations $\Lambda$ of the gauge potential $A^{\mu}(x)$
and the (complex)  scalar field $\phi (x)$. In the continuum,
the fields  $A^{\mu}(x)$  and $\phi (x)$ transform under gauge
transformations as
\beq    gA^{\mu}(x)\rightarrow \Lambda^{-1}(x)gA^{\mu}(x)\Lambda(x)-
i\Lambda^{-1}(x)\partial^{\mu}\Lambda(x) \;\;\;(\mbox{$g=$ coupling
constant})\label{gpot} \eeq
\beq \phi(x) \rightarrow \Lambda(x)\phi(x) \label{phi} \eeq
In the lattice  formulation,  each  of  the  four  components  of
the $A^{\mu}$ field corresponds  to  a  group-valued  variable
$U(\linkxax)$ defined on links $\link$ of the lattice. The index
$\nu$ specifies the direction of the link connecting sites with
coordinates $x^{\rho}$ and $x^{\rho}+a\delta^{\rho}_{\nu}$; often
such coordinates are written more briefly as $x$ and
$x+a\delta_{\nu}$. Under a local gauge transformation, the
$U(\linkxax)$ transform as
\begin{align} U(\linkxax)\rightarrow %\Lambda^{-1}(x)U(\linkxy)\Lambda(y)=
&\Lambda^{-1}(x)U(\linkxax)\Lambda(x+a\delta_{\nu})\approx\nonumber\\
&\approx\Lambda^{-1}(x)U(\linkxax)(\Lambda(x)+\partial^{\rho}\Lambda(x)a\delta^{\rho}_{\nu})
\label{lgt}\\
&=\Lambda^{-1}(x)U(\linkxax)\Lambda(x)(1+\partial^{\rho}(\log
\Lambda(x))a\delta^{\rho}_{\nu}) \approx \nonumber\\
&\approx \Lambda^{-1}(x)U(\linkxy)\Lambda(x)\exp(\partial^{\rho} (\log
\Lambda(x))a\delta^{\rho}_{\nu})\nonumber
\end{align}
That this corresponds to the transformation (\ref{gpot}) for
the continuum fields $A^{\rho}$ is readily verified: write
$U(\linkxax)=\exp(igA^{\rho}(x)a\delta_{\rho}^{\nu})\approx
1+igA^{\rho}(x)a\delta_{\rho}^{\nu}$  in which case the gauge
transformation above is
\begin{align*}
&\Lambda^{-1}(x)(1+igA^{\rho}(x)a\delta^{\nu}_{\rho})\Lambda(x)(1+
\partial^{\rho}(\log \Lambda(x))a\delta^{\nu}_{\rho})\approx \\
&\approx 1+\Lambda^{-1}(x)(igA^{\rho}(x)a\delta^{\nu}_{\rho})\Lambda(x)+
\partial^{\rho}(\log \Lambda(x))a\delta^{\nu}_{\rho}= \\
&=1+\Lambda^{-1}(x)(igA^{\rho}(x)a\delta^{\nu}_{\rho})\Lambda(x)+
\Lambda^{-1}(x)\Lambda(x)
\frac{1}{\Lambda(x)}\partial^{\rho}(\Lambda(x))a\delta^{\nu}_{\rho}\\
&=1+i[\Lambda^{-1}(x)gA^{\rho}(x)\Lambda(x)-i\Lambda^{-1}(x)\partial^{\rho}(\Lambda(x))]a\delta^{\nu}_{\rho} \end{align*}
which corresponds to the transformation rule~(\ref{gpot}).
On the lattice, the group-valued field $\phi$ is defined on
lattice sites; the transformation rule is as in (\ref{phi})
above.
%XXXXXXXXXXXXXXXXXXXXXXXXXXXXXXXXXXXXXXXXXXXXXXXXXXXXXXXXXXX
%second insertion her

\subsection{``Phase'' classification according to symmetry properties
of vacuum}\label{phclass}

We are interested in the case in which the gauge field
$U(\linkxax)$ takes values in a non-simple gauge group such as
$G=SMG$.
% having many subgroups and invariant subgroups
%(including discrete subgroups), it is possible for \dofx
%corresponding say to different  subgroups to take  group values
%according  to distributions  that characterise qualitatively
%different physical  behaviours along   the different subgroups.
%Some \dofx can have a fluctuation pattern characteristic of a
%Higgsed phase; some of the  \dofx  having fluctuation  patterns
%characteristic of  an un-Higgsed  phase can  be  further
%classified according  to whether  they  have Coulomb-like  or
%confinement-like patterns of  fluctuation. The point  is that a
%``phase'', which of course corresponds to a region  in the
%action parameter space, can, for a non-simple gauge group, be
%described in terms of characteristics that differ along
%different subgroups.
The gauge field for the SMG has 12 \dof : if we allow a slight
simplification one can say that 8 of these are associated with
$SU(3)$ \dof, 3 with $SU(2)$ \dof $\:\:$ and one with the
$U(1)$ \dofperiod  It is possible for these \dofx to take
values in various structures all of which are determined for
each choice $(K,H)$ such that $K \subseteq SMG$ and $H
\triangleleft K$. The various structures are the subgroup $K$,
the invariant subgroup $H$, the homogeneous space $SMG/K$ and
the factor group $K/H$. For gauge field \dofx there is a
correspondence between distributions that characterize
qualitatively different physical behaviors (e.g., quantum
fluctuation patterns) and which structures the gauge field
\dofx take values in (e.g., {\it elements} of $K\subseteq G$
and $H\triangleleft K$ and {\it cosets} of $G/K$ and $K/H$. As
already hinted, there is a one-to-one correspondence between
the possible phases for the gauge field theory and the possible
combinations $(K,H)$ with $K\subseteq G$ and $H\triangleleft
K$. Discrete subgroups must be included among the possible
subgroups. The choice of the pair $(K,H)$ specifies which \dofx
are in a Higgsed phase and also whether the un-Higgsed \dofx
are in a confined or Coulomb-like phase. Now I need to reveal
how $K\subseteq G$ and $H\triangleleft K$ are defined.

%The fluctuation patterns for the various \dofx corresponding to
%these subgroups  can be classified
The subgroups $K\subseteq G$ and $H\triangleleft K$ are defined by
the transformation properties of the vacuum
%under the two classes of
%gauge transformations $\Lambda_{Const}$ and $\Lambda_{Linear}$;
%following a partial fixing of the gauge, the phases of  the vacuum
%are to be classified\cite{frogniel,nonabel}
according to whether or not there is spontaneous breakdown of
gauge symmetry under gauge transformations corresponding to the
sets of gauge functions $\Lambda_{Const}$ and
$\Lambda_{Linear}$ that are respectively constant and linear in
the spacetime coordinates\cite{frogniel,nonabel}:

\beq \Lambda_{Const}\in \{\Lambda:\br^4\rightarrow
G|\exists\alpha[\forall x\in \br^4 [\Lambda(x)= e^{i\alpha}]]
\}\eeq
and
\beq\Lambda_{Linear}\in \{\Lambda:\br^4\rightarrow
G|\exists\alpha_{\mu}[\forall x\in \br^4[\Lambda(x)=
e^{i\alpha_{\mu}x^{\mu}}]]\}. \label{lin} \eeq

Here $\alpha=\alpha^at^a$ and $\alpha_{\mu}=\alpha^a_{\mu}t^a$
where $a$ labels the Lie algebra generators in the case of
non-Abelian subgroups. The $t^a$ denote a basis of  the Lie
algebra satisfying the commutation      relations
$[t^a,t^b]=c^{ab}_ct^c$  where the $c^{ab}_c$  are   the
structure constants.

Spontaneous  symmetry   breakdown   is   manifested   as
non-vanishing values for gauge variant quantities.  However,
according to Elitzur's theorem, such quantities cannot survive
under the  full gauge symmetry. Hence a partial fixing of the
gauge is necessary  before  it makes sense to talk about the
spontaneous breaking of symmetry.  We choose the Lorentz gauge
for the reason that  this  still  allows  the freedom of making
gauge transformations of the types $\Lambda_{Const}$ and
$\Lambda_{Linear}$ to be used in classifying the lattice
artifact ``phases'' of the vacuum. On  the  lattice, the choice
of the Lorentz gauge amounts  to  the condition
$\prod_{\;\linkx \;
\begin{tiny}  \mbox{emanating   from   }
\end{tiny}     \sitex     }     U(\link)=1$     for     all
sites $\sitejx $.
%\begin{large}$\cdot$\end{large}.

%It will  be seen that the set of possible ``phases''
%corresponds one-to-one to the set of  all  possible  subgroup
%pairs  $(K,H)$
%START FOOTNOTE
%\footnote{In this classification scheme it has been assumed
%that the action energetically favours $U(\Box)\approx \bunit$;
%however, a vacuum also having fluxes corresponding to
%nontrivial elements of the centre could be favoured if for
%instance there are negative values for coefficients of
%plaquette terms in the action. Such terms would lead to new
%partially confining phases that were Coulomb-like
%%w.r.t. a discrete subgroup
%but for which fluctuations in the  \dofx are centred at a
%nontrivial element of the centre instead of at the identity.}
%STOP FOOTNOTE
%consisting  of  a subgroup $K\subseteq SMG$ and invariant
%subgroup $H\triangleleft K$. Each ``phase'' $(K,H)$ in general
%corresponds to a partitioning of the \dofx (these latter can be
%labelled by a Lie algebra basis). Let us now define the
%subgroup $K$ and the invariant subgroup $H$
%XXXXXXXXXXXXXXXXXXXXXXXXXXXXXXXXXXXXXXXXXXX third insert below
By definition the  \dofx  belonging  to  the subgroup $K$ exhaust
the un-Higgsed degrees of freedom if, after fixing the gauge in
accord with say the Lorentz condition,  $K\subseteq G$ is the
maximal subgroup of gauge transformations belonging to the set
$\Lambda_{Const}$ that leaves the vacuum invariant
%\footnote{The
%vacuum invariance referred  to really   means   the invariance
%of   the   coefficients $\langle D^{(\mu)}_{ij}(U(\linkxax))
%\rangle_{vac}$ in an expansion in (matrix elements   of)
%continuous   unitary   irreducible representations
%$D^{(\mu)}_{ij}(U(\linkxax))$.  The  expansion referred  to  is
%that corresponding to  some  link  variable probability
%density  function $P(U(\link^{\!\!\!\!l_0}))  = \int
%\prod_{\link  \;\;  \not  =  \;\; \linklo} d^{{\cal
%H}aar}U(\link) e^S.$}.
For  the  vacuum  of  field variables
defined on sites (denoted by  $\langle  \phi(\sitex)\rangle $),
invariance under transformations $\Lambda_{Const.}$ is possible
only  if $\langle\phi(\sitex)\rangle=0$. For the vacuum of
field variables defined on links (denoted by $\langle
U(\linkxax)\rangle$, invariance  under  transformations
$\Lambda_{Const}$ requires that $\langle U(\linkxax)\rangle $
takes values in the centre of the subgroup $K$. Conventionally,
the idea of Higgsed \dofx pertains to field variables defined
on sites. With the above criterion using $\Lambda_{Const}$, the
notion of Higgsed \dofx is generalised to also include link
variables.

If $K\subseteq G$ is the maximal subgroup for which the
transformations $\Lambda_{Const}$ leave the vacuum invariant, the
gauge field variables taking values in the homogeneous  space $G/K$
(see for example \cite{dubrovin,nakahara}) are by definition Higgsed
in the vacuum. For these degrees of freedom, gauge symmetry is
spontaneously   broken   in   the   vacuum under   gauge
transformations $\Lambda_{Const}$ (i.e., global gauge
transformations).

%XXXXXXXXXXXXXXXXXXXXXXXXXXXXXXXXXXXXXXXXX fourth insertion
In the vacuum, the un-Higgsed \dofx -  taking  values  in  the
subgroup $K$~- can be in a confining phase or a Coulomb-like phase
according  to the  way  these  \dofx  transform  under gauge
transformations $\Lambda_{Linear} $ having linear gauge functions.

Degrees  of  freedom  taking  values   in   the   invariant
subgroup $H\triangleleft K$ are by definition confined in the
vacuum if $H$ is the  maximal  invariant   subgroup   of
gauge transformations $\Lambda_{Linear}$  that  leaves the
vacuum invariant; i.e.,  $h$ consists  of  the  set  of
elements $h=\exp\{i\alpha^1_at_a\}$ such that the  gauge
transformations with linear gauge function $\Lambda_{Linear}$
exemplified by\footnote{In the quantity $x^1/a$, $a$ denotes
the lattice constant; modulo lattice artifacts, rotational
invariance  allows the  (arbitrary)  choice  of $x^1$ as the
axis $x^{\mu}$  that  we use.}$\Lambda_{Linear}\stackrel
{def.}{=}h^{x^1/a}$ leave the vacuum invariant.

If $H\triangleleft K$ is the maximal invariant  subgroup  of
\dofx that  are confined in the vacuum, \dofx taking as values
the cosets belonging to the factor group $K/H$ are by
definition in a Coulomb phase (again, in the Lorentz gauge).
For \dofx corresponding to this set of cosets, there is
invariance of the vacuum expectation value under coset
representatives of the type $\Lambda_{Const}$ while gauge
symmetry is spontaneously broken in the  vacuum  under  coset
representatives of the type $\Lambda_{Linear}$.

Having now formal criteria for distinguishing the different phases
of the vacuum, it would be useful to elaborate a bit further on what
is meant by having  a  phase  associated  with a  subgroup  -
invariant subgroup  pair  $(K_i\subseteq  G, H_j \lhd  K_i)$.  A
phase  is  a characteristic region of action parameter space. Where
does an  action parameter space come from and what makes a region of
it characteristic of a given phase $(K_i\subseteq G, H_j \lhd K_i)$?
An action parameter space comes about by choosing  a  functional
form  of the plaquette action. This will normally be a sum of terms
each of which is a product of an action parameter (action parameters
are  related to coupling constants) multiplied by a sum over lattice
plaquettes each term of which is the trace of group-valued plaquette
variable in one of the desired representations (e.g., the
fundamental representation, the adjoint representation, etc.).

Having an action allows the calculation of the partition
function and subsequently the free energy. As each phase
$(K_i\subseteq G, H_j \lhd K_i)$ corresponds to different micro
physical patterns of fluctuations  along various group
structures and homogeneous spaces as described above,
%(recall
%that we are dealing with phases separated by first order phase
%transitions),
the partition function and hence the free energy is a different
function  of the plaquette action parameters for each phase
$(K_i\subseteq G, H_j \lhd K_i)$.  Have in mind that transitions
between these ``lattice artefact'' phase are first order. A region
of plaquette action parameter space is in a given phase
$(K_i\subseteq G, H_j \lhd K_i)$ if the free energy $- \log
Z_{K_i\subseteq G, H_j \triangleleft K_i}$ associated with this
phase has the largest value of  all free energy functions.
%(i.e.,
%one free energy function for each phase $(L\subseteq  G, M\lhd L)$)
%in this region of plaquette action parameter space. The realized
%phase in a region of action parameter space is the phase having the
%largest free energy in this region.
One should imagine that at any given point in the action parameter
space all free energy funcions  (one for each possible phase
$(K,H)$) are defined (and have values). The realized phase at the
point in question is determined by which of these free energy
functions is the largest.

In seeking the multiple point, we  seek  the point or surface
in parameter space where ``all'' (or a maximum number of)
phases $(K_i\subseteq  G,  H_j  \lhd  K_i)$  ``touch''  one
another. MPP claims that action parameter values (which are
simply related to coupling constants) at the multiple point are
those realized in Nature.

%XXXXXXXXXXXXXXXXXXXXXXXXXXXXXXXXXXXXXXXXXXXXXXXXXXXXXXXXX
\section{The Higgsed phase}
On a lattice consisting of sites $(\sitex)$ and
site-connecting links $(\linkxy)$ denote by $\phi(\sitex)$ a
scaler field variable defined on lattice sites. We want to
describe the conditions to be fulfilled if the field variable
$\phi(\sitex)$ is a Higgsed degree of freedom. The appropriate
mathematical structure is that of a homogeneous space. If $K
\subseteq SMG$ ($K$ not an invariant subgroup of SMG) is the
subgroup of not-Higgsed gauge degrees of freedom, the Higgsed
degrees of freedom $\phi(\sitex)$ take values in the
homogeneous space $ SMG/K $

It might be be useful with a reminder about the mathematical
structure of a homogeneous space. For the purpose of exposition
it is expedient to use the example of the group  $G=SO(3)$
instead of the $G=SMG$  and the subgroup $SO(2)\subset SO(3)$
(instead of the unspecified $K \subseteq SMG$). So we  consider
the homogeneous space $SO(3)/SO(2)$. In this case the cosets
(i.e. elements) of $SO(3)/SO(2)$ are in one-to-one
correspondence with the points on a $S_2$ sphere: for an
arbitrary coset $h \in SO(3)/SO(2)$, the orbit of the action of
$SO(3)$ on $h$ is just $S_2$. The homogeneous space
$SO(3)/SO(2)$ is mapped onto itself under the action of
$SO(3)$:

$$ h_2 \stackrel{g \in SO(3)}{\longrightarrow} h_1 \;\;\; (h_1,h_2 \in SO(3)/SO(2);$$
Note that there is no multiplication (i.e., composition) rule
for the cosets (i.e., elements) of a homogeneous space. For example,
$h_1 \cdot h_2$ for $h_1,h_2 \in SO(3)/SO(2)$ is \underline{not}
meaningful. It can be shown that the action the group $G$ on the
homogeneous space $G/K$ is {\it transitive} which means that for any
two cosets $h_1,h_2 \in G/K$ there exists at least one element $g
\in G$ such that $h_1 = gh_2$. In the example with $G=SO(3)$ and
$G/K = SO(3)/SO(2)$ this means that for any two points $h_1$ and
$h_2$ on $S_2 \cong SO(3)/SO(2)$ there is at least one element $g
\in SO(3)$ such that $h_1=gh_2$. The set of such elements $g$:

$$ \{g \in SO(3)|gh_2=h_1\}  $$
is the coset of $SO(3)/SO(2)$ associated with $h_1$ (here
$h_2$ can be thought of as a (arbitrarily chosen) basis coset
from which all other cosets of $SO(3)/SO(2)$ can be obtained by
the appropriate action of $SO(3)$).

Any element $g^{\prime}$ belonging to the coset $\{g \in
SO(3)|gh_2=h_1\}$ is a {\it representative} of this coset
associated with $h_1$. Other representatives of this same coset
are obtained by letting $g^{\prime}$ act on the $SO(2)\subset
SO(3)$ that leaves the basis coset $h_2$ invariant (denote the
latter by $SO(2)_{h_2\,inv}$). In fact {\it all} of the
representatives of the coset $\{g \in SO(3)|gh_2=h_1\}$ are
given by $g^{\prime}\cdot SO(2)_{h_2\,inv}$. So when
$g^{\prime}$ is a representative of the coset associated with
$h_1$, so is $g^{\prime}\cdot k$ when $k \in SO(2)_{h_2\,inv}$.
It goes without saying that a representative of a coset always
belongs to the coset that it represents.
%XXXXXXXXXXXXXXXXXXXXXXXXXXXXXXXXXXXXXXXXXXXXX fifth insertion

To get a feeling for it means to have a Higgsed phase, think of
having an $S_2$ situated at each site $\sitex$ of the
(space-time) lattice. In this picture, the variable
$\phi(\sitex)$ at each site $\sitex$ corresponds to a point on
the $S_2$ at this site. A priori there is no special point in
this homogeneous space $SO(3)/SO(2)\stackrel{-}{\simeq} S_2$
which implies $\langle \phi(\site) \rangle=0$. The Higgs
mechanism comes into play when, for all sites on the lattice,
the vacuum
%distribution of the
value of $\phi(\site)$ - modulo parallel transport between
sites by link variables -
%clusters about some
is (in a classical approximation) the same coset of
$SO(3)/SO(2)$ or in other words the same point on all the (site
situated) $S_2$'s (modulo parallel transport) inasmuch as
$SO(3)/SO(2)\cong  S_2$. With one point of $S_2$ singled out
globally - call it $h_2$  - it is obvious
% point in
%and we conclude
%START FOOTNOTE
%\footnote{There are technical problems here. This conclusion
%presumably requires the validity of the ``cluster decomposition
%principle'' \cite{streater}}
%END FOOTNOTE
that $\langle \phi(\site) \rangle\neq 0$. The symmetry of the
homogeneous space $SO(3)/SO(2)$ is broken globally down to the
$SO(2) \subset SO(3)$ that leaves the point $h_2 \in S_2$
invariant. This is just the {\it isotropy} group of $h_2 \in
S_2$ which we can - using a notation defined above - denote as
$SO(2)_{h_2 \:inv.}$. After a Higgsning corresponding to
singling out $h_2 \in S_2$ we can think of $h_2$ as the axis
about which the symmetry remaining after this Higgsning are
just the rotations $SO(2)_{h_2 \:inv.}$.

In a quantum field theoretic description of a Higgsed phase
corresponding to $h_2 \in S_2$  where we allow for quantum
fluctuations, we expect a {\it clustering} of the values of
$\phi(\site)$  about the coset $h_2 \in S_2$
%In the Higgsed $situation, the point of $S_2$ about which there is a
%clustering of the values of $\phi(\site)$
for all sites of the lattice (modulo parallel transport).
%But
%in the presence of quantum fluctuations the coset $h_2 \in S_2$
%defines the axis of rotations of the $SO(2)$ gauge symmetry -
%i.e., $SO(2)_{h_2\;\; inv}$ surviving the spontaneous breakdown
%of the $SO(3)$ symmetry by Higgsing about which quantum
%fluctuations are centered.
This brings us to a technical problem\cite{hbnpriv}: the average
value of such quantum fluctuations is expected to be $h_2$: $\langle
\phi(\site) \rangle = h_2$. But the average value of for example two
cosets of in the neighborhood of the coset corresponding $h_2$ does
not lie in $S_2 \cong SO(3)/SO(2) $ but rather in the interior of
$S_2$ (the convex closure). In order to have such average values in
our target space we need the convex closure  \footnote{
%Even if one were to succeed in embedding a
%homogeneous space in an affine space in a natural manner, such
%an embedding would not in general be convex. Therefore it would
%generally be necessary to construct the convex closure (e.g.,
%in a vector space)
If we want for example to include averages of the cosets of
%field
%variables. As an example, think of
the homogeneous space $SO(3)/SO(2)$ (which we know is metrically
equivalent to an $S_2$ sphere), it would generally be necessary to
construct the convex closure (e.g., in a vector space). In this
case, one could obtain the complex closure as a ball in the linear
embedding space $\br^3$. Alternatively, we can imagine supplementing
the $SO(3)/SO(2)$ manifold with the necessary (strictly speaking
non-existent) points needed in order to render averages on the $S_2$
meaningful. Either procedure eliminates the problem that an average
taken on a non-convex envelope is generally unstable in the
following way: e.g., think of the ``north pole'' of an $S_2$ about
which quantum fluctuations are initially clustered (the Higgsed
situation); if the fluctuations become so large that that they are
concentrated near the equator, the average on an $S_2$ will jump
discontinuously back and forth between the north and south poles
depending respectively on whether the fluctuations are concentrated
just north of or just south of the equator). It is interesting to
note that by including the points in the ball enclosed by an $S_2$,
it is possible for $\langle \phi \rangle$ to have a value lying in
the symmetric point (i.e., center) when quantum fluctuations are
large enough. This point, corresponding to $\langle \phi \rangle=0$,
is of course unique in not leading to spontaneous breakdown under
rotations of the $S_2$. This scenario describes an inverse Higgs
mechanism\cite{hbnpriv}.} of $S_2$.

The Higgs mechanism outlined above can be provoked if there is a
term in the action of the form

\beq \kappa dist^2(\phi(\sitejx), U(\linkjxy)\phi(\sitejy))
\eeq
where $\kappa$ is a parameter and $dist^2(\phi(\sitejx),
U(\linkjxy)\phi(\sitejy))$ is the suitably defined squared distance
on the $S_2$ at the site $\sitejx$  between the point
$\phi(\sitejx)$ and the point $\phi(\sitejy)$ after the latter is
``parallel transported'' to $\sitejx$ using the link variable
$U(\linkjxy)\in G$. This is the so-called Manton
action\cite{manton}.

In terms of of elements $g\in SO(3)$,

\beq dist^2(\phi(\sitejx),
U(\linkjxy)\phi(\sitejy))\stackrel{def}{=} \eeq
\[ inf\{dist^2(g_x\cdot SO(2), U(\linkjxy)g_y\cdot SO(2)|\]
\[ \mbox{
$g_x$ \& $g_y$ \begin{footnotesize} are reps. of respectively the
cosets
\end{footnotesize}$\phi(\sitejx)$ \& $\phi(\sitejy)$}\} \]
 In order to provoke the Higgs mechanism, not only must the
parameter $\kappa$ be sufficiently large to ensure that it doesn't
pay not to have clustered values of the variables $\phi(\cdot)$. It
is also necessary that ``parallel transport'' be well defined so
that it makes sense to talk about the values of $\phi(\site)$ being
organised (i.e., clustered) at some coset of $SO(3)/SO(2)$. This
would obviously not be the case if the theory were confined. In
confinement, $\langle U(\linkjxy) \rangle=0$ and parallel transport
is meaningless. In the continuum theory, this would correspond to
having large curvature (i.e., large $F_{\mu\nu}$) which in turn
would make parallel transport very path dependent
%START FOOTNOTE
%\footnote{Even when confinement is absent, there are technical
%difficulties in defining parallel transport over large
%spacetime distances; presumably it is necessary to average over
%a bundle of spacetime parallel paths.}.
%END FOOTNOTE

\section{The un-Higgsed Phases}

The un-Higgsed gauge field \dofx (i.e., link variables) take
values that correspond to the Lie algebra of the subgroup $K
\subseteq G$. The confined \dofx take as values the elements of
the invariant subgroup $H \triangleleft K$. The Coulomb-like
\dofx take as values the cosets of the factor group $K/H$.

\subsection{Confined degrees of freedom}

The confined phase is characterized by large quantum
fluctuations in the group-valued link variables so that at
least crudely speaking the whole confined subgroup $H$ is
accessed. So roughly speaking all elements $h\in H$ are visited
with nearly the same probability. In other words the
distribution of quantum fluctuations for confined link
variables is not strongly clustered in a small part of the
group space (e.g. at the group identity or in the center of the
group). Since the distribution of confined \dofx is essentially
flat (i.e., without much characteristic structure) the effect
of gauge transformations is not noticeable. The subgroup $H$ is
therefore essentially invariant under all classes of gauge
transformations including the for us interesting types of gauge
transformation $\Lambda_{Const}$ and $\Lambda_{Linear}$.

%The scheme to be used for vacuum phase classification utilises
\subsection{The Coulomb-like phase}

The claim above is that Coulomb-like link variable \dofx take
as values the cosets of the factor group $K/H$. Recall that by
definition of a factor group all of the elements of $H$ are
identified (i.e., not distinguishable from one another) and the
(invariant) subgroup $H$ becomes the identity element in the
coset space. That elements of $H$ are not distinguished from
one another is consistent with the intuitive properties of
having confinement along the subgroup $H$ as sketched above: a
consequence of having large quantum fluctuations along $H$ is
that all elements of $H$ enter into the fluctuation pattern
(which is a manifestation of the underlying  physics)  with
essentially the same weight (as opposed to e.g., a Coulomb-like
phase in which the fluctuation pattern is more or less tightly
clustered around the group identity).

The transformation  properties  of  the  vacuum  that  are
appropriate for having a Coulomb-like phase are suggested by
examining the requirements\cite{frogniel} for getting a massless
gauge particle as the Nambu-Goldstone boson accompanying  the
spontaneous breakdown of gauge symmetry. To this end  we need to
examine  the Goldstone Theorem

% the $\langle
%U(\linkxax)\rangle $. The transformation properties of the
%vacuum $\langle U(\linkxax)\rangle $ that we use to define a
%Coulomb phase in the classification scheme given above should
%be consistent with these requirements.

As already pointed out, a gauge choice must be made in order
that spontaneous breakdown of gauge symmetry is at all
possible. Otherwise Elitzur's Theorem insures that all gauge
variant quantities vanish identically. Once a gauge choice is
made - the Lorenz gauge is  strongly suggested inasmuch as  we
want, in order to classify phases, to retain the freedom to
make  gauge  transformations of the types $\Lambda_{Const}$ and
$\Lambda_{Linear}$ - the symmetry  under the remaining gauge
symmetry must somehow be broken in order to get a
Nambu-Goldstone boson that, according to the Nambu-Goldstone
Theorem, is present for each generator of  a spontaneously
broken  continuous gauge symmetry.

Recalling from~(\ref{lgt}) that a link variable $U(\linkxy)$
transforms under gauge transformations as

\beq  U(\linkxax)  \rightarrow
\Lambda^{-1}(x)U(\linkxax)\Lambda(x)\cdot \underbrace{\exp(
\partial^{\rho}(\log         \Lambda(x))\cdot
a\delta^{\nu}_{\rho}}_{\mbox{\footnotesize gradient part of
transf.}} \label{utransf},\eeq
it is seen that, for the special case of an  Abelian gauge
group, a gauge function that is linear in the coordinates (or
higher order  in the coordinates) is required for  spontaneous
breakdown  because  the only possibility for spontaneously
breaking the  symmetry  comes  from the ``gradient'' part of
the transformation (\ref{utransf}). So the needed  spontaneous
breakdown of gauge symmetry is garanteed if gauge symmetry for
gauge transformations of the type
 $\Lambda_{Linear}$ is spontaneously broken (i.e., the vacuum is
 not invariant under this class of gauge transformations).
 Let $Q_{\nu}$ denote the generator of such gauge transformations.

%However,  we need also to take into account  the  requirement
%of  translational invariance.

However the proof  of  the Nambu-Goldstone  Theorem also
requires the assumption  of translational   invariance. This
amounts to the requirement that the vacuum be invariant under
gauge transformations
%having gauge functions such that these
%transformations correspond to gauge transformations
generated by the commutator  of  the momentum operator  with the
generator  of the spontaneously broken symmetry which is just
$Q_{\nu}$ as defined above.
%Let
%$Q_{\nu}$ denote the generator of such transformations.
%For
%such transformations.
Then the requirement of translational
symmetry is   equivalent to requiring that the vacuum $\langle
U(\linkxax)\rangle  $  is annihilated by the commutator
$[P_{\mu},Q_{\nu}]=ig_{\mu\nu}Q$ where $Q$  denotes the
generator of gauge transformations with constant gauge
functions. So the condition for having translational invariance
translates  into the requirement  that the  vacuum $\langle
U(\linkxax)\rangle  $  be invariant under gauge transformations
with constant  gauge functions.  An examination of
(\ref{utransf}) verifies that this is always true  for Abelian
gauge groups and also for  non-Abelian groups if  the  vacuum
expectation value $\langle U(\linkxax)\rangle $ lies in the
centre of the group (which just means that the vacuum is not
``Higgsed'').

%We show now that these two requirements. i.e.,

%\begin{enumerate}

%\item spontaneous breakdown via the  gradient  in
 %   (\ref{utransf})

%\item translational  invariance

%\end{enumerate}

%\nin are satisfied for gauge transformations with gauge
%functions that are linear in the coordinates., the first
%requirement is   obviously satisfied.  The  second requirement

%\marginpar{\tiny It is unusual to talk
%about Higgs fields with matter fields}

%Note that while gauge transformations with gauge  functions
%quadratic (and higher order) in  the  coordinates  would
%suffice for  giving spontaneous symmetry breaking, such  gauge
%functions would  preclude translational invariance.

% (and of course are not allowed if the Lorentz
%gauge is chosen) WRONG: $\Box \lambda=0 is  possible  without  having
%\lambda linear or constant

%These
%transformation properties of the vacuum  will  emerge in
%Section~\ref{lingf} as the defining features  of  a
%Coulomb-like phase.

\section{Summary and Concluding Remarks}

We have presented a formalism that can be used to define the various
possible phases for a non-simple gauge group in the context of
lattice gauge theory (LGT). Specifically we are interested in the
non-simple $SMG$. These phases are normally said to be artefacts of
the unphysical lattice regulator. As we assume that a lattice is one
way to implement what we take to be a fundamental ontological
(roughly Planck scale) regulator, the ``artefact'' phases take on a
physical meaning.

The various phases are realized by adjusting intensive parameters
(which are closely related to the couplings) in the action. These
span the so-called action parameter space which is the space in
which the boundaries separating the various possible phases can be
constructed in a way analogous to the way that temperature and
pressure span the space in which the boundaries separating the
solid, liquid and gaseous phases of $H_2O$ can be drawn. In LGT a
typical term in the action is the product of such an intensive
parameter with the trace in some representation of a gauge group
element defined on a lattice plaquette. In each action term these
traces are summed over the plaquettes of the lattice

In this contribution we have developed the formalism for
distinguishing the  possible phases of a non-simple gauge group $G$
each of which corresponds to a pair of subgroups $(K,H)$ such that
$K \subseteq G$ and $H \triangleleft K$.

For each phase $(K,H)$ the free energy $- \log Z_{K_i\subseteq G,
H_j \triangleleft K_i}$ is defined for the entire action parameter
space. At any point in this space, the phase realized is that for
the free enery function has the largest value.

The point in the action parameter space at which the maximum
number of different phases come together - the multiple point -
corresponds according to the MPP to the parameter values
(couplings) realized in Nature. At this point the free energy
functions for all the phases that come together at the multiple
point are of course all equal.

The  \dofx belonging  to  the subgroup $K$ are the un-Higgsed
degrees of freedom if, after fixing the gauge in accord with
say the Lorentz condition,  $K\subseteq G$ is the maximal
subgroup of gauge transformations belonging to the set
$\Lambda_{Const}$ that leaves the vacuum invariant. The field
variables taking values in the homogeneous space $G/K$ are by
definition Higgsed in the vacuum. For these degrees of freedom,
gauge symmetry is spontaneously broken in the vacuum under
global gauge transformations $\Lambda_{Const}$.

Degrees  of  freedom  taking  values   in   the   invariant
subgroup $H\triangleleft K$ are by definition confined in the
vacuum if $H$ is the  maximal  invariant   subgroup   of gauge
transformations $\Lambda_{Linear}$  that  leaves the vacuum
invariant.

The \dofx in a  Coulomb-like phase take as values the cosets of
the factor group $K/H$. The symmetry properties of the vacuum
for a Coulomb-like phase are dictated by the requirements of
the Goldstone Theorem. The conditions to be fulfilled in order
that the Nambu-Goldstone boson accompanying a spontaneous
breakdown of gauge symmetry can be identified with a massless
gauge particle (the existence of which is the characteristic
feature  of  a Coulomb-like phase) suggests that the  Coulomb
phase  vacuum is invariant  under gauge transformations having
a  constant gauge function but spontaneously broken under gauge
transformations having linear gauge functions.

Summarizing one can say that each phase corresponds to a
partitioning of the \dofx (these latter can be labelled by a Lie
algebra basis) - some that are Higgsed, others that are un-Higgsed;
of the latter, some \dofx can be confining, others Coulomb-like. It
is useful to think of a group element $U$ of the gauge group as
being parameterized  in  terms of three  sets of coordinates
corresponding  to  three   different structures that are appropriate
to the symmetry properties  used to define a given phase $(K,H)$ of
the vacuum. These  three  sets  of coordinates, which are definable
in terms of the gauge group $G$, the subgroup $K$, and the invariant
subgroup $H\triangleleft K$, are the {\em homogeneous space} $G/K$,
the {\em  factor  group} $K/H$, and $H$ itself:

\beq U=U(g,k,h) \;\;\mbox{with }\;g\in G/K, k\in K/H, h\in H. \eeq
The coordinates $g\in G/K$ will  be  seen to
 correspond to
Higgsed \dof, the coordinates $k\in  K/H$  to un-Higgsed,
Coulomb-like \dofx and the coordinates $h\in H$ to un-Higgsed,
confined \dof.

%% G. Bregar, SNMB, article
%%\documentclass[preprint,aps]{revtex4}
%%\usepackage[pdftex]{graphicx}
%%% \usepackage[dvips]{graphicx}
%% \begin{document}
%%  \newcommand{\Qed}{\rule{2.5mm}{3mm}}
%% \newcommand{\balpha}{\mbox{\boldmath {$\alpha$}}}
%% \def\Tr{{\rm Tr}}
%% \def\(#1)#2{{\stackrel{#2}{(#1)}}}
%% \def\[#1]#2{{\stackrel{#2}{[#1]}}}
%% \def\A{{\cal A}}
%% \def\B{{\cal B}}
%% \def\Sb#1{_{\lower 1.5pt \hbox{$\scriptstyle#1$}}}
%% \draft 
%%%\documentstyle[preprint,aps]{revtex}
%%%\begin{document}
%%%\draft
\title{Does Dark Matter Consist of Baryons of New Stable Family Quarks?}
\author{G. Bregar and N.S. Manko\v c Bor\v stnik}
\institute{%
Department of Physics, FMF, University of
Ljubljana, \\
Jadranska 19, 1000 Ljubljana}

\authorrunning{G. Bregar and N.S. Manko\v c Bor\v stnik}
\titlerunning{Does Dark Matter Consist of Baryons of New Stable Family Quarks?}
 \maketitle

\begin{abstract} 
We investigate the possibility that the dark matter consists of clusters of the   
heavy family quarks and leptons with zero Yukawa couplings to the lower families. 
Such a family is predicted by the {\it approach unifying spin and charges} as the 
fifth family.   
We make a rough estimation of properties of  baryons  of this new family members, 
of their behaviour during the evolution of the universe and when scattering on 
the ordinary matter and study possible limitations on the family properties due 
to the cosmological and direct experimental evidences. This paper will be published
in  October 2009 in Phys. Rev D. We add it here since in the discussion sections 
the derivations and conclusions of this paper are commented.
\end{abstract}

\section{Introduction}
\label{gbsnmbintroduction}

Although the origin of the dark matter is 
unknown, its gravitational interaction with the known matter  and other cosmological 
observations 
require from the candidate for the dark matter constituent that: 
i. The scattering amplitude of a cluster of constituents with the ordinary matter 
and among the dark matter clusters themselves must be small enough, % to be in agreement 
% with the observations 
so that no effect of such scattering has been observed, except possibly 
in the DAMA/NaI~\cite{gbsnmbrita0708} and not (yet?) in the CDMS and other experiments~\cite{gbsnmbcdms}. 
ii. Its  density distribution (obviously different from the ordinary matter density 
distribution) causes that all the stars  within a galaxy rotate approximately with the same
velocity (suggesting that the density is approximately spherically symmetrically distributed, 
descending with the second power of the distance from the center, it is extended also far 
out of the galaxy,  manifesting the gravitational lensing by galaxy clusters).  
iii. The dark matter constituents must be stable in comparison with the age of our universe,  
having obviously for many orders of magnitude different time scale for forming (if at all)  
solid matter than the ordinary matter. 
iv. The dark matter constituents had to be formed during the evolution of our 
universe %(with the inflationary development included) 
so that they contribute today the main part of the matter ((5-7) times as much as the 
ordinary matter).

There are several candidates for the massive dark matter constituents in the literature, like, for example,
WIMPs (weakly interacting massive particles), the references can be found 
in~\cite{gbsnmbdodelson,gbsnmbrita0708}. 
In this paper we discuss the possibility that the dark matter constituents are 
clusters of a stable (from the point of view of the age of the universe)
family of quarks and leptons. Such a family is predicted 
by the approach unifying spin and  charges~\cite{gbsnmbpn06,gbsnmbn92,gbsnmbgmdn07}, proposed by one 
of the authors of this paper: N.S.M.B. This approach is showing a new way beyond the standard model of the 
electroweak  and colour interactions by answering the open questions of this model like: Where do 
the families originate?, Why do only the left handed quarks and leptons carry the weak charge, 
while the right handed ones 
do not? Why do particles carry the observed $SU(2), U(1)$ and $SU(3)$ charges? Where does the Higgs field 
originate from?, and others. 

 There are several 
attempts in the literature trying to understand the origin of families. All of them, however, 
in one or another way (for example through choices of appropriate groups) simply postulate that there are at 
least three families, as 
does the standard model of the electroweak and colour interactions. 
Pro\-posing the (right) mechanism for generating families is to our understanding  
the most promising guide to   
physics beyond the standard model. 

{\it The approach unifying spin and charges is 
offering the mechanism for the appearance of families.} It introduces the {\it second 
kind}~\cite{gbsnmbpn06,gbsnmbn92,gbsnmbn93,gbsnmbhn02hn03} 
of the Clifford algebra objects,  which generates families as the  
{\it equivalent representations 
to the Dirac spinor representation}. The references~\cite{gbsnmbn93,gbsnmbhn02hn03} show that there are 
two, only two, kinds of the Clifford algebra objects, one used by Dirac to describe the spin of fermions. 
The second kind forms the equivalent representations with respect to the Lorentz group for spinors~\cite{gbsnmbpn06} and  
the families do form the equivalent representations with respect to the Lorentz group. 
The approach, in which fermions carry two kinds of spins (no charges), predicts from the simple starting action 
 more than the observed three families. It predicts 
two times  four families with masses several orders of magnitude bellow the 
unification scale of the three observed charges. 

Since due to the approach 
(after assuming a particular, but to our opinion trustable, way of a nonperturbative breaking of the starting symmetry) 
the fifth family decouples in the Yukawa couplings from 
the lower four families (whose the fourth family quark's  mass is predicted to be at around $250$ GeV or 
above~\cite{gbsnmbpn06,gbsnmbgmdn07}),    
the fifth family quarks and leptons are stable as required by the condition iii.. 
Since the masses of all the members of the fifth family lie, due to the approach, much above the known three  
and the predicted fourth family masses,  
%(the fourth family might according to the first very rough estimates of the approach be even seen at the LHC), 
the baryons made out of the fifth family form small enough  clusters (as we shall see in section~\ref{gbsnmbproperties}) so that 
 their scattering amplitude among themselves and with the ordinary matter  
is small enough and also the number of clusters (as we shall see in  section~\ref{gbsnmbevolution})  is low enough to fulfil 
the conditions i. and iii.. 
Our study of the behaviour of the fifth family quarks in the cosmological evolution 
(section~\ref{gbsnmbevolution}) shows that also the condition 
iv. is fulfilled, if the fifth family masses are large enough.
%%%%%******

Let us add that there are several assessments about masses of a possible (non stable) fourth family of quarks and leptons, 
which follow from the analyses of the existing experimental data and the cosmological observations. 
Although most of physicists have  doubts about the existence of any more 
than the three observed families, the analyses clearly show that neither the experimental electroweak data~\cite{gbsnmbokun,gbsnmbpdg}, 
nor the cosmological observations~\cite{gbsnmbpdg} forbid the existence of more than three 
%%%%
%%%%KOMENTAR ZELI REFEREE
%%%%
families, as long as the masses of the fourth family quarks are higher than a few hundred 
GeV and the masses of the fourth family leptons above one hundred GeV ($\nu_4$ could be above $50$ GeV). 
 We studied in the ¨references~\cite{gbsnmbpn06,gbsnmbgmdn07,gbsnmbn09} possible (non perturbative)  breaks of the  symmetries
 of the simple starting Lagrangean which, by predicting the Yukawa couplings, leads at low energies first 
to twice four families with no Yukawa couplings between these two groups of families. One group 
obtains at the last break masses of several hundred TeV or higher, 
while the lower four families stay massless and mass protected~\cite{gbsnmbn09}.
For one choice of the next break~\cite{gbsnmbgmdn07} the fourth family members ($u_{4}, d_{4}, \nu_4, e_4$)  
obtain the masses at
($224$ GeV (285 GeV), $285$ GeV (224 GeV), $84$ GeV, $170$ GeV), respectively. 
For the other choice  of the next break %, more 
%trustable in our understanding, 
we could not determine the fourth family masses, but  when assuming the values for these masses we predicted mixing 
matrices in dependence on the masses. All these studies were done on the tree level. We are 
studying now symmetries of the Yukawa couplings if we go beyond the tree level. 
Let us add that the last experimental data~\cite{gbsnmbmoscow09} from the HERA experiments require that there is  
no $d_4$ quark with the mass lower than $250$ GeV. 

Our stable  fifth family baryons, which might form the dark matter, also   
do not contradict the so far observed experimental data---as it is the measured 
(first family) baryon number and its ratio to the photon  energy density, 
as long as the fifth family quarks are 
 heavy enough ($> 1$ TeV). (This would be true for any stable heavy family.) 
Namely, all the measurements, which connect the baryon and the photon 
energy density, relate to the moment(s) in the history of 
the universe, when baryons of the first family where formed ($k_b T$ bellow the binding energy 
of the three first family quarks dressed into constituent mass of $m_{q_1}c^2 \approx 300$ MeV, that is bellow 
$10$ MeV) and the electrons and nuclei   formed  atoms ($k_b \,T 
 \approx 1$ eV). The chargeless (with respect to the colour and electromagnetic 
 charges) clusters of the fifth family were 
 formed long before (at $ k_b T\approx E_{c_5}$ (see Table~\ref{gbsnmbTableI.})), contributing the equal amount of the
 fifth family baryons and anti-baryons to the dark matter, provided that there is no 
 fifth family baryon---anti-baryon asymmetry   
 %%%%
 %%%%POSTAVI TO tudi V ZAKLJUCEK
 (if the asymmetry  is nonzero the colourless baryons or anti-baryons are formed also at the 
 early stage of the colour phase transition at around $1$ GeV). %(section~\ref{gbsnmbevolution}). 
 %%%%
 They  manifest 
 after decoupling from the plasma (with their small number density and  small cross 
 section) (almost) only their gravitational  interaction.

In this paper we estimate the properties of the fifth family members
($u_5$, $d_5$, $\nu_5$, $e_5$), as well 
as of the clusters of these members, in particular the fifth family neutrons, under the assumptions that:\\
%I. The fifth family neutrons and anti-neutrons are candidates to form the dark matter.\\
I. Neutron is the lightest fifth family baryon.\\ 
II. There is no fifth family baryon---anti-baryon asymmetry.\\
The assumptions are made since we are not yet able to derive the properties of the family from the starting 
Lagrange density of the approach. The results of the present paper's study are helpful 
to better understand  steps needed to come from the approach's starting Lagrange density to the 
low energy effective one.

From  the approach unifying spin and charges we  learn: \\
i. The stable fifth family members have masses higher than $\approx 1$ TeV  and smaller than  
$ \approx 10^6$ TeV. \\
ii. The stable fifth family members have the properties of the lower four families; 
that is the same family members with the same 
(electromagnetic, weak and colour) charges 
and interacting correspondingly with the same gauge fields.

We estimate the masses of the fifth family quarks by studying 
their behaviour in the evolution of the universe, their formation of %colour and electromagnetic 
chargeless 
(with respect to the electromagnetic and colour interaction) clusters and the properties of these clusters 
when scattering on the ordinary (made mostly of the first family members) matter and among themselves.  
We use a simple (the hydrogen-like) model~\cite{gbsnmbgnBled07} to estimate
the size and the binding energy of the fifth family baryons, assuming that the fifth family 
quarks are heavy enough to interact mostly by exchanging one gluon. 
We solve the Boltzmann equations for the fifth family quarks (and anti-quarks) 
forming the colourless clusters in the expanding universe, starting  in the energy 
region when the fifth family members are ultrarelativistic, up to $\approx 1$ GeV when the colour phase transition starts. 
In this energy interval the  one gluon exchange is the dominant interaction among quarks and the plasma. 
We conclude that the quarks and anti-quarks, which   
succeed to form neutral (colourless and electromagnetic chargeless) clusters,  have the properties 
of the dark matter constituents if their masses are within the interval of a few TeV $< m_{q_5} c^2< $ a few hundred TeV, 
while the  rest of the coloured fifth family objects annihilate within the colour phase 
transition period %starting at around $1$ GeV 
with their anti-particles for the zero fifth family baryon number asymmetry. 
%%%%
%%%%IZRACUNATI STEVILO NEVTRONOV IN ANTINEVTRONOVPRI NEKI PREDPOSTAVLJENI ASIMETRIJI, DENIMO ISTI KOT PRI PRVI DRUZINI. 
%%%%

We estimate also the behaviour of 
our fifth family clusters if hitting the DA\-MA/NaI---DAMA-LIBRA~\cite{gbsnmbrita0708} and CDMS~\cite{gbsnmbcdms} 
experiments presenting the limitations the DAMA/NaI experiments put on our fifth family quarks when    
%which exhibit annual modulation and 
recognizing that CDMS has not found any event (yet). 
%%%%
%%%%KOMENTIRAJ LIMITO, KI JO POSTAVI CDMS. POVEJ V POGLAVJU O DYNAMIKI, KAKO SE OBNASAJO GRUCE
%%%%TO  JE TUDI ZELJA REFEREEJA
%%%%

%%%%
%%%%POSTAVI V ZAKLJUCEK
%%%%
The fifth family baryons are not the objects (WIMPS), which would interact with only the weak interaction, 
since their decoupling from the rest of the 
plasma in the expanding universe is determined by the colour force and  
their interaction with the ordinary matter is determined with the  
fifth family "nuclear force" (this is the force among clusters of the fifth family quarks, 
manifesting much smaller cross section than does the ordinary, mostly first family, 
"nuclear force") as long as their mass is not higher than $10^{4} $ TeV, when the weak interaction starts to 
dominate as commented 
in the last paragraph of section~\ref{gbsnmbdynamics}.

\section{Properties of clusters of the heavy family}
\label{gbsnmbproperties}

Let us  study the properties of the fifth family of quarks and leptons as 
predicted by the approach unifying spin and charges, with masses several orders of 
magnitude greater than those of the known three families,  decoupled in the Yukawa couplings from 
the lower mass families and with 
the charges and their couplings to the gauge fields of the known families (which all seems, 
due to our estimate predictions of the approach, reasonable assumptions).   
Families distinguish among themselves (besides in masses) 
in the family index (in the quantum number, which in the approach is determined  
by the second kind of the Clifford algebra objects' operators~\cite{gbsnmbpn06,gbsnmbn92,gbsnmbn93} 
$\tilde{S}^{ab}=\frac{i}{4}(\tilde{\gamma}^a \tilde{\gamma}^b - \tilde{\gamma}^b 
\tilde{\gamma}^a)$, anti-commuting with the Dirac $\gamma^a$'s), and 
(due to the Yukawa couplings)  in their masses.  

For a heavy enough family the properties of baryons (protons $p_5$ $(u_5 u_5 d_5)$, 
neutrons $n_5$ $(u_5 d_5 d_5)$, $\Delta_{5}^{-}$, $\Delta_{5}^{++}$) 
made out of  
quarks $u_5$ and $d_5$ can be estimated by using the non relativistic Bohr-like model 
with the $\frac{1}{r}$ %(radial) 
dependence of the potential  
between a pair of quarks  $V= - \frac{2}{3} \frac{\hbar c \,\alpha_c}{r}$, where $\alpha_c$ is in this case the 
colour coupling constant. 
Equivalently goes for anti-quarks. % (with the factor $\frac{4}{3}$ instead of $\frac{2}{3}$). 
This is a meaningful approximation as long as the   
one gluon exchange is the dominant contribution to the interaction among quarks,  
that is as long as excitations  of a cluster are not influenced by  the linearly rising 
part of the potential~\footnote{Let us tell that a simple bag model evaluation does not 
contradict such a simple Bohr-like model.}. The electromagnetic 
and weak interaction contributions are  of the order of $10^{-2}$ times smaller. 
%~\footnote{
%A simple bag model %~\cite{gbsnmbkuti}, 
%with the potential $V(r)=0 $ for $r<R$ 
%and $V(r)= \infty $ otherwise, supports our rough estimation. It, namely, 
%predicts for the lowest energy $E$ (the mass) of a cluster of three quarks:  
%$E= 3 \,m_{q_5} c^2 (1+ (x \hbar c /m_{q_5}c^2 R)^2),$ with $\tan x = x/ [1-
%(m_{q_5}c^2 R/\hbar c) - \sqrt{x^2 + (m_{q_5}c^2 R/\hbar c)^2} \,], $ where $2.04< x <\pi$ 
%for $0 < (m_{q_5}c^2 R/ \hbar c) < \infty$. For  $R$ taken from our Bohr-like model 
%$R= r_{c_5}$,  $(m_{q_5}c^2 R/ \hbar c) \approx 8$, 
%for example, is $x$ close to $3$ and rises very slowly to $\pi$. Accordingly 
%the mass of the three quark cluster is close to three masses of the quark even in the bag model, while 
% one gluon exchange suggests the binding energy for one pair of quarks of the order of 
% $\frac{1}{20} m_{q_5}c^2$}.
%
Which one of $p_5$, $n_5$, or maybe $\Delta_{5}^-$ or $\Delta_{5}^{++}$,  
is a stable fifth family baryon, depends on the ratio of the bare masses 
$m_{u_5}$ and  $m_{d_5}$, as well as on the  weak and the 
electromagnetic interactions among quarks. 
If $m_{d_5}$ is appropriately 
smaller than $m_{u_5}$ so that the  
weak and electromagnetic interactions favor the neutron $n_5$, then $n_5$ is 
a colour singlet electromagnetic chargeless stable cluster of quarks, with 
the weak charge $-1/2$. %with the lowest mass among 
%the nucleons of the fifth family. 
If $m_{d_5}$ is larger (enough, due to the stronger electromagnetic repulsion among 
the two $u_5$ than among the two $d_5$) than $m_{u_5}$, the proton $p_5$ which is 
a colour singlet stable nucleon with the weak charge $1/2$,  %and the electromagnetic charge 
needs the electron $e_5$ or $e_1$ or $\bar{p}_1$ to form  a stable  electromagnetic %(and weak) 
chargeless cluster (in the last case it could also be the weak singlet and would accordingly  manifest 
the ordinary nuclear force only).  
An atom made out of only fifth family members might be lighter or not than $n_5$, 
depending on the masses of the fifth family members. 

Neutral (with respect to the electromagnetic and colour charge) fifth family 
particles that constitute the dark matter can be $n_5,\nu_5$ 
or  charged baryons like  $p_5, \Delta^{++}_5$, $\Delta^{-}_5$, forming neutral atoms with
$e^{-}_5$ or $\bar{e}^{+}_5$, correspondingly, or (as said above) $p_{5} \bar{p}_1$ . 
We treat the case that $n_5$ as well as $\bar{n}_5$ 
form the major part of the dark matter, assuming %in this letter 
that $n_5$ (and $\bar{n}_5$) are stable baryons (anti-baryons). Taking  $m_{\nu_5}< m_{e_5}$ 
also $\nu_5$ contributes to the dark matter. We shall comment this in section~\ref{gbsnmbdirectmeasurements}.

In the Bohr-like model % or the a little more sophisticated hydrogen-like model in which 
%we let the widths of the hydrogen wave functions to be adapted as explained in Appendix, 
we obtain if neglecting more than one gluon exchange contribution
%(****GREGOR, %POISKATI REFERENCE IN POPRAVITI TABELO!!!!VSTAVITI v FOOTNOTE PRAVI IZRAZ, 
%Napisati Appendix****)
%
\begin{eqnarray}
\label{gbsnmbbohr}
E_{c_{5}}\approx -3\; \frac{1}{2}\; \left( \frac{2}{3}\, \alpha_c \right)^2\; \frac{m_{q_5}}{2} c^2,
\quad r_{c_{5}} \approx  \frac{\hbar c}{ \frac{2}{3}\;\alpha_c \frac{m_{q_5}}{2} c^2}. 
\end{eqnarray}
The mass of the cluster is approximately $m_{c_5}\, c^2 \approx  
3 m_{q_5}\, c^2(1- (\frac{1}{3}\, \alpha_c)^2)$. We use the  factor of $\frac{2}{3}$ 
for a  two quark pair potential and of $\frac{4}{3}$ for a quark and an anti-quark pair potential. %~\cite{gbsnmbruhula} 
If treating correctly the three quarks' (or anti-quarks') center of mass motion in the 
hydrogen-like model, allowing  the hydrogen-like functions to adapt the width as presented in 
Appendix,
%\ref{gbsnmbbetterhf},   
the factor $-3\; \frac{1}{2}\; (\frac{2}{3})^2\; \frac{1}{2}$ in Eq.~\ref{gbsnmbbohr} is replaced by
$0.66$, and the mass of the cluster is accordingly $3 m_{q_5} c^2(1-0.22\, \alpha_{c}^2)$, while 
the average radius takes the values as presented in  Table~\ref{gbsnmbTableI.}.

Assuming that the coupling constant   
of the colour charge  $\alpha_c$   runs with the kinetic energy $- E_{c_{5}}/3$ and taking into account 
the number of families which contribute to the running coupling constant in dependence on the kinetic energy 
(and correspondingly on the mass of the fifth family quarks)
%of a quark %as in %ref.~\cite{gbsnmbgreiner}  
%
%($\,\alpha_c(E^2)=\frac{\alpha_c(M^2)}{1+\frac{\alpha_c(M^2)}{4 \pi} (11-\frac{2 N_F}{3}) 
%\textrm{ln}(\frac{E^2}{M^2}) }$, with $\alpha_{c}((91 \; \textrm{GeV})^2)=0.1176(20)$,
%the number of flavours $N_F=8$ or less, depending on the temperature (****ALI JE UPOSTEVANO?****)) 
%
we estimate  the properties of a baryon as presented on Table~\ref{gbsnmbTableI.} (the table 
is calculated from the hydrogen-like model presented in Appendix), 
\begin{table}
\begin{center}  
\begin{tabular}{||c||c|c|c|c|}
\hline
$\frac{m_{q_5} c^2}{{\rm TeV}}$ & $\alpha_c$ & $\frac{E_{c_5}}{m_{q_5} c^2}$ & 
$\frac{r_{c_5}}{10^{-6}{\rm fm}}$ & $\frac{\Delta m_{ud} c^2}{{\rm GeV}}$ 
\\
\hline
\hline
$1   $ & 0.16   & -0.016   & $3.2\, \cdot 10^3$   & 0.05		             \\
\hline
$10  $ & 0.12   & -0.009   & $4.2\, \cdot 10^2$   & 0.5           \\
\hline
$10^2$ & 0.10   & -0.006   & $52$            & 5           \\
\hline
$10^3$ & 0.08   & -0.004   & $6.0$           & 50           \\
\hline
$10^4$ & 0.07   & -0.003   & $0.7$           & $5 \cdot 10^2$           \\
\hline
$10^5$ & 0.06   & -0.003   & $0.08$          & $5 \cdot 10^3$            \\
\hline
\end{tabular}
\end{center}
\caption{\label{gbsnmbTableI.} 
%Na desni v vsaki deljeni celici je primer, ko je $N_F=8$, na levi pa ko je 
%$N_F=6$.
The properties of a cluster of the fifth family quarks
within the extended Bohr-like (hydrogen-like) model from Appendix. 
$m_{q_5}$ in TeV/c$^2$ is the assumed fifth family quark mass,
$\alpha_c$ is the coupling constant 
of the colour interaction at $E\approx (- E_{c_{5}}/3)\;$ (Eq.\ref{gbsnmbbohr})   
which is the kinetic energy 
of  quarks in the baryon, 
$r_{c_5}$ is the corresponding average  radius. Then  $\sigma_{c_5}=\pi r_{c_5}^2 $  
is the corresponding scattering cross section.} % for a chosen quark mass.} 
\end{table}

The binding energy is approximately  $\frac{1}{ 100}$  of the mass 
of the cluster (it is $\approx \frac{\alpha_{c}^2}{3}$).  The baryon $n_5$ ($u_{5} d_{5} d_{5}$) 
is lighter than the baryon $p_{5}$,   ($u_{q_5} d_{q_5} d_{q_5}$) 
if $\Delta m_{ud}=(m_{u_5}-m_{d_5})$ is smaller than $\approx (0.05,0.5,5, 50, 500, 5000)$ GeV  
for the six  values of the $m_{q_5} c^2$ on Table~\ref{gbsnmbTableI.}, respectively. %(****PREVERITI!****) 
We see  from Table~\ref{gbsnmbTableI.} that the ''nucleon-nucleon'' 
force among the fifth family baryons leads to many orders of 
magnitude smaller cross section than in the case 
of the first family nucleons ($\sigma_{c_5}= \pi r_{c_5}^2$ is from $10^{-5}\,{\rm fm}^2$  for 
$m_{q_5} c^2 = 1$ TeV to $10^{-14}\, {\rm fm}^2$  for $m_{q_5} c^2 = 10^5$ TeV). 
Accordingly is the scattering cross section between two  fifth family baryons   
determined by the weak interaction as soon as the mass   exceeds  several GeV.

If a cluster of the heavy (fifth family) quarks and leptons and  of the 
ordinary (the lightest) family is made, 
then, since ordinary family   dictates the radius and the excitation energies  
of a cluster, its 
properties are not far from the properties of the ordinary hadrons and atoms, except that such a  
cluster has the mass dictated by the heavy family members. 
\section{Evolution of the abundance of the fifth family members in the universe}
\label{gbsnmbevolution}

We assume that there is no fifth family baryon---anti-baryon asymmetry
and that the  neutron is the lightest baryon made out of the 
fifth family quarks. Under these assumptions and with the knowledge from 
our rough estimations~\cite{gbsnmbgmdn07} that the fifth family masses are within  the interval    
from  $1$ TeV to $10^6$ TeV  
%
% (We only have a rough estimation that this could be the case).
%%%%
%%%%
%KOMENTIRAJ NA KONCU, PRIMER, KO TA PRIVZETEK SPROSTIMO
%%%%
%%%%
we study the behaviour of our fifth family quarks and anti-quarks in the expanding  
(and accordingly cooling down~\cite{gbsnmbdodelson}) 
universe in the plasma of all other fields (fermionic and bosonic) from the 
period, when the fifth family members carrying all the three charges (the colour, 
weak and electromagnetic) are ultra relativistic and is their number 
(as there are the numbers of all the other fermions and bosons in the 
ultra relativistic regime) determined by the temperature. We follow the fifth family quarks and 
anti-quarks first through the 
freezing out period, when the fifth family quarks and anti-quarks 
start to have too large mass to be formed out of the plasma (due to  the plasma's too low temperature), 
then through the period when  first  the clusters of 
di-quarks and di-anti-quarks and then the colourless neutrons and anti-neutrons ($n_5$  and $\bar{n}_5$) 
are formed.  The fifth family neutrons being tightly bound into the colourless objects do not feel   
the colour phase transition  when it  
starts bellow $ k_b  T\approx 1$ GeV ($k_b$ is the Boltzmann constant) and %$n_5$  and $\bar{n}_5$
decouple accordingly from the rest of quarks and anti-quarks and gluons and manifest 
today as the dark matter constituents. 
We take the quark mass as a free parameter in the interval 
from $1$ TeV to $10^6$ TeV and determine the mass from the observed dark matter density.  
%assuming that there are $n_5$ and 
%$\bar{n}_5$ which form the today's dark matter.  

At the colour phase transition, however, the coloured fifth family quarks and anti-quarks 
annihilate  to the today's unmeasurable density: Heaving much larger mass (of the order of $10^{5} $ times larger),  
%as we shall evaluate from the dark matter density for our fifth family quarks)  and accordingly 
and correspondingly much larger momentum (of the order of $10^3$ times larger) as well as much 
larger binding energy (of the order 
of $10^5$ times larger)
than the first family quarks when they are  "dressed" into constituent mass,   
the coloured fifth family quarks succeed in the colour phase transition region  to annihilate with the 
corresponding anti-quarks to the non measurable extend, 
if it is no fifth family baryon asymmetry. 
%%%%
%%%%
%NAPRAVI RACUN
%KOMENTIRAJ, CE ASIMETRIJA JE, TEDAJ PAC PRISPEVAJO K TEMNI MASI. TUDI TEDAJ JE NJIHOVA MASA 
% DOLOCENA S TEM, DA MORAJO BITI DOVOLJ TEZKI, KER SICER IYJAVA NE VELJA.
%%%%
%%%%

In the freezing out period almost up to the colour phase transition
the kinetic energy of quarks is high enough so that  
the one gluon exchange 
dominates in the colour interaction of quarks with the plasma, while the (hundred times) weaker weak 
%interaction and even much weaker 
and electromagnetic 
interaction can be neglected. 

The quarks and anti-quarks start to freeze out when the temperature of the plasma falls close to  
$m_{q_5}\,c^2/k_b $. They are forming clusters (bound states) 
when the temperature falls close to  the binding energy (which is due to Table~\ref{gbsnmbTableI.} 
$\approx \frac{1}{100} m_{q_5} c^2$). 
When the three quarks (or three anti-quarks) of the 
fifth family form a colourless baryon (or anti-baryon), they decouple from the rest of plasma due to small 
scattering cross section manifested by the average radius presented in Table~\ref{gbsnmbTableI.}.

%To estimate the behaviour of our stable heavy family quarks and anti-quarks in the expanding 
%universe we would need to know the mass of the fifth family quarks and the fifth family baryon asymmetry. 
%%We take the quark mass as a free parameter and determine it from the observed dark matter density, 
%assuming that there are $n_5$ and 
%$\bar{n}_5$ which form the today's dark matter.  

Recognizing that at the temperatures ($ 10^6$ TeV $> k_b T >1 $ GeV) the one gluon exchange gives the 
dominant contribution to the interaction among quarks of any family, it is not difficult to estimate 
the thermally averaged scattering cross sections (as the function of the temperature) for 
the fifth family quarks and   anti-quarks to 
scatter:\\    
$\;\;\;\;$ i. into all the 
relativistic quarks and anti-quarks of lower mass families ($<\sigma v>_{q\bar{q}}$),\\  
$\;\;\;$ ii.  into gluons ($<\sigma v>_{gg}$), \\
$\;\;$ iii.  into (annihilating) bound states of a fifth family quark and an anti-quark mesons $\;\;\;$
($<\sigma v>_{(q\bar{q})_b}$), \\ 
%with the summation over $2 \times 8$ gluons) 
$\;\;$ iv. into bound states of two fifth family quarks and into the fifth family baryons 
($<\sigma v>_{c_5}$)  %which here 
%we evaluate only with the assumptions that these two cross section can be treated together  
(and equivalently into two anti-quarks and 
into anti-baryons).

%%%%
%%%%
%KAR SLEDI JE ZE BILO POVEDANO IN NE PONAVLJAJ
%%%%
%%%%
%A rough estimation of the probability for the first family quarks and anti-quarks   to annihilate  
%at the colour phase transition ($  k_b T\approx 1$ GeV) shows that the fifth family quarks do 
%annihilate with the fifth family anti-quarks, leaving (almost, negligible small) number of stable 
%mesons of the first and the fifth family members. 
%%iv.) the ''Thompson'' scattering cross section for gluons on quarks or anti-quarks ($\sigma_{T}$) 

The one gluon exchange scattering cross sections are namely (up to the stren\-gth of the  coupling constants 
and up to the numbers of the order one 
determined by the corresponding groups) 
equivalent to the corresponding cross sections for the one photon exchange scattering cross sections, and  
we use correspondingly also the  expression  for scattering of an electron and a proton into 
the bound  state of a hydrogen when treating the scattering of two quarks into the bound states.
%So we use the 
%adapted cross sections from the electromagnetic interactions%when there are no  expressions available 
%in the literature for the one gluon exchange
We take the roughness  
of such estimations into account by two parameters: 
The parameter $\eta_{c_5}$ takes care of scattering 
of two quarks (anti-quarks) into three 
colourless quarks (or anti-quarks), which are the fifth family baryons (anti-baryons) 
and about the uncertainty with which this 
cross section is estimated. $\eta_{(q \bar{q})_b}$ takes care of the roughness of the used formula 
for $<\sigma v>_{(q\bar{q})_b}$.

%$\eta_{c_5}$ and $\eta_{(q \bar{q})_b}$ (defined bellow), which 
%define accordingly the accuracy with which the fifth family mass is estimated and correspondingly 
%the acceptable fifth family mass interval. 

The following expressions for the thermally averaged cross sections are used%in the Boltzmann equations 
\begin{eqnarray}
\label{gbsnmbsigmasq}
< \sigma v>_{q\bar{q}} &=&  \frac{16 \,\pi}{9} 
\;\left( \frac{\alpha_{c} \hbar c}{m_{q_{5}}\,c^2}\right)^2 \, c ,\nonumber\\
< \sigma v>_{gg} &=&  \frac{37 \,\pi}{108}\;\left( \frac{\alpha_{c} 
\hbar c}{m_{q_{5}}\,c^2}\right)^2\, c, \nonumber\\
< \sigma v>_{c_5} &=& \eta_{c_5}\; 10 \;\left( \frac{\alpha_{c} \hbar c}{m_{q_5}\,c^2} \right)^2\, c\; 
\sqrt{\frac{ E_{c_5}}{ k_b T}} \ln{\frac{E_{c_5}}{ k_b T}}, \nonumber\\
<\sigma  v>_{(q \bar{q})_b}&=& \eta_{(q \bar{q})_b} \;10 \;\left(
\frac{\alpha_{c} \hbar c}{m_{q_5}\,c^2}\right)^2\, c\; 
\sqrt{\frac{ E_{c_5}}{ k_b T}} \ln{\frac{E_{c_5}}{ k_b T}}, \nonumber\\
\sigma_{T } &=&  \frac{8 \pi}{3} \left(\frac{\alpha_{c} \hbar c }{m_{q_5} \, c^2}\right)^2,
\end{eqnarray}
where $v$ is the relative velocity between the fifth family  quark and its anti-quark, or between two quarks and 
$E_{c_5}$ is the binding energy for a cluster (Eq.~\ref{gbsnmbbohr}). 
%$< \sigma v> \;$ is the thermally averaged scattering cross section 
%times the  relative velocity: 
%i. $< \sigma v>_{q\bar{q}} $ for all the pairs  
%of the fifth family  quarks and anti-quarks into all the lower mass (of the four families') 
%quarks and anti-quarks, which are, while scattering takes place, ultra relativistic. 
%ii. $< \sigma v>_{gg}$  for scattering into gluons.  
%iii. $ <\sigma v>_{c_5}$ for two quarks (or two anti-quarks) to scatter into a bound state of 
%two quarks (anti-quarks) and from two to three quarks (anti-quarks) colourless clusters. 
%We use the equivalent expression as for scattering of an electron and a proton into 
%the bound  state of a hydrogen. 
%The parameter $\eta_{c_5}$ takes care of scattering 
%of two quarks (anti-quarks) into three 
%colourless quarks (or anti-quarks), which are the fifth family baryons (anti-baryons) and about the uncertainty with which this 
%cross section is estimated. 
%
%
%iv. $<\sigma v>_{(q\bar{q})_b}$ for  scattering into a bound state of the fifth family quark and anti-quark,  
%which annihilate in the time $\tau_{(q\bar{q})_b} < 10^{-28}$ s. 
%$\eta_{(q \bar{q})_b}$ takes care of the roughness of the used formula. 
%
$\sigma_{T }$ is the Thompson-like scattering cross section of gluons on quarks (or anti-quarks).  
%responsible for destroying the bound states of baryons. %  an acceptable approximation, when 
%the $\hbar \omega$ of a gluon is much smaller than the $m_{q_5} c^2$ and high enough that it scatters 
%elastically. 
%%%%%

To see how many fifth family quarks and anti-quarks of a chosen mass  form the fifth family 
baryons and anti-baryons today we solve the coupled systems of 
Boltzmann equations presented bellow as  a function of time (or temperature). 
The value of the fifth family quark mass which predicts the today observed dark matter is 
 the mass we are looking for. Due to the inaccuracy of the estimated 
scattering cross sections 
entering into the Boltzmann equations we %only can 
tell the interval within which the mass lies. 
We follow in our derivation of the Boltzmann equations (as much as possible) 
%, when estimating the number density of the fifth family quarks $n_{q_5}$ and 
%anti-quarks $n_{\bar{q}_5}$ clustered into baryons (with the number density $n_{c_5}$) and anti-baryons 
%($n_{\bar{c}_5}$), which to our prediction form  the dark matter today, 
the ref.~\cite{gbsnmbdodelson}, chapter 3. 
%$n_{q_{5}}$ is the number density of all the  fifth family quarks of any colour and spin and 
% correspondingly is assumed for the other number densities. 

Let $T_0$ be the today's black body radiation temperature, $T(t)$ the actual (studied) temperature, 
$a^2(T^0) =1$ and $a^2(T)= a^2(T(t))$ is the metric tensor component in 
the expanding flat universe---the Friedmann-Robertson-Walker metric:  
${\rm diag}\,  g_{\mu \nu} = 
(1, - a(t)^2, - a(t)^2, - a(t)^2),\;$  $(\frac{\dot{a}}{a})^2= \frac{8 \pi G}{3} \rho$, 
with $\rho= \frac{\pi^2}{15} \, g^*\, T^4$, $ \, T=T(t)$,  
$g^*$  measures the number of degrees of freedom of those of the  
four family members (f) and  gauge bosons (b), which are at the treated temperature $T$ 
ultra-relativistic ($g^*= \sum_{i\in {\rm b}} \,g_i + \frac{7}{8} \sum_{i\in {\rm f}} \,g_i$). 
$H_0 \, \approx 1.5\,\cdot 10^{-42} \,\frac{{\rm GeV} c}{\hbar c} $ is the present Hubble constant 
and $G = \frac{\hbar c }{ (m_{pl}^2)}$, $m_{pl} c^2 =  
1.2 \cdot 10^{19}$ GeV. 

Let us write down the Boltzmann equation, which treats in the expanding universe 
the number density of all the fifth 
family quarks as a function of time $t$. The fifth family quarks scatter with  anti-quarks into 
all the other relativistic quarks (with the number density $n_{q}$) and anti-quarks ($n_{\bar{q}}$ 
($< \sigma v>_{q\bar{q}}$) and into gluons 
($< \sigma v>_{gg}$). At the beginning, when the quarks are becoming non-relativistic and   
start to freeze out, the  formation of bound states is negligible. One finds~\cite{gbsnmbdodelson} 
the Boltzmann equation for the fifth family quarks $n_{q_5}$ (and  equivalently  for 
anti-quarks $n_{\bar{q}_5}$)
\begin{eqnarray}
\label{gbsnmbboltzq1}
a^{-3}\frac{d( a^3 n_{q_5})}{dt} &=& < \sigma v>_{q\bar{q}}\; n^{(0)}_{q_5} n^{(0)}_{\bar{q}_5}\,
\left( - \frac{n_{q_5} n_{\bar{q}_5}}{n^{(0)}_{q_5} n^{(0)}_{\bar{q}_5}} + 
 \frac{n_{q} n_{\bar{q}}}{n^{(0)}_{q} n^{(0)}_{\bar{q}}} \right) + \nonumber\\ 
&&< \sigma v>_{gg} \; 
n^{(0)}_{q_5} n^{(0)}_{\bar{q}_5}\,
\left( - \frac{n_{q_5} n_{\bar{q}_5}}{n^{(0)}_{q_5} n^{(0)}_{\bar{q}_5}} +  
\frac{n_{g} n_{g}}{n^{(0)}_{g} n^{(0)}_{g}} \right). 
\end{eqnarray}
Let us tell that $n^{(0)}_{i} = g_i\, (\frac{m_i c^2  k_b T}{(\hbar c)^2})^{\frac{3}{2}} 
e^{-\frac{m_i c^2}{ k_b T}}$ for 
$m_i c^2 >>  k_b T$  %(which is our case 
and   $\frac{g_i}{\pi^2}\, (\frac{ k_b T}{\hbar c})^3$ 
for $m_i c^2 << k_b T$.
Since the ultra-relativistic quarks and anti-quarks of the lower families are in the  
thermal equilibrium with the plasma and so 
are gluons, it follows $\frac{n_{q} n_{\bar{q}}}{n^{(0)}_{q} n^{(0)}_{\bar{q}}}=1= 
\frac{n_{g} n_{g}}{n^{(0)}_{g} n^{(0)}_{g}}$. Taking into account that $(a\, T)^3 \, g^*(T)$ is a constant  
it is appropriate~\cite{gbsnmbdodelson} to introduce a new parameter $x=\frac{m_{q_5}c^2}{k_b T}$  and 
the quantity $Y_{q_5}= 
n_{q_5}\, (\frac{\hbar c}{k_b T})^3$, $Y^{(0)}_{q_5}= 
n^{(0)}_{q_5}\, (\frac{\hbar c}{k_b T})^3$.  When taking into account that  the number of 
quarks is the same as the number of anti-quarks, and that 
$\frac{dx}{dt} = \frac{h_m \,m_{q_5}c^2}{x} $, with $h_m = \sqrt{\frac{4 \pi^3 g^*}{45}}\, 
\frac{c}{\hbar c \, m_{pl} c^2}$, Eq.~\ref{gbsnmbboltzq1} transforms into $\frac{dY_{q_5}}{dx} = 
\frac{\lambda_{q_5}}{x^2}\, (Y^{(0)2}_{q_5} - Y^{2}_{q_5}), $ with $\lambda_{q_5} = \frac{(<\sigma v>_{q\bar{q}} + 
<\sigma v>_{gg}) \, m_{q_5} c^2}{h_{m}\, (\hbar c)^3}$. It is this equation which  we are solving 
(up to the region of $x$ when the clusters of quarks 
and anti-quarks start to be formed) to see 
the behaviour of the fifth family quarks as a function of the temperature.  

When the temperature of the expanding universe falls close enough to the binding energy of the cluster 
of the fifth family quarks (and anti-quarks), the bound states of quarks (and anti-quarks) and the 
clusters of fifth family baryons (in our case neutrons $n_{5}$) (and anti-baryons $\bar{n}_{5}$---anti-neutrons) 
start to form. 
%%%%%
%%%%%
To a fifth family di-quark  ($q_5 + q_5 \rightarrow $ di-quark + gluon) 
a third quark clusters (di-quark $+ q_5 \rightarrow c_5 +$ gluon) to form the colourless fifth 
family neutron (anti-neutron), in an excited state (contributing gluons back into the plasma in the thermal bath 
when going into the ground state), all in thermal equilibrium. Similarly goes with the anti-quarks clusters.
We take into account both processes approximately within the same 
equation of motion by correcting the averaged amplitude $< \sigma v>_{c_5} $ for quarks to scatter into 
a bound state of di-quarks with the parameter $\eta_{c_5}$, 
as explained above.  
%%%%%
%%%%%
The corresponding Boltzmann equation for the number of baryons $n_{c_5}$ then reads
\begin{eqnarray}
\label{gbsnmbboltzc}
a^{-3}\frac{d( a^3 n_{c_5})}{dt} &=& < \sigma v>_{c_5}\; n^{(0)^2}_{q_5}\,
\left( \left( \frac{n_{q_5}}{n^{(0)}_{q_5}} \right)^2 -  
\frac{n_{c_5}}{ n^{(0)}_{c_5}} \right).
\end{eqnarray}
Introducing  again $Y_{c_5}= n_{c_5}\, (\frac{\hbar c}{k_b T})^3$, 
$Y^{(0)}_{c_5}= n^{(0)}_{c_5}\, (\frac{\hbar c}{k_b T})^3$ and 
$\lambda_{c_5} = \frac{<\sigma v>_{c_5}  \, m_{q_5} c^2}{h_m\, (\hbar c)^3}$, with the 
same $x$ and $h_m$ as above, we obtain the equation 
 $\frac{dY_{c_5}}{dx} = 
\frac{\lambda_{c_5}}{x^2}\, (Y^{2}_{q_5} - Y_{c_5} \,Y^{(0)}_{q_5}\, 
\frac{Y^{(0)}_{q_5}}{Y^{(0)}_{c_5}} )$. 

The number density of the fifth family quarks $n_{q_5}$  (%and anti-quarks %
and correspondingly  $Y_{q_5}$), which has above the 
temperature of the binding energy of the clusters of the fifth family quarks (almost) 
reached the decoupled value, starts to decrease again due  to the formation of the clusters of the 
fifth family quarks (and anti-quarks) as well as due to forming the bound state of 
the fifth family quark with an anti-quark, which annihilates into gluons.  
It follows  
\begin{eqnarray}
\label{gbsnmbboltzq2}
a^{-3}\frac{d( a^3 n_{q_5})}{dt} &=& 
< \sigma v>_{c_5}\; n^{(0)}_{q_5}\, n^{(0)}_{q_5} 
\left[ -\left( \frac{n_{q_5}}{n^{(0)}_{q_5}} \right)^2 + \frac{n_{c_5}}{ n^{(0)}_{c_5}} -  
\frac{\eta_{(q\bar{q})_b}}{\eta_{c_5}} \;\left( \frac{n_{q_5}}{n^{(0)}_{q_5}} \right)^2 \right] + \nonumber\\ 
&& < \sigma v>_{q\bar{q}}\; n^{(0)}_{q_5} n^{(0)}_{\bar{q}_5}\,
\left(- \frac{n_{q_5} n_{\bar{q}_5}}{n^{(0)}_{q_5} n^{(0)}_{\bar{q}_5}} + 
 \frac{n_{q} n_{\bar{q}}}{n^{(0)}_{q} n^{(0)}_{\bar{q}}} \right) + \nonumber\\ 
&&< \sigma v>_{gg} \; 
n^{(0)}_{q_5} n^{(0)}_{\bar{q}_5}\,
\left(- \frac{n_{q_5} n_{\bar{q}_5}}{n^{(0)}_{q_5} n^{(0)}_{\bar{q}_5}} +  
\frac{n_{g} n_{g}}{n^{(0)}_{g} n^{(0)}_{g}} \right), 
\end{eqnarray}
with $\eta_{(q\bar{q})_b}$ and $\eta_{c_5}$ defined in Eq.~\ref{gbsnmbsigmasq}.
Introducing the above defined $Y_{q_5}$  and $Y_{c_5}$ the Eq.~\ref{gbsnmbboltzq2} transforms into 
$\frac{dY_{q_5}}{dx} = 
\frac{\lambda_{c_5}}{x^2}\, (- Y^{2}_{q_5} + Y_{c_5} \,Y^{(0)}_{q_5}\, 
\frac{Y^{(0)}_{q_5}}{Y^{(0)}_{c_5}} ) + \frac{\lambda_{(q\bar{q})_b}}{x^2}\, (-  Y^{2}_{q_5})
+ \frac{\lambda_{q_5}}{x^2}\, (Y^{(0)2}_{q_5} - Y^{2}_{q_5})$, 
with $\lambda_{(q\bar{q})_b} = \frac{<\sigma v>_{(q\bar{q})_b}  \, m_{q_5} c^2}{h_m\, (\hbar c)^3}$ 
(and with the same $x$ and $h_m$ as well as $\lambda_{c_5}$ and $\lambda_{q_5}$
as defined above). We solve this equation together with the above equation 
for $Y_{c_5} $. 

%Let us look also at the Thompson scattering of gluons on the bound states, 
%destroying clusters, which 
%starts to be negligible when the rate for gluons to scatter off the quarks ($n_{q_5} \,\sigma_T \,c$) 
%starts to be smaller than the expansion rate ($H= \sqrt{\frac{8\pi^3 \,g^*}{45}} 
%\; \frac{( k_b T)^2 \,c}{\hbar c\, m_{pl} c^2}$), with $g^*$ defined above. 
%Recognizing that the binding energy of Table~\ref{gbsnmbTableI.} is approximately  $\frac{1}{100}$ the mass 
%of the fifth family quarks we get the requirement that the bound states can be formed when 
%$n_{q_5} << 3. 10^{-25} (\frac{m_{q_5} \, c^2}{{\rm GeV}})^4$ 
%fm$^{-3}$, which for $m_{q_5}\, c^2 = 1$ TeV gives $n_{q_5} << 3. 10^{-13}$ fm$^{-3}$ and  for 
%$m_{q_5}\, c^2 = 10$ TeV gives $n_{q_5} << 3. 10^{-9}$ fm$^{-3}$. One can easily check from the 
%solutions of the Boltzmann equations that this requirements are fulfilled. %(****GREGOR, PREVERITE*****)

\begin{figure}[h]
\begin{center}
\includegraphics[width=13cm,angle=0]{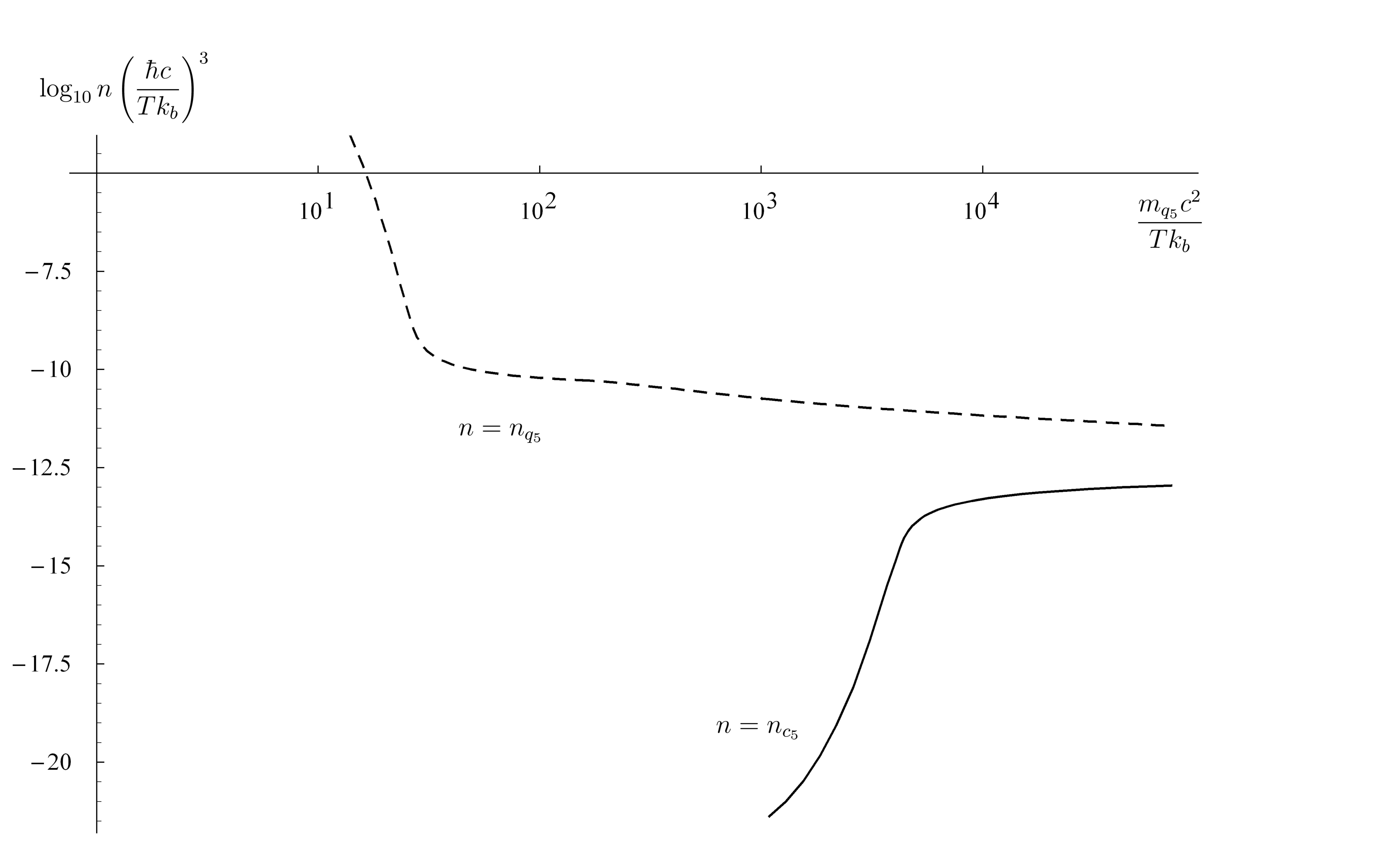}
\caption{The dependence of the two number densities $n_{q_5}$ (of the fifth family quarks) and $n_{c_5}$ (of 
the fifth family clusters) as function of $\frac{m_{q_5} \, c^2}{ k_b \, T}$ is presented 
for the special values $m_{q_5} c^2= 71 \,{\rm TeV}$, $\eta_{c_5} = \frac{1}{50}$ and $\eta_{(q\bar{q})_b}=1$.
We take $g^*=91.5$. In the treated energy (temperature $ k_b T$) interval the one gluon exchange gives the  main 
contribution to the scattering cross sections of Eq.(\ref{gbsnmbsigmasq}) entering into the Boltzmann equations for 
$n_{q_5}$ and $n_{c_5}$. In the figure we make a choice of the  parameters within the estimated  intervals.}
\end{center}
\label{gbsnmbDiagramI.}
\end{figure}

Solving the  Boltzmann  equations (Eqs.~\ref{gbsnmbboltzq1},~\ref{gbsnmbboltzc},~\ref{gbsnmbboltzq2}) we obtain 
 the number density of the fifth family quarks  $n_{q_5}$ (and 
anti-quarks) and the number density of the fifth family baryons $n_{c_5}$ (and anti-baryons)
 as a function of the parameter $x=\frac{m_{q_5} c^2}{k_b T}$   and the two parameters $\eta_{c_5}$ 
 and $\eta_{(q\bar{q})_b}$. The evaluations are made, as we explained above, 
with the approximate expressions for the thermally averaged cross sections  from Eq.(~\ref{gbsnmbsigmasq}), 
corrected  by the parameters  
$\eta_{c_5}$  and  $\eta_{(q \bar{q})_b}$ (Eq.~\ref{gbsnmbsigmasq}). 
%%%%%%%%%%%%%%%%%%
%%%%%%%%%%%%%%%%%
We made a rough estimation of the two intervals, within which the parameters 
$\eta_{c_5}$  and  $\eta_{(q \bar{q})_b}$ (Eq.~\ref{gbsnmbsigmasq}) seem to be acceptable. 
More accurate evaluations 
of the cross sections are under consideration.
%%%%%%%%%%%%%%%
%%%%%%%%%%%%%%%
In fig.~\ref{gbsnmbDiagramI.} both number densities (multiplied by $(\frac{\hbar \, c}{ k_b T})^3$, which is 
$Y_{q_5}$ and $Y_{c_5}$, respectively for the quarks and the clusters of quarks) as a function 
of  $ \frac{m_{q_5} \, c^2}{ k_b T}$ for $\eta_{(q\bar{q})_3}=1$ and $\eta_{c_5}=\frac{1}{50}$ are presented. 
%%%%%%%%%%%%%%
%%%%%%%%%%%%%%%%
The particular choice of the parameters $\eta_{(q\bar{q})_3}$ and $\eta_{c_5}$ in fig.~\ref{gbsnmbDiagramI.}
is made as a typical example.
%%%%%%%%%%%%%%%%%%%
%%%%%%%%%%%%%
The calculation is performed up to $ k_b T=1$ GeV (when the colour phase transition starts and the one 
gluon exchange stops to be the acceptable approximation).

Let us repeat how the $n_5$ and $\bar{n}_5
$ evolve in the evolution of our universe. 
 The quarks and anti-quarks are at high temperature 
($\frac{m_{q_5} c^2}{k_b T}<< 1$) in thermal equilibrium with the plasma (as are also all the other 
families and bosons of  lower masses). 
As the temperature of the plasma (due to the expansion of the universe) drops close to the 
mass of the fifth family 
quarks, quarks and anti-quarks scatter into all the other (ultra) relativistic fermions and bosons, but 
can not be created any longer from the plasma (in the average).  
At the temperature close to  the binding energy of the quarks in a cluster, the clusters of  the fifth family 
($n_{c_5}, n_{\bar{c}_5}$)
baryons start to be formed. We evaluated the number density  $n_{q_5} (T) \,
(\frac{\hbar c}{ k_b T})^3 = Y_{q_5} $  
of the fifth family quarks (and anti-quarks) and the number density of the fifth family baryons 
$n_{c_5} (T) \,(\frac{\hbar c}{ k_b T})^3 = Y_{c_5} $ for several choices  of  
$m_{q_5}, \eta_{c_5}$ and $\eta_{(q \bar{q})_b}$ up to $ k_b T_{lim}= 1$ GeV $=\frac{m_{q_5} c^2}{x_{lim}} $.  
%The fifth family quark mass then follows from Eq.~\ref{gbsnmbdm}. 

From the calculated decoupled number density of baryons and anti-baryons of the fifth family quarks  
(and anti-quarks) $n_{c_5}(T_1)$ at temperature $ k_b T_1=1$ GeV, where we stopped our 
calculations  as a function of the quark mass and of the 
two parameters $\eta_{c_5}$ and $\eta_{(q\bar{q})_b}$, the today's mass density of the dark matter 
follows (after taking into account that when once the $n_{5}$ and $\bar{n}_{5}$ decouple, their number 
stays unchanged but due to the expansion of the universe their density decreases according to 
$a^{3}_1 n_{c_5}(T_1)= 
a^{3}_2 n_{c_5}(T_2)$, with the today's $a_0=1$ and the temperature $T_0=2.725^0 $ K) leading to~\cite{gbsnmbdodelson}
\begin{eqnarray}
\label{gbsnmbdm}
\rho_{dm} &=& \Omega_{dm} \rho_{cr}= 2 \, m_{c_5}\, n_{c_5}(T_1) \, 
\left(\frac{T_0}{T_1} \right)^3 \frac{g^*(T_1)}{g^*(T_0)},
\end{eqnarray}
where we take into account that $g^*(T_1) (a_1 T_1)^3= g^*(T_0)(a_0 T_0)^3$, 
with $T_0 = 2.5 \,\cdot 10^{-4}\,\frac{{\rm eV}}{k_b}$, $g^*(T_0)= 2 + \frac{7}{8}\,\cdot3 \,\cdot \,
(\frac{4}{11})^{4/3}$, $g^*(T_1)= 2 + 2\,\cdot 8 + \frac{7}{8}\, (5\cdot3\cdot2\cdot2 + 6\cdot 
2\cdot2)$ and  
 $\rho_{cr} \,c^2 \,\approx \frac{3\, H^{2}_{0}\, c^2}{8 \pi G} %\approx 7\cdot 10^{-45} {\rm GeV} (fm)^{-3}
 \approx 5.7\, \cdot 10^3 \frac{{\rm eV}}{{\rm cm}^3}$, factor $2$ counts 
 baryons and anti-baryons. % (since the spin  of baryons is taken into account in $n_{c_5}$).  

The  intervals for the acceptable  parameters $\eta_{c_5}$ and $\eta_{(q \bar{q})_b}$  (determining  
the inaccuracy, with which the scattering cross sections were evaluated) influence 
the value of $n_{c_5}$ and  
determine the interval, within which one expects the fifth family mass.
\begin{table}
\begin{center} \begin{tabular}{|c||c|c|c|c|c|}
\hline
$\frac{m_{q_5} c^2}{{\rm TeV}}$&$\eta_{(q\bar{q})_b}=\frac{1}{10}$&$\eta_{(q\bar{q})_b}=\frac{1}{3}$&$\eta_{(q\bar{q})_b}=1$& 
$\eta_{(q\bar{q})_b}=3$&$\eta_{(q\bar{q})_b}=10$\\
\hline\hline
$\eta_{c_5}=\frac{1}{50}$ & 21 & 36 & 71& 159&417\\
\hline
$\eta_{c_5}=\frac{1}{10} $ & 12 & 20 & 39&  84&215 \\
\hline
$\eta_{c_5}=\frac{1}{3} $ & 9  & 14 & 25&  54&134 \\
\hline
$\eta_{c_5}=1 $           & 8  & 11 & 19&  37 &88 \\
\hline
$\eta_{c_5}= 3$ &           7  & 10 & 15&  27 &60 \\
\hline
$\eta_{c_5}=10$ &           7* &  8*& 13&  22 &43 \\
\hline
\end{tabular}
\end{center}
\caption{\label{gbsnmbTableII.} The fifth family quark mass is presented (Eq.(\ref{gbsnmbdm})), calculated for different 
choices of $\eta_{c_5}$ (which takes care of the inaccuracy with which  a colourless cluster of three 
quarks (anti-quarks) cross section was estimated  and of $\eta_{(q\bar{q})_b}$ (which takes care of the inaccuracy
with which the cross section for the 
annihilation of a bound state of quark---anti-quark was taken into account) from Eqs.~(\ref{gbsnmbdm}, \ref{gbsnmbboltzc}, \ref{gbsnmbboltzq1}).
* denotes non stable calculations.}
\end{table}
We read from Table~\ref{gbsnmbTableII.} the mass interval for the fifth family quarks' mass, 
which fits Eqs.~(\ref{gbsnmbdm}, \ref{gbsnmbboltzc}, \ref{gbsnmbboltzq1}):
\begin{eqnarray}
\label{gbsnmbmassinterval}
10 \;\; {\rm TeV} < m_{q_5}\, c^2 < {\rm a\, few} \cdot 10^2 {\rm TeV}.
\end{eqnarray}
From this mass interval we estimate from Table~\ref{gbsnmbTableI.} the cross section for the 
fifth family neutrons $\pi (r_{c_5})^2$:
\begin{eqnarray}
\label{gbsnmbsigma}
10^{-8} {\rm fm}^2 \, < \sigma_{c_5} < \, 10^{-6} {\rm fm}^2.
\end{eqnarray}
(It is  at least $10^{-6} $ smaller than the cross section for the first family neutrons.)

Let us comment on the fifth family quark---anti-quark annihilation at the colour phase transition, 
which starts at 
approximately $1$ GeV. When the colour phase transition starts, the quarks start to "dress" into 
constituent mass, which brings to them 
 $\approx 300$ MeV/$c^2$, since to the force many gluon exchanges start to contribute. 
 The scattering cross sections, which were up to 
 the phase transition dominated by one gluon exchange, rise now to the value of a few ${\rm fm}^2$ and more,
 say $(50 {\rm fm})^2$. 
 Although the colour phase transition is not yet 
 well understood even for the first family quarks, the evaluation of what happens to the 
 fifth family quarks and anti-quarks and coloured clusters of the fifth family 
 quarks or anti-quarks can still be done as follows. 
 At the interval, when the temperature $ k_b T$ is 
considerably above the binding energy of the "dressed" first family quarks and anti-quarks  
into mesons or of the binding energy of the 
three first family quarks 
or anti-quarks into the first family baryons or anti-baryons, which is $\approx $ a few MeV 
(one must be more careful with the mesons), 
the first family quarks  and anti-quarks 
move in the plasma like being free. (Let us remind the reader 
that the nuclear interaction can be derived as the interaction among the  clusters of quarks~\cite{gbsnmbbmmn}.) 
25 years ago there were several proposals to treat nuclei as  clusters of dressed 
quarks instead of as clusters of baryons. Although this idea was not very fruitful (since even models 
with nuclei as  bound states of $\alpha$ 
particles work many a time reasonably) it also was not far from the reality. 
Accordingly it is meaningful to accept the description 
of plasma at temperatures above a few $ 10$ MeV/$k_b$ as the plasma of less or more "dressed" quarks 
%(where "dressing" matters only the 
%first family quarks)   
with the very large scattering amplitude (of   $\approx (50{\rm fm})^2$). 
The fifth family quarks and anti-quarks, heaving much higher mass (several ten  thousands 
GeV/$c^2$ to be compared with  $\approx 300$ MeV/$c^2$) than the first family quarks and accordingly 
much higher momentum,  "see" the first family quarks as  a "medium"  in which they (the fifth family
quarks) scatter among themselves. The fifth family quarks and anti-quarks, 
having much higher binding energy when forming a meson among themselves  than when forming  
mesons with the first family 
quarks and anti-quarks (few thousand GeV  to be compared with few MeV or few $10$ MeV)  
and correspondingly very high annihilation probability and 
also pretty low velocities ($\approx 10^{-3} c$),  have during the scattering 
enough time to annihilate with their anti-particles.
The ratio of the scattering time between two coloured quarks (of any kind) and the Hubble time 
is of the order of $\approx 10^{-18}$ and therefore although the number of the fifth family 
quarks and anti-quarks is  of the order of $10^{-13}$ smaller than the number of the quarks and anti-quarks 
of the first family (as show the solutions of the Boltzmann equations 
presented in fig.~\ref{gbsnmbDiagramI.}),  the fifth family quarks and anti-quarks have in the first period 
of the colour phase transition 
(from $\approx$ GeV to $\approx 10$ MeV) enough opportunity to scatter often enough among themselves to deplete (their  
annihilation time is for several orders of magnitude smaller than
  the time needed to pass by). 
 More detailed calculations, which are certainly needed, are under considerations. Let us still do 
 rough estimation about the number of the coloured fifth family quarks (and anti-quarks).
 Using the expression for the thermally averaged cross section for scattering of a  quark and an 
 anti-quark and annihilating  ($<\sigma  v>_{(q \bar{q})_b} $ from Eq.(\ref{gbsnmbsigmasq})) 
 and correcting the  part which determines 
 the scattering cross section by replacing it with $ \eta\, ( 50 {\rm fm})^2 c\; $ (which takes into account the scattering in the 
 plasma during the colour phase transition in the expanding universe) we obtain the  expression 
 $ <\sigma  v>_{(q \bar{q})_b}= \eta_{(q \bar{q})_b} \, \eta\, ( 50 {\rm fm})^2 c\; 
\sqrt{\frac{ E_{c_5}}{ k_b T}} \ln{\frac{E_{c_5}}{ k_b T}}$, which is almost independent of the velocity of 
the fifth family 
quarks (which slow down when the temperature lowers). We shall assume that the temperature is 
lowering as it would be no phase transition and correct this fact with the parameter $\eta$, which could  
for a few orders of magnitude (say $10^2$) enlarge the depleting probability.  
Using this expression for $<\sigma  v>_{(q \bar{q})_b}$ in 
the expression for $\lambda= \frac{<\sigma  v>_{(q \bar{q})_b} \; m_{q_5} c^2}{h_m (\hbar c)^3}$, we 
obtain for a factor up to $10^{19}$ larger $\lambda$ than it was the one dictating the freeze out 
procedure of $q_5$ and $\bar{q}_{5}$ before the phase transition. 
Using then the equation $\frac{dY_{q_5}}{dx} = 
\frac{\lambda_{c_5}}{x^2}\, (- Y^{2}_{q_5})$ and integrating it from $Y_1$ which is the value 
from the fig.~\ref{gbsnmbDiagramI.}
at $1$ GeV up to the value when $ k_b T\approx 20 $ MeV, when the first family quarks start to bindd into baryons, 
we obtain in the approximation that $\lambda $ is independent of $x$ (which is not really the case) that 
$\frac{1}{Y(20 {\rm MeV})}= 10^{32} \frac{1}{2\cdot 10^5} $ or $Y(20 {\rm MeV}) = 10^{-27}$ and correspondingly
$n_{q_5}(T_0)= \eta^{-1} 10^{-24} cm^{-3}$. Some of these fifth family quarks can form the mesons or baryons and anti-baryons 
with the first family quarks $q_1$ when they start to form baryons and mesons. They would form the 
anomalous hydrogen in the ratio: $\frac{n_{ah}}{n_{h}} \approx  \eta^{-1} \cdot 10^{-12} $, where 
$n_{ah}$ determines the number of the anomalous (heavy) hydrogen atoms and ${n_{h}}$ the number 
of the hydrogen atoms, with $\eta$ which might be bellow $10^{2}$.  
The  best measurements in the context of such baryons with the masses of a few hundred TeV/${\rm c}^2$  
which we were able to find were done $25$ years ago~\cite{gbsnmbsuperheavy}. The authors declare 
that their measurements manifest that such a ratio should be $\frac{n_{ah}}{n_{h}}< 10^{-14}$ for the mass interval 
between $10$ TeV/$c^{2}$ to $10^{4}$ TeV/$c^{2}$. Our evaluation presented above is very rough and more careful 
treating the problem might easily lead to lower values than required. On the other side we can not say 
how trustable is the value for the above ratio %(which confidence level it has) 
for the masses of a few hundreds 
TeV. Our evaluations are very approximate and if $\eta= 10^{2}$ we  conclude that the evaluation 
agrees with measurements.  
%%%%
%%%%
%(****TO BE CALCULATED AND CHECKED BY GREGOR, ALSO THE REFERENCE AND COMMENT ARE MISSING and will be added by him.****) 
%%%%
%%%%
  
% The de Broglie wavelength is $\approx 10 $ times the average distance among neighbour quarks.
%(The same would happen to all the lower 
%families' quarks and anti-quarks (going due to the Yukawa couplings to the first family members), 
%if there would be no   quark---anti-quark asymmetry, but mostly at the very end of this period, 
%when the temperature falls bellow $1$ MeV).  

%%POPRAVI CONCLUSION AND INTRODUCTION TER POVEJ OCENO, KOLIKO DOGODKOV BI IZMERILI CDMS PRI MASSI 100 TeV

%%END EVOLUTION

%
\section{Dynamics of a heavy family baryons in our galaxy}
\label{gbsnmbdynamics}

%

%%%
There are experiments~\cite{gbsnmbrita0708,gbsnmbcdms} which are trying to directly measure the dark matter clusters. Let us 
make a short introduction into these measurements, treating our fifth family clusters in particular.
The density of the dark 
matter $\rho_{dm}$ in the Milky way can be evaluated from the measured rotation velocity  
of  stars and gas in our galaxy, which appears to be approximately independent of the distance $r$ from the 
center of our galaxy. For our Sun this velocity 
is $v_S \approx (170 - 270)$ km/s. $\rho_{dm}$ is approximately spherically symmetric distributed 
and proportional to $\frac{1}{r^2}$.  Locally (at the position of our Sun) $\rho_{dm}$ 
is known within a factor of 10 to be 
$\rho_0 \approx 0.3 \,{\rm GeV} /(c^2 \,{\rm cm}^3)$, 
we put $\rho_{dm}= \rho_0\, \varepsilon_{\rho},$ 
with $\frac{1}{3} < \varepsilon_{\rho} < 3$. 
The local velocity distribution of the dark matter cluster $\vec{v}_{dm\, i}$, in the 
velocity  class 
$i$ of clusters, can only be estimated, 
results depend strongly on the model. Let us illustrate this dependence.  
%It is taken usually  as zero in the coordinate system fixed on 
%the center of our  galaxy. 
In a simple model that all the clusters at any radius $r$ from the center 
of our galaxy travel in all possible circles around the center so that the paths are 
spherically symmetrically distributed, the velocity of a cluster at the position of 
the Earth is equal to $v_{S}$, the velocity of our Sun in the absolute value,
but has all possible orientations perpendicular to the radius $r$ with  equal probability.
In the model %~\cite{gbsnmbgnBled07} 
that the clusters only oscillate through the center of the galaxy, 
the velocities of the dark matter clusters at the Earth position have values from 
zero to the escape velocity, each one weighted so that all the contributions give  
$ \rho_{dm} $. %Also the model  that clusters make all possible paths 
%from the oscillatory one to the circle, weighted so that they reproduce the $\rho_{dm}$,
%seems acceptable. 
Many other possibilities are presented in the references cited in~\cite{gbsnmbrita0708}. 

The velocity of the Earth around the center of the galaxy is equal to:  
$\vec{v}_{E}= \vec{v}_{S} + \vec{v}_{ES} $, with $v_{ES}= 30$ km/s and 
$\frac{\vec{v}_{S}\cdot \vec{v}_{ES}}{v_S v_{ES}}\approx \cos \theta \, \sin \omega t, \theta = 60^0$. 
Then the velocity with which the dark matter cluster of the $i$- th  velocity class  
hits the Earth is equal to:  
$\vec{v}_{dmE\,i}= \vec{v}_{dm\,i} - \vec{v}_{E}$. % where the index $i$ points out 
%that  the distribution in the velocity, which is very model dependent, is in 
%the class $i$. 
$\omega $ 
determines the rotation of our Earth around the Sun.

One finds for the flux %per unit time and unit surface 
of the  
%(any heavy with the small enough cross section) 
dark matter clusters hitting the Earth:    
$\Phi_{dm} = \sum_i \,\frac{\rho_{dm \,i}}{m_{c_5}}  \,
|\vec{v}_{dm \,i} - \vec{v}_{E}|  $ 
to be approximately  (as long as $\frac{v_{ES}}{|\vec{v}_{dm \,i}- \vec{v}_S|}$ is small%than $1$
) equal to  
\begin{eqnarray}
\label{gbsnmbflux}
\Phi_{dm}\approx \sum_i \,\frac{\rho_{dm \,i}}{m_{c_5}}  \,
\{|\vec{v}_{dm \,i} - \vec{v}_{S}| - \vec{v}_{ES} \cdot \frac{\vec{v}_{dm\, i}- \vec{v}_S}{
|\vec{v}_{dm \,i}- \vec{v}_S|} \}.
\end{eqnarray}
Further terms are neglected. %The flux is very much model dependent. 
We shall approximately take that
$$\sum_i \, |\vec{v_{dm \,i}}- \vec{v_S}| \,\rho_{dm \,i} \approx \varepsilon_{v_{dmS}} 
\, \varepsilon_{\rho}\,  v_S\, \rho_0 ,$$ 
%with $\rho_0 = 0.3 \, {\rm GeV}/(c^2\, cm^3), $ 
%while we estimate $\frac{1}{4} < \varepsilon_{\rho} < 4$,   
%$ \frac{1}{3} < \varepsilon_{v_{dmS}} < 3$ 
and correspondingly 
$ \sum_i \, \vec{v}_{ES}  \cdot \frac{\vec{v}_{dm \,i}- \vec{v}_S}{
|\vec{v}_{dm \,i}- \vec{v}_S|} \approx v_{ES} \varepsilon_{v_{dmS}}
\cos \theta \, \sin \omega t $, % with $v_{ES} = 30$ km/s  
%$\theta = 60^0$, 
(determining the annual modulations observed by DAMA~\cite{gbsnmbrita0708}). 
Here $\frac{1}{3} < \varepsilon_{v_{dmS}} < 
3$ and $\frac{1}{3} < \frac{\varepsilon_{v_{dmES}}}{\varepsilon_{v_{dmS}}} < 3$ are
estimated with respect to experimental and (our) theoretical evaluations.  

Let us evaluate the cross section for our heavy dark matter baryon to elastically
(the excited states of nuclei,  
which we shall treat, I and Ge, are at $\approx 50$ keV 
or higher and are very narrow, while the average recoil energy of Iodine is expected to be 
$30$ keV) 
scatter  on an ordinary nucleus with $A$ nucleons 
$\sigma_{A} = 
\frac{1}{\pi \hbar^2} <|M_{c_5 A}|>^2 \, m_{A}^2$. 
For our heavy dark matter cluster %with a small cross section $\sigma_{c_{5}}$  
is  $m_{A}  $  approximately the mass of the ordinary nucleus~\footnote{Let us illustrate 
what is happening when a very heavy ($10^4$ times or more heavier than the ordinary nucleon) cluster 
hits the nucleon. Having the "nuclear force" cross section of $10^{-8}$ ${\rm fm}^2$ or smaller, 
it "sees" with this cross section a particular quark, which starts to move. But since at this velocities 
the quark is tightly bound into a nucleon and nucleon into the nucleus, the hole nucleus is forced to move 
with the moving  quark.}. 
%%%%
%%%%KOMENTIRAJ, KAKO ZADENE TEZAK DELEC KVARK, NAPNE STRUNE IN PREMAKNE CELO JEDRO
%%%%
In the case of a 
coherent scattering (if recognizing that $\lambda= \frac{h}{p_A}$ is for a nucleus large enough 
to make scattering coherent when the mass of  the cluster is 
 $1$ TeV or more and its velocity 
$\approx v_{S}$), the cross section is  almost independent of the recoil 
velocity of the nucleus. 
For the case that the ''nuclear force'' as manifesting  in the cross section $\pi\, (r_{c_5})^2$ 
in Eq.(\ref{gbsnmbbohr}) 
brings the main contribution~\footnote{The very heavy colourless cluster of three quarks,  
hitting with the relative velocity $\approx 200 \,{\rm km}/{\rm s}$ the nucleus of the first 
family quarks, ''sees'' the (light) quark  $q_1$ of the 
nucleus through the cross section $\pi\, (r_{c_5})^2$.
But since the quark $q_{1}$ is at these velocities strongly bound to the proton and the 
proton to the nucleus,  the hole nucleus takes the momentum.} 
the cross section  is  proportional to $(3A)^2$ 
(due to the square of the matrix element) times $(A)^2$ (due to the mass of the nuclei 
$m_A\approx 3 A \,m_{q_1}$, with $m_{q_1}\, c^2 \approx \frac{1 {\rm GeV}}{3}$).  
When $m_{q_5}$ is  heavier than $10^4 \, {\rm TeV}/c^2$ (Table~\ref{gbsnmbTableI.}), 
the weak interaction dominates and $\sigma_{A}$ is proportional to $(A-Z)^2 \, A^2$, 
since to $Z^0$ boson exchange only neutron gives an appreciable contribution. 
Accordingly we have,  when the ''nuclear force'' dominates,
$\sigma_A \approx \sigma_{0} \, A^4 \, \varepsilon_{\sigma}$, with 
$\sigma_{0}\, \varepsilon_{\sigma}$, which is $\pi r_{c_5}^2 \, 
\varepsilon_{\sigma_{nucl}} $ and with 
 $\frac{1}{30} < \varepsilon_{\sigma_{nucl}} < 30$.  
$\varepsilon_{\sigma_{nucl}}$
takes into account the roughness 
with which we treat our  heavy baryon's properties and the scattering procedure.  
When the weak interaction dominates, $ \varepsilon_{\sigma}$ is smaller and we have $  
 \sigma_{0}\, \varepsilon_{\sigma}=(\frac{m_{n_1} G_F}{\sqrt{2 \pi}} 
\frac{A-Z}{A})^2 \,\varepsilon_{\sigma_{weak}}  $
($=( 10^{-6} \,\frac{A-Z}{ A} \, {\rm fm} )^2 \,\varepsilon_{\sigma_{weak}} $), 
$ \frac{1}{10}\, <\,  \varepsilon_{\sigma_{weak}} \,< 1$. The weak force is pretty accurately 
 evaluated, but the way how we are averaging is not.

\section{Direct measurements of the fifth family  baryons as dark matter constituents} 
\label{gbsnmbdirectmeasurements}

We are making very rough estimations of what the  
 DAMA~\cite{gbsnmbrita0708} and CDMS~\cite{gbsnmbcdms} experiments are measuring, provided that the 
 dark matter clusters are made out 
 of our (any) heavy family quarks as discussed above. 
 We are looking for limitations these two experiments might put on 
 properties of our heavy family members. 
 We discussed about our estimations and their relations to the measurements 
 with R. Bernabei~\cite{gbsnmbprivatecommRBJF} and 
 J. Filippini~\cite{gbsnmbprivatecommRBJF}. 
 Both pointed out (R.B. in particular) that the two experiments can hardly be compared, 
 and that our very approximate estimations may be right only within the orders of magnitude. 
 % detailed taking into 
 %account  the way how do the dark matter constituents scatter on the nuclei and with 
 %which velocity do they scatter (in ref.~\cite{gbsnmbrita0708} such studies were done),   
 %as well as how does a particular  experiment measure events,  is essential and that 
 %results depend  significantly on the details, so that too rough treating might change the 
 %results for orders of magnitude. 
 We are completely aware of how rough our estimation is, % and we 
 %do agree with  their comments, 
 yet we conclude that, since the number of measured events  is  proportional to 
 $(m_{c_5})^{-3}$ %the third power of the clusters' mass, 
 for masses $\approx 10^4$ TeV or smaller (while for 
 higher masses, when the weak interaction dominates, it is proportional to  
 $(m_{c_5})^{-1}$) that even such rough  estimations   
 may in the case of our heavy baryons say whether both experiments
 do at all measure our (any) heavy family clusters, if one experiment 
 clearly sees  the dark matter signals and the 
 other does not (yet?) and we accordingly estimate the mass of our cluster. 
 % Let us point out that the number of events an experiment  on the Earth might 
 %recognize as triggered by our heavy  dark matter cluster  is proportional to 
 %$1/m_{c_5}^3$, so that accordingly the  mass estimated from the measured events 
 % depends on the third route of the number of events. 
 
 Let $N_A$ be the number of nuclei of a type $A$ in the %measurement 
 apparatus  
 (of either DAMA~\cite{gbsnmbrita0708}, which has $4\cdot 10^{24}$ nuclei per kg of $I$, 
 with $A_I=127$,  
  and  $Na$, with $A_{Na}= 23$ (we shall neglect $Na$), 
 or of CDMS~\cite{gbsnmbcdms}, which has $8.3 \cdot 10^{24}$ of $Ge$ nuclei 
 per kg,  with $A_{Ge}\approx 73$). 
 At velocities  of a dark matter cluster  $v_{dmE}$ $\approx$ $200$ km/s  
 are the $3A$ scatterers strongly bound in the nucleus,    
 so that the whole nucleus with $A$ nucleons elastically scatters on a 
 heavy dark matter cluster.  
Then the number of events per second  ($R_A$) taking place 
in $N_A$ nuclei   is  due to the flux $\Phi_{dm}$ and the recognition that the cross section 
is at these energies almost independent 
of the velocity %(and depends accordingly only  on $A$ of the nucleus),  
equal to
\begin{eqnarray}
\label{gbsnmbra}
R_A = \, N_A \,  \frac{\rho_{0}}{m_{c_5}} \;
\sigma(A) \, v_S \, \varepsilon_{v_{dmS}}\, \varepsilon_{\rho} \, ( 1 + 
\frac{\varepsilon_{v_{dmES}}}{\varepsilon_{v_{dmS}}} \, \frac{v_{ES}}{v_S}\, \cos \theta
\, \sin \omega t).
\end{eqnarray}
Let $\Delta R_A$ mean the amplitude of the annual modulation of $R_A$ 
\begin{eqnarray}
\label{gbsnmbanmod}
\Delta R_A &=& R_A(\omega t = \frac{\pi}{2}) - R_A(\omega t = 0) = N_A \, R_0 \, A^4\, 
\frac{\varepsilon_{v_{dmES}}}{\varepsilon_{v_{dmS}}}\, \frac{v_{ES}}{v_S}\, \cos \theta,
\end{eqnarray}
where $ R_0 = \sigma_{0} \, \frac{\rho_0}{m_{c_5}} \,  v_S\, \varepsilon$, 
$R_0$ is for the case that the ''nuclear force'' 
dominates $R_0 \approx  \pi\, (\frac{3\, \hbar\, c}{\alpha_c \, m_{q_5}\, c^2})^2\, 
\frac{\rho_0}{m_{q_5}} \, v_S\, \varepsilon$, with 
$\varepsilon = 
\varepsilon_{\rho} \, \varepsilon_{v_{dmES}} \varepsilon_{\sigma_{nucl}} $.  $R_0$ is therefore 
proportional to $m_{q_5}^{-3}$. 
We estimated  $10^{-4} < \varepsilon < 10$,   %(****ALI LAHKO TO NAPAKO ZMANJSAMO?****) 
which demonstrates both, the uncertainties in the knowledge about the dark matter dynamics 
in our galaxy and our approximate treating of the dark matter properties.  
(When for $m_{q_5} \, c^2 > 10^4$ TeV the weak interaction determines the cross section  
$R_0 $ is in this case proportional to $m_{q_5}^{-1}$.) 
We estimate that an experiment with $N_A$ scatterers  should  measure the amplitude
$R_A \varepsilon_{cut\, A}$, with $\varepsilon_{cut \, A}$ determining  the efficiency  of 
a particular experiment to detect a dark matter cluster collision. 
For small enough $\frac{\varepsilon_{v_{dmES}}}{\varepsilon_{v_{dmS}}}\, 
\frac{v_{ES}}{v_S}\, \cos \theta$ we have 
\begin{eqnarray}
R_A \, \varepsilon_{cut \, A}  \approx  N_{A}\, R_0\, A^4\, 
 \varepsilon_{cut\, A} = \Delta R_A \varepsilon_{cut\, A} \,
 \frac{\varepsilon_{v_{dmS}}}{\varepsilon_{v_{dmES}}} \, \frac{v_{S}}{v_{ES}\, \cos \theta}. 
\label{gbsnmbmeasure}
\end{eqnarray}
If DAMA~\cite{gbsnmbrita0708}   is measuring 
our  heavy  family baryons %with weak enough scattering cross section 
%(mostly)  
 scattering mostly on $I$ (we neglect the same number of $Na$,  with $A =23$),  
then the average $R_I$ is 
\begin{eqnarray}
\label{gbsnmbridama}
R_{I} \varepsilon_{cut\, dama} \approx  \Delta R_{dama} % \; \varepsilon_{cut\, dama}\,
\frac{\varepsilon_{v_{dmS}}}{\varepsilon_{v_{dmES}}}\,
\frac{v_{S}  }{v_{ES}\, \cos 60^0 } ,
\end{eqnarray}
with $\Delta R_{dama}\approx 
\Delta R_{I}  \, \varepsilon_{cut\, dama}$, this is what we read from their papers~\cite{gbsnmbrita0708}.  
In this rough estimation 
most of unknowns about the dark matter properties, except the local velocity of our Sun,  
the cut off procedure ($\varepsilon_{cut\, dama}$) and 
$\frac{\varepsilon_{v_{dmS}}}{\varepsilon_{v_{dmES}}}$,
(estimated to be $\frac{1}{3} < \frac{\varepsilon_{v_{dmS}}}{\varepsilon_{v_{dmES}}} < 3$), 
 are hidden in $\Delta R_{dama}$. If we assume that the 
Sun's velocity is 
$v_{S}=100, 170, 220, 270$ km/s,  we find   $\frac{v_S}{v_{ES} \cos \theta}= 7,10,14,18, $ 
respectively. (The recoil energy of the nucleus $A=I$ changes correspondingly %in the average 
with the square of   $v_S $.)
DAMA/NaI, DAMA/LIBRA~\cite{gbsnmbrita0708} publishes %with $4. 10^{24}$ scatterers  per kg 
%$\varepsilon_{cut \,I}$ \, 
$ \Delta R_{dama}= 0.052  $ counts per day and per kg of NaI. 
Correspondingly  is $R_I \, \varepsilon_{cut\, dama}  = 
 0,052 \, \frac{\varepsilon_{v_{dmS}}}{\varepsilon_{v_{dmES}}}\, \frac{v_S}{v_{SE} \cos \theta} $ 
counts per day and per kg. 
CDMS should then in $121$ days with 1 kg of Ge ($A=73$) detect   
$R_{Ge}\, \varepsilon_{cut\, cdms}$
$\approx \frac{8.3}{4.0} \, 
 (\frac{73}{127})^4 \; \frac{\varepsilon_{cut\,cdms}}{\varepsilon_{cut \,dama}}\, 
 \frac{\varepsilon_{v_{dmS}}}{\varepsilon_{v_{dmES}}}\;
 \frac{v_S}{v_{SE} \cos \theta} \;  0.052 \cdot 
 121 \;$ events, 
which is for the above measured velocities equal to $(10,16,21,25)
\, \frac{\varepsilon_{cut\, cdms}}{\varepsilon_{cut\,dama}}\;
\frac{\varepsilon_{v_{dmS}}}{\varepsilon_{v_{dmES}}}$. CDMS~\cite{gbsnmbcdms} 
has found no event.

The approximations we made might cause that the expected  numbers 
($10$, $16$, $21$, $25$) multiplied by $\frac{\varepsilon_{cut\,Ge}}{\varepsilon_{cut\,I}}\;
\frac{\varepsilon_{v_{dmS}}}{\varepsilon_{v_{dmES}}}$  
are too high (or too low!!) for a factor let us say $4$ or $10$. 
%(But they also might be too low for the same fastor!)  
If in the near future  
CDMS (or some other experiment) 
will measure the above predicted events, then there might be  heavy 
family clusters which form the dark matter. In this case the DAMA experiment   
puts the limit on our heavy family masses (Eq.(\ref{gbsnmbmeasure})). 

Taking into account all the uncertainties   mentioned above, with the uncertainty with
the ''nuclear force'' cross section included (we evaluate these uncertainties  to be 
$10^{-4}  <\,\varepsilon^{"}\,< 3\cdot 10^3$), we can estimate the mass range of the fifth family quarks 
from the DAMA experiments:  
$$(m_{q_5}\, c^2)^3= \frac{1}{\Delta R_{dama}} 
N_I\,A^4\, \pi \,(\frac{3 \,\hbar c}{\alpha_c})^2 \,
\rho_0\, c^2\, v_{ES} \,\cos \theta\, \varepsilon^{"}= (0.3\, \cdot 10^7)^3 \, 
\varepsilon^{"} (\frac{0.1}{\alpha_c})^{2} \textrm{GeV}.$$ 
The lower mass limit, which follows from the DAMA experiment,  is accordingly  
$m_{q_5}\, c^2> 200$ TeV. 
Observing that 
for $m_{q_5} \, c^2> 10^4$ TeV 
the weak force starts to dominate, we estimate the upper limit $m_{q_5}\, c^2< 10^5$ TeV. 
Then
%In the case that the weak 
%interaction determines the $n_5$ cross section we find for the mass range    
$200 {\rm\; TeV} < m_{q_5} \, c^2 < 10^5$ TeV. 

Let us at the end evaluate the total number of our fifth family neutrons ($n_5$) which in $\delta t= 121$ days 
strike $1$ kg of Ge and which CDMS experiment could detect, that is  $R_{Ge} \delta t 
\varepsilon_{cut_{Ge}}= N_{Ge} \sigma_0 \frac{\rho_0}{m_{c_5}}\,v_{S}\, A^{4}_{Ge}\, \varepsilon  \varepsilon_{cut+{Ge}}$ 
(Eq.~\ref{gbsnmbmeasure}), 
with $N_{Ge} = 8.3 \cdot 10^{24}$/kg, with the cross section from Table~\ref{gbsnmbTableI.}, with $A_{Ge} = 73$ 
and $1$ kg of Ge, while  $10^{-5} < \varepsilon \varepsilon_{cut_{Ge}}< 5\cdot 10$. The  coefficient 
$\varepsilon \varepsilon_{cut_{Ge}}$ determines all the uncertainties: about the scattering amplitudes of the 
fifth family neutrons on the Ge nuclei (about the scattering amplitude of one 
$n_5$ on the first family quark, about the degree of coherence when scattering on the nuclei, about the 
local density of the dark matter, about the local velocity of the dark matter and about the efficiency of the 
experiment). Quite a part of these uncertainties were hidden in the number of events the DAMA/LIBRA 
experiments measure, when we compare both experiments. 
If we assume that the fifth family quark mass ($m_{q_5}$) is several hundreds TeV, as evaluated (as  
the upper bound (Eq.~\ref{gbsnmbmassinterval})) when considering the cosmological history of our fifth family neutrons,
we get for the number of events the CDMS experiment should measure: 
$\varepsilon \varepsilon_{cut_{Ge}} \cdot 10^{4}$. If we take $\varepsilon \varepsilon_{cut_{Ge}}= 10^{-5}$,
the CDMS experiment should continue to measure 10 times as long as they did.

Let us see how many events CDMS should measure if the dark matter clusters would interact weakly 
with the Ge nuclei and  if the 
weak interaction would determine also their freezing out procedure, that is if any kind of WIMP would 
form the dark matter. One easily sees from the 
Boltzmann equations for the freezing out procedure for $q_5$ that since the weak massless boson exchange 
is approximately hundred times weaker than the 
one gluon exchange which determines the freeze out procedure of the fifth family quarks,  
the mass of such an object should be hundred times smaller, which means a few TeV. Taking into account the 
expression for the weak interaction of such an object with Ge nuclei, which leads to $10^{-2}$ smaller 
cross section for scattering of one such weakly interacting particle on one proton (see derivations in the previous 
section), we end up with the 
number of events which the CDMS experiment should measure: $\varepsilon \varepsilon_{cut_{Ge}} 5 \cdot 10^3$. 
Since the weak interaction with the matter is much better known that the (''fifth family nuclear force'') 
interaction of the colourless clusters of $q_5$ ($n_5$), the 
$\varepsilon $ is smaller. Let us say $\varepsilon$ is $5 \cdot10^{-4}$. Accordingly, even in the case of 
weakly interacting 
dark matter particles the CDMS should continue to measure to see some events.

\section{ Concluding remarks}
\label{gbsnmbconclusion}

We estimated in this paper the possibility that a new  stable  family, 
predicted by the approach unifying spin and 
charges~\cite{gbsnmbpn06,gbsnmbn92,gbsnmbgmdn07} to have the same charges and the same couplings 
to the corresponding gauge fields as the known  families,    
%having the matrix elements of the Yukawa couplings to the lower mass 
%families equal to zero, 
%
forms baryons which are the dark matter constituents. The approach (proposed by S.N.M.B.)
 is to our knowledge the only proposal  
in the literature so far which offers the mechanism for generating families, 
if we do not count those which in one or another way just assume more 
than three families. 
Not  being able so far to derive  from the approach precisely enough the fifth family masses and also not 
(yet) the baryon asymmetry,
we assume that the neutron is the lightest fifth family baryon and that there is no baryon---anti-baryon asymmetry.
We comment what changes if the asymmetry exists.
%%%%
%POPRAVITI MORDA
%%%%
%%%%
We evaluated under these assumptions the properties of the fifth family members  in the expanding universe, 
their clustering into the fifth family 
neutrons, the scattering of these neutrons on ordinary matter and find the limit on the properties of the stable fifth family quarks 
due to the 
cosmological observations and the direct experiments provided that these neutrons constitute the dark matter.

We use the simple hydrogen-like model to evaluate the 
properties of these heavy baryons and their interaction %of heavy baryons %among themselves and 
among themselves and with the ordinary  nuclei. We take into account that for masses of the order 
of $1$ TeV/$c^2$ or larger the one gluon exchange determines the force among the constituents of 
the fifth family baryons. Studying the interaction of these baryons with the ordinary matter we 
find out  that %in the latter case 
for massive enough fifth family quarks ($m_{q_5}> 10^4$ TeV) the weak interaction 
starts to dominate over the 
''nuclear interaction'' which the fifth family neutron manifests.  
The non relativistic 
fifth family baryons interact among themselves with the weak force only.

%
%We assume %further (with no justification yet) 
%that in the evolution of our universe $q_5$ and $\bar{q}_5$ were formed with no asymmetry. 
We study  
the freeze out procedure of the fifth family quarks and anti-quarks and the formation of  
baryons and anti-baryons up to the temperature  $ k_b T= 1$ GeV,  when the colour phase transition 
starts which to our estimations depletes almost all the fifth family quarks and anti-quarks while the colourless
% (neutral with respect to the colour and electromagnetic charge) 
fifth family neutrons with very small scattering cross section decouples long before (at $ k_b T= 100$ GeV). 

The cosmological evolution 
suggests for the mass limits the range $10$ TeV $< m_{q_5} \, c^2 < {\rm a \, few} \cdot 10^2$ TeV 
and for the  scattering cross sections 
$ 10^{-8}\, {\rm fm}^2\, < \sigma_{c_5}\, <   10^{-6} \,{\rm fm}^2  $. 
The measured density of  the  dark matter 
does not put much limitation on the properties of heavy enough clusters.

The DAMA experiments~\cite{gbsnmbrita0708} limit (provided that they measure 
our heavy fifth family clusters) the quark mass 
to:  $ 200 \,{\rm TeV} < m_{q_{5}}c^2 < 10^5\, {\rm TeV}$.   
%In the case that the weak interaction determines the $n_5$ cross section we find     
%$10\; {\rm TeV} < m_{q_5} \, c^2< 10^5$ TeV.    
The estimated cross section for the dark matter cluster to 
(elastically, coherently and nonrelativisically) scatter on the (first family) nucleus is in this case 
determined on the lower mass limit by the ''fifth family nuclear force'' of the fifth family clusters %and is equal to 
($ (3\cdot 10^{-5}\,A^2\, {\rm fm} )^2$) % while 
and on the higher mass limit by the weak force %determines %the cross 
%section, which is equal to  
($ ( A (A-Z)\, 10^{-6} \, {\rm fm} )^2 $). 
Accordingly we conclude that if the DAMA experiments are measuring our fifth family neutrons,  
the mass of the fifth family quarks is a few hundred  TeV $/c^2$. 

Taking into account all the uncertainties in connection with the dark matter clusters (the local density of the 
dark matter and its local velocity) including the scattering cross sections  of our fifth family neutrons on the 
ordinary nuclei as well as the experimental errors, we do expect that CDMS will in a few years 
measure our fifth family baryons.

Let us point out that  the stable fifth family neutrons are not the WIMPS, which would interact with the weak force 
only: the 
cosmological behaviour (the freezing out procedure) of these clusters are dictated by the colour force, while their 
interaction with the ordinary matter is determined by the "fifth family nuclear force" if 
they have masses smaller than $10^4$ TeV/$c^2$.

In the ref.~\cite{gbsnmbmbb}~\footnote{ The referee of PRL suggested 
that we should comment on the paper~\cite{gbsnmbmbb}.} the authors 
study the limits on a scattering cross section of 
a heavy dark matter cluster of particles and anti-particles (both of approximately the same amount) 
with the ordinary matter, estimating the energy 
flux produced by the annihilation of such pairs of clusters. %They assume (approximately) the same number of 
%particles and antiparticles in the dark matter. 
They treat the conditions under which 
would  %in a stationary case  
the heat flow  following  
from the annihilation of dark matter  particles and anti-particles in the Earth core start to be noticeable. 
Using their limits we conclude that our fifth family baryons of the mass of a few hundreds TeV/${c^2} $ 
have  for a factor more than $100$ too small scattering amplitude with the ordinary matter to cause a measurable 
heat flux on the Earth's surface. 
On the other hand could the measurements~\cite{gbsnmbsuperheavy} tell whether the fifth family members do deplete 
at the colour phase transition of our universe enough to be in agreement with them. Our very rough estimation 
show that the fifth family  members are on the allowed limit, but they are too rough to be taken as 
a real limit.

%Let us add that our Earth would for $m_{q_5}\,c^2 \approx 1 {\rm TeV}$ or lower contain a mass part 
%$10^{-9}$ or lower of the dark matter clusters, and that  
%the mean time between two collisions among the dark matter clusters in our galaxy would be 
%from $10^{22}$ years on. 

Our estimations predict that, if the DAMA experiments %~\cite{gbsnmbrita0708} 
observe the events due to our (any) 
heavy family members, (or any heavy enough family clusters with 
small enough cross section),  
the CDMS experiments~\cite{gbsnmbcdms} will in the near future observe  
a few events as well. %(limiting now the heavy family mass
%to be $8. 10^3$ TeV or higher).  
%(in particular if clusters scatter elastically and coherently, which 
%is happening if their velocities are low enough).  
%The CDMS experiment limits up to now the heavy family mass
%to be $8. 10^3$ TeV or higher. 
%
If CDMS will not confirm the heavy family events, then we must conclude, 
trusting the DAMA experiments, that either our 
fifth family clusters have much higher cross section due to the possibility that $u_5$ is lighter than 
$d_5$  so that their velocity slows down when 
scattering on nuclei of the earth above the measuring apparatus 
bellow the threshold of the CDMS experiment (and that there must be in this case the fifth family 
quarks---anti-quarks asymmetry)~\cite{gbsnmbmaxim}) 
while the DAMA experiment still observes them, 
%  
%other possibilities ($p_5, \Delta_{5}^{++}, \Delta_{5}^{-}, \bar{\Delta}_{5}^{++} $) might fit 
%the data, % ~\cite{gbsnmbmaxim} 
or the fifth family clusters (any heavy stable family clusters) are not what forms the dark matter.

Let us comment again the question whether it is at all possible (due to electroweak experimental
data) that there exist more than three up to now observed families, that is, whether the approach 
unifying spin and charges  by predicting the fourth and the stable fifth  
family (with neutrinos included) contradict the observations. In the 
ref.~\cite{gbsnmbmdnbled06} the properties 
of all the members of the fourth family were studied (for  one particular choice of breaking the starting 
symmetry). The predicted fourth family neutrino mass is at around $100$ GeV/$c^2$ or higher, therefore it 
does not due to the detailed analyses of the electroweak data done by the Russian group~\cite{gbsnmbokun} 
contradict any experimental data. %(%****DODAJ REFERENCO STRAN DATA PRL****).  
The stable fifth family neutrino has due to our calculations %~\cite{gbsnmbmdnbled06} 
considerably higher mass. Accordingly none of these two neutrinos contradict  
the electroweak data. They also do not 
contradict the nucleosynthesis, since to the nucleosynthesis only the neutrinos with masses 
bellow the electron mass contribute. 
The fact that the fifth family baryons might form the dark matter does not contradict  
 the measured (first family) baryon number and its ratio to the photon 
 energy density as well, as long as the fifth family quarks are 
 heavy enough ($>$1 TeV). All the measurements, which connect the baryon and the photon 
 energy density, relate to the moment(s) in the history of 
 the universe, when the baryons (of the first family) where formed ($m_1 c^2 \approx 
  k_b T = 1$ GeV and lower)  and the electrons and nuclei were  forming  atoms ($k_b \,T 
 \approx 1$ eV). The chargeless (with respect to the colour and electromagnetic 
 charges, not with respect to the weak charge) clusters of the fifth family were 
 formed long before (at $ k_b T\approx E_{c_5}$ (Table~\ref{gbsnmbTableI.})). They  manifest 
 after decoupling from the plasma (with their small number density and  small cross 
 section) (almost) only their gravitational  interaction.

%If   future results from CDMS and 
%DAMA and other experiments will confirm our heavy family clusters with no light family quarks contributing,   
%then we shall soon know, what is the origin of the dark matter.  
%If a possible answer is a complicated mechanism %(like in ~\cite{gbsnmbmaxim}),  
%then, since there might be  many possible complicated scenarios for the 
%dark matter origin, it might be very difficult to make the right choice among them, and 
%we shall not find out very soon what is the dark matter constituted out of.

  Let the reader recognize that the fifth family baryons are not the objects---WIMPS---which 
  would interact with only the weak interaction, 
 since their decoupling from the rest of the 
 plasma in the expanding universe is determined by the colour force and  
 their interaction with the ordinary matter is determined with the  
 fifth family "nuclear force" (the force among the fifth family nucleons, 
 manifesting much smaller cross section than does the ordinary 
 "nuclear force") as long as their mass is not higher than $10^{4} $ TeV, when the weak interaction starts to 
 dominate as commented in %the last paragraph of 
 section~\ref{gbsnmbdynamics}.

Let us conclude this paper with the recognition:   
 If the approach unifying spin and charges is the right way beyond the 
 standard model of the electroweak and colour interaction,  
 then more than three 
 families of quarks and leptons do exist, and the stable 
 (with respect to the age of the universe) fifth family of quarks and leptons 
 is the candidate to form the dark matter. The assumptions we made (i. The fifth 
 family neutron is the lightest fifth family baryon, ii. There is no fifth family baryon asymmetry),
 could be derived from the approach unifying spins and charges and we are working on these problems. 
 The fifth family baryon anti-baryon asymmetry does not very much  change the conclusions of 
 this paper as long as the fifth family quarks's mass is  a few hundreds TeV or higher.

 \section{ Appendix: Three fifth family quarks' bound states}
 \label{gbsnmbbetterhf}

 We look for the ground  state solution of the Hamilton equation  $H\,
 |\psi\rangle= E_{c_5}\,|\psi\rangle $ for a cluster of three heavy quarks with 
 \begin{eqnarray}
  H=\sum_{i=1}^3 \,\frac{p_{i}^2}{2 \,m_{q_5}} 
  -\frac{2}{3}\,  \, \sum_{i<j=1}^3
   \frac{\hbar c \; \alpha_c}{|\vec{x}_i-\vec{x}_j|},   
 \end{eqnarray}
 in the center of mass motion
 \begin{eqnarray}
  \vec{x}=\vec{x}_2 - \vec{x}_1,\quad  \vec{y}=\vec{x}_3 - \frac{\vec{x}_1+\vec{x}_2}{2},\quad
  \vec{R}=\frac{\vec{x}_1+\vec{x}_2+\vec{x}_3}{3}, 
 \end{eqnarray}
 assuming the anti-symmetric colour part  ($|\psi\rangle_{c,\, \cal{A} }$), 
 symmetric spin and weak charge part  ($|\psi\rangle_{w  \, {\rm spin},\, \cal{S}  }$)
 and symmetric space part ($|\psi\rangle_{{\rm space}, \, \cal{S}}$). 
 For the space part we take the hydrogen-like wave functions 
 $ \psi_a(\vec{x})=$$\frac{1}{\sqrt{\pi a^3}} \; e^{-|\vec{x}|/a}$ and 
  $\psi_b(\vec{y})=$$\frac{1}{\sqrt{\pi b^3}} \; e^{-|\vec{y}|/b}$, allowing $a$ and $b$ to 
  adapt variationally.
  Accordingly 
$$\langle\, \vec{x}_1, \vec{x}_2, \vec{x}_3|\psi\rangle_{{\rm space}\, \cal{S}}=
  \mathcal{N}
 \left( \psi_a(\vec{x}) \psi_{b}(\vec{y}) + \textrm{symmetric
     permutations} \right).$$ 
It follows 
\begin{multline}
 \langle\, \vec{x}_1, \vec{x}_2, \vec{x}_3|\psi\rangle_{{\rm space}\, \cal{S}}=\\
 \mathcal{N} \,  \bigl(  2 \psi_a(\vec{x}) \psi_{b}(\vec{y})   +
  2 \psi_a(\vec{y}-\frac{\vec{x}}{2}) \psi_{b}(\frac{\vec{y}}{2}+\frac{3 \vec{x}}{4}))  +
 2 \psi_a(\vec{y}+\frac{\vec{x}}{2})
 \psi_{b}(\frac{\vec{y}}{2}-\frac{3 \vec{x}}{4}) \bigr) .
\end{multline}
 
 The Hamiltonian in the center of mass motion reads 
 $H=\frac{p_x^2}{2 (\frac{m_{q_5}}{2})}+\frac{p_y^2}{2 (\frac{2m_{q_5}}{3})}+\frac{p_R^2}{2 
 \cdot 3 m_{q_5}}
 -\frac{2}{3} \hbar c \; \alpha_c \left(\frac{1}{x}+\frac{1}{|\vec{y}+\frac{\vec{x}}{2}|}+
 \frac{1}{|\vec{y}-\frac{\vec{x}}{2}|} \right).
 $
 Varying the expectation value of the Hamiltonian with respect to $a$ and $b$ 
 it follows: $\frac{a}{b}=1.03, \, \frac{a\, \alpha_c\, m_{q_5}\, c^2}{\hbar c} = 1.6$. 
 
 Accordingly we get for the binding energy  $ E_{c_5}=0.66\; m_{q_5}\, c^2 \alpha_{c}^2$ and for the size 
 of the cluster $\sqrt{\langle |\vec{x}_2-\vec{x}_1|^2 \rangle} = 2.5\, \frac{\hbar c}{\alpha_c m_{q_5 \, c^2}}
 $.

 To estimate  the  mass difference between $u_5$ and $d_5$
 for which $u_5 d_5 d_5$ is stable we treat the electromagnetic ($\alpha_{elm}$) and weak ($\alpha_w $) 
 interaction as a small correction 
 to the above calculated binding energy: $H'=  \alpha_{elm\,w}  \; \hbar c \,
 \left(\frac{1}{x}+\frac{1}{|\vec{y}+\frac{\vec{x}}{2}|}+
 \frac{1}{|\vec{y}-\frac{\vec{x}}{2}|} \right)$. $\alpha_{elm\,w} $ stays for electromagnetic and 
 weak coupling constants.
 For $m_{q_5}= 200$ TeV we take $\alpha_{elm\,w} =\frac{1}{100}$, then 
 $|m_{u_5}- m_{d_5}|< \frac{1}{3}\,  E_{c_5} \frac{(\frac{3}{2} \alpha_{elm\,w})^2}{\alpha_c^2}
 = 0.5\,\cdot 10^{-4} \; m_{q_5}\,  c^2 $.

 %
 %\appendix{Boltzmann equations for fifth family quarks and clusters}
%\label{gbsnmbbeq}
%

%
\section*{Acknowledgments} 

The authors  would like to thank  all the participants 
of the   workshops entitled 
"What comes beyond the Standard models", 
taking place  at Bled annually (usually) in  July, starting in 1998, and in particular  
H. B. Nielsen, since all the open problems were there very openly discussed.

%pn06,n92,n93,n07bled,hn02hn03

%% V. Dvoeglazov, first contribution, 11.09.2009
%\documentclass{article}
%
%\documentclass[aps]{revtex4}
%
%\begin{document}
\author{V.V. Dvoeglazov}
\title{P, C and T for Truly  Neutral Particles\thanks{Presented at the QTRF-5, V\"axj\"o. Sweden, June 14-18, 2009  and at the ICSSUR09, Olomouc, Czech Republic, June 22-26, 2009.
The extended version  is contributed to the 12th International Workshop 'What Comes Beyond 
the Standard Models', 14. - 24. July 2009, Bled.}}
\institute{%
Universidad de Zacatecas\\Ap. Postal 636, Suc. 3 Cruces, C. P. 98064\\Zacatecas, Zac., M\'exico}

\titlerunning{P, C and T for Truly  Neutral Particles}
\authorrunning{V.V. Dvoeglazov}
\maketitle

\begin{abstract}
We  present a realization of a quantum field theory, envisaged many years ago by Gelfand, Tsetlin, Sokolik and  
Bilenky. Considering the special case of the $(1/2,0)\oplus (0,1/2)$ field and developing the Majorana construct for neutrino we show that 
a fermion and its antifermion can have the same properties with respect to the intrinsic parity ($P$) operation. The transformation laws for $C$
and $T$ operations have also been given. The construct can be applied to explanation of the present situation in neutrino physics.
The case of the $(1,0)\oplus (0,1)$ field is also considered.
\end{abstract}

During the 20th century various authors introduced {\it self/anti-self} charge-con\-jugate 4-spinors
(including in the momentum representation), see~\cite{vd1Majorana,vd1Bilenky,vd1Ziino,vd1Ahluwalia}. Later, Lounesto, 
Dvoeglazov, Kirchbach {\it etc} studied these spinors, they found dynamical equations, gauge transformations 
and other specific features of them. Recently, in~\cite{vd1Kirchbach} it was claimed that ``for imaginary $C$ parities, the neutrino mass can drop out from the single $\beta$ decay trace and reappear in $0\nu\beta\beta$,...  in principle experimentally testable signature for a non-trivial impact of Majorana framework in experiments with polarized  sources" (see also Summary of the cited paper). Thus, phase factors can have physical significance in quantum mechanics. So, the aim of my talk is to remind what several researchers presented in the 90s concerning with the neutrino description.

The definitions are:
\begin{equation}
C= e^{i\theta_c} \begin{pmatrix}0&0&0&-i\\
0&0&i&0\\
0&i&0&0\\
-i&0&0&0\end{pmatrix} {\cal K} = -e^{i\theta_c} \gamma^2 {\cal K}
\end{equation}
is the anti-linear operator of charge conjugation. We  define the {\it self/anti-self} charge-conjugate 4-spinors 
in the momentum space\footnote{In~\cite{vd1Kirchbach}  a bit different notation was used referring to~\cite{vd1Bilenky}.}
\begin{eqnarray}
C\lambda^{S,A} (p^\mu) &=& \pm \lambda^{S,A} (p^\mu)\,,\\
C\rho^{S,A} (p^\mu) &=& \pm \rho^{S,A} (p^\mu)\,,
\end{eqnarray}
where
\begin{equation}
\lambda^{S,A} (p^\mu)=\begin{pmatrix}\pm i\Theta \phi^\ast_L (p^\mu)\\
\phi_L (p^\mu)\end{pmatrix}
\end{equation}
and
\begin{equation}
\rho^{S,A} (p^\mu)=\begin{pmatrix}\phi_R (p^\mu)\\ \mp i\Theta \phi^\ast_R (p^\mu)\end{pmatrix}\,.
\end{equation}
The Wigner matrix is
\begin{equation}
\Theta_{[1/2]}=-i\sigma_2=\begin{pmatrix}0&-1\\
1&0\end{pmatrix}\,,
\end{equation}
and $\phi_L$, $\phi_R$ are the Ryder (Weyl) left- and right-handed 2-spinors
\begin{eqnarray}
\phi_R (p^\mu) &=&\Lambda_R ({\bf p}
\leftarrow {\bf 0}) \phi_R ({\bf 0})= \exp (+ {\mathbf \sigma}\cdot {\mathbf \varphi}/2) \phi_R ({\bf 0})\,,\\
\phi_L (p^\mu) &=&\Lambda_L {\bf p}
\leftarrow {\bf 0}) \phi_L ({\bf 0})= \exp (- {\mathbf \sigma}\cdot {\mathbf \varphi}/2) \phi_L ({\bf 0})\,,
\end{eqnarray}
with ${\mathbf \varphi} = {\bf n} \varphi$ being the boost parameters:
\begin{equation}
cosh \varphi =\gamma = \frac{1}{\sqrt{1-v^2/c^2}}\,,\, sinh \varphi =\beta \gamma =\frac{v/c}{\sqrt{1-v^2/c^2}}\,,\,
tanh \varphi =v/c\,.
\end{equation}
As we have shown the 4-spinors  $\lambda$ and $\rho$ are NOT the eigenspinors of helicity. Moreover, 
$\lambda$ and $\rho$ are NOT the eigenspinors of the parity $P=\begin{pmatrix}0&1\\ 1&0\end{pmatrix}R$, as opposed to the Dirac case.

Such definitions of 4-spinors differ, of course, from the original Majorana definition in x-representation:
\begin{equation}
\nu (x) = \frac{1}{\sqrt{2}} (\Psi_D (x) + \Psi_D^c (x))\,,
\end{equation}
\begin{equation}
\nu (x) = \int \frac{d^3 {\bf p}}{(2\pi)^3 2E_p}  \sum_\sigma [ u_\sigma ({\bf p}) a_\sigma ({\bf p}) e^{-ip\cdot x}
+ v_\sigma ({\bf p}) [\lambda a_\sigma^\dagger ({\bf p})] e^{+ip\cdot x}]\,,
\end{equation}
\begin{equation}
a_\sigma ({\bf p}) =\frac{1}{\sqrt{2}} (b_\sigma ({\bf p}) + d_\sigma^\dagger ({\bf p}))\,,
\end{equation}
$C \nu (x) = \nu (x)$ that represents the positive real $C-$ parity field operator. However, the momentum-space Majorana-like spinors 
open various possibilities for description of neutral  particles 
(with experimental consequences, see~\cite{vd1Kirchbach}).

The 4-spinors of the second kind  $\lambda^{S,A}_{\uparrow\downarrow}
(p^\mu)$ and $\rho^{S,A}_{\uparrow\downarrow} (p^\mu)$
are~\cite{vd1Dvoeglazov2}:
\begin{gather}
  \lambda^S_\uparrow (p^\mu) = \frac{1}{2\sqrt{E+m}}
\begin{pmatrix}ip_l\\ i (p^- +m)\\ p^- +m\\ -p_r\end{pmatrix} ,
\lambda^S_\downarrow (p^\mu)= \frac{1}{2\sqrt{E+m}}
\begin{pmatrix}-i (p^+ +m)\\ -ip_r\\ -p_l\\ (p^+ +m)\end{pmatrix},\\
  \lambda^A_\uparrow (p^\mu) = \frac{1}{2\sqrt{E+m}}
\begin{pmatrix}-ip_l\\ -i(p^- +m)\\ (p^- +m)\\ -p_r\end{pmatrix} ,
\lambda^A_\downarrow (p^\mu) = \frac{1}{2\sqrt{E+m}}
\begin{pmatrix}i(p^+ +m)\\ ip_r\\ -p_l\\ (p^+ +m)\end{pmatrix},\\
  \rho^S_\uparrow (p^\mu) = \frac{1}{2\sqrt{E+m}}
\begin{pmatrix}p^+ +m\\ p_r\\ ip_l\\ -i(p^+ +m)\end{pmatrix} ,
\rho^S_\downarrow (p^\mu) = \frac{1}{2\sqrt{E+m}}
\begin{pmatrix}p_l\\ (p^- +m)\\ i(p^- +m)\\ -ip_r\end{pmatrix},\\
  \rho^A_\uparrow (p^\mu) = \frac{1}{2\sqrt{E+m}}
\begin{pmatrix}p^+ +m\\ p_r\\ -ip_l\\ i (p^+ +m)\end{pmatrix} ,
\rho^A_\downarrow (p^\mu) = \frac{1}{2\sqrt{E+m}}
\begin{pmatrix}p_l\\ (p^- +m)\\ -i(p^- +m)\\ ip_r\end{pmatrix}
\end{gather}
with $p_r = p_x + i p_y$, $p_l = p_x -ip_y$, $p^{\pm} = p_0 \pm p_z$.
The indices $\uparrow\downarrow$ should be referred to either the chiral helicity 
quantum number introduced 
in the 60s, $\eta=-\gamma^5 h$ or to the $\hat S_3$ operator quantum numbers.
While 
\begin{equation}
Pu_\sigma ({\bf p}) = + u_\sigma ({\bf p})\,,
Pv_\sigma ({\bf p}) = - v_\sigma ({\bf p})\,,
\end{equation}
we have
\begin{equation}
P\lambda^{S,A} ({\bf p}) = \rho^{A,S} ({\bf p})\,,
P \rho^{S,A} ({\bf p}) = \lambda^{A,S} ({\bf p})\,,
\end{equation}
for the Majorana-like momentum-space 4-spinors
on the first quantization level.
In this basis one has
\begin{eqnarray}
\rho^S_\uparrow (p^\mu) \,&=&\, - i \lambda^A_\downarrow (p^\mu)\,,\,
\rho^S_\downarrow (p^\mu) \,=\, + i \lambda^A_\uparrow (p^\mu)\,,\\
\rho^A_\uparrow (p^\mu) \,&=&\, + i \lambda^S_\downarrow (p^\mu)\,,\,
\rho^A_\downarrow (p^\mu) \,=\, - i \lambda^S_\uparrow (p^\mu)\,.
\end{eqnarray}

The normalization of the spinors $\lambda^{S,A}_{\uparrow\downarrow}
(p^\mu)$ and $\rho^{S,A}_{\uparrow\downarrow} (p^\mu)$ are the following ones:
\begin{eqnarray}
\overline \lambda^S_\uparrow (p^\mu) \lambda^S_\downarrow (p^\mu) \,&=&\,
- i m \,,\,
\overline \lambda^S_\downarrow (p^\mu) \lambda^S_\uparrow (p^\mu) \,= \,
+ i m \,,\\
\overline \lambda^A_\uparrow (p^\mu) \lambda^A_\downarrow (p^\mu) \,&=&\,
+ i m \,,\,
\overline \lambda^A_\downarrow (p^\mu) \lambda^A_\uparrow (p^\mu) \,=\,
- i m \,,\\
\overline \rho^S_\uparrow (p^\mu) \rho^S_\downarrow (p^\mu) \, &=&  \,
+ i m\,,\,
\overline \rho^S_\downarrow (p^\mu) \rho^S_\uparrow (p^\mu)  \, =  \,
- i m\,,\\
\overline \rho^A_\uparrow (p^\mu) \rho^A_\downarrow (p^\mu)  \,&=&\,
- i m\,,\,
\overline \rho^A_\downarrow (p^\mu) \rho^A_\uparrow (p^\mu) \,=\,
+ i m\,.
\end{eqnarray}
All other conditions are equal to zero.

First of all,  one must derive dynamical equations for the
Majorana-like spinors in order to see what dynamics do the neutral
particles have. One can use the generalized form of the Ryder
relation for zero-momentum 
spinors:
\begin{equation}
\label{vd1rbug12} \left [\phi_{_L}^h
({\bf 0})\right ]^* = (-1)^{1/2-h}\, e^{-i(\vartheta_1^L
+\vartheta_2^L)} \,\Theta_{[1/2]} \,\phi_{_L}^{-h} ({\bf 0})\,,
\end{equation}

Relations for zero-momentum right spinors are obtained with the substitution $L \leftrightarrow R$. $h$ is the helicity quantum number for
the left- and right 2-spinors. Hence, implying that $\lambda^S (p^\mu)$
(and $\rho^A (p^\mu)$) answer for positive-frequency solutions; $\lambda^A
(p^\mu)$ (and $\rho^S (p^\mu)$), for negative-frequency solutions, one can
obtain the dynamical coordinate-space equations~\cite{vd1Dvoeglazov1}
\begin{eqnarray}
i \gamma^\mu \partial_\mu \lambda^S (x) - m \rho^A (x) &=& 0 \,,
\label{vd111}\\
i \gamma^\mu \partial_\mu \rho^A (x) - m \lambda^S (x) &=& 0 \,,
\label{vd112}\\
i \gamma^\mu \partial_\mu \lambda^A (x) + m \rho^S (x) &=& 0\,,
\label{vd113}\\
i \gamma^\mu \partial_\mu \rho^S (x) + m \lambda^A (x) &=& 0\,.
\label{vd114}
\end{eqnarray}
These are NOT the Dirac equations.

They can be written in the 8-component form as follows:
\begin{eqnarray}
\left [i \Gamma^\mu \partial_\mu - m\right ] \Psi_{_{(+)}} (x) &=& 0\,,
\label{vd1psi1}\\
\left [i \Gamma^\mu \partial_\mu + m\right ] \Psi_{_{(-)}} (x) &=& 0\,,
\label{vd1psi2}
\end{eqnarray}
with
\begin{eqnarray}
\Psi_{(+)} (x) = \begin{pmatrix}\rho^A (x)\\
\lambda^S (x)\end{pmatrix}\, ,\,
\Psi_{(-)} (x) = \begin{pmatrix}\rho^S (x)\\
\lambda^A (x)\end{pmatrix}\,, \mbox{and}\,\,
\Gamma^\mu =\begin{pmatrix}0 & \gamma^\mu\\
\gamma^\mu & 0\end{pmatrix}
\end{eqnarray}
One can also re-write the equations into the two-component form.
Similar formulations have been presented by M. Markov~\cite{vd1Markov} long ago, and
A. Barut and G. Ziino~\cite{vd1Ziino}. The group-theoretical basis for such doubling has 
been first given
in the papers by Gelfand, Tsetlin and Sokolik~\cite{vd1Gelfand} and other authors.

Hence, the Lagrangian is
\begin{eqnarray}
&&{\cal L}= \frac{i}{2} \left[\bar \lambda^S \gamma^\mu \partial_\mu \lambda^S - (\partial_\mu \bar \lambda^S ) \gamma^\mu \lambda^S +
\right.\nonumber\\
&&\left. \bar \rho^A \gamma^\mu \partial_\mu \rho^A - (\partial_\mu \bar \rho^A ) \gamma^\mu \rho^A +\right.\nonumber\\
&&\left.\bar \lambda^A \gamma^\mu \partial_\mu \lambda^A - (\partial_\mu \bar \lambda^A ) \gamma^\mu \lambda^A +\right.\nonumber\\
&&\left.\bar \rho^S
\gamma^\mu \partial_\mu \rho^S - (\partial_\mu \bar \rho^S ) \gamma^\mu \rho^S -\right.\nonumber\\
&&\left. - m (\bar\lambda^S \rho^A +\bar \rho^A \lambda^S -\bar\lambda^A \rho^S -\bar\rho^S \lambda^A )
\right ]\,.
\end{eqnarray}

The connection with the Dirac spinors has been found. 
For instance~\cite{vd1Ahluwalia,vd1Dvoeglazov1},
\begin{eqnarray}
\begin{pmatrix}\lambda^S_\uparrow (p^\mu) \\ \lambda^S_\downarrow (p^\mu) \\
\lambda^A_\uparrow (p^\mu) \\ \lambda^A_\downarrow (p^\mu)\end{pmatrix} = \frac{1}{2} \begin{pmatrix} & i & -1 & i\\ -i & 1 & -i & -1\\ 1 & -i & -1 & -i\\ i&
1& i& -1\end{pmatrix} \begin{pmatrix}u_{+1/2} (p^\mu) \\ u_{-1/2} (p^\mu) \\
v_{+1/2} (p^\mu) \\ v_{-1/2} (p^\mu)\end{pmatrix}\,.\label{vd1connect}
\end{eqnarray}
See also ref.~\cite{vd1Gelfand,vd1Ziino}.

The sets of $\lambda$ spinors and of $\rho$ spinors are claimed to be
{\it bi-orthonormal} sets each in the mathematical sense,  provided
that overall phase factors of 2-spinors $\theta_1 +\theta_2 = 0$ or $\pi$.
For instance, on the classical level $\bar \lambda^S_\uparrow
\lambda^S_\downarrow = 2iN^2 \cos ( \theta_1 + \theta_2 )$.
Corresponding commutation relations for this
type of states have also been earlier proposed.

\begin{itemize}
\item
The Lagrangian for $\lambda$ and $\rho$-type  $j=1/2$ states was
given.

\item
While in the massive case there are four $\lambda$-type spinors, two
$\lambda^S$ and two $\lambda^A$ (the $\rho$ spinors are connected by
certain relations with the $\lambda$ spinors for any spin case),  in a
massless case $\lambda^S_\uparrow$ and $\lambda^A_\uparrow$ identically
vanish, provided that one takes into account that $\phi_L^{\pm 1/2}$ are
 eigenspinors of ${\mathbf \sigma}\cdot \hat {\bf n}$.

\item
It was noted the possibility of the generalization of the concept of the
Fock space, which leads to the ``doubling" Fock space~\cite{vd1Gelfand,vd1Ziino}.

\end{itemize}

It was shown~\cite{vd1Dvoeglazov1} that the covariant derivative (and, hence, the
 interaction) can be introduced in this construct in the following way:
\begin{equation}
\partial_\mu \rightarrow \nabla_\mu = \partial_\mu - ig {\text{\L}}^5 B_\mu\quad,
\end{equation}
where ${\text{\L}}^5 = \mbox{diag} (\gamma^5 \quad -\gamma^5)$, the $8\times 8$
matrix. With respect to the transformations
\begin{eqnarray}
\lambda^\prime (x)
\rightarrow (\cos \alpha -i\gamma^5 \sin\alpha) \lambda
(x)\quad,\label{vd1g10}\\
\overline \lambda^{\,\prime} (x) \rightarrow
\overline \lambda (x) (\cos \alpha - i\gamma^5
\sin\alpha)\quad,\label{vd1g20}\\
\rho^\prime (x) \rightarrow  (\cos \alpha +
i\gamma^5 \sin\alpha) \rho (x) \quad,\label{vd1g30}\\
\overline \rho^{\,\prime} (x) \rightarrow  \overline \rho (x)
(\cos \alpha + i\gamma^5 \sin\alpha)\quad\label{vd1g40}
\end{eqnarray}
the spinors retain their properties to be self/anti-self charge conjugate
spinors and the proposed Lagrangian~\cite[p.1472]{vd1Dvoeglazov1} remains to be invariant.
This tells us that while self/anti-self charge conjugate states has
zero eigenvalues of the ordinary (scalar) charge operator but they can
possess the axial charge (cf.  with the discussion of~\cite{vd1Ziino} and
the old idea of R. E. Marshak and others).

In fact, from this consideration one can recover the Feynman-Gell-Mann
equation (and its charge-conjugate equation). They are re-written in the
two-component forms:
%\begin{eqnarray} 
\begin{gather}
\begin{cases}\left [\pi_\mu^- \pi^{\mu\,-}
-m^2 -\frac{g}{2} \sigma^{\mu\nu} F_{\mu\nu} \right ] \chi (x)=0\,, & \\
\left [\pi_\mu^+ \pi^{\mu\,+} -m^2
+\frac{g}{2} \widetilde\sigma^{\mu\nu} F_{\mu\nu} \right ] \phi (x)
=0\,, & \end{cases}\label{vd1iii}
\end{gather}
%\end{eqnarray}
where one now has $\pi_\mu^\pm =
i\partial_\mu \pm gA_\mu$, \, $\sigma^{0i} = -\widetilde\sigma^{0i} =
i\sigma^i$, $\sigma^{ij} = \widetilde\sigma^{ij} = \epsilon_{ijk}
\sigma^k$ and $\nu^{^{DL}} (x) =\mbox{column} (\chi \quad \phi )$.

Next, because the transformations
\begin{eqnarray}
\lambda_S^\prime (p^\mu) &=& \begin{pmatrix}\Xi &0\\ 0&\Xi\end{pmatrix} \lambda_S (p^\mu)
\equiv \lambda_A^\ast (p^\mu)\quad,\quad\\
\lambda_S^{\prime\prime} (p^\mu) &=& \begin{pmatrix}i\Xi &0\\ 0&-i\Xi\end{pmatrix} \lambda_S
(p^\mu) \equiv -i\lambda_S^\ast (p^\mu)\quad,\quad\\
\lambda_S^{\prime\prime\prime} (p^\mu) &=& \begin{pmatrix}0& i\Xi\\
i\Xi &0\end{pmatrix} \lambda_S (p^\mu) \equiv i\gamma^0 \lambda_A^\ast
(p^\mu)\quad,\quad\\
\lambda_S^{IV} (p^\mu) &=& \begin{pmatrix}0& \Xi\\
-\Xi&0\end{pmatrix} \lambda_S (p^\mu) \equiv \gamma^0\lambda_S^\ast
(p^\mu)\quad
\end{eqnarray}
with the $2\times 2$ matrix $\Xi$ defined as ($\phi$ is the azimuthal
angle  related to ${\bf p} \rightarrow {\bf 0}$)
\begin{equation}
\Xi = \begin{pmatrix}e^{i\phi} & 0\\ 0 &
e^{-i\phi}\end{pmatrix}\quad,\quad \Xi \Lambda_{R,L} (0 \leftarrow
p^\mu) \Xi^{-1} = \Lambda_{R,L}^\ast (0 \leftarrow
 p^\mu)\,\,\, ,
\end{equation}
and corresponding transformations for
$\lambda^A$ do {\it not} change the properties of bispi\-nors to be in the
self/anti-self charge conjugate spaces, the Majorana-like field operator
($b^\dagger \equiv a^\dagger$) admits additional phase (and, in general,
normalization) $SU(2)$ transformations:
\begin{equation} \nu^{ML\,\,\prime}
(x^\mu) = \left [ c_0 + i({\bf \tau}\cdot  {\bf c}) \right
]\nu^{ML\,\,\dagger} (x^\mu) \quad, \end{equation} where $c_\alpha$ are
arbitrary parameters. The ${\bf \tau}$ matrices are defined over the
field of $2\times 2$ matrices and the Hermitian
conjugation operation is assumed to act on the $c$- numbers as the complex
conjugation. One can parametrize $c_0 = \cos\phi$ and ${\bf c} = {\bf n}
\sin\phi$ and, thus, define the $SU(2)$ group of phase transformations.
One can select the Lagrangian which is composed from both field
operators (with $\lambda$ spinors and $\rho$ spinors)
and which remains to be
invariant with respect to this kind of transformations.  The conclusion
is:  a non-Abelian construct is permitted, which is based on
the spinors of the Lorentz group only (cf. with the old ideas of T. W.
Kibble and R. Utiyama) .  This is not surprising because both $SU(2)$
group and $U(1)$ group are  the sub-groups of the extended Poincar\'e group
(cf.~\cite{vd1Ryder}).

The Dirac-like and the Majorana-like field operators can
be built from both $\lambda^{S,A} (p^\mu)$ and $\rho^{S,A} (p^\mu)$,
or their combinations. For 
instance,
\begin{eqnarray}
\Psi (x^\mu) &\equiv& \int {d^3 {\bf p}\over (2\pi)^3} {1\over 2E_p}
\sum_\eta \left [ \lambda^S_\eta (p^\mu) \, a_\eta ({\bf p}) \,\exp
(-ip\cdot x) +\right.\nonumber\\
&&\left. \lambda^A_\eta (p^\mu)\, b^\dagger_\eta ({\bf p}) \,\exp
(+ip\cdot x)\right ]\quad.\label{vd1oper}
\end{eqnarray}

The anticommutation relations are the following ones (due to the {\it bi-ortho\-normality}):
\begin{eqnarray}
[a_{\eta{\prime}} (p^{\prime^\mu}), a_\eta^\dagger (p^\mu) ]_\pm = (2\pi)^3 2E_p \delta ({\bf p} -{\bf p}^\prime) \delta_{\eta,-\eta^\prime}
\end{eqnarray}
and 
\begin{eqnarray}
[b_{\eta{\prime}} (p^{\prime^\mu}), b_\eta^\dagger (p^\mu) ]_\pm = (2\pi)^3 2E_p \delta ({\bf p} -{\bf p}^\prime) \delta_{\eta,-\eta^\prime}
\end{eqnarray}
Other (anti)commutators are equal to zero: ($[ a_{\eta^\prime} (p^{\prime^\mu}), 
b_\eta^\dagger (p^\mu) ]=0$).

In the Fock space the operations of the charge conjugation and space
inversions can be defined through unitary operators such that:
\begin{equation}
U^c_{[1/2]} \Psi (x^\mu) (U^c_{[1/2]})^{-1} = {\cal C}_{[1/2]}
\Psi^\dagger_{[1/2]} (x^\mu),
U^s_{[1/2]} \Psi (x^\mu) (U^s_{[1/2]})^{-1} = \gamma^0
\Psi (x^{\prime^{\,\mu}}),
\end{equation}
the time reversal operation, through {\it an antiunitary}
operator\footnote{Let us remind that the operator of hermitian conjugation does not act on $c$-numbers on the left side of the equation (\ref{vd1tr}).
This fact is conected with the properties of the antiunitary operator:
$\left [ V^{^T} \lambda A (V^{^T})^{-1}\right ]^\dagger =
\left [\lambda^\ast V^{^T} A (V^{^T})^{-1}\right ]^\dagger =
\lambda \left [ V^{^T} A^\dagger (V^{^T})^{-1} \right ]$.}
\begin{equation}
\left [V^{^T}_{[1/2]}  \Psi (x^\mu)
(V^{^T}_{[1/2]})^{-1} \right ]^\dagger = S(T) \Psi^\dagger
(x^{{\prime\prime}^\mu}) \,,\label{vd1tr}
\end{equation}
with
$x^{\prime^{\,\mu}} \equiv (x^0, -{\bf x})$ and $x^{{\prime\prime}^{\,\mu}}
=(-x^0,{\bf x})$.  We  further assume the vacuum state to be assigned an
even $P$- and $C$-eigenvalue and, then, proceed as in ref.~\cite{vd1Itzykson}.
As a result we have the following properties of creation (annihilation)
operators in the Fock space:
\begin{eqnarray}
U^s_{[1/2]} a_\uparrow ({\bf p}) (U^s_{[1/2]})^{-1} &=& - ia_\downarrow
(-  {\bf p})\,,\\
U^s_{[1/2]} a_\downarrow ({\bf p}) (U^s_{[1/2]})^{-1} &=& + ia_\uparrow
(- {\bf p})\,,\\
U^s_{[1/2]} b_\uparrow^\dagger ({\bf p}) (U^s_{[1/2]})^{-1} &=&
+ i b_\downarrow^\dagger (- {\bf p})\,,\\
U^s_{[1/2]} b_\downarrow^\dagger ({\bf p}) (U^s_{[1/2]})^{-1} &=&
- i b_\uparrow (- {\bf p})\,,
\end{eqnarray} 
what signifies that the states created by the operators $a^\dagger
({\bf p})$ and $b^\dagger ({\bf p})$ have very different properties
with respect to the space inversion operation, comparing with
Dirac states (the case was also regarded in~\cite{vd1Ziino}):
\begin{eqnarray}
U^s_{[1/2]} \vert {\bf p},\uparrow >^+ = + i \vert -{\bf p},
\downarrow >^+,
U^s_{[1/2]} \vert {\bf p},\uparrow >^- = + i
\vert -{\bf p}, \downarrow >^-\\
U^s_{[1/2]} \vert {\bf p},\downarrow >^+ = - i \vert -{\bf p},
\uparrow >^+,
U^s_{[1/2]} \vert {\bf p},\downarrow >^- =  - i
\vert -{\bf p}, \uparrow >^-
\end{eqnarray}

For the charge conjugation operation in the Fock space we have
two physically different possibilities. The first one, {\it e.g.},
\begin{eqnarray}
U^c_{[1/2]} a_\uparrow ({\bf p}) (U^c_{[1/2]})^{-1} = + b_\uparrow
({\bf p}),
U^c_{[1/2]} a_\downarrow ({\bf p}) (U^c_{[1/2]})^{-1} = + b_\downarrow
({\bf p}),\\
U^c_{[1/2]} b_\uparrow^\dagger ({\bf p}) (U^c_{[1/2]})^{-1} =
-a_\uparrow^\dagger ({\bf p}),
U^c_{[1/2]} b_\downarrow^\dagger ({\bf p})
(U^c_{[1/2]})^{-1} = -a_\downarrow^\dagger ({\bf p}),
\end{eqnarray}
in fact, has some similarities with the Dirac construct.
The action of this operator on the physical states are
\begin{eqnarray}
U^c_{[1/2]} \vert {\bf p}, \, \uparrow >^+ &=& + \,\vert {\bf p},\,
\uparrow >^- \,,\,
U^c_{[1/2]} \vert {\bf p}, \, \downarrow >^+ = + \, \vert {\bf p},\,
\downarrow >^- \,,\\
U^c_{[1/2]} \vert {\bf p}, \, \uparrow >^-
&=&  - \, \vert {\bf p},\, \uparrow >^+ \,,\,
U^c_{[1/2]} \vert
{\bf p}, \, \downarrow >^- = - \, \vert {\bf p},\, \downarrow >^+ \,.
\end{eqnarray}
But, one can also construct the charge conjugation operator in the
Fock space which acts, {\it e.g.}, in the following manner:
\begin{eqnarray}
\widetilde U^c_{[1/2]} a_\uparrow ({\bf p}) (\widetilde U^c_{[1/2]})^{-1}
= - b_\downarrow ({\bf p}), \widetilde U^c_{[1/2]}
a_\downarrow ({\bf p}) (\widetilde U^c_{[1/2]})^{-1} = - b_\uparrow
({\bf p}),\\
\widetilde U^c_{[1/2]} b_\uparrow^\dagger ({\bf p})
(\widetilde U^c_{[1/2]})^{-1} = + a_\downarrow^\dagger ({\bf
p}),
\widetilde U^c_{[1/2]} b_\downarrow^\dagger ({\bf p})
(\widetilde U^c_{[1/2]})^{-1} = + a_\uparrow^\dagger ({\bf p}),
\end{eqnarray}
and, therefore,
\begin{eqnarray}
\widetilde U^c_{[1/2]} \vert {\bf p}, \, \uparrow >^+ &=& - \,\vert {\bf
p},\, \downarrow >^- \,,\,
\widetilde U^c_{[1/2]} \vert {\bf p}, \, \downarrow
>^+ = - \, \vert {\bf p},\, \uparrow >^- \,,\\
\widetilde U^c_{[1/2]} \vert
{\bf p}, \, \uparrow >^- &=& + \, \vert {\bf p},\, \downarrow >^+
\,,\,
\widetilde U^c_{[1/2]} \vert {\bf p}, \, \downarrow >^- = + \, \vert {\bf
p},\, \uparrow >^+ \,.
\end{eqnarray}
This is due to corresponding algebraic structures of self/anti-self charge-conjugate spinors.

Investigations of several important cases, which are different from the
 above ones, are required a separate paper. Next, it is
 possible a situation when the operators of the space inversion and 
charge conjugation commute each other in the Fock space. For instance,
\begin{eqnarray}
U^c_{[1/2]} U^s_{[1/2]} \vert {\bf
p},\, \uparrow >^+ &=& + i U^c_{[1/2]}\vert -{\bf p},\, \downarrow >^+ =
+ i \vert -{\bf p},\, \downarrow >^- \,,\\
U^s_{[1/2]} U^c_{[1/2]} \vert {\bf
p},\, \uparrow >^+ &=& + U^s_{[1/2]}\vert {\bf p},\, \uparrow >^- = + i
\vert -{\bf p},\, \downarrow >^- \,.
\end{eqnarray}
The second choice of the charge conjugation operator answers for the case
when the $\widetilde U^c_{[1/2]}$ and $U^s_{[1/2]}$ operations
anticommute:
\begin{eqnarray}
\widetilde U^c_{[1/2]} U^s_{[1/2]} \vert {\bf p},\, \uparrow >^+ &=&
+ i \widetilde U^c_{[1/2]}\vert -{\bf
p},\, \downarrow >^+ = -i \, \vert -{\bf p},\, \uparrow >^- \,,\\
U^s_{[1/2]} \widetilde U^c_{[1/2]} \vert {\bf p},\, \uparrow >^+ &=& -
U^s_{[1/2]}\vert {\bf p},\, \downarrow >^- = + i \, \vert -{\bf p},\,
\uparrow >^- \,.
\end{eqnarray} 

Next, one can compose states which would have somewhat similar
properties to those which we have become accustomed.
The states $\vert {\bf p}, \,\uparrow >^+ \pm
i\vert {\bf p},\, \downarrow >^+$ answer for positive (negative) parity,
respectively.  But, what is important, {\it the antiparticle states}
(moving backward in time) have the same properties with respect to the
operation of space inversion as the corresponding {\it particle states}
(as opposed to $j=1/2$ Dirac particles).  
The states which are 
eigenstates of the charge conjugation operator in the Fock space are
\begin{equation}
U^c_{[1/2]} \left ( \vert {\bf p},\, \uparrow >^+ \pm i\,
\vert {\bf p},\, \uparrow >^- \right ) = \mp i\,  \left ( \vert {\bf p},\,
\uparrow >^+ \pm i\, \vert {\bf p},\, \uparrow >^- \right ) \,.
\end{equation}
There is no any simultaneous sets of states which would be ``eigenstates" of the 
operator of the space inversion and of the charge conjugation 
$U^c_{[1/2]}$.

Finally, the time reversal {\it anti-unitary} operator in 
the Fock space should be defined in such a way that the formalism to be
 compatible with the $CPT$ theorem. If we wish the Dirac states to transform 
as 
$V(T) \vert {\bf p}, \pm 1/2 > = \pm \,\vert -{\bf p}, \mp 1/2 >$ we
 have to choose (within a phase factor), ref.~\cite{vd1Itzykson}:
\begin{equation}
S(T) = \begin{pmatrix}\Theta_{[1/2]} &0\\ 0 &
\Theta_{[1/2]}\end{pmatrix}\,.
\end{equation}

Thus, in the first relevant case we obtain for the $\Psi
(x^\mu)$ field, Eq.  (\ref{vd1oper}):
\begin{eqnarray}
V^{^T} a^\dagger_\uparrow ({\bf p}) (V^{^T})^{-1} &=& a^\dagger_\downarrow
(-{\bf p})\,,\,
V^{^T} a^\dagger_\downarrow ({\bf p}) (V^{^T})^{-1} = -
a^\dagger_\uparrow (-{\bf p})\,, \\
V^{^T} b_\uparrow ({\bf p}) (V^{^T})^{-1} &=& b_\downarrow
(-{\bf p})\,,\,
V^{^T} b_\downarrow ({\bf p}) (V^{^T})^{-1} = -
b_\uparrow (-{\bf p})\,.
\end{eqnarray}

The analogs of the above equations in the $(1,0)\oplus (0,1)$ representation space are:
\begin{equation}
C_{[1]} =e^{i\theta_c}\begin{pmatrix}0&\Theta_{[1]}\\
-\Theta_{[1]}&0\end{pmatrix}\,,\quad \Theta_{[1]}=\begin{pmatrix}0&0&1\\
0&-1&0\\
1&0&0\end{pmatrix}\,,
\end{equation}
\begin{equation}
P=e^{i\theta_s}\begin{pmatrix}1&0\\ 0&1\end{pmatrix} R= e^{i\theta_s}\gamma_{00} R\,,
\end{equation}
\begin{equation}
\Gamma^5 =\begin{pmatrix}1&0\\ 0&-1\end{pmatrix}\,.
\end{equation}
One can define the $\Gamma^5 C$ self/anti-self charge conjugate 6-component objects.
\begin{eqnarray}
\Gamma^5 C_{[1]}  \lambda ( p^\mu ) &=&\pm \lambda ( p^\mu )\,,\\
\Gamma^5 C_{[1]}  \rho ( p^\mu ) &=&\pm \rho ( p^\mu )\,.
\end{eqnarray}
The $C_{[1]}$ matrix is constructed from dynamical equations for charged spin-1 particles.
No self/anti-self charge-conjugate states are possible.
They are also NOT the eigenstates of the parity operator (except for $\lambda_{\rightarrow}$):
\begin{eqnarray}
P\lambda^S_\uparrow =+\lambda^S_\downarrow\,, P\lambda^S_\rightarrow =-\lambda^S_\rightarrow\,, P\lambda^S_\downarrow =+\lambda^S_\uparrow\,, \\
P\lambda^A_\uparrow =-\lambda^A_\downarrow\,, P\lambda^A_\rightarrow =+\lambda^A_\rightarrow\,, P\lambda^A_\downarrow =+\lambda^A_\uparrow\,.
\end{eqnarray}
The dynamical equations are
\begin{eqnarray}
\gamma_{\mu\nu} p^\mu p^\nu  \lambda^S_{\uparrow\downarrow} - m^2 \lambda^S_{\downarrow\uparrow}&=&0\,,\\
\gamma_{\mu\nu} p^\mu p^\nu  \lambda^A_{\uparrow\downarrow} + m^2 \lambda^A_{\downarrow\uparrow}&=&0\,,\\\
\gamma_{\mu\nu} p^\mu p^\nu  \lambda^S_{\rightarrow} + m^2 \lambda^S_{\rightarrow}&=&0\,,\\
\gamma_{\mu\nu} p^\mu p^\nu  \lambda^A_{\rightarrow} -m^2 \lambda^A_{\rightarrow}&=&0\,.
\end{eqnarray}
Under the appropriate choice of the basis and phase factors we have
\begin{eqnarray}
&&\rho^S_{\uparrow\downarrow}=+\lambda^S_{\downarrow\uparrow}\,,
\rho^A_{\uparrow\downarrow}=-\lambda^A_{\downarrow\uparrow}\\
&&\rho^S_{\rightarrow}=-\lambda^S_{\rightarrow}\,,
\rho^A_{\rightarrow}=+\lambda^S_{\rightarrow}\,.
\end{eqnarray}

On the secondary quantization level we obtained similar results as in the spin-1/2 case.

The conclusions are:

\begin{itemize}

\item
The momentum-space Majorana -like spinors are considered in the $(j,0)\oplus (0,j)$ representation space.

\item
They have different properties from the Dirac spinors even on the classical level.

\item
It is convenient to work in the 8-dimensional space. Then, we can impose the Gelfand-Tsetlin-Sokolik 
(Bargmann-Wight\-man-Wigner) prescription of 2-dimensional representation of the inversion group.

\item
Gauge transformations are different. The axial charge is possible.

\item
Experimental differencies have been recently discussed  (the possibility of observation 
of the phase factor/eigenvalue of the $C$-parity), see~\cite{vd1Kirchbach}.

\item
(Anti)commutation relations are assumed to be different from the Dirac case (and the $2(2j+1)$ case) due to the bi-orthonormality of the states (the spinors are self-orthogonal).

\item
The $(1,0)\oplus (0,1)$ case has also been considered. The $\Gamma^5 C$-self/anti-self conjugate objects have been introduced. The results  are similar to 
the $(1/2,0)\oplus (0,1/2)$ representation. The 12-dimensional formalism was introduced.

\item
The field operator can describe both charged and neutral states.

\end{itemize}

%}

%% V. Dvoeglazov, second contribution, 11.09.2009
%%
\author{V.V. Dvoeglazov}
\title{Relativistic Equations for Spin Particles: What Can We Learn From Noncommutativity?\thanks{Talk given at the XXVIII WGMP09, Bia{\l}owie\`{z}a, Poland, June 28-July 4, 2009.
The extended version  is contributed to the 12th International Workshop 'What Comes Beyond 
the Standard Models', 14. - 24. July 2009, Bled.}}
\institute{%
Universidad de Zacatecas, 
Apartado Postal 636, Suc. 3 Cruces\\
Zacatecas 98064, Zac., M\'exico\\
URL: http://planck.reduaz.mx/\~ valeri/ \\
e-mail address: valeri@fisica.uaz.edu.mx} 

\authorrunning{V.V. Dvoeglazov}
\titlerunning{Relativistic Equations for Spin Particles}
\maketitle

\begin{abstract}
We  derive relativistic equations for charged and neutral  spin particles.
The approach for higher-spin particles is based on generalizations of the Bargmann-Wigner  formalism.
Next, we study , what new physical information can the introduction of non-commutativity give us. Additional non-commutative parameters can provide a suitable basis for explanation of the origin of mass.
\end{abstract}

\section{Introduction}

In the spin-1/2 case the Klein-Gordon equation  can be written 
for the two-compo\-nent spinor ($c=\hbar =1$)
\begin{equation}
(E I^{(2)} - {\sigma}\cdot {\bf p})
(E I^{(2)} + {\sigma}\cdot {\bf p})\Psi^{(2)} = m^2 \Psi^{(2)}\,,
\end{equation}
or, in the 4-component form
\begin{equation}
[i\gamma_\mu \partial_\mu +m_1 +m_2 \gamma^5 ] \Psi^{(4)} = 0\,.
\end{equation}
There exist various generalizations
of the Dirac formalism. For instance, the Barut generalization
is based on
\begin{equation}
[i\gamma_\mu \partial_\mu +  a (\partial_\mu \partial_\mu)/m - \kappa ] \Psi =0\,,
\end{equation}
which can describe states of different masses. If one fixes the parameter $a$ by the requirement that the equation gives the state with the classical anomalous magnetic moment, then $m_2 =
m_1 (1+\frac{3}{2\alpha})$, i.e., it gives the muon mass. Of course, one can propose a generalized equation:
\begin{equation}
[i\gamma_\mu \partial_\mu +  a +b \partial_\mu \partial_\mu + \gamma_5 (c+d \partial_\mu \partial_\mu ) ] \Psi =0
\,,
\end{equation}
and, perhaps, even that of higher orders in derivatives.

In the spin-1 case  we have
\begin{equation}
(E I^{(3)} - {\bf S}\cdot {\bf p})
(E I^{(3)} + {\bf S}\cdot {\bf p}){\Psi}^{(3)} 
- {\bf p} ({\bf p}\cdot {\Psi}^{(3)})= m^2 \Psi^{(3)}\,,
\end{equation}
that lead to (\ref{vd21}-\ref{vd24}), when $m=0$. We can continue writing down equations for higher spins in a similar fashion.

In Ref.~\cite{vd2Hua,vd2Hua1}
I derived the Maxwell-like equations with the additional gradient of a scalar field ${\chi}$ from the first principles.\footnote{Cf. $'chi$-field with the $S=0$ field in the $(1/2,1/2)$ representation, ref.~\cite{vd2Weinb}.}
Here they are:
\begin{eqnarray}
&&{\nabla}\times {\bf
E}=-\frac{1}{c}\frac{\partial {\mathbf{B}}}{\partial t} + {\nabla} {Im} \chi \,, \label{vd21}\\
&&{\nabla }\times {\mathbf{B}}=\frac{1}{c}\frac{\partial {\mathbf{E}}}{\partial t}  +{\nabla} {Re} \chi\,,\label{vd22}\\
&&{\nabla}\cdot {\mathbf{E}}=-\frac{1}{c} \frac{\partial}{\partial
t} {Re}\chi \,,\label{vd23}\\
&&{\nabla }\cdot {\mathbf{B}}= \frac{1}{c} \frac{\partial}{\partial t} {Im} \chi \,.  \label{vd24}
\end{eqnarray}
The $\chi$ may depend on the ${\mathbf{E}}, {\mathbf{B}}$, so we can have the non-linear electrodynamics.
Of course, similar equations can be obtained 
in the massive case $m\neq 0$, i.e., within the Proca-like theory.

On this basis we are ready to generalize the BW formalism~\cite{vd2bw-hs,vd2Lurie}. Why is that convenient? In Ref.~\cite{vd2dv-hpa,vd2wig} I presented the mapping between the Weinberg-Tucker-Hammer (WTH) equation, Ref.~\cite{vd2WTH,vd2WTH1}, and the equations for antisymmetric tensor  (AST) fields. The equation for a 6-component field function is\footnote{ In order to have solutions satisfying the Einstein dispersion relations $E^2 -{\bf p}^2 =m^2$ we have to assume $B/(A+1)= 1$, or $B/(A-1)=1$.}
\begin{equation}
[\gamma_{\alpha\beta} p_\alpha p_\beta +A p_\alpha p_\alpha +Bm^2 ]
\Psi^{(6)} =0\,.
\end{equation}
Corresponding equations for the AST fields are:
\begin{eqnarray}
&&\partial_\alpha\partial_\mu F_{\mu\beta}^{(1)}
-\partial_\beta\partial_\mu F_{\mu\alpha}^{(I1)}
+ \frac{A-1}{2} \partial_\mu \partial_\mu F_{\alpha\beta}^{(1)}
-\frac{B}{2} m^2 F_{\alpha\beta}^{(1)} = 0\,,\label{vd2wth1}\\
&&\partial_\alpha\partial_\mu F_{\mu\beta}^{(2)}
-\partial_\beta\partial_\mu F_{\mu\alpha}^{(2)}
- \frac{A+1}{2} \partial_\mu \partial_\mu F_{\alpha\beta}^{(2)}
+\frac{B}{2} m^2 F_{\alpha\beta}^{(2)}=0\label{vd2wth2}
\end{eqnarray}
depending on the parity properties of $\Psi^{(6)}$ (the first case corresponds 
to the eigenvalue $P=-1$; the second one, to $P=+1$). 

We have noted:

\begin{itemize}

\item
One can derive equations for the dual tensor $\tilde F_{\alpha\beta}$,
which are similar to equations (\ref{vd2wth1},\ref{vd2wth2}), Ref.~\cite{vd2dv-cl,vd2dv-hpa}.

\item
In the Tucker-Hammer case ($A=1$, $B=2$), the first equation gives the Proca theory $\partial_\alpha \partial_\mu 
F_{\mu\beta} -
\partial_\beta \partial_\mu F_{\mu\alpha} = m^2 F_{\alpha\beta}$. In the second case one finds something different, $\partial_\alpha \partial_\mu F_{\mu\beta} -
\partial_\beta \partial_\mu F_{\mu\alpha} = (\partial_\mu \partial_\mu - m^2 ) F_{\alpha\beta}$.

\item
If $\Psi^{(6)}$ has no definite parity, e.~g., $\Psi^{(6)} = 
\mbox{column} ({\bf E}+i{\bf B}\,\,\, {\bf B}+i{\bf E}\, )$, the equation
for the AST field will contain both the tensor and the dual tensor:
\begin{equation}
\partial_\alpha \partial_\mu F_{\mu\beta}
-\partial_\beta \partial_\mu F_{\mu\alpha}
=\frac{1}{2} \partial^2 F_{\alpha\beta} +
[-\frac{A}{2} \partial^2 + \frac{B}{2} m^2] \tilde
F_{\alpha\beta}.\label{vd2pv1}
\end{equation}

\item
Depending on the relation between $A$ and $B$ and on which
parity solution  do we consider, the WTH equations may describe different mass states. For instance, when $A=7$ and $B=8$ we have the second mass state $(m^{\prime})^2 = 4m^2/3$.

\end{itemize}

We tried to find relations between the generalized WTH theory
and other spin-1 formalisms.  Therefore, we Have been forced to modify the Bargmann-Wigner formalism~\cite{vd2dv-cl,vd2dv-ps}. For instance, we introduced the sign operator in the Dirac equations which are the inputs for the formalism for symmetric 2-rank spinor:
\begin{eqnarray}
\left [ i\gamma_\mu \partial_\mu + \epsilon_1 m_1 +\epsilon_2 m_2 \gamma_5
\right ]_{\alpha\beta} \Psi_{\beta\gamma} &=&0\,,\\
\left [ i\gamma_\mu
\partial_\mu + \epsilon_3 m_1 +\epsilon_4 m_2 \gamma_5 \right ]_{\gamma\beta}
\Psi_{\alpha\beta} &=&0\,,
\end{eqnarray}
In general we have 16 possible combinations, but 4 of them give the same
sets of the Proca-like equations. We obtain~\cite{vd2dv-cl}:
\begin{eqnarray} 
&&\partial_\mu A_\lambda - \partial_\lambda A_\mu + 2m_1 A_1 F_{\mu \lambda}
+im_2 A_2 \epsilon_{\alpha\beta\mu\lambda} F_{\alpha\beta} =0\,,\\
&&\partial_\lambda
F_{\mu \lambda} - \frac{m_1}{2} A_1 A_\mu -\frac{m_2}{2} B_2 \tilde
A_\mu=0\,,
\end{eqnarray} 
with 
$A_1 = (\epsilon_1 +\epsilon_3) /2$,
$A_2 = (\epsilon_2 +\epsilon_4 )/ 2$,
$B_1 = (\epsilon_1 -\epsilon_3 )/ 2$,
and
$B_2 = (\epsilon_2 -\epsilon_4 )/ 2$. See the additional constraints in the cited paper~\cite{vd2dv-cl}.
So, we have the dual tensor and the pseudovector potential
in the Proca-like sets. The pseudovector potential is the same as that
which enters in the Duffin-Kemmer set for the spin 0. 

Moreover, it appears that the properties of the polarization
vectors with respect to parity operation depend on the choice of the spin basis.
For instance, in Ref.~\cite{vd2GR,vd2dv-cl} the momentum-space polarization vectors have been 
listed in the helicity basis.
Berestetski\u{\i}, Lifshitz and Pitaevski\u{\i} claimed too, Ref.~\cite{vd2BLP}, that
the helicity states cannot be the parity states. If one applies common-used
relations between fields and potentials it appears that the ${\bf E}$ and ${\bf B}$ fields have no usual properties with respect to space inversions.

Thus, the conclusions of the previous works are:
\begin{itemize}

\item
The mapping exists between the WTH formalism for $S=1$ and the AST fields of four kinds (provided that the solutions of the WTH equations are of the definite
parity).

\item
Their massless limits contain additional solutions comparing with the Maxwell equations. This was related to the possible theoretical existence of the Ogievetski\u{\i}-Polubarinov-Kalb-Ramond notoph, Ref.~\cite{vd2Og,vd2Og1,vd2Og2}.

\item
In some particular cases ($A=0, B=1$) massive solutions of different parities are naturally divided into the classes of causal and tachyonic solutions.

\item
If we want to take into account the solutions of the WTH equations of
different parity properties, this induces us to generalize the BW, Proca and Duffin-Kemmer formalisms.

\item
In the $(1/2,0)\oplus (0,1/2)$, $(1,0)\oplus (0,1)$ etc. representations
it is possible to introduce the parity-violating frameworks. The corresponding solutions are the mixing of various polarization states.

\item
The sum of the Klein-Gordon equation with the $(S,0)\oplus (0,S)$ equations may change the theoretical content even on the free level. For instance, the higher-spin equations may actually describe various spin and mass states.

\item
The mappings exists between the WTH solutions of undefined parity
and the AST fields, which contain both tensor and dual tensor. They are eight.

\item
The 4-potentials and electromagnetic fields~\cite{vd2dv-cl,vd2GR} in the helicity
basis have different parity properties comparing with the standard basis of the polarization vectors.

\item
In the previous talk~\cite{vd2dv-pl} I presented a theory in the $(1/2,0)\oplus (0,1/2)$ representation in the helicity basis. Under the space inversion operation,
different helicity states transform each other, $Pu_h (-{\bf p}) = -i u_{-h} ({\bf p})$, $Pv_h (-{\bf p}) = +i v_{-h} ({\bf p})$.

\end{itemize}

\section{The 4-Vector Field}

Next, we show that the equation for the 4-vector field can be presented
in a matrix form. Recently, S. I. Kruglov proposed, Refs.~\cite{vd2krug1}, 
a general form of the Lagrangian for  4-potential field $B_\mu$, which also contains the spin-0 state. 
Initially, we have 
\begin{equation}
\alpha \partial_\mu \partial_\nu B_\nu +\beta \partial_\nu^2 B_\mu +\gamma m^2 B_\mu =0\, ,\label{vd2eq-pot}
\end{equation}
provided that derivatives commute.
When $\partial_\nu B_\nu =0$ (the Lorentz gauge) we obtain spin-1 states only.
However, if it is not equal to zero we have a scalar field and an axial-vector potential. We can also verify this 
statement by consideration of the dispersion relations of the equation (\ref{vd2eq-pot}). One obtains 4+4 states 
(two of them may differ in mass from others).

Next, one can fix one of the constants $\alpha,\beta,\gamma$ 
without loosing any physical content. For instance, when $\alpha=-2$
one gets the equation
\begin{equation}
\left [ \delta_{\mu\nu} \delta_{\alpha\beta} - \delta_{\mu\alpha}\delta_{\nu\beta} - \delta_{\mu\beta} \delta_{\nu\alpha}\right ] \partial_\alpha \partial_\beta B_\nu + A \partial_\alpha^2 \delta_{\mu\nu} B_\nu - Bm^2  B_\mu =0\,,\label{vd2eq1-m}
\end{equation} 
where  $\beta= A+1$ and $\gamma=-B$. In the matrix form the equation (\ref{vd2eq1-m}) reads:
\begin{equation}
\left [ \gamma_{\alpha\beta} \partial_\alpha \partial_\beta +A \partial_\alpha^2 - Bm^2 \right ]_{\mu\nu} B_\nu = 0\,,
\end{equation}
with
\begin{equation}
[\gamma_{\alpha\beta}]_{\mu\nu} = \delta_{\mu\nu}\delta_{\alpha\beta}
-\delta_{\mu\alpha}\delta_{\nu\beta} - \delta_{\mu\beta} \delta_{\nu\alpha}\,.
\end{equation}
They are the analogs of the Barut-Muzinich-Williams (BMW) $\gamma$-matrices
for bivector fields.\footnote{One can also define the analogs of the BMW $\gamma_{5,\alpha\beta}$ matrices 
\begin{equation}
\gamma_{5,\alpha\beta} = \frac{i}{6} [\gamma_{\alpha\kappa}, \gamma_{\beta\kappa} ]_{-, \mu\nu} = i [\delta_{\alpha\mu} \delta_{\beta\nu} - \delta_{\alpha\nu}\delta_{\beta\mu} ]\,.
\end{equation}
As opposed to $\gamma_{\alpha\beta}$ matrices they are totally antisymmetric.
They are related to  boost and rotation generators of this representation.
The $\gamma$-matrices are pure real; $\gamma_5$-matrices are pure imaginary.
In the $(1/2,1/2)$ representation, we need 16 matrices to form the complete set.}
It is easy to prove by the textbook method~\cite{vd2Itzyk} that $\gamma_{44}$
can serve as the parity matrix.

{\it Lagrangian and the equations of motion.}
Let us try
\begin{equation}
{\cal L} = (\partial_\alpha B_\mu^\ast) [\gamma_{\alpha\beta} ]_{\mu\nu} (\partial_\beta B_\nu)
+ A (\partial_\alpha B_\mu^\ast) (\partial_\alpha B_\mu) + Bm^2
B_\mu^\ast B_\mu\,.
\end{equation}
On using the Lagrange-Euler equation
we have
\begin{equation}
[\gamma_{\nu\beta}]_{\kappa\tau} \partial_\nu \partial_\beta B_\tau + A \partial_\nu^2 B_\kappa - Bm^2  B_\kappa =0\,.\label{vd2equat}
\end{equation}
It may be presented in the form of (\ref{vd2eq-pot}). 

{\it Masses.} We are convinced that in the case of spin 0, we have $B_\mu \rightarrow
\partial_\mu \chi$; in the case of spin 1 we have $\partial_\mu B_\mu =0$.

\begin{equation}
(\delta_{\mu\nu}\delta_{\alpha\beta}
-\delta_{\mu\alpha}\delta_{\nu\beta} - \delta_{\mu\beta} \delta_{\nu\alpha})
\partial_\alpha \partial_\beta \partial_\nu \chi = - \partial^2 \partial_\mu \chi\,.
\end{equation}
Hence, from (\ref{vd2equat}) we have
\begin{equation}
[(A-1) \partial^2_\nu - Bm^2 ] \partial_\mu \chi=0\,.
\end{equation}
If $A-1=B$ we have the spin-0 particles with masses $\pm m$ with the correct relativistic dispersion.

In another case
\begin{equation}
[\delta_{\mu\nu}\delta_{\alpha\beta}
-\delta_{\mu\alpha}\delta_{\nu\beta} - \delta_{\mu\beta} \delta_{\nu\alpha}]
\partial_\alpha \partial_\beta B_\nu  =  \partial^2 B_\mu \,.
\end{equation}
Hence,
\begin{equation}
[(A+1) \partial^2_\nu  - Bm^2] B_\mu =0\,.
\end{equation}
If $A+1 =B$ we have the spin-1 particles with masses $\pm m$ with the correct relativistic dispersion.

The equation (\ref{vd2equat}) can be transformed in  two equations:
 \begin{eqnarray}
\left [\gamma_{\alpha\beta} \partial_\alpha \partial_\beta + (B+1)\partial_\alpha^2 - Bm^2 \right ]_{\mu\nu} B_\nu &=&0,\,\mbox{spin 0 with} \pm m,\nonumber\\
&&\\
\left [\gamma_{\alpha\beta} \partial_\alpha \partial_\beta + (B-1)\partial_\alpha^2  - Bm^2 \right ]_{\mu\nu}  B_\nu &=&0,\, \mbox{spin 1 with} \pm m.\nonumber\\
\end{eqnarray}
 
The first one has the solution with spin 0 and masses $\pm m$. However, it has also the spin-1
solution with the {\it different} masses, $[\partial_\nu^2 +(B+1)\partial^2_\nu - Bm^2 ] B_\mu =0$:
\begin{equation}
\tilde m = \pm \sqrt{\frac{B}{B+2}} m\,.
\end{equation}
The second one has the solution with spin 1 and masses $\pm m$. But, it also has 
the {\it spin-0} solution with the {\it different} masses, $[ -\partial_\nu^2 + (B-1) \partial^2_\nu - Bm^2 ] \partial_\mu \chi =0$. So,
$\tilde m = \pm \sqrt{\frac{B}{B-2}}m$.
One can come to the same conclusion by checking the dispersion relations 
from $\mbox{Det} [\gamma_{\alpha\beta} p_\alpha p_\beta - Ap_\alpha p_\alpha +Bm^2] = 0$\,. When $\tilde m^2 = \frac{4}{3} m^2$, we have $B=-8, A=-7$, that is compatible with our consideration of bi-vector fields, Ref.~\cite{vd2wig}.
Thus, one can form the Lagrangian with the particles of spines 1, masses $\pm m$, the particle with the mass $\sqrt{\frac{4}{3}} m$, spin 1, for which the particle is equal to the antiparticle, by choosing the appropriate creation/annihilation operators; and the particles with spines 0 with masses $\pm m$ and 
$\pm \sqrt{\frac{4}{5}} m$ (some of them may be neutral).

{\it Energy-momentum tensor.}
According to Ref.~\cite{vd2Lurie}, it is defined as
 \begin{eqnarray}
&&T_{\mu\nu} = - \sum_{\alpha}^{} \left [ \frac{\partial {\cal L}}{\partial (\partial_\mu B_\alpha)} \partial_\nu B_\alpha
+\partial_\nu B_\alpha^\ast \frac{\partial {\cal L}}{\partial (\partial_\mu B_\alpha^\ast)}\right ]
+{\cal L} \delta_{\mu\nu}\\
&& P_\mu = -i \int T_{4\mu} d^3 {\bf x}\,.
\end{eqnarray}
\begin{eqnarray}
&&T_{\mu\nu} =  -(\partial_\kappa B_\tau^\ast) [\gamma_{\kappa\mu}]_{\tau\alpha} (\partial_\nu B_\alpha) - (\partial_\nu B_\alpha^\ast) [\gamma_{\mu\kappa}]_{\alpha\tau} (\partial_\kappa B_\tau)-\nonumber\\
&-& A [(\partial_\mu B_\alpha^\ast) (\partial_\nu B_\alpha) + (\partial_\nu B_\alpha^\ast)  (\partial_\mu B_\alpha)] + {\cal L}\delta_{\mu\nu} =\nonumber\\
&=& - (A+1) [(\partial_\mu B_\alpha^\ast) (\partial_\nu B_\alpha) + (\partial_\nu B_\alpha^\ast)  (\partial_\mu B_\alpha)] +
\left [ (\partial_\alpha B_\mu^\ast) (\partial_\nu B_\alpha) + \right . \nonumber\\
&+&\left . (\partial_\nu B_\alpha^\ast)  (\partial_\alpha B_\mu) \right ] +
[(\partial_\alpha B_\alpha^\ast) (\partial_\nu B_\mu) + (\partial_\nu B_\mu^\ast)  (\partial_\alpha B_\alpha)] + {\cal L} \delta_{\mu\nu}\,.
\end{eqnarray}
Remember that after substitutions of  the explicite forms of the $\gamma$'s, the Lagrangian is
\begin{eqnarray}
&&{\cal L} = (A+1) (\partial_\alpha B_\mu^\ast) (\partial_\alpha B_\mu ) - (\partial_\nu B_\mu^\ast)  (\partial_\mu B_\nu)- (\partial_\mu B_\mu^\ast) (\partial_\nu B_\nu)\nonumber\\ 
&+& Bm^2   B_\mu^\ast B_\mu\,,
\end{eqnarray}
and the third term cannot be removed by the standard substitution ${\cal L} \rightarrow {\cal L}^\prime +\partial_\mu \Gamma_\mu$\,, $\Gamma_\mu = B_\nu^\ast \partial_\nu B_\mu - B_\mu^\ast \partial_\nu B_\nu$ 
to get the textbook Lagrangian ${\cal L}^\prime = (\partial_\alpha B_\mu^\ast) (\partial_\alpha B_\mu ) +m^2 B_\mu^\ast B_\mu$\,.

{\it The current vector} is defined
 \begin{eqnarray}
&&J_{\mu} = -i \sum_{\alpha}^{} [\frac{\partial {\cal L}}{\partial (\partial_\mu B_\alpha)} 
B_\alpha
- B_\alpha^\ast \frac{\partial {\cal L}}{\partial (\partial_\mu B_\alpha^\ast)} ]\,,\\
&& Q = -i \int J_{4} d^3 {\bf x}\,.
\end{eqnarray}
\begin{eqnarray}
&&J_{\lambda} =  -i \left \{ (\partial_\alpha B_\mu^\ast) [\gamma_{\alpha\lambda}]_{\mu\kappa}  B_\kappa - B_\kappa^\ast [\gamma_{\lambda\alpha}]_{\kappa\mu} (\partial_\alpha B_\mu)\right.\nonumber\\
&&\left. + A (\partial_\lambda B_\kappa^\ast) B_\kappa - A B_\kappa^\ast (\partial_\lambda B_\kappa) \right \} \nonumber\\
&&= - i \left \{ (A+1) [(\partial_\lambda B_\kappa^\ast) B_\kappa
 -  B_\kappa^\ast (\partial_\lambda B_\kappa) ] +  [ B_\kappa^\ast (\partial_\kappa B_\lambda) -
(\partial_\kappa B_\lambda^\ast) B_\kappa ]  \right . \nonumber\\
&&  \left . +[B_\lambda^\ast (\partial_\kappa B_\kappa) - (\partial_\kappa B_\kappa^\ast) B_\lambda ] \right \} \,.
\end{eqnarray}
Again, the second term and the last term cannot be removed at the same time by adding the total derivative to the Lagrangian. These terms correspond to the contribution of the scalar (spin-0) portion.

{\it Angular momentum.}
Finally,
 \begin{eqnarray}
&&{\cal M}_{\mu\alpha,\lambda} = x_\mu T_{\{\alpha\lambda\}} - x_\alpha T_{\{\mu\lambda\}} + {\cal S}_{\mu\alpha,\lambda} =\nonumber\\
&=& x_\mu T_{\{\alpha\lambda\}} - x_\alpha T_{\{\mu\lambda\}} -  i \left \{\sum_{\kappa\tau}^{}
\frac{\partial {\cal L}}{\partial (\partial_\lambda B_\kappa)} {\cal T}_{\mu\alpha,\kappa\tau} B_\tau+\right.\\
&+&\left. B_\tau^\ast {\cal T}_{\mu\alpha,\kappa\tau} \frac{\partial {\cal L}}{ 
\partial (\partial_\lambda B_\kappa^\ast)}\right \}\nonumber\\
&& {\cal M}_{\mu\nu} = -i \int {\cal M}_{\mu\nu,4} d^3 {\bf x}\,,
\end{eqnarray}
where ${\cal T}_{\mu\alpha,\kappa\tau} \sim [\gamma_{5,\mu\alpha}]_{\kappa\tau}$\,.

{\it The field operator.} Various-type field operators are possible in this representation. Let us remind the textbook procedure to get them.
During the calculations below we have to present $1=\theta (k_0) +\theta (-k_0)$
in order to get positive- and negative-frequency parts.
\begin{eqnarray}
&&A_\mu (x) = \frac{1}{(2\pi)^3} \int d^4 k \,\delta (k^2 -m^2) e^{+ik\cdot x}
A_\mu (k) =\nonumber\\
&=&\frac{1}{(2\pi)^3} \int \frac{d^3 {\bf k}}{2E_k} \theta(k_0)  
[A_\mu (k) e^{+ik\cdot x}  + A_\mu (-k) e^{-ik\cdot x} ]
\nonumber\\
&=&\frac{1}{(2\pi)^3} \sum_{\lambda}^{}\int \frac{d^3 {\bf k}}{2E_k}   
[\epsilon_\mu (k,\lambda) a_\lambda (k) e^{+ik\cdot x}  + \epsilon_\mu (-k,\lambda) 
a_\lambda (-k) e^{-ik\cdot x} ]\,.\nonumber\\
\end{eqnarray}
Moreover, we should transform the second part to $\epsilon_\mu^\ast (k,\lambda) b_\lambda^\dagger (k)$ as usual. In such a way we obtain the charge-conjugate states. Of course, one can try to get $P$-conjugates or $CP$-conjugate states too. 
We set
\begin{equation}
\sum_{\lambda}^{} \epsilon_\mu (-k,\lambda) a_\lambda (-k) = 
\sum_{\lambda}^{} \epsilon_\mu^\ast (k,\lambda) b_\lambda^\dagger (k)\,,
\label{vd2expan}
\end{equation}
multiply both parts by $\epsilon_\nu [\gamma_{44}]_{\nu\mu}$, and use the normalization conditions for polarization vectors.

In the $(\frac{1}{2}, \frac{1}{2})$ representation we can also expand
the second term  in the different way:
\begin{equation}
\sum_{\lambda}^{} \epsilon_\mu (-k, \lambda) a_\lambda (-k) =
\sum_{\lambda}^{} \epsilon_\mu (k, \lambda) a_\lambda (k)\,.
\end{equation}
From the first definition we obtain (the signs $\mp$
depends on the value of $\sigma$):
\begin{equation}
b_\sigma^\dagger (k) = \mp \sum_{\mu\nu\lambda}^{} \epsilon_\nu (k,\sigma) 
[\gamma_{44}]_{\nu\mu} \epsilon_\mu (-k,\lambda) a_\lambda (-k)\,,
\end{equation}
The second definition is $\Lambda^2_{\sigma\lambda} = \mp \sum_{\nu\mu}^{} \epsilon^{\ast}_\nu (k, \sigma) [\gamma_{44}]_{\nu\mu}
\epsilon_\mu (-k, \lambda)$.
The field operator will only destroy particles. 

{\it Propagators.} From Ref.~\cite{vd2Itzyk} it is known for the real vector field:
\begin{eqnarray}
&&<0\vert T(B_\mu (x) B_\nu (y)\vert 0> =\\
&& -i \int \frac{d^4 k}{(2\pi)^4} e^{ik (x-y)} 
(\frac{\delta_{\mu\nu} +k_\mu k_\nu/\mu^2}{k^2 +\mu^2 +i\epsilon} 
- \frac{k_\mu k_\nu/\mu^2}{k^2 +m^2 +i\epsilon})\,.\nonumber
\end{eqnarray}
If $\mu=m$ (this depends on relations between $A$ and  $B$) we have the cancellation of divergent parts. Thus, we can overcome the well-known  difficulty of the Proca theory with the massless limit. 

If $\mu\neq m$ we can still have a {\it causal} theory, but in this case we need more than one equation, and should apply the method proposed 
in Ref.~\cite{vd2dv-hpa}.
The reasons were that the Weinberg equation propagates both causal and tachyonic solutions. 

{\it Indefinite metrics.}
Usually, one considers the hermitian field operator in the pseudo-Euclidean metric for the electromagnetic potential:
\begin{equation}
A_\mu = \sum_{\lambda}^{} \int \frac{d^3 {\bf k}}{(2\pi)^3 2E_k} 
[\epsilon_\mu (k,\lambda) a_\lambda ({\bf k}) +\epsilon_\mu^\ast (k,\lambda)
a_\lambda^\dagger ({\bf k})]
\end{equation}
with {\it all} four polarizations to be independent ones. Next, one introduces the Lorentz condition in the weak form
\begin{equation}
[a_{0_t} ({\bf k}) - a_0 ({\bf k})] \vert \phi> =0
\end{equation} 
and the indefinite metrics in the Fock space, Ref.~\cite{vd2Bogol}:
$a_{0_t}^\ast = -a_{0_t}$ and $\eta a_\lambda = -a^\lambda \eta$, $\eta^2 =1$,
in order to get the correct sign in the energy-momentum vector
and to not have the problem with the vacuum average.

We observe:
1) that the indefinite metric problems may appear even on the massive level
in the Stueckelberg formalism; 2) The Stueckelberg theory has a good massless limit for propagators, and it reproduces the handling of the indefinite metric in the massless limit (the electromagnetic 4-potential case); 3) we generalized the Stueckelberg formalism (considering, at least, two equations); instead of charge-conjugate solutions we may consider the $P-$  or $CP-$ conjugates. The potential field becomes to be the complex-valued field, that may justify the introduction of the anti-hermitian amplitudes.

\section{The Spin-2 Case}

The general scheme for derivation of higher-spin equations
was given in~\cite{vd2bw-hs}. A field of rest mass $m$ and spin $j \geq \frac{1}{2}$ is represented by a completely symmetric multispinor of rank $2j$.
The particular cases $j=1$ and $j=\frac{3}{2}$ have been given in the
textbooks, e.~g., ref.~\cite{vd2Lurie}. The spin-2 case can also be of some
interest because it is generally believed that the essential features of
the gravitational field are  obtained from transverse components of the
$(2,0)\oplus (0,2)$  representation of the Lorentz group. Nevertheless,
questions of the redandant components of the higher-spin relativistic
equations have not yet been understood in detail.

In this section we use the commonly-accepted procedure
for the derivation  of higher-spin equations.
We begin with the equations for the 4-rank symmetric spinor:
\begin{eqnarray}
&&\left [ i\gamma^\mu \partial_\mu - m \right ]_{\alpha\alpha^\prime}
\Psi_{\alpha^\prime \beta\gamma\delta} = 0\, ,
\left [ i\gamma^\mu \partial_\mu - m \right ]_{\beta\beta^\prime}
\Psi_{\alpha\beta^\prime \gamma\delta} = 0\\
&&\left [ i\gamma^\mu \partial_\mu - m \right ]_{\gamma\gamma^\prime}
\Psi_{\alpha\beta\gamma^\prime \delta} = 0\, ,
\left [ i\gamma^\mu \partial_\mu - m \right ]_{\delta\delta^\prime}
\Psi_{\alpha\beta\gamma\delta^\prime} = 0.
\end{eqnarray} 
The massless limit (if one needs) should be taken in the end of all
calculations.

We proceed expanding the field function in the set of symmetric matrices
(as in the spin-1 case). The total function is
\begin{eqnarray}
\lefteqn{\Psi_{\{\alpha\beta\}\{\gamma\delta\}}
= (\gamma_\mu R)_{\alpha\beta} (\gamma^\kappa R)_{\gamma\delta}
G_\kappa^{\quad \mu} + (\gamma_\mu R)_{\alpha\beta} (\sigma^{\kappa\tau}
R)_{\gamma\delta} F_{\kappa\tau}^{\quad \mu} + } \nonumber\\
&+& (\sigma_{\mu\nu} R)_{\alpha\beta} (\gamma^\kappa R)_{\gamma\delta}
T_\kappa^{\quad \mu\nu} + (\sigma_{\mu\nu} R)_{\alpha\beta}
(\sigma^{\kappa\tau} R)_{\gamma\delta} R_{\kappa\tau}^{\quad\mu\nu};
\end{eqnarray}
and the resulting tensor equations are:
\begin{eqnarray}
&&\frac{2}{m} \partial_\mu T_\kappa^{\quad \mu\nu} =
-G_{\kappa}^{\quad\nu}\, ,
\frac{2}{m} \partial_\mu R_{\kappa\tau}^{\quad \mu\nu} =
-F_{\kappa\tau}^{\quad\nu}\, ,\\
&& T_{\kappa}^{\quad \mu\nu} = \frac{1}{2m} \left [
\partial^\mu G_{\kappa}^{\quad\nu}
- \partial^\nu G_{\kappa}^{\quad \mu} \right ] \, ,\\
&& R_{\kappa\tau}^{\quad \mu\nu} = \frac{1}{2m} \left [
\partial^\mu F_{\kappa\tau}^{\quad\nu}
- \partial^\nu F_{\kappa\tau}^{\quad \mu} \right ] \, .
\end{eqnarray}
The constraints are re-written to
\begin{eqnarray}
&&\frac{1}{m} \partial_\mu G_\kappa^{\quad\mu} = 0\, ,\quad
\frac{1}{m} \partial_\mu F_{\kappa\tau}^{\quad\mu} =0\, ,\\
&& \frac{1}{m} \epsilon_{\alpha\beta\nu\mu} \partial^\alpha
T_\kappa^{\quad\beta\nu} = 0\, ,\quad
\frac{1}{m} \epsilon_{\alpha\beta\nu\mu} \partial^\alpha
R_{\kappa\tau}^{\quad\beta\nu} = 0\, .
\end{eqnarray}
However, we need to make symmetrization over these two sets
of indices $\{ \alpha\beta \}$ and $\{\gamma\delta \}$. The total
symmetry can be ensured if one contracts the function $\Psi_{\{\alpha\beta
\} \{\gamma \delta \}}$ with {\it antisymmetric} matrices
$R^{-1}_{\beta\gamma}$, $(R^{-1} \gamma^5 )_{\beta\gamma}$ and
$(R^{-1} \gamma^5 \gamma^\lambda )_{\beta\gamma}$ and equate
all these contractions to zero (similar to the $j=3/2$ case
considered in ref.~\cite[p. 44]{vd2Lurie}.  
We  encountered with
the known difficulty of the theory for spin-2 particles in
the Minkowski space.
We explicitly showed that all field functions become to be equal to zero.
Such a situation cannot be considered as a satisfactory one (because it
does not give us any physical information) and can be corrected in several
ways. We  modified the formalism~\cite{vd2dv-ps}. The field function is now presented as
\begin{equation}
\Psi_{\{\alpha\beta\}\gamma\delta} =
\alpha_1 (\gamma_\mu R)_{\alpha\beta} \Psi^\mu_{\gamma\delta} +
\alpha_2 (\sigma_{\mu\nu} R)_{\alpha\beta} \Psi^{\mu\nu}_{\gamma\delta}
+\alpha_3 (\gamma^5 \sigma_{\mu\nu} R)_{\alpha\beta}
\widetilde \Psi^{\mu\nu}_{\gamma\delta}\, ,
\end{equation}
with
\begin{align}
\Psi^\mu_{\{\gamma\delta\}} &= \beta_1 (\gamma^\kappa R)_{\gamma\delta}
G_\kappa^{\quad\mu} + \beta_2 (\sigma^{\kappa\tau} R)_{\gamma\delta}
F_{\kappa\tau}^{\quad\mu} +\beta_3 (\gamma^5 \sigma^{\kappa\tau}
R)_{\gamma\delta} \widetilde F_{\kappa\tau}^{\quad\mu},\\
\Psi^{\mu\nu}_{\{\gamma\delta\}} &=\beta_4 (\gamma^\kappa
R)_{\gamma\delta} T_\kappa^{\quad\mu\nu} + \beta_5 (\sigma^{\kappa\tau}
R)_{\gamma\delta} R_{\kappa\tau}^{\quad\mu\nu} +\beta_6 (\gamma^5
\sigma^{\kappa\tau} R)_{\gamma\delta}
\widetilde R_{\kappa\tau}^{\quad\mu\nu},\\
\widetilde \Psi^{\mu\nu}_{\{\gamma\delta\}} &=\beta_7 (\gamma^\kappa
R)_{\gamma\delta} \widetilde T_\kappa^{\quad\mu\nu} + \beta_8
(\sigma^{\kappa\tau} R)_{\gamma\delta}
\widetilde D_{\kappa\tau}^{\quad\mu\nu}
+\beta_9 (\gamma^5 \sigma^{\kappa\tau} R)_{\gamma\delta}
D_{\kappa\tau}^{\quad\mu\nu}.
\end{align}
Hence, the function $\Psi_{\{\alpha\beta\}\{\gamma\delta\}}$
can be expressed as a sum of nine terms:
\begin{eqnarray}
&&\Psi_{\{\alpha\beta\}\{\gamma\delta\}} =
\alpha_1 \beta_1 (\gamma_\mu R)_{\alpha\beta} (\gamma^\kappa
R)_{\gamma\delta} G_\kappa^{\quad\mu} +\alpha_1 \beta_2
(\gamma_\mu R)_{\alpha\beta} (\sigma^{\kappa\tau} R)_{\gamma\delta}
F_{\kappa\tau}^{\quad\mu} + \nonumber\\
&+&\alpha_1 \beta_3 (\gamma_\mu R)_{\alpha\beta}
(\gamma^5 \sigma^{\kappa\tau} R)_{\gamma\delta} \widetilde
F_{\kappa\tau}^{\quad\mu} +
+ \alpha_2 \beta_4 (\sigma_{\mu\nu}
R)_{\alpha\beta} (\gamma^\kappa R)_{\gamma\delta} T_\kappa^{\quad\mu\nu}
+\nonumber\\
&+&\alpha_2 \beta_5 (\sigma_{\mu\nu} R)_{\alpha\beta} (\sigma^{\kappa\tau}
R)_{\gamma\delta} R_{\kappa\tau}^{\quad \mu\nu}
+ \alpha_2
\beta_6 (\sigma_{\mu\nu} R)_{\alpha\beta} (\gamma^5 \sigma^{\kappa\tau}
R)_{\gamma\delta} \widetilde R_{\kappa\tau}^{\quad\mu\nu} +\nonumber\\
&+&\alpha_3 \beta_7 (\gamma^5 \sigma_{\mu\nu} R)_{\alpha\beta}
(\gamma^\kappa R)_{\gamma\delta} \widetilde
T_\kappa^{\quad\mu\nu}+
\alpha_3 \beta_8 (\gamma^5
\sigma_{\mu\nu} R)_{\alpha\beta} (\sigma^{\kappa\tau} R)_{\gamma\delta}
\widetilde D_{\kappa\tau}^{\quad\mu\nu} +\nonumber\\
&+&\alpha_3 \beta_9
(\gamma^5 \sigma_{\mu\nu} R)_{\alpha\beta} (\gamma^5 \sigma^{\kappa\tau}
R)_{\gamma\delta} D_{\kappa\tau}^{\quad \mu\nu}\, .
\label{vd2ffn1}
\end{eqnarray}
The corresponding dynamical
equations are given by the set
\begin{eqnarray}
&& \frac{2\alpha_2
\beta_4}{m} \partial_\nu T_\kappa^{\quad\mu\nu} +\frac{i\alpha_3
\beta_7}{m} \epsilon^{\mu\nu\alpha\beta} \partial_\nu
\widetilde T_{\kappa,\alpha\beta} = \alpha_1 \beta_1
G_\kappa^{\quad\mu}\,; \label{vd2b}\\
&&\frac{2\alpha_2 \beta_5}{m} \partial_\nu
R_{\kappa\tau}^{\quad\mu\nu} +\frac{i\alpha_2 \beta_6}{m}
\epsilon_{\alpha\beta\kappa\tau} \partial_\nu \widetilde R^{\alpha\beta,
\mu\nu} +\frac{i\alpha_3 \beta_8}{m}
\epsilon^{\mu\nu\alpha\beta}\partial_\nu \widetilde
D_{\kappa\tau,\alpha\beta} - \nonumber\\
&-&\frac{\alpha_3 \beta_9}{2}
\epsilon^{\mu\nu\alpha\beta} \epsilon_{\lambda\delta\kappa\tau}
D^{\lambda\delta}_{\quad \alpha\beta} = \alpha_1 \beta_2
F_{\kappa\tau}^{\quad\mu} + \frac{i\alpha_1 \beta_3}{2}
\epsilon_{\alpha\beta\kappa\tau} \widetilde F^{\alpha\beta,\mu}\,; \\
&& 2\alpha_2 \beta_4 T_\kappa^{\quad\mu\nu} +i\alpha_3 \beta_7
\epsilon^{\alpha\beta\mu\nu} \widetilde T_{\kappa,\alpha\beta}
=  \frac{\alpha_1 \beta_1}{m} (\partial^\mu G_\kappa^{\quad \nu}
- \partial^\nu G_\kappa^{\quad\mu})\,; \\
&& 2\alpha_2 \beta_5 R_{\kappa\tau}^{\quad\mu\nu} +i\alpha_3 \beta_8
\epsilon^{\alpha\beta\mu\nu} \widetilde D_{\kappa\tau,\alpha\beta}
+i\alpha_2 \beta_6 \epsilon_{\alpha\beta\kappa\tau} \widetilde
R^{\alpha\beta,\mu\nu}-\nonumber\\
&-& \frac{\alpha_3 \beta_9}{2} \epsilon^{\alpha\beta\mu\nu}
\epsilon_{\lambda\delta\kappa\tau} D^{\lambda\delta}_{\quad \alpha\beta}
= \nonumber\\
&=& \frac{\alpha_1 \beta_2}{m} (\partial^\mu F_{\kappa\tau}^{\quad \nu}
-\partial^\nu F_{\kappa\tau}^{\quad\mu} ) + \frac{i\alpha_1 \beta_3}{2m}
\epsilon_{\alpha\beta\kappa\tau} (\partial^\mu \widetilde
F^{\alpha\beta,\nu} - \partial^\nu \widetilde F^{\alpha\beta,\mu} )\, .
\label{vd2f}
\end{eqnarray}
The essential constraints can be found in Ref.~\cite{vd2aaca}.
They are  the results of contractions of the field function (\ref{vd2ffn1})
with three antisymmetric matrices, as above. Furthermore,
one should recover the above relations in the particular
case when $\alpha_3 = \beta_3 =\beta_6 = \beta_9 = 0$ and
$\alpha_1 = \alpha_2 = \beta_1 =\beta_2 =\beta_4
=\beta_5 = \beta_7 =\beta_8 =1$.

As a discussion we note that in such a framework we have already physical
content because only certain combinations of field functions
would be equal to zero. In general, the fields
$F_{\kappa\tau}^{\quad\mu}$, $\widetilde F_{\kappa\tau}^{\quad\mu}$,
$T_{\kappa}^{\quad\mu\nu}$, $\widetilde T_{\kappa}^{\quad\mu\nu}$, and
$R_{\kappa\tau}^{\quad\mu\nu}$,  $\widetilde
R_{\kappa\tau}^{\quad\mu\nu}$, $D_{\kappa\tau}^{\quad\mu\nu}$, $\widetilde
D_{\kappa\tau}^{\quad\mu\nu}$ can  correspond to different physical states
and the equations above describe oscillations one state to another.
Furthermore, from the set of equations (\ref{vd2b}-\ref{vd2f}) one
obtains the {\it second}-order equation for symmetric traceless tensor of
the second rank ($\alpha_1 \neq 0$, $\beta_1 \neq 0$):
\begin{equation} \frac{1}{m^2} \left [\partial_\nu
\partial^\mu G_\kappa^{\quad \nu} - \partial_\nu \partial^\nu
G_\kappa^{\quad\mu} \right ] =  G_\kappa^{\quad \mu}\, .
\end{equation}
After the contraction in indices $\kappa$ and $\mu$ this equation is
reduced to the set
\begin{eqnarray}
&&\partial_\mu G_{\quad\kappa}^{\mu} = F_\kappa\,,  \\
&&\frac{1}{m^2} \partial_\kappa F^\kappa = 0\, ,
\end{eqnarray}
i.~e.,  to the equations connecting the analogue of the energy-momentum
tensor and the analogue of the 4-vector potential. 
Further investigations may provide additional foundations to
``surprising" similarities of gravitational and electromagnetic
equations in the low-velocity limit.

\section{Noncommutativity}

The questions of "non-commutativity" see, for instance, in Ref.~\cite{vd2Bled}.
The assumption that operators of
coordinates do {\it not} commute $[\hat{x}_{\mu },\hat{x}_{\nu }]_{-} = i\theta_{\mu\nu}$ (or, alternatively, $[\hat{x}_{\mu },\hat{x}_{\nu }]_{-}= iC_{\mu\nu}^\beta x_\beta$)
has been first made by H. Snyder~\cite{vd2snyder}. Later it was shown that such an anzatz may lead to non-locality. Thus, the Lorentz symmetry may be
broken. Recently, some attention has again been paid to this 
idea~\cite{vd2noncom} in the context of ``brane theories''.

On the other hand, the famous Feynman-Dyson proof of Maxwell equations~\cite{vd2FD}
contains intrinsically the non-commutativity of velocities. While 
$$[ x^i, x^j ]_-=0$$ 
therein, but  
$$[ \dot x^i (t), 
\dot x^j (t) ]_- = \frac{i\hbar}{m^2} \epsilon^{ijk} B_k \neq 0$$ 
(at the same time with $[x^i, \dot x^j]_- = \frac{i\hbar}{m} \delta^{ij}$) that also may be considered as a contradiction with
the well-accepted theories. Dyson wrote in a very clever way: ``Feynman in 1948 was not alone in trying to build theories outside the framework of conventional physics... All these radical programms, including Feynman's, failed... I venture to disagree with Feynman now, as I often did while he was alive..."

Furthermore, it was recently shown that notation and terminology, 
which physicists used when speaking about partial
derivative of many-variables functions, are sometimes 
confusing, see the discussion in~\cite{vd2eld}. Some authors clai\-med~\cite{vd2chja}:
``this equation [cannot be correct] because the partial differentiation would involve increments of the functions ${\bf r} (t)$ in the form ${\bf r} (t) +\Delta {\bf r} (t)$ and we do not know how we must interpret this increment because we have two options: {\it either} $\Delta {\bf r} (t) = {\bf r} (t) - {\bf r}^\ast (t)$, {\it or} $\Delta {\bf r} (t) = {\bf r} (t) - {\bf r} (t^\ast)$. Both are different processes because the first one involves changes in the functional form of the functions ${\bf r} (t)$, while the second involves changes in the position along the path defined by ${\bf r} = {\bf r} (t)$ but preserving the same functional form." 

Another well-known physical example of the situation, when we have both explicite and implicite dependences of the function which derivatives act upon, is the field of an accelerated charge~ \cite{vd2landau}.
First, Landau and Lifshitz wrote that the functions depended on the retarded time $t^{\prime }$
and only through $t^{\prime }+R(t^{\prime })/c=t$ they depended implicitly
on $x,y,z,t$. However, later they used
the explicit dependence of $R$ and fields on the space coordinates of the
observation point too. Otherwise, the ``simply" retarded fields do not satisfy the Maxwell equations.  So, actually
the fields and the potentials are the functions of the following forms:
$A^\mu (x, y, z, t' (x,y,z,t)), {\bf E} (x, y, z, t' (x,y,z,t)), {\bf B} (x, y, z, t' 
(x,y,z,t))$. 

In~\cite{vd2Bled} I studied the case when we deal with explicite and implicite dependencies  $f ({\bf p}, E ({\bf p}))$. It is well known that the energy in the
relativism is connected with the 3-momentum as $E=\pm \sqrt{{\bf p}^2 +m^2}$
; the unit system $c=\hbar=1$ is used. In other words, we must choose the
3-dimensional hyperboloid from the entire Minkowski space and the energy is 
{\it not} an independent quantity anymore. Let us calculate the commutator
of the whole derivative $\hat\partial /\hat\partial E$ and $\hat\partial / 
\hat\partial p_i$.\footnote{
In order to make distinction between differentiating the explicit function
and that which contains both explicit and implicit dependencies, the `whole
partial derivative' may be denoted as $\hat\partial$.} In the general case
one has 
\begin{equation}
{\frac{\hat\partial f ({\bf p}, E({\bf p})) }{\hat\partial p_i}} \equiv {
\frac{\partial f ({\bf p}, E({\bf p})) }{\partial p_i}} + {\frac{\partial f (
{\bf p}, E({\bf p})) }{\partial E}} {\frac{\partial E}{\partial p_i}}\, .
\end{equation}
Applying this rule, we surprisingly find 
\begin{eqnarray}
&&[{\frac{\hat\partial }{\hat\partial p_i}},{\frac{\hat\partial }{\hat
\partial E}}]_- f ({\bf p},E ({\bf p})) = {\frac{\hat\partial }{\hat\partial
p_i}} {\frac{\partial f }{\partial E}} -{\frac{\partial }{\partial E}} ({
\frac{\partial f}{\partial p_i}} +{\frac{\partial f}{\partial E}}{\frac{
\partial E}{\partial p_i}}) =  \nonumber \\
&=& - {\frac{\partial f }{\partial E}} {\frac{\partial
}{\partial E}}({\frac{\partial E}{\partial p_i}})\,.  \label{vd2com}
\end{eqnarray}
So, if $E=\pm \sqrt{m^2+{\bf p}^2}$ 
and one uses the generally-accepted 
representation form of $\partial E/\partial p_i
=  p^i/E$,
one has that the expression (\ref{vd2com})
appears to be equal to $(p_i/E^2) {\frac{\partial f({\bf p}, E ({\bf p}))}{
\partial E}}$. 
On the other hand, the commutator 
\begin{equation}
[{\frac{\hat\partial}{\hat\partial p_i}}, {\frac{\hat\partial}{\hat\partial
p_j}}]_- f ({\bf p},E ({\bf p})) = {\frac{1}{E^3}} {\frac{
\partial f({\bf p}, E ({\bf p}))}{\partial E}} [p_i, p_j]_-\,.
\end{equation}
This may be considered to be zero unless we would trust to the genious
Feynman. He postulated that the velocity (or, of course, the 3-momentum)
commutator is equal to $[p_i,p_j]\sim i\hbar\epsilon_{ijk} B^k$, i.e., to
the magnetic field.

Furthermore, since the energy derivative corresponds to the operator of time
and the $i$-component momentum derivative, to $\hat x_i$, we put forward the
following anzatz in the momentum representation: 
\begin{equation}
[\hat x^\mu, \hat x^\nu]_- = \omega ({\bf p}, E({\bf p})) \,
F^{\mu\nu}_{\vert\vert}{\frac{\partial }{\partial E}}\,,
\end{equation}
with some weight function $\omega$ being different for different choices of
the antisymmetric tensor spin basis. In the modern literature, the idea of the broken Lorentz invariance by this method is widely discussed, see e.g.~\cite{vd2amelino}. 

Let us turn now to the application of the presented ideas to the Dirac case.
Recently, we analized Sakurai-van der Waerden method of derivations of the Dirac
(and higher-spins too) equation~\cite{vd2Dvoh}. We can start from
\begin{equation}
(E I^{(2)}-{\mathbf \sigma}\cdot {\bf p}) (E I^{(2)}+ {\mathbf\sigma}\cdot
{\bf p} ) \Psi_{(2)} = m^2 \Psi_{(2)} \,,
\end{equation}
or
\begin{equation}
(E I^{(4)}+{\mathbf \alpha}\cdot {\bf p} +m\beta) (E I^{(4)}-{\mathbf\alpha}\cdot
{\bf p} -m\beta ) \Psi_{(4)} =0.\label{vd2f4}
\end{equation}
Of course, as in the original Dirac work, we have
\begin{equation}
\beta^2 = 1\,,\quad
\alpha^i \beta +\beta \alpha^i =0\,,\quad
\alpha^i \alpha^j +\alpha^j \alpha^i =2\delta^{ij} \,.
\end{equation}
For instance, their explicite forms can be chosen 
\begin{eqnarray}
\alpha^i =\begin{pmatrix}
\sigma^i& 0\cr
0&-\sigma^i\cr
\end{pmatrix}\,,\quad
\beta = \begin{pmatrix}0&1_{2\times 2}\cr
1_{2\times 2} &0\cr
\end{pmatrix}\,,
\end{eqnarray}
where $\sigma^i$ are the ordinary Pauli $2\times 2$ matrices.
We also postulate the non-commuta\-tivity
\begin{equation}
[E, {\bf p}^i]_- = \Theta^{0i} = \theta^i,,
\end{equation}
as usual. Therefore the equation (\ref{vd2f4}) will {\it not} lead
to the well-known equation $E^2 -{\bf p}^2 = m^2$. Instead, we have
\begin{equation}
\left \{ E^2 - E ({\mathbf \alpha} \cdot {\bf p})
+({\mathbf \alpha} \cdot {\bf p}) E - {\bf p}^2 - m^2 - i {\mathbf\sigma}\times I_{(2)}
[{\bf p}\otimes {\bf p}] \right \} \Psi_{(4)} = 0
\end{equation}
For the sake of simplicity, we may assume the last term to be zero. Thus we come to
\begin{equation}
\left \{ E^2 - {\bf p}^2 - m^2 -  ({\mathbf \alpha}\cdot {\mathbf \theta})
\right \} \Psi_{(4)} = 0\,.
\end{equation} 
However, let us make the unitary transformation. It is known~\cite{vd2Berg}
that one can\footnote{Of course, the certain relations for the components ${\bf a}$ should be assumed. Moreover, in our case ${\bf \theta}$ should not depend on $E$ and ${\bf p}$. Otherwise, we must take the noncommutativity $[E, {\bf p}^i]_-$ again.}
\begin{equation}
U_1 ({\mathbf \sigma}\cdot {\bf a}) U_1^{-1} = \sigma_3 \vert {\bf a} \vert\,.\label{vd2s3}
\end{equation}
For ${\mathbf \alpha}$ matrices we re-write (\ref{vd2s3}) to
\begin{eqnarray}
U_1 ({\mathbf \alpha}\cdot {\mathbf \theta}) U_1^{-1} = \vert {\mathbf \theta} \vert
\begin{pmatrix}
1&0&0&0\cr
0&-1&0&0\cr
0&0&-1&0\cr
0&0&0&1\cr
\end{pmatrix} = \alpha_3 \vert {\mathbf\theta}\vert\,.
\end{eqnarray}
applying the second unitary transformation:
\begin{eqnarray}
U_2 \alpha_3 U_2^\dagger =
\begin{pmatrix}
1&0&0&0\cr
0&0&0&1\cr
0&0&1&0\cr
0&1&0&0\cr
\end{pmatrix} \alpha_3 \begin{pmatrix}
1&0&0&0\cr
0&0&0&1\cr
0&0&1&0\cr
0&1&0&0\cr
\end{pmatrix} = \begin{pmatrix}
1&0&0&0\cr
0&1&0&0\cr
0&0&-1&0\cr
0&0&0&-1\cr
\end{pmatrix}\,.
\end{eqnarray}
The final equation is
\begin{equation}
[E^2 -{\bf p}^2 -m^2 -\gamma^5_{chiral} \vert {\mathbf \theta}\vert ] \Psi^\prime_{(4)} = 0\,.
\end{equation}
In the physical sense this implies the mass splitting for a Dirac particle over the non-commutative space, $m_{1,2}= \sqrt{m^2 \pm \vert {\mathbf \theta}\vert}$. This procedure may be attractive for explanation of the mass creation and the mass splitting for fermions.

\section{Conclusions}

\begin{itemize}

\item
The $(1/2,1/2)$ representation contains both the  spin-1 and spin-0
states (cf. with the Stueckelberg formalism).

\item
Unless we take into account the fourth state (the ``time-like" state, or
the spin-0 state) the set of 4-vectors is {\it not} a complete set in a mathematical sense.

\item
We cannot remove terms like $(\partial_\mu B^\ast_\mu)(\partial_\nu B_\nu)$ 
terms from the Lagrangian and dynamical invariants unless apply the Fermi 
method, i.~e., manually. The Lorentz condition applies only to the spin 1 states.

\item
We have some additional terms in the expressions of the energy-mo\-men\-tum vector (and, accordingly, of the 4-current and the Pauli-Lunbanski vectors), which are the consequence of the impossibility to apply the Lorentz condition for spin-0 states.

\item
Helicity vectors are not eigenvectors of the parity operator. Meanwhile, the parity is a ``good" quantum number, $[{\cal P}, {\cal H}]_- =0$ in the Fock space.

\item
We are able to describe the states of different masses in this representation from the beginning.

\item
Various-type field operators can be constructed in the $(1/2,1/2)$ representation space. For instance, they can contain $C$, $P$ and $CP$ conjugate states.
Even if $b_\lambda^\dagger =a_\lambda^\dagger$ 
we can have complex 4-vector fields.
We found the relations between creation, annihilation operators for different types of the field operators $B_\mu$.

\item
Propagators have good behavious in the massless limit as opposed to those of the Proca theory.

\item
The spin-2 case can be considered on an equal footing with the spin-1 case.

\item The postulate of non-commutativity leads to the mass spliting for leptons.

\end{itemize}

%% Albino Hernandez Galeana, 14.11.2009, contribution
%%
\newcommand{\AHGgh}{\mbox{$SU(3)$}}
\title{Radiative Charged Fermion Masses and Quark Mixing $(V_{CKM})_{4\times 4}$ in a
$SU(3)$ Gauged Flavor Symmetry Model}
\author{A. Hern\'andez-Galeana}
\institute{% 
Departamento de F\'{\i}sica,   Escuela Superior de
F\'{\i}sica y Matem\'aticas, I.P.N., \\
U. P. "Adolfo L\'opez Mateos". C. P. 07738, M\'exico, D.F.,
M\'exico}

\authorrunning{A. Hern\'andez-Galeana}
\titlerunning{Radiative Charged Fermion Masses and Quark Mixing\ldots}
\maketitle

\begin{abstract}
We report the analysis on charged fermion masses and quark mixing,
within the context of a non-supersymmetric $SU(3)$ gauged family
symmetry model with hierarchical one loop radiative mass
generation mechanism for light fermions, mediated by the massive
bosons associated to the $\AHGgh$ family symmetry that is
spontaneously broken, meanwhile the top and bottom quarks as well
as the tau lepton are generated at tree level by the
implementation of {\bf Dirac See-saw} mechanisms through the
introduction of new vector like fermions. A quantitative analysis
shows that this model is successful to accommodate a realistic
spectrum of masses and mixing in the quark sector as well as the
charged lepton masses. Furthermore, the above scenario enable us
to suppress within current experimental bounds the tree level
$\Delta F=2$ processes for $K^o-\bar{K^o}$ and $D^o-\bar{D^o}$
meson mixing mediated by these extra horizontal gauge bosons.
\end{abstract}

%%
%%Keywords: Fermion masses and mixing, Flavor symmetry.
%%
%%\vspace{5mm}
%%
%%{\footnotesize  PACS: 14.60.Pq, 12.15.Ff, 12.60.-i}
%%
%%

%\pagebreak

%\tableofcontents

%%\pagebreak

\section{ Introduction }

The known hierarchical spectrum of quark masses and mixing as well
as the charged lepton masses has suggested to many model building
theorists that light fermion masses could be generated from
radiative corrections\cite{ahg2earlyradm}, while the mass of the top and
bottom quarks as well as that of the tau lepton are generated at
tree level. This may be understood as a consequence of the breaking
of a symmetry among families ( a horizontal symmetry ). This
symmetry may be discrete \cite{ahg2modeldiscrete}, or continuous,
\cite{ahg2modelcontinuous}. The radiative generation of the light
fermions may be mediated by scalar particles as it is proposed, for
instance, in references \cite{ahg2modelrad,ahg2medscalars} or
also through vectorial bosons as it happens for instance in
"Dynamical Symmetry Breaking" (DSB) theories like " Extended
Technicolor ", \cite{ahg2medgaugebosons}.

%\vspace{2mm}

In this report we address the problem of fermion masses and quark
mixing within a non-supersymmetric $SU(3)$ gauged flavor symmetry
model introduced by the author in \cite{ahg2su3models}. In this model
we introduced a radiative hierarchical mass generation mechanism
in which the masses of the top and bottom quarks as well as for
the tau lepton are generated at tree level by the implementation
of "Dirac See-saw" mechanisms induced by the introduction of a new
generation\footnote{Recently, some authors have pointed out
interesting features regarding the possibility of the existence of
a fourth generation\cite{ahg2fourthge}} of $SU(2)_L$ weak singlet
vector like fermions, where as light families get mass through one
loop radiative corrections, mediated by the massive bosons
associated to the $\AHGgh$ family symmetry that is spontaneously
broken.

\section{Model with $SU(3)$ flavor symmetry}

\subsection{Fermion content}

We define the gauge group symmetry $G\equiv SU(3) \otimes G_{SM} $ , where $SU(3)$ is a flavor
symmetry among families and $G_{SM}\equiv SU(3)_C \otimes SU(2)_L
\otimes U(1)_Y$ is the "Standard Model" gauge group of elementary
particles. The content of fermions assume the ordinary quarks and
leptons assigned under the  $G$ as: $\Psi_q^o = ( 3 , 3 , 2 ,
\frac{1}{3} )_L  \;,\; \Psi_l^o = ( 3 , 1 , 2 , -1 )_L
\;,\;\Psi_u^o = ( 3 , 3, 1 , \frac{4}{3} )_R \;,\; \Psi_d^o = (3,
3 , 1 , -\frac{2}{3} )_R \;,\; \Psi_e^o = (3 , 1 , 1,-2)_R $,
where the last entry correspond to the hypercharge $Y$, and the
electric charge is defined by $Q = T_{3L} + \frac{1}{2} Y$. The
model also includes two types of extra fermions: Right handed
neutrinos $\Psi_\nu^o = ( 3 , 1 , 1 , 0 )_R$, and the $SU(2)_L$
singlet vector like fermions \begin{eqnarray} U_{L,R}^o= ( 1 , 3 , 1 ,
\frac{4}{3} )  \qquad , \qquad D_{L,R}^o = ( 1 , 3 , 1 ,-
\frac{2}{3} ) \label{ahg2eq1} \\ N_{L,R}^o= ( 1 , 1 , 1 , 0 ) \qquad , \qquad
E_{L,R}^o= ( 1 , 1 , 1 , -2 ) \label{ahg2eq2}\end{eqnarray}
The above fermion content and its assignment under the group $G$
make the model anomaly free. After the definition of the gauge
symmetry and the assignment of the ordinary fermions in the
canonical form under the standard model group and in the most
simply non trivial way under the $SU(3)$ family symmetry, the
introduction of the right-handed neutrinos becomes a necessity to
cancel anomalies, while the vector like fermions has been
introduced to give masses at tree level only to the heaviest
family of known fermions through Dirac See-saw mechanisms. These
vectorial fermions play a crucial role to
implement a hierarchical spectrum for quarks and charged lepton masses.

\section{Spontaneous Symmetry breaking}

 The "Spontaneous Symmetry Breaking" (SSB) is proposed to be achieved in the
form:
\begin{equation} G \stackrel{\Lambda_1}{\longrightarrow} SU(2)\otimes G_{SM}
\stackrel{\Lambda_2}{\longrightarrow} G_{SM}
\stackrel{\Lambda_3}{\longrightarrow} SU(3)_C \otimes U(1)_Q 
\label{ahg2eq3}
\end{equation}
in order the model had the possibility to
be consistent with the known low energy physics,  here
$\Lambda_1$, $\Lambda_2$ and $\Lambda_3$ are the scales of SSB.

\subsection{Electroweak symmetry breaking}

To achieve the spontaneous breaking of the electroweak symmetry to
$U(1)_Q$,  we introduce the scalars: $ \Phi = ( 3 , 1 , 2 , -1 )$
and  $\Phi^{\prime} = ( 3 , 1 , 2 , +1 ) $, with the
VEV´s: $\langle \Phi \rangle^T = ( \langle \Phi_1 \rangle ,
\langle \Phi_2 \rangle , \langle \Phi_3 \rangle )$, $\langle
\Phi^{\prime} \rangle^T = ( \langle \Phi^{\prime}_1 \rangle ,
\langle \Phi^{\prime}_2 \rangle , \langle \Phi^{\prime}_3 \rangle
)$; where $T$ means transpose, and
\begin{equation} \qquad \langle \Phi_i \rangle = \frac{1}{\sqrt[]{2}} \left(
\begin{array}{c} v_i
\\ 0  \end{array} \right) \qquad , \qquad
\langle \Phi^{\prime}_i \rangle = \frac{1}{\sqrt[]{2}} \left(
\begin{array}{c} 0
\\ V_i  \end{array} \right) \:.
\label{ahg2eq4}
\end{equation}
Assuming $(v_1, v_2, v_3) \neq (V_1, V_2, V_3)$ with
$v_1^2+v_2^2+v_3^2=V_1^2+V_2^2+V_3^2 $, the contributions from
$\langle \Phi \rangle$ and $\langle \Phi^{\prime} \rangle$ yield
the $W$ gauge boson mass $\frac{1}{2} g^2 (v_1^2+v_2^2+v_3^2)
W^{+} W^{-} $. Hence, if we define as usual $M_W=\frac{1}{2} g v$,
we may write $ v=\sqrt{2} \sqrt{v_1^2+v_2^2+v_3^2} \thickapprox
246$ GeV.

\subsection{$SU(3)$ flavor symmetry breaking}

With the purpose to implement a hierarchical spectrum for charged
fermion mas\-ses, and simultaneously
to achieve the SSB of $SU(3)$, we introduce the scalar fields:
$\eta_i,\;i=1,2,3$ transforming as $(3 , 1 , 1 , 0)$ under the
gauge group and taking the "Vacuum Expectation Values" (VEV's):
\begin{equation} \langle \eta_3 \rangle^T = ( 0 , 0, {\cal{V}}_3) \quad , \quad
\langle \eta_2 \rangle^T = ( 0 , {\cal{V}}_2,0) \quad , \quad
\langle \eta_1 \rangle^T = ( {\cal{V}}_1,0,0) \:, 
\label{ahg2eq5}
\end{equation}
The above scalar fields and
VEV's break completely the $SU(3)$ flavor symmetry. The corresponding
$SU(3)$ 
gauge bosons are defined in Eq.(\ref{ahg2eq12}) through their
couplings to fermions. To simplify
computations we impose a $SU(2)$ global symmetry in the gauge
boson masses. So, we assume ${\cal{V}}_1={\cal{V}}_2 \equiv
{\cal{V}}$ in order to cancel mixing between $Z_1$ and $Z_2$ gauge
bosons. Thus, a natural hierarchy among the VEV´s consistent with
the proposed sequence of SSB in Eq.(\ref{ahg2eq3}) is $
{\cal{V}}_3\:>>\:{\cal{V}} \; \gg
\;\sqrt{v_1^2+v_2^2+v_3^2}=\frac{v}{\sqrt{2}}\simeq
\frac{246\:\text{GeV}}{\sqrt{2}} \backsimeq 173.9 \:\text{GeV} \approx \;m_t $.
Hence, neglecting tiny contributions from electroweak symmetry
breaking, we obtain the gauge bosons masses\footnote{Note
that the $SU(2)$ global symmetry and the hierarchy of the scales
of SSB yield a spectrum of $SU(3)$ gauge boson masses without
mixing}
\begin{multline} g_H^2 \left\{ \frac{1}{2} ({\cal{V}})^2
[\:Z_1^2+(Y_1^1)^2+(Y_1^2)^2\:] + \frac{1}{6}\:[\:2
({\cal{V}}_3)^2+({\cal{V}})^2\:]
 \:Z_2^2          \right. \\ \left. + \frac{1}{4}
(\:({\cal{V}}_3)^2+({\cal{V}})^2\:) [\:(Y_2^1)^2+(Y_2^2)^2
+(Y_3^1)^2+(Y_3^2)^2\:] \right\}.
\label{ahg2eq6}
\end{multline}
Thus, we may define the horizontal boson masses \begin{equation}
\begin{array}{rcl}(M_{Z_1})^2=(M_{Y_1^1})^2=(M_{Y_1^2})^2 & = & M_1^2
\equiv g_H^2 {{\cal{V}}}^2 \:, \\
(M_{Y_2^1})^2=(M_{Y_2^2})^2=(M_{Y_3^1})^2=(M_{Y_3^2})^2 & = & M_2^2
\equiv
\frac{g_H^2}{2} ({{\cal{V}}_3}^2+{{\cal{V}}}^2 )   \\
(M_{Z_2})^2 & = & 4/3 M_2^2 - 1/3 M_1^2 \end{array} \:, 
\label{ahg2eq7}
\end{equation}
with the hierarchy $ M_{Z_2} \gtrsim M_2 > M_1 \gg M_W$. 

\section{ Fermion masses}

\subsection{Dirac See-saw mechanisms}

Now we describe briefly the procedure to get the masses for fermions. 
The analysis is presented explicitly for the charged
lepton sector, with a completely analogous procedure for the $u$
and $d$ quark sectors. With the fields of particles introduced in
the model, we may write the gauge invariant Yukawa couplings:
\begin{equation}
h \bar{\Psi}_l^o \Phi^\prime E_R^o \;\;+\;\; h_3 \bar{\Psi}_e^o
\eta_3 E_L^o \;\;+\;\; h_2 \bar{\Psi}_e^o \eta_2 E_L^o \;\;+\;\;
h_1 \bar{\Psi}_e^o \eta_1 E_L^o \;\;+\;\ M \bar{E}_L^o E_R^o \;\;+
h.c 
\label{ahg2eq8}
\end{equation}
where $M$ is a free mass parameter because its mass term
is gauge invariant, $h$, $h_1$, $h_2$ and $h_3$ are Yukawa
coupling constants. When the involved scalar fields acquire VEV's we
get, in the gauge basis ${\Psi^{o}_{L,R}}^T = ( e^{o} , \mu^{o} ,
\tau^{o}, E^o )_{L,R}$, 
the mass terms $\bar{\Psi}^{o}_L {\cal{M}}^o \Psi^{o}_R + h.c $ where
\begin{equation} {\cal M}^o = \begin{pmatrix} 0 & 0 & 0 & h \:v_1\\ 0 & 0 & 0 & h \:v_2\\
0 & 0 & 0 & h \:v_3\\ - h_1 {\cal{V}} & - h_2 {\cal{V}} & h_3
{\cal{V}}_3 & M \end{pmatrix} \equiv \begin{pmatrix} 0 & 0 & 0 & a_1\\ 0 & 0 & 0 & a_2\\
0 & 0 & 0 & a_3\\ - b_1 & - b_2 & b_3 & c
\end{pmatrix} \;. 
\label{ahg2eq9}
\end{equation}
Notice that ${\cal{M}}^o$ has the same structure of a
See-saw mass matrix, but in this case for Dirac fermion masses.
So, we name ${\cal{M}}^o$ as a {\bf "Dirac See-saw"} mass matrix.
${\cal{M}}^o$ is diagonalized by applying a biunitary
transformation $\Psi^{o}_{L,R} = V^{o}_{L,R} \;\chi_{L,R}^o$. The
orthogonal matrices $V^{o}_L$ and $V^{o}_R$ are obtained
explicitly in the Appendix. From $V_L^o$ and $V_R^o$, and using the
relationships defined in this appendix, one
computes
%\begin{equation} 
\begin{align}
{V^{o}_L}^T {\cal{M}}^{o} \;V^{o}_R &=Diag(0,0,-
\sqrt{\lambda_-},\sqrt{\lambda_+})  \label{ahg2eq10}\\
{V^{o}_L}^T {\cal{M}}^{o} {{\cal{M}}^{o}}^T \;V^{o}_L =
{V^{o}_R}^T {{\cal{M}}^{o}}^T {\cal{M}}^{o} \;V^{o}_R &=
Diag(0,0,\lambda_-,\lambda_+)  \:. \label{ahg2eq11}
\end{align}
%\end{equation}
$\lambda_-$ and $\lambda_+$ are the nonzero
eigenvalues, Eq. (\ref{ahg2eq37}). We see from Eqs.(\ref{ahg2eq10},\ref{ahg2eq11}) that at tree level the
See-saw mechanism yields two massless eigenvalues associated to
the light fermions. The eigenvalue $\sqrt{\lambda_+}$ is
associated with the fourth very heavy fermion, and
$\sqrt{\lambda_-}$ is of the order of the heaviest ordinary
fermion(tau mass).

\subsection{One loop contribution to fermion masses}

\begin{figure}[htp]
\begin{center}
\includegraphics{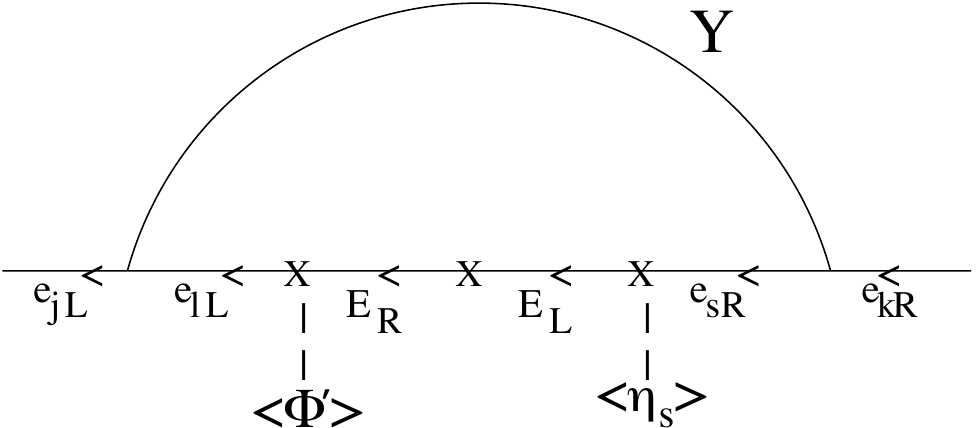}
\end{center}
\caption{\label{ahg2Fig1} Generic one loop diagram contribution to the
mass term $m_{jk} \;{\bar{e}}_{jL}^o  e_{kR}^o$}
\end{figure}

Subsequently the masses for the light fermions arise through one
loop radiative corrections. After the breakdown of the electroweak
symmetry we can construct the generic one loop mass diagram of
Fig.~\ref{ahg2Fig1} . The vertices in this diagram come from the $SU(3)$ flavor
symmetry interaction Lagrangian
\begin{multline} i {\cal{L}}_{int} = \frac{g_{H}}{2} \left\{ (\bar{e^{o}}
\gamma_{\mu} e^{o}- \bar{\mu^{o}} \gamma_{\mu} \mu^{o})Z_1^\mu +
\frac{1}{\sqrt{3}}(\bar{e^{o}} \gamma_{\mu} e^{o}+ \bar{\mu^{o}}
\gamma_{\mu} \mu^{o} - 2 \bar{\tau^{o}}
\gamma_{\mu} \tau^{o})Z_2^\mu    \right.              \\
+ (\bar{e^{o}} \gamma_{\mu} \mu^{o}+ \bar{\mu^{o}} \gamma_{\mu}
e^{o})Y_1^{1 \mu}+(-i\bar{e^{o}} \gamma_{\mu}
\mu^{o}+ i\bar{\mu^{o}} \gamma_{\mu} e^{o})Y_1^{2 \mu}   \\
+(\bar{e^{o}} \gamma_{\mu} \tau^{o}+ \bar{\tau^{o}} \gamma_{\mu}
e^{o})Y_2^{1 \mu}+(-i\bar{e^{o}} \gamma_{\mu}
\tau^{o}+ i\bar{\tau^{o}} \gamma_{\mu} e^{o})Y_2^{2 \mu}  \\
+ \left. (\bar{\mu^{o}} \gamma_{\mu} \tau^{o}+ \bar{\tau^{o}}
\gamma_{\mu} \mu^{o})Y_3^{1 \mu}+(-i\bar{\mu^{o}} \gamma_{\mu}
\tau^{o}+ i\bar{\tau^{o}} \gamma_{\mu} \mu^{o})Y_3^{2 \mu}
\right\} \:,\label{ahg2eq12}\end{multline}
where $g_H$ is the $\AHGgh$ coupling constant, $Z_1$,
$Z_2$ and $Y_i^j\;,i=1,2,3\;,j=1,2$ are the eight gauge bosons.
The crosses in the internal fermion line mean the mixing, and the
mass $M$, generated by the Yukawa couplings in Eq.(\ref{ahg2eq8}), after the
scalar fields take VEV's. The one loop diagram of Fig.~\ref{ahg2Fig1} gives
the generic contribution
\begin{equation} c_Y \frac{\alpha_H}{\pi} \sum_{i=3,4} m_i^o
\:(V_L^o)_{ji}(V_R^o)_{ki} f(M_Y, m_i^o) \qquad , \qquad \alpha_H
\equiv \frac{g_H^2}{4 \pi} \label{ahg2eq13}\end{equation}
to the mass term $m_{jk} \:{\bar{e}}_{jL}^o e_{kR}^o$,
where $M_Y$ is the gauge boson mass, $c_Y$ is a factor coupling
constant, Eq.(\ref{ahg2eq12}), $m_3^o=-\sqrt{\lambda_-}$ and
$m_4^o=\sqrt{\lambda_+}$ are the See-saw mass eigenvalues,
Eq.(\ref{ahg2eq10}), and $f(a,b)=\frac{a^2}{a^2-b^2}
\ln{\frac{a^2}{b^2}}$. Using again the results of Appendix, we compute
\begin{equation} \sum_{i=3,4} m_i^o \:(V_L^o)_{ji}(V_R^o)_{ki} f(M_Y, m_i^o)=
\frac{a_j \:\beta_k \:M}{\lambda_+ - \lambda_-}\:
F(M_Y,\sqrt{\lambda_-},\sqrt{\lambda_+}) \:,\label{ahg2eq14}\end{equation}
with
$F(M_Y,\sqrt{\lambda_-},\sqrt{\lambda_+})\equiv \frac{M_Y^2}{M_Y^2
- \lambda_+} \ln{\frac{M_Y^2}{\lambda_+}} - \frac{M_Y^2}{M_Y^2 -
\lambda_-} \ln{\frac{M_Y^2}{\lambda_-}}\:$, $\beta_1= -b_1$,
$\beta_2= -b_2$ and $\beta_3= b_3$. Adding up all the one loop
$\AHGgh$ gauge boson contributions, we get in the gauge basis the
mass terms $\bar{\Psi^{o}_L} {\cal{M}}_1^o  \:\Psi^{o}_R + h.c.$,
\begin{equation} {\cal{M}}_1^o = \left( \begin{array}{ccrc} R_{11} & R_{12} & R_{13}  & 0\\
R_{21} & R_{22} & R_{23} & 0\\ R_{31} & R_{32} & R_{33} & 0\\
0 & 0 & 0 & 0
\end{array} \right) \:\frac{\alpha_H}{\pi}\;,\label{ahg2eq15}
\end{equation}
%\vspace{1mm}
\begin{align} R_{11} &=-\frac{1}{4} F_1 (m_{11} + 2 m_{22}) - \frac{1}{12}
F_{Z_2} m_{11} + \frac{1}{2} F_2 m_{33}  \:,\nonumber\\
R_{22} &=-\frac{1}{4} F_1 (2 m_{11} + m_{22}) -
\frac{1}{12} F_{Z_2} m_{22} + \frac{1}{2} F_2 m_{33} \:,\nonumber\\
R_{12} &=(\frac{1}{4} F_1 - \frac{1}{12} F_{Z_2}
)m_{12}\quad ,\quad R_{21}=(\frac{1}{4} F_1 - \frac{1}{12} F_{Z_2}
)m_{21} \;,\label{ahg2eq16}\\
R_{33} &=\frac{1}{3} F_{Z_2} m_{33} -
\frac{1}{2} F_2 (m_{11} + m_{22}) \quad ,\quad  R_{13}= -
\frac{1}{6} F_{Z_2} m_{13} \:,\nonumber\\
R_{31} &= \frac{1}{6} F_{Z_2} m_{31} \quad ,\quad R_{23}= -
\frac{1}{6} F_{Z_2} m_{23}  \quad ,\quad R_{32}= \frac{1}{6} F_{Z_2}
m_{32} \;, \nonumber 
\end{align}
$F_{1,2} \equiv
F(M_{1,2},\sqrt{\lambda_-},\sqrt{\lambda_+}) \; , \;F_2 \equiv
F(M_{Z_2},\sqrt{\lambda_-},\sqrt{\lambda_+})$, with $M_1 \:,\:M_2$
and $M_{Z_2}$ being the horizontal boson masses defined in Eq.(\ref{ahg2eq7}),
\begin{equation} m_{jk}=\frac{a_j \:b_k \:M}{\lambda_+ - \lambda_-} = \frac{a_j
\:b_k}{a \:b} \:\sqrt{\lambda_-}\:c_{\alpha} c_{\beta} \:,\label{ahg2eq17}\end{equation}
$\cos\alpha \equiv c_{\alpha}\:,\;\cos\beta \equiv
c_{\beta}\:,\;\sin\alpha \equiv s_{\alpha}\:,\;\sin\beta \equiv
s_{\beta}$. So, up to one loop contribution we obtain the fermion
masses
\begin{equation} \bar{\Psi}^{o}_L {\cal{M}}^{o} \:\Psi^{o}_R +
\bar{\Psi^{o}_L} {\cal{M}}_1^o \:\Psi^{o}_R = \bar{\chi_L^o}
\:{\cal{M}}_1 \:\chi_R^o \label{ahg2eq18}\end{equation} 
with ${\cal{M}}_1\equiv  \left[ Diag(0,0,-
\sqrt{\lambda_-},\sqrt{\lambda_+})+ {V_L^o}^T {\cal{M}}_1^o
\:V_R^o \right]$; explicitly
\begin{equation} {\cal{M}}_1= \left( \begin{array}{rrrr} q_{11}&q_{12}&c_\beta \:q_{13}&s_\beta \:q_{13} \\
q_{21}& q_{22} & c_\beta \:q_{23} & s_\beta \:q_{23}\\
c_\alpha \:q_{31}& c_\alpha \:q_{32} & -\sqrt{\lambda_-}+c_\alpha
c_\beta
\:q_{33} & c_\alpha s_\beta \:q_{33} \\
s_\alpha \:q_{31}& s_\alpha \:q_{32} & s_\alpha c_\beta \:q_{33} &
\sqrt{\lambda_+}+s_\alpha s_\beta \:q_{33}
\end{array} \right) \;,\label{ahg2eq19}\end{equation}
where the mass entries $q_{ij}\: ;i,j=1,2,3$ are written as:
\begin{equation} \begin{array}{lll} q_{11}=-c_1
\frac{H}{q}\:,&q_{12}=\frac{b_3}{b} c_1 \epsilon
\frac{H}{q}\:,&q_{13}=\frac{b^\prime}{b} c_1 \epsilon
\frac{H}{q} \:,\\
                                                       \\
q_{21}=-\frac{a_3}{a} c_1 \epsilon \frac{H}{q}\:,&q_{22}=c_2
\left[-\frac{H}{q}+u q (\frac{\Delta}{2}+J) \right]\:,&    \\
                                                       \\
 q_{31}=-\frac{a^\prime}{a} c_1 \epsilon
\frac{H}{q}\:,& q_{32}=c_2 \left[-\frac{a^\prime}{a_3}
\frac{H}{q}+u q (\frac{a^\prime}{a_3}
\frac{\Delta}{2}-\frac{a_3}{a^\prime} J)
 \right], &  \end{array}  \label{ahg2eq20} \end{equation}

\begin{equation} q_{23}=c_2 \left[-\frac{b^\prime}{b_3} \frac{H}{q}+u q
(\frac{b^\prime}{b_3} \frac{\Delta}{2}-\frac{b_3}{b^\prime} J)
 \right] \:, \nonumber \end{equation}

\begin{equation} q_{33}=c_2 \left[-u H+J+\frac{1}{6} u^2 q^2 \Delta
-\frac{1}{3} \left(u^2 q^2
F_1+(1+\frac{{a^\prime}^2}{a_3^2}+\frac{{b^\prime}^2}{b_3^2})
\right) F_{Z_2} \right] \:,\nonumber \end{equation}

%\vspace{2mm}

\begin{equation} c_1=\frac{1}{2} c_\alpha c_\beta \frac{a_3 \:b_3}{a\:b}
\frac{\alpha_H}{\pi} \quad , \quad c_2= \frac{a_3 \:b_3}{a\:b}
\:c_1 \quad , \quad u=\frac{\eta_+}{a_3\:b_3} \quad , \quad
\epsilon=\frac{\eta_-}{\eta_+}        \nonumber \end{equation}

\begin{equation} \eta_-=a_1 \:b_2 - a_2 \:b_1 \; , \; \eta_+=a_1 \:b_1 + a_2
\:b_2 \; , \; \frac{a^\prime \:b^\prime}{a_3 \:b_3}=u \:q \:,\label{ahg2eq21}\end{equation}

\begin{equation} q=\sqrt{1+\epsilon^2} \quad , \quad H=F_2 - u \:F_1 \quad ,
\quad J=F_{Z_2} - u \:F_2 \quad , \quad \Delta=F_{Z_2} - F_1\:.
\nonumber \end{equation}
The diagonalization of ${\cal{M}}_1$, Eq.(\ref{ahg2eq19}),  gives the
physical masses for fermions in each sector u, d and e. Using a
new biunitary transformation $\chi_{L,R}^o=V_{L,R}^{(1)}
\psi_{L,R}$; $\bar{\chi_L^o} \;{\cal{M}}_1\;\chi_R^o= \bar{\psi_L}
\:{V_L^{(1)}}^T {\cal{M}}_1\: V_R^{(1)} \:\psi_R $, with
${\Psi_{L,R}}^T = ( f_1 , f_2 , f_3 , F )_{L,R}$ being the mass
eigenfields, that is
\begin{equation} {V^{(1)}_L}^T {\cal{M}}_1 \:{\cal M}_1^T \;V^{(1)}_L =
{V^{(1)}_R}^T {\cal M}_1^T \:{\cal{M}}_1 \;V^{(1)}_R =
Diag(m_1^2,m_2^2,m_3^2,M_F^2) \:,\label{ahg2eq22}\end{equation}
$m_1^2=m_e^2$, $m_2^2=m_\mu^2$, $m_3^2=m_\tau^2$ and
$M_F^2=M_E^2$ for charged leptons. Thus, the final transformation
from massless to mass fermions eigenfields in this
scenario reads
\begin{equation} \Psi_L^o = V_L^{o} \:V^{(1)}_L \:\Psi_L \qquad \mbox{and}
\qquad \Psi_R^o = V_R^{o} \:V^{(1)}_R \:\Psi_R \label{ahg2eq23}\end{equation}

\subsection{Quark Mixing and $( V_{CKM} )_{4\times 4}$ }

The interaction of quarks ${f_{uL}^o}^T=(u^o,c^o,t^o)_L$ and ${f_{dL}^o}^T=(d^o,cso,b^o)_L$ with the $W$ charged gauge
boson is\footnote{Recall that vector like quarks, Eq.(\ref{ahg2eq1}), are $SU(2)_L$
weak singlets, and so, they do not couple to $W$ boson in the
interaction basis.}
\begin{equation} \bar{f^o}_{u L} \gamma_\mu f_{d L}^o {W^+}^\mu = \bar{\psi}_{u
L}\;{V_{u L}^{(1)}}^T\;[(V_{u L}^o)_{3\times 4}]^T \;(V_{d
L}^o)_{3\times 4} \;V_{d L}^{(1)}\;\gamma_\mu \psi_{d L}
\;{W^+}^\mu \:,\label{ahg2eq24}\end{equation}
and therefore, the non-unitary $V_{CKM}$ of dimension
$4\times 4$ is identified as
\begin{equation} (V_{CKM})_{4\times 4}\equiv {V_{u L}^{(1)}}^T\;[(V_{u L}^o)_{3\times
4}]^T \;(V_{d L}^o)_{3\times 4} \;V_{d L}^{(1)} \:.\label{ahg2eq25}\end{equation}
Assuming the relationship
$\frac{v_1}{v_1^2+v_2^2}=\frac{V_1}{V_1^2+V_2^2}$, we may write 
\begin{equation} V_o \equiv [(V_{u L}^o)_{3\times 4}]^T \;(V_{d L}^o)_{3\times
4} = \left(
\begin{array}{cccc}
       1  &0  &0  &0  \\
       0  & C_o& - c_\alpha^d \:S_o& - s_\alpha^d \:S_o\\
       0  & c_\alpha^u \:S_o & c_\alpha^u \:c_\alpha^d \:C_o & c_\alpha^u \:s_\alpha^d \:C_o  \\
      0  & s_\alpha^u \:S_o & s_\alpha^u \:c_\alpha^d \:C_o&s_\alpha^u \:s_\alpha^d \:C_o
\end{array} \right) \:,\label{ahg2eq26}\end{equation}

\begin{eqnarray} C_o=\frac{1+r_u \:r_d}{\sqrt{(1+r_u^2)(1+r_d^2)}} \quad ,
\quad S_o=\frac{r_u - r_d}{\sqrt{(1+r_u^2)(1+r_d^2)}} \quad ,
\quad  \nonumber \\
                 \label{ahg2eq27}\\
C_o^2+S_o^2=1 \quad , \quad r_u=(\frac{a^\prime}{a_3})_u \quad ,
\quad r_d=(\frac{a^\prime}{a_3})_d    \nonumber    \end{eqnarray}

\section{Numerical results}

Using the strong hierarchy of masses for quarks and
charged leptons and the results in\cite{ahg2prd2007}, we report here
the magnitudes of quark masses and mixing coming from the analysis of
a small region of the parameter space in
this model. For this numerical analysis we use the input global
parameters $\frac{\alpha_H}{\pi}=.12$, $M_1=620$ TeV and
$M_2=6500$ TeV .

%\vspace{1mm}

\subsection{Sector d:}

Parameter space: $(\sqrt{\lambda_-})_d=
4.31175$ GeV, $(\sqrt{\lambda_+})_d =1.06618 \times 10^6$ GeV,
$r_d=3$, $u_d=1.73906$, $\epsilon=7.51219$,
$s_\alpha^d=1\:\times 10^{-5}$, and $s_\beta^d=.970762$ lead to the
down quark masses: $m_d=4.4$ MeV, $m_s=75$ MeV, $m_b=4.2$
GeV, and the mixing matrix
\begin{equation} V_{d L}^{(1)}= \left(
\begin{array}{rrrr}
.9741  &.2257  &-.0037  & 5.98 \times 10^{-8} \\
-.2256  &.9741  &.0018 &  -3.03 \times 10^{-8} \\
.0040&-.0009 &.9999 & 4.26 \times 10^{-7}\\
-6.69 \times 10^{-8}& 1.64 \times 10^{-8}& -4.26 \times 10^{-7} & 1
\end{array} \right) \:.\label{ahg2eq28}\end{equation}

\subsection{Sector u:}

Parameter space: $ (\sqrt{\lambda_-})_u= 180.463$ GeV, 
$(\sqrt{\lambda_+})_u =6.48273 \times 10^6$ GeV, $r_u=2.568$,
$u_u=1.23019$, $\epsilon=0$, $s_\alpha^u=1\:\times
10^{-5}$ and $s_\beta^u=.941119$ yield the up quark masses
$m_u=2.4$ MeV, $m_c=1.25$ GeV, $m_t=172$ GeV, and
the mixing
\begin{equation} V_{u L}^{(1)}= \left(
\begin{array}{rrrr}
1&0&0 &0 \\
0&.9999  &.0095 &  -7.17 \times 10^{-7} \\
0&-.0095 &.9999 & 3.65 \times 10^{-6}\\
0& 7.52 \times 10^{-7}& -3.64 \times 10^{-6} & 1
\end{array} \right) \:.\label{ahg2eq29}\end{equation}

\subsection{$(V_{CKM})_{4\times 4}$}

The above up and down quark mixing matrices $V_{u L}^{(1)}$ and
$V_{d L}^{(1)}$, and the matrix $V_o$, Eq.(\ref{ahg2eq26}), defined by
the See-saw mixing angles $s_\alpha^d$, $s_\beta^d$, $s_\alpha^u$,
$s_\beta^u$, and the values of parameters $r_u$ and $r_d$, yield the quark mixing
\begin{equation} (V_{CKM})_{4 \times 4}=\left(
\begin{array}{rrrr}
.9741  &.2257  &-.0037  &5.98 \times 10^{-8} \\
-.2253  &.9733  &.0418 &  2.67 \times 10^{-8} \\
.0130&-.0399 &.9991 & 1.42 \times 10^{-6}\\
3.69 \times 10^{-7}& -1.37 \times 10^{-6}& 1.35 \times 10^{-5} &
1.94\times 10^{-11}
\end{array} \right) \label{ahg2eq30}\end{equation}
Notice that except the $(V_{CKM})_{31}$ matrix element,
all the others entries are within the allowed range of values
reported in PDG\cite{ahg2PDG}.

\subsection{Charged Leptons:}

For this sector, the parameter space: $(\sqrt{\lambda_-})_e=
4.0986$ GeV, $(\sqrt{\lambda_+})_e =1.62719\times
10^{7}$ GeV, $r_e=r_d=3$, $u_e=1.18259$, $\epsilon=0 $,
$\alpha^e=1\:\times10^{-5}$ and $\beta^e=.0251802$,
reproduce the known charged lepton masses: $m_e=.51099$ MeV,
$m_\mu=105.658$ MeV, $m_\tau=1776.84$ MeV.

\subsection{FCNC's in $K^o-\bar{K^o}$ meson mixing}

The $SU(3)$ horizontal gauge bosons contribute to new FCNC's, in
particular they mediate $\Delta F=2$ processes at tree level. Here
we compute their leading contribution to $K^o-\bar{K^o}$ meson
mixing. In the previous scenario the $(V_{CKM})_{12}$ and
$(V_{CKM})_{13}$ mixing angles come completely from the down quark
sector, and hence, the effective hamiltonian from the tree level diagrams
mediated by the $SU(2)$ horizontal gauge bosons of mass $M_1$ to
the ${\cal O}_{LL}(\Delta S=2)=(\bar{d}_L \gamma_\mu
s_L)(\bar{d}_L \gamma^\mu s_L)$ operator is
\begin{equation} {\cal H }_{eff} = C_{\bar{d} s}\:{\cal O}_{LL} \quad , \quad
C_{\bar{d} s} \approx \frac{g_H^2}{4}  \frac{1}{M_1^2}
\frac{r_d^4}{(1+r_d^2)^2} s_{12}^2 \:,\label{ahg2eq31}\end{equation}
and then contribute to the $K^o-\bar{K^o}$ mass difference as
\begin{equation} \Delta m_K \approx \frac{2 \pi^2}{3}   \frac{\alpha_H}{\pi}
\frac{r_d^4}{(1+r_d^2)^2} s_{12}^2 \frac{F_K^2}{M_1^2} B_K(\mu)
M_K \:.\label{ahg2eq32}\end{equation}
Using the input values $s_{12}=.2257$, $F_K=160$ MeV,
$M_K=497.614$ MeV and $B_K=.8$, one gets
\begin{equation} \Delta m_K \thickapprox 0.8637 \times {10}^{-12}\text{MeV}  
\label{ahg2eq33}
\end{equation}
which is consistent with the present experimental
bounds\cite{ahg2PDG}. The quark mixing alignment in
Eqs.(\ref{ahg2eq28}--\ref{ahg2eq30}) avoids tree level contributions
to $D^0-\bar{D^o}$ mixing mediated by these $SU(2)$ horizontal
gauge bosons.

\section{Conclusions}

We have reported a detailed analysis on charged fermion masses and
mixing, within the $SU(3)$ gauged flavor symmetry model with radiative mass
generation for light fermions, introduced by the author in
Ref.\cite{ahg2su3models}. A quantitative
analysis shows that this hierarchical mechanisms enables us to
accommodate a realistic spectrum of masses and mixing for quarks
and the charged leptons masses. A crucial feature to achieve these
results has been the introduction of just one $SU(2)_L$ weak
singlet vector like fermion for each sector u, d and e, and in
this sense this is a simple and economical model. Moreover, tree
level FCNC's processes mediated by the $SU(3)$ massive gauge
bosons like $ K^0-\bar{K^o}$ and $D^0-\bar{D^o}$ are suppressed
within current experimental limits.

%\vspace{5mm}

\section*{Acknowledgments}

It is my pleasure to thank the organizers and colleagues, in
particular to Maxim Y. Khlopov and N. Manko\v c-Bor\v stnik for this
stimulating Workshop in Bled, Slovenia. The author thanks S.F.
King for useful discussions and acknowledge the hospitality of the
University of Southampton,U.K., where part of this work was done.
This work was partially supported by the "Instituto Polit\'ecnico
Nacional", (Grants from EDI and COFAA) and "Sistema Nacional de
Investigadores" (SNI) in Mexico.

\section*{Appendix: Diagonalization of the generic Dirac See-saw mass matrix}

\begin{equation} {\cal M}=
\begin{pmatrix} 0 & 0 & 0 & a_1\\ 0 & 0 & 0 & a_2\\ 0 & 0 & 0 &
a_3\\ - b_1 & - b_2 & b_3 & c \end{pmatrix} \label{ahg2eq34}\end{equation}
Using a biunitary transformation $\Psi^{o}_L =
V^{o}_L \;\chi_L^o$ and  $\Psi^{o}_R = V^{o}_R \;\chi_R^o $ to
diagonalize ${\cal{M}}^o$, where the orthogonal matrices $V^{o}_L$ and
$V^{o}_R$ may be written explicitly as
\begin{equation} V^{o}_L = \left( \begin{array}{ccrr}
\frac{a_2}{a^\prime} & \frac{a_1 a_3}{a a^\prime} & \frac{a_1}{a}
\cos{\alpha} & \frac{a_1}{a} \sin{\alpha}\\ -\frac{a_1}{a^\prime}
& \frac{a_2 a_3}{a a^\prime} & \frac{a_2}{a} \cos{\alpha} &
\frac{a_2}{a} \sin{\alpha}\\ 0 & -\frac{a^\prime}{a} &
\frac{a_3}{a} \cos{\alpha} & \frac{a_3}{a} \sin{\alpha}\\ 0 & 0 &
-\sin{\alpha} & \cos{\alpha}
\end{array} \right) \;,\label{ahg2eq35}\end{equation}

\begin{equation} V^{o}_R = \left( \begin{array}{ccrr} \frac{b_2}{b^\prime} &
\frac{b_1 b_3}{b b^\prime} & - \frac{b_1}{b} \cos{\beta} & -
\frac{b_1}{b} \sin{\beta}\\ -\frac{b_1}{b^\prime} & \frac{b_2 b_3}{b
b^\prime} & - \frac{b_2}{b} \cos{\beta} & - \frac{b_2}{b}
\sin{\beta}\\ 0 & \frac{b^\prime}{b} & \frac{b_3}{b} \cos{\beta} &
\frac{b_3}{b} \sin{\beta}\\ 0 & 0 & -\sin{\beta} & \cos{\beta}
\end{array} \right) \;,\label{ahg2eq36}\end{equation}
where $a^\prime=\sqrt{a_1^2+a_2^2}\;, \;
b^\prime=\sqrt{b_1^2+b_2^2} \;, \;a=\sqrt{{a^\prime}^2+a_3^2} \; ,
\; b=\sqrt{{b^\prime}^2+b_3^2} \;,$
\begin{equation} \lambda_{\pm } = \frac{1}{2} \left( B \pm \sqrt{B^2 -4D} \right)
\label{ahg2eq37}
\end{equation}
are the nonzero eigenvalues of
${\cal{M}}^{o} {{\cal{M}}^{o}}^T$ (${{\cal{M}}^{o}}^T
{\cal{M}}^{o}$),
\begin{eqnarray} B = a^2 + b^2 + c^2 =
\lambda_{-}+\lambda_{+}\quad &, \quad D= a^2
b^2=\lambda_{-}\lambda_{+} \;,\label{ahg2eq38}\end{eqnarray}

\begin{eqnarray} \cos{\alpha} =\sqrt{\frac{\lambda_+ -
a^2}{\lambda_+ - \lambda_-}} \;, \; \sin{\alpha} = \sqrt{\frac{a^2
- \lambda_-}{\lambda_+ - \lambda_-}} \nonumber \\
                                     \label{ahg2eq39}\\
\cos{\beta} =\sqrt{\frac{\lambda_+ - b^2}{\lambda_+ - \lambda_-}}
\;, \; \sin{\beta} = \sqrt{\frac{b^2 - \lambda_-}{\lambda_+ -
\lambda_-}} \nonumber \end{eqnarray}

\begin{eqnarray} \cos{\alpha}\: \cos{\beta}=
\frac{c\:\sqrt{\lambda_+}}{\lambda_+ - \lambda_-} \quad , \quad
\cos{\alpha} \:\sin{\beta}=
\frac{b\:c^2\:\sqrt{\lambda_+}}{(\lambda_+ - b^2)(\lambda_+ -
\lambda_-)}        \nonumber \\
                            \label{ahg2eq40} \\
\sin{\alpha} \:\sin{\beta}= \frac{c\:\sqrt{\lambda_-}}{\lambda_+ -
\lambda_-} \quad , \quad \sin{\alpha} \:\cos{\beta}=
\frac{a\:c^2\:\sqrt{\lambda_+}}{(\lambda_+ - a^2)(\lambda_+ -
\lambda_-)} \nonumber \end{eqnarray}
Note that in the space parameter $ a^2 \ll
c^2 \:,\:b^2 \; , \; \frac{\lambda_-}{\lambda_+} \ll 1$, and hence
we may approach the eigenvalues as
\begin{equation} \lambda_- \approx \frac{D}{B} \approx \frac{a^2\:b^2}{c^2+b^2}
\qquad , \qquad \lambda_+ \approx c^2+b^2+a^2 -
\frac{a^2\:b^2}{c^2+b^2} \label{ahg2eq41}\end{equation}

%%\vspace{10mm}
%%
%%\begin{figure}[htp]
%%\begin{center}
%%\includegraphics{1lazo.eps}
%%\end{center}
%%\caption{\label{ahg2Fig1} Generic one loop diagram contribution to the
%%mass term $m_{jk} \;\bar{e^o}_{jL} {e^o}_{kR}$}
%%\end{figure}
%%

%%\end{document}

%%

%%\end{document}

%% Maxim Khlopov, last version of the contribution, 24.11.2009
%%\documentclass[11pt]{article}
%%\usepackage{amssymb}
%%\usepackage{amsmath}
%%\usepackage[english]{babel}
%%\usepackage{graphicx}
%%
%%\oddsidemargin=2.0cm %
%%\voffset=-0.5cm %
%%\textheight=20cm %
%%\textwidth=12.5cm %
%%
%%\bibliographystyle{unsrt}
% A useful Journal macro
\def\Journal#1#2#3#4{{#1} {\bf #2}, #3 (#4)}
%%
% Some useful journal names
\def\NCA{\em Nuovo Cimento}
\def\RNC{\em Rivista Nuovo Cimento}
\def\NIM{\em Nucl. Instrum. Methods}
\def\NIMA{{\em Nucl. Instrum. Methods} A}
\def\NPB{{\em Nucl. Phys.} B}
\def\PLB{{\em Phys. Lett.}  B}
\def\PRL{\em Phys. Rev. Lett.}
\def\PRD{{\em Phys. Rev.} D}
\def\ZPC{{\em Z. Phys.} C}
\def\GaC{\em Gravitation and Cosmology}
\def\GaCS{{\em Gravitation and Cosmology} Supplement}
\def\JETP{\em JETP}
\def\JETPL{\em JETP Lett.}
\def\PAN{\em Phys.Atom.Nucl.}
\def\CQG{\em Class. Quantum Grav.}
\def\APJ{\em Astrophys. J.}
\def\SCI{\em Science}
\def\MPLA{{\em Mod. Phys. Lett.}  A}
\def\IJTP{\em Int. J. Theor. Phys.}
\def\NJP{\em New J. of Phys.}
\def\JHEP{\em JHEP}
\def\EPHJ{\em Eur.Phys.J}
%%
% Some other macros used in the sample text
%%\def\st{\scriptstyle}
%%\def\sst{\scriptscriptstyle}
%%\def\mco{\multicolumn}
%%\def\epp{\epsilon^{\prime}}
%%\def\vep{\varepsilon}
%%\def\ra{\rightarrow}
%%\def\ppg{\pi^+\pi^-\gamma}
%%\def\vp{{\bf p}}
%%\def\ko{K^0}
%%\def\kb{\bar{K^0}}
%%\def\al{\alpha}
%%\def\ab{\bar{\alpha}}
\def\s{{\,\rm s}}
\def\eV{\,{\rm eV}}
\def\GeV{\,{\rm GeV}}
\def\TeV{\,{\rm TeV}}
\def\cm{{\,\rm cm}}
%%\def\K{{\,\rm K}}
%%\def\kpc{{\,\rm kpc}}
%%\def\beq{\begin{equation}}
%%\def\eeq{\end{equation}}
%%\def\bea{\begin{eqnarray}}
%%\def\eea{\end{eqnarray}}
%%\def\CPbar{\hbox{{\rm CP}\hskip-1.80em{/}}}
%%\begin{document}
\title{Low Energy Binding of Composite Dark Matter with Nuclei as a Solution for the Puzzles of Dark Matter Searches}
\author{M.Yu. Khlopov$^{1,2,3}$, A.G. Mayorov$^{1}$ and E.Yu. Soldatov$^{1}$}
\institute{%
$^{1}$Moscow Engineering Physics Institute (National Nuclear Research University), 115409 Moscow, Russia \\
$^{2}$ Centre for Cosmoparticle Physics "Cosmion" 115409 Moscow, Russia \\
$^{3}$ APC laboratory 10, rue Alice Domon et L\'eonie Duquet \\75205
Paris Cedex 13, France}

\authorrunning{M.Yu. Khlopov, A.G. Mayorov and E.Yu. Soldatov}
\titlerunning{Low Energy Binding of Composite Dark Matter with Nuclei}
\maketitle

\begin{abstract}
Positive results of dark matter searches in experiments DAMA/NaI and
DAMA/ LIBRA taken together with negative results of other groups can imply nontrivial particle physics solutions for cosmological dark matter. Stable particles with charge -2 bind with primordial helium in O-helium "atoms" (OHe), representing a specific
Warmer than Cold nuclear-interacting form of dark matter. Slowed down in the
terrestrial matter, OHe is elusive for direct methods of underground
Dark matter detection like those used in CDMS experiment, but its
low energy binding with nuclei can lead to annual variations of energy
release in the interval of energy 2-6 keV in DAMA/NaI and DAMA/LIBRA
experiments. Schrodinger equation for system of nucleus and OHe is considered and reduced to an equation of relative motion in a spherically symmetrical potential, formed by the Yukawa tail of nuclear scalar isoscalar attraction potential, acting on He beyond the nucleus, and dipole Coulomb repulsion between the nucleus and OHe at distances from the nuclear surface, smaller than the size of OHe. The values of coupling strength and mass of meson, mediating scalar isoscalar nuclear potential, are rather uncertain. Within these uncertainties and in the approximation of rectangular potential wells we find a range of these parameters, at which  the sodium and/or iodine nuclei have a few keV binding energy with OHe. At nuclear parameters, reproducing DAMA results, the energy release predicted for detectors with chemical content other than NaI differ in the most cases from the one in DAMA detector. In particular, it is shown that in the case of CDMS germanium state has binding energy with OHe beyond the range of 2-6 keV and its formation should not lead to ionization in the energy range of DAMA signal.
%The estimated rate of annual modulations of radiative capture of Na by OHe to this level is consistent with the number of events registered in DAMA/NaI and DAMA/LIBRA experiments.
Due to dipole Coulomb barrier, transitions to more energetic levels of Na(I)+OHe system with much higher energy release are suppressed in the correspondence with the results of DAMA experiments. The proposed explanation inevitably leads to prediction of abundance of anomalous Na and I, corresponding to the signal, observed by DAMA.
%We discuss a possibility of similar effects for other chemical content of underground set-ups.

\end{abstract}
\section{Introduction}
The widely shared belief is that the dark matter, corresponding to
$25\%$ of the total cosmological density, is nonbaryonic and
consists of new stable particles. One can formulate the set of
conditions under which new particles can be considered as candidates
to dark matter (see e.g. \cite{mkpa2book,mkpa2Cosmoarcheology,mkpa2Bled07} for
review and reference): they should be stable, saturate the measured
dark matter density and decouple from plasma and radiation at least
before the beginning of matter dominated stage. The easiest way to
satisfy these conditions is to involve neutral weakly interacting
particles. However it is not the only particle physics solution for
the dark matter problem. In the composite dark matter scenarios new
stable particles can have electric charge, but escape experimental
discovery, because they are hidden in atom-like states maintaining
dark matter of the modern Universe.

It offers new solutions for the
physical nature of the cosmological dark matter. The main problem
for these solutions is to suppress the abundance of positively
charged species bound with ordinary electrons, which behave as
anomalous isotopes of hydrogen or helium. This problem is
unresolvable, if the model predicts stable particles with charge -1,
as it is the case for tera-electrons \cite{mkpa2Glashow,mkpa2Fargion:2005xz}.
To avoid anomalous isotopes overproduction, stable particles with
charge -1 should be absent, so that stable negatively charged
particles should have charge -2 only.

Elementary particle frames for heavy stable -2 charged species are provided by:

\begin{itemize}
\item[(a)] stable "antibaryons" $\bar U \bar U \bar U$ formed by anti-$U$ quark of fourth generation
\cite{mkpa2I,mkpa2lom,mkpa2Khlopov:2006dk,mkpa2Q} 
\item[(b)] AC-leptons \cite{mkpa2Khlopov:2006dk,mkpa25,mkpa2FKS}, predicted in the
extension \cite{mkpa25} of standard model, based on the approach of
almost-commutative geometry \cite{mkpa2bookAC}.  
\item[(c)] Technileptons and anti-technibaryons
\cite{mkpa2KK} in the framework of walking
technicolor models (WTC) \cite{mkpa2Sannino:2004qp}. 
\item[(d)] Finally, stable
charged clusters $\bar u_5 \bar u_5 \bar u_5$ of (anti)quarks $\bar
u_5$ of 5th family can follow from the approach, unifying spins and
charges \cite{mkpa2Norma}.
\end{itemize}

In the asymmetric case, corresponding to excess of -2 charge
species, $X^{--}$, as it was assumed for $(\bar U \bar U \bar
U)^{--}$ in the model of stable $U$-quark of a 4th generation, as
well as can take place for $(\bar u_5 \bar u_5 \bar u_5)^{--}$ in
the approach \cite{mkpa2Norma} their positively charged partners
effectively annihilate in the early Universe. Such an asymmetric
case was realized in \cite{mkpa2KK} in the framework of WTC, where it was
possible to find a relationship between the excess of negatively
charged anti-techni-baryons $(\bar U \bar U )^{--}$ and/or
technileptons $\zeta^{--}$ and the baryon asymmetry of the Universe.
The relationship between baryon asymmetry and excess of -2 charge stable species
is supported by sphaleron transitions at high temperatures and can be realized in all the models,
in which new stable species belong to non-trivial representations of electroweak SU(2) group.

 After it is formed
in the Standard Big Bang Nucleosynthesis (SBBN), $^4He$ screens the
$X^{--}$ charged particles in composite $(^4He^{++}X^{--})$ {\it
O-helium} ``atoms''
 \cite{mkpa2I}.
 For different models of $X^{--}$ these "atoms" are also
called ANO-helium \cite{mkpa2lom,mkpa2Khlopov:2006dk}, Ole-helium
\cite{mkpa2Khlopov:2006dk,mkpa2FKS} or techni-O-helium \cite{mkpa2KK}. We'll call
them all O-helium ($OHe$) in our further discussion, which follows
the guidelines of \cite{mkpa2I2}.

In all these forms of O-helium $X^{--}$ behave either as leptons or
as specific "heavy quark clusters" with strongly suppressed hadronic
interaction. Therefore O-helium interaction with matter is
determined by nuclear interaction of $He$. These neutral primordial
nuclear interacting objects contribute to the modern dark matter
density and play the role of a nontrivial form of strongly
interacting dark matter \cite{mkpa2Starkman,mkpa2McGuire:2001qj}. The active
influence of this type of dark matter on nuclear transformations
seems to be incompatible with the expected dark matter properties.
However, it turns out that the considered scenario of nuclear-interacting O-helium
Warmer than Cold Dark Matter is not easily
ruled out \cite{mkpa2I,mkpa2FKS,mkpa2KK,mkpa2Khlopov:2008rp} and challenges the
experimental search for various forms of O-helium and its charged
constituents.

Here we
concentrate on its effects in underground detectors. We present
qualitative confirmation of the earlier guess \cite{mkpa2I2,mkpa2KK2} that
the positive results of dark matter searches in DAMA/NaI (see for
review \cite{mkpa2Bernabei:2003za}) and DAMA/LIBRA \cite{mkpa2Bernabei:2008yi}
experiments can be explained by O-helium, resolving the controversy
between these results and negative results of other experimental
groups.

\section{OHe in the terrestrial matter}
The evident consequence of the O-helium dark matter is its
inevitable presence in the terrestrial matter, which appears opaque
to O-helium and stores all its in-falling flux.

After they fall down terrestrial surface, the in-falling $OHe$
particles are effectively slowed down due to elastic collisions with
matter. Then they drift, sinking down towards the center of the
Earth with velocity \beq V = \frac{g}{n \sigma v} \approx 80 S_3
A^{1/2} \cm/\s. \label{mkpa2dif}\eeq Here $A \sim 30$ is the average
atomic weight in terrestrial surface matter, $n=2.4 \cdot 10^{24}/A \cm^{-3}$
is the number density of terrestrial atomic nuclei, $\sigma v$ is the rate
of nuclear collisions, $m_o \approx M_X+4m_p=S_3 \TeV$ is the mass of O-helium, $M_X$ is the mass of the $X^{--}$ component of O-helium, $m_p$ is the mass of proton and $g=980~ \cm/\s^2$.

Near the Earth's surface, the O-helium abundance is determined by
the equilibrium between the in-falling and down-drifting fluxes.

The in-falling O-helium flux from dark matter halo is
$$
  F=\frac{n_{0}}{8\pi}\cdot |\overline{V_{h}}+\overline{V_{E}}|,
$$
where $V_{h}$-speed of Solar System (220 km/s), $V_{E}$-speed of
Earth (29.5 km/s) and $n_{0}=3 \cdot 10^{-4} S_3^{-1} \cm^{-3}$ is the
local density of O-helium dark matter. For qualitative estimation we don't take into account here velocity dispersion and distribution of particles in the incoming flux that can lead to significant effect.

At a depth $L$ below the Earth's surface, the drift timescale is
$t_{dr} \sim L/V$, where $V \sim 400 S_3 \cm/\s$ is given by
Eq.~(\ref{mkpa2dif}). It means that the change of the incoming flux,
caused by the motion of the Earth along its orbit, should lead at
the depth $L \sim 10^5 \cm$ to the corresponding change in the
equilibrium underground concentration of $OHe$ on the timescale
$t_{dr} \approx 2.5 \cdot 10^2 S_3^{-1}\s$.

The equilibrium concentration, which is established in the matter of
underground detectors at this timescale, is given by
\begin{equation}
    n_{oE}=\frac{2\pi \cdot F}{V} = n_{0}\frac{n \sigma v}{4g} \cdot
    |\overline{V_{h}}+\overline{V_{E}}|,
\end{equation}
where, with account for $V_{h} > V_{E}$, relative velocity can be
expressed as
$$
    |\overline{V_{o}}|=\sqrt{(\overline{V_{h}}+\overline{V_{E}})^{2}}=\sqrt{V_{h}^2+V_{E}^2+V_{h}V_{E}sin(\theta)} \simeq
$$
$$
\simeq V_{h}\sqrt{1+\frac{V_{E}}{V_{h}}sin(\theta)}\sim
V_{h}(1+\frac{1}{2}\frac{V_{E}}{V_{h}}sin(\theta)).
$$
Here $\theta=\omega (t-t_0)$ with $\omega = 2\pi/T$, $T=1yr$ and
$t_0$ is the phase. Then the concentration takes the form
\begin{equation}
    n_{oE}=n_{oE}^{(1)}+n_{oE}^{(2)}\cdot sin(\omega (t-t_0))
    \label{mkpa2noE}
\end{equation}

So, there are two parts of the signal: constant and annual
modulation, as it is expected in the strategy of dark matter search
in DAMA experiment \cite{mkpa2Bernabei:2008yi}.

Such neutral $(^4He^{++}X^{--})$ ``atoms" may provide a catalysis of
cold nuclear reactions in ordinary matter (much more effectively
than muon catalysis). This effect needs a special and thorough
investigation. On the other hand, $X^{--}$ capture by nuclei,
heavier than helium, can lead to production of anomalous isotopes,
but the arguments, presented in \cite{mkpa2I,mkpa2FKS,mkpa2KK} indicate that their
abundance should be below the experimental upper limits.

It should be noted that the nuclear cross section of the O-helium
interaction with matter escapes the severe constraints
\cite{mkpa2McGuire:2001qj} on strongly interacting dark matter particles
(SIMPs) \cite{mkpa2Starkman,mkpa2McGuire:2001qj} imposed by the XQC experiment
\cite{mkpa2XQC}. Therefore, a special strategy of direct O-helium  search
is needed, as it was proposed in \cite{mkpa2Belotsky:2006fa}.

%%%%%%%%%%%%%%%%%%%%%%%%%%%%%%%%%%%%%%%%%%%%%%%%%%%%%%%%%%%%%%%%%%%%%%%%
In underground detectors, $OHe$ ``atoms'' are slowed down to thermal
energies and give rise to energy transfer $\sim 2.5 \cdot 10^{-4}
\eV A/S_3$, far below the threshold for direct dark matter
detection. It makes this form of dark matter insensitive to the
severe CDMS constraints \cite{mkpa2Akerib:2005kh}. However, $OHe$ induced
processes in the matter of underground detectors can result in observable effects.

\section{Low energy bound state of O-helium with nuclei}

In the essence, our explanation of the results of experiments DAMA/NaI and DAMA/LIBRA is based on the idea that OHe,
slowed down in the terrestrial matter and present in the matter of DAMA detectors, can form a few keV bound state with
nucleus, in which OHe is situated \textbf{beyond} the nucleus. Formation of such bound state leads to the corresponding energy release and ionization signal, detected in DAMA experiments.

\subsection{Low energy bound state of O-helium with nuclei}

We assume the following picture: at the distances larger, than its size,
OHe is neutral and it feels only Yukawa exponential tail of nuclear attraction,
due to scalar-isoscalar nuclear potential. It should be noted that scalar-isoscalar
nature of He nucleus excludes its nuclear interaction due to $\pi$ or $\rho$ meson exchange,
so that the main role in its nuclear interaction outside the nucleus plays $\sigma$ meson exchange,
on which nuclear physics data are not very definite. When the distance from the surface of nucleus becomes
smaller than the size of OHe, the mutual attraction of nucleus and OHe is changed by dipole Coulomb repulsion. Inside the nucleus strong nuclear attraction takes place. In the result the spherically symmetric potential appears,given by
\begin{equation}
U=-\frac{A_{He} A g^2 exp(-\mu r)}{r} + \frac{Z_{He} Z e^2 r_o \cdot F(r)}{r^2}.
\label{mkpa2epot}
\end{equation}
Here $A_{He}=4$, $Z_{He}=2$ are atomic weight and charge of helium, $A$ and $Z$ are respectively atomic weight and charge of nucleus, $\mu$ and $g^2$ are the mass and coupling of scalar-isoscalar meson - mediator of nuclear attraction, $r_o$ is the size of OHe and $F(r)$ is its electromagnetic formfactor, which strongly suppresses the strength of dipole electromagnetic interaction outside the OHe "atom".
%Qualitatively, this potential has the form, presented on Fig. \ref{mkpa2pic3}.

Schrodinger equation for this system is reduced
(taking apart the equation for the center of mass) to the equation of relative motion for the reduced mass
\begin{equation}
            m=\frac{Am_p m_o}{Am_p+m_o},
            \label{mkpa2m}
 \end{equation}
where $m_p$ is the mass of proton.

In the case of orbital momentum \emph{l}=0 the wave functions depend only on \emph{r}.

\begin{figure}
    \begin{center}
        \includegraphics[width=4in]{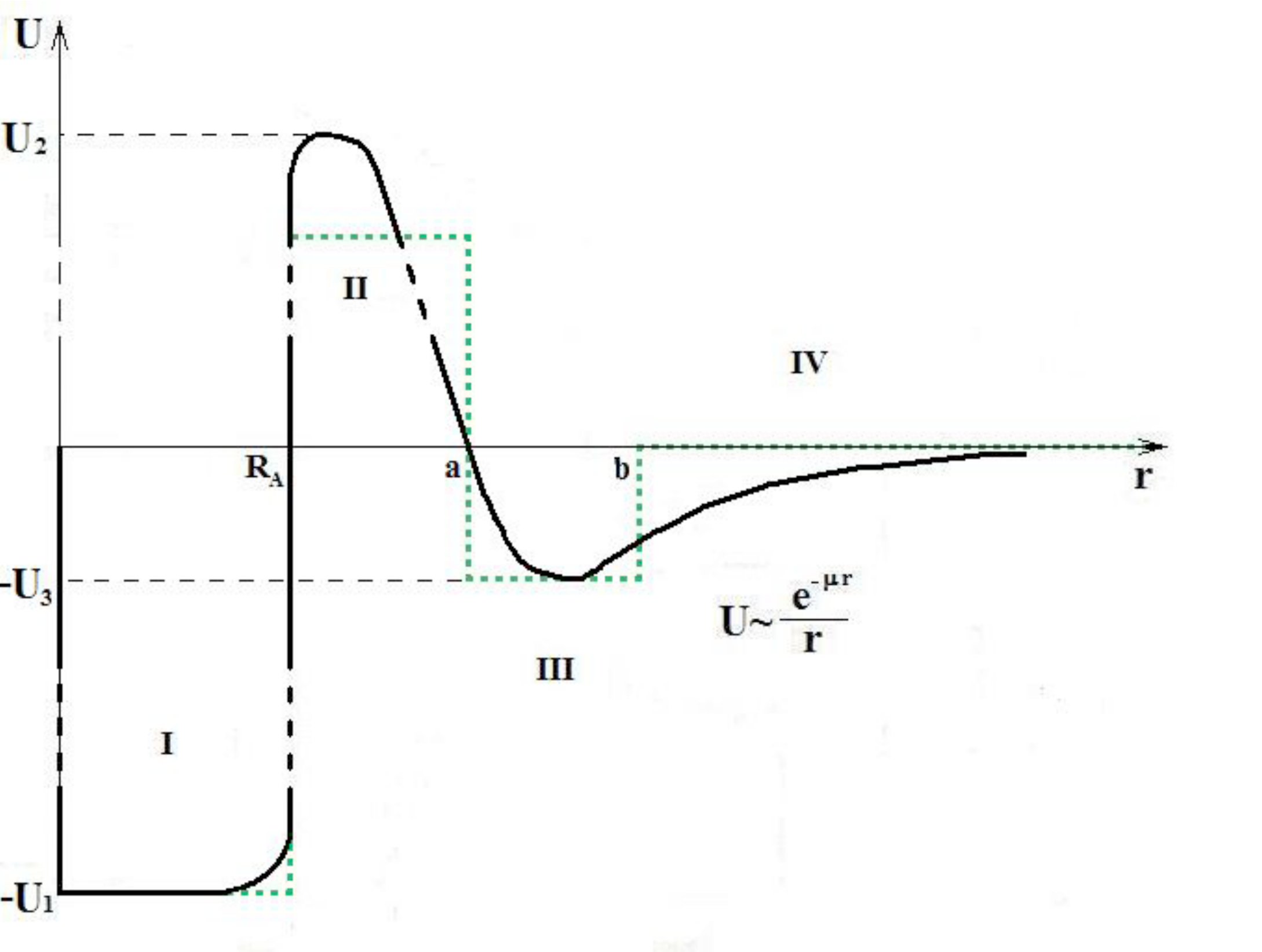}\\
        \caption{The approximation of rectangular well for potential of OHe-nucleus system.}\label{mkpa2pic23}
    \end{center}
\end{figure}

To simplify the solution of Schrodinger equation we approximate the potential (\ref{mkpa2epot})
by a rectangular potential that
consists of a deep potential well within the
radius of nucleus $R_A$, of a rectangular dipole Coulomb potential barrier outside its surface up to the
 radial layer $a=R_A+r_o$, where it is suppressed by the OHe atom formfactor, and of the outer potential well of the width $\sim 1/\mu$, formed by the tail of Yukawa nuclear interaction. It leads to the approximate potential, given by

\begin{equation}
    \left\{
        \begin{aligned}
        r<R_A: U=U_{1}=-\frac{4Ag^{2}exp(-\mu R_A)}{R_A},  \\
        R_A<r<a: U=U_{2}=\frac{\int_{R_A}^{R_A+r_o} \frac{2Z \alpha 4\pi(ro/x)}{x} dx}{r_o},  \\
        a<r<b: U=U_{3}=\frac{4Ag^{2}exp(-\mu (R_A+r_o) )}{R_A+r_o},  \\
        b<r: U=U_{4}=0,
        \end{aligned}
            \right.
            \label{mkpa2Pot1}
 \end{equation}

    presented on Fig. \ref{mkpa2pic23}.

Solutions of Schrodinger
equation for each of the four regions, indicated on Fig. \ref{mkpa2pic23}, are considered in Appendix.
In the result of their sewing one obtains the condition for the existence of a low-energy level in OHe-nucleus system,
\begin{equation}
sin(k_3 b + \delta)=\sqrt{\frac{1}{2mU_3}} \cdot k_3,
\label{mkpa2e21}
\end{equation}
where $k_3$ and $\delta$ are, respectively, the wave number and phase of the wave function in the region III (see Appendix for details).

\begin{figure}
    \begin{center}
        \includegraphics[width=3in]{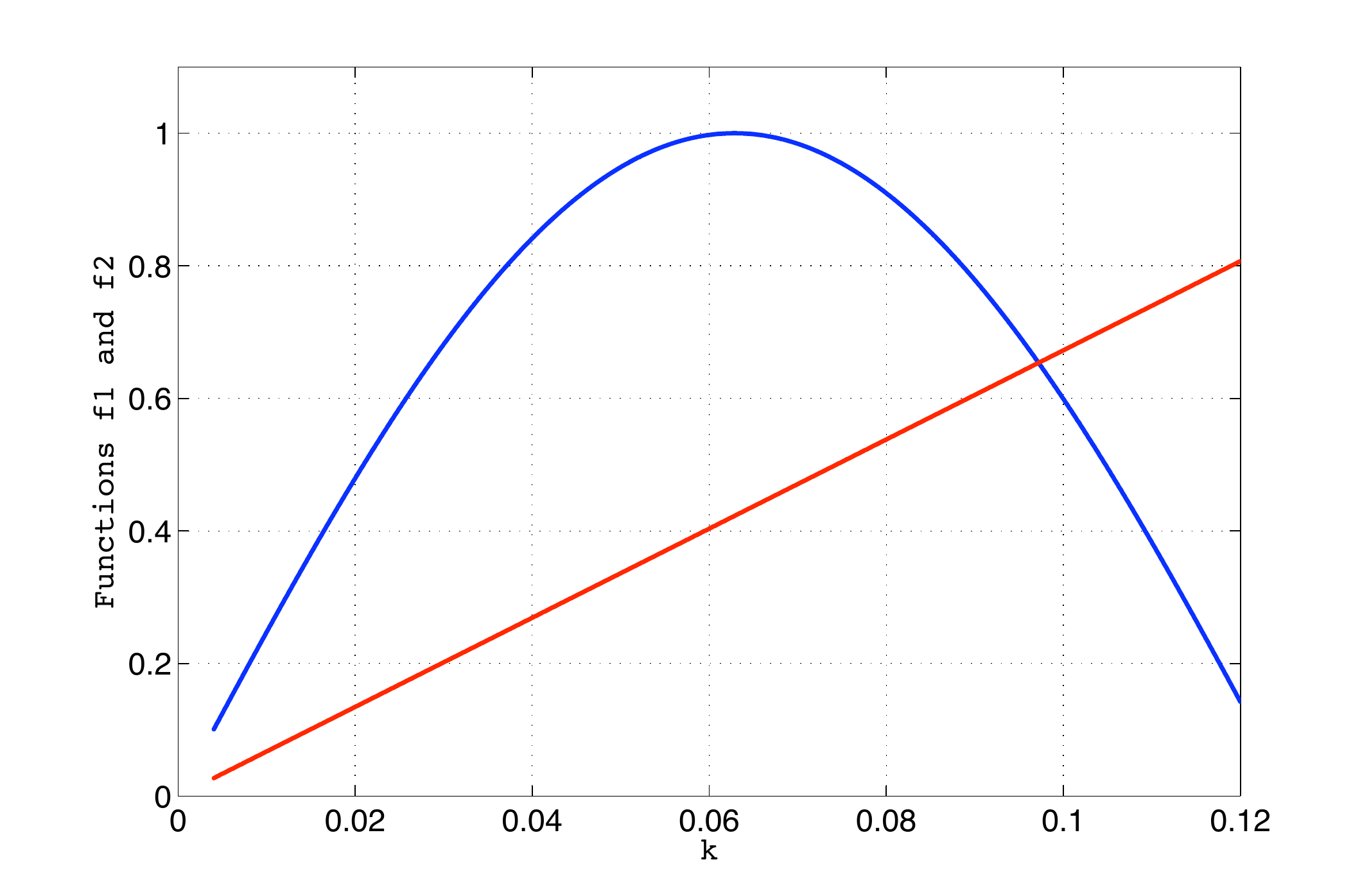}\\
        \caption{Graphical solution of transcendental equation.}\label{mkpa2F12}
    \end{center}
\end{figure}

With the use of the potential (\ref{mkpa2Pot1}) in the Eq.(\ref{mkpa2e21}), intersection of the two lines gives graphical
solution presented on Fig. \ref{mkpa2F12}.

Based on this solution one obtains from Eq.(\ref{mkpa2e18})
the energy levels of a bound state in the considered potential well.

\begin{figure}
    \begin{center}
        \includegraphics[width=4in]{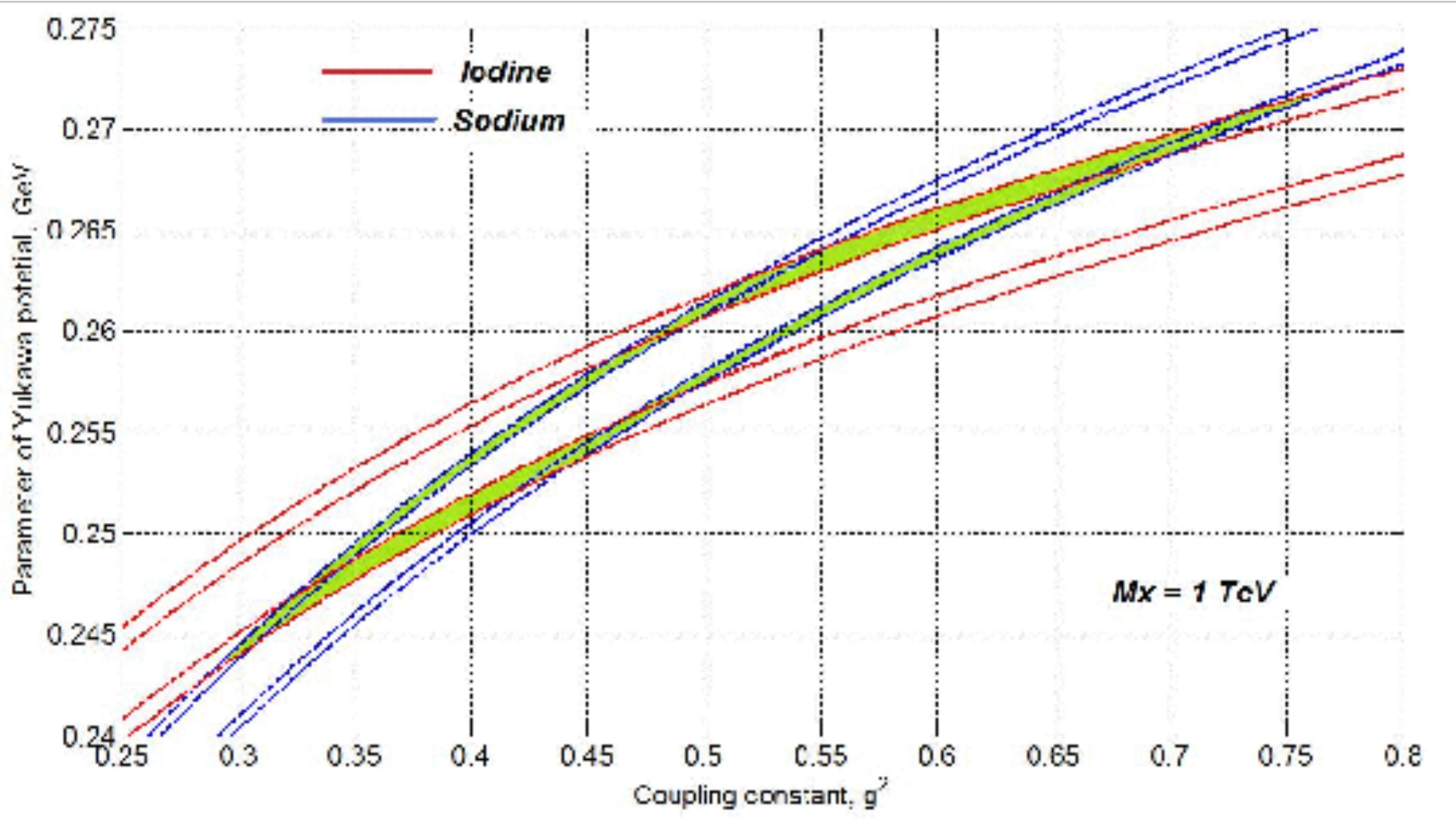}\\
        \caption{The region of parameters $\mu$ and $g^2$, for which Na and I have a level in the interval 2-6 keV. For each nucleus two narrow strips determine the region of parameters, at which the bound system of this element with OHe has a level in 2-6 keV energy range. The outer line of strip corresponds to the level of 6 keV and the internal line to the level of 2 keV. The region of intersection of strips correspond to existence of 2-6 keV levels in both OHe-Na and OHe-I systems, while the piece of strip between strips of other nucleus corresponds to the case, when OHe bound state with this nucleus has 2-6 keV level, while the binding energy of OHe with the other nuclei is less than 2 keV by absolute value.}\label{mkpa2NaI}
    \end{center}
\end{figure}

The energy of this bound state and its existence strongly depend on the parameters $\mu$ and $g^2$ of nuclear potential (\ref{mkpa2epot}). On the Fig. \ref{mkpa2NaI} the region of these parameters, giving 2-6 keV energy level in OHe bound states with sodium and iodine are presented. In these calculations the mass of OHe was taken equal to $m_o=1 TeV$.

The rate of radiative capture of OHe by nuclei should be accurately calculated with the use of exact form of wave functions, obtained for the OHe-nucleus bound state. This work is now in progress. One can use the analogy with the radiative capture of neutron by proton, considered in textbooks (see e.g. \cite{mkpa2LL4})  with the following corrections:
 \begin{itemize}
\item
  There is only E1 transition in the case of OHe capture.
\item
  The reduced masses of n-p and OHe-nucleus systems are different
\item
  The existence of dipole Coulomb barrier leads to a suppression of the cross section of OHe radiative capture.
\end{itemize}
 With the account for these effects our first estimations give the rate of OHe radiative capture, reproducing the level of signal, detected by DAMA.

Formation of OHe-nucleus bound system leads to energy release of its binding energy, detected as ionization signal in DAMA experiment. In the context of our approach the existence of annual modulations of this signal in the range 2-6 keV and absence of such effect at energies above 6 keV means that binding energy of Na-OHe and I-OHe systems should not exceed 6 keV, being in the range 2-6 keV for at least one of these elements. These conditions were taken into account for determination of nuclear parameters, at which the result of DAMA can be reproduced. At these values of $\mu$ and $g^2$ energy of OHe binding with other nuclei can strongly differ from 2-6 keV. In particular, energy release at the formation of OHe bound state with thallium can be larger than 6 keV. However, taking into account that thallium content in DAMA detector is 3 orders of magnitude smaller, than NaI, such signal is to be below the experimental errors.

It should be noted that the results of DAMA experiment exhibit also absence of annual modulations at the energy of MeV-tens MeV. Energy release in this range should take place, if OHe-nucleus system comes to the deep level inside the nucleus (in the region I of Fig. \ref{mkpa2pic23}). This transition implies tunneling through dipole Coulomb barrier and is suppressed below the experimental limits.
%Preliminary results give the energy level of
%for $\mu= 320 \MeV$ and $g^2=2$, $\mu= 350 \MeV$ and $g^2=4$, $\mu= 380 \MeV$ and $g^2=10$ or for $\mu= 460 \MeV$ and $g^2=100$.

\subsection{Energy levels in other nuclei}
\begin{figure}
    \begin{center}
        \includegraphics[width=4in]{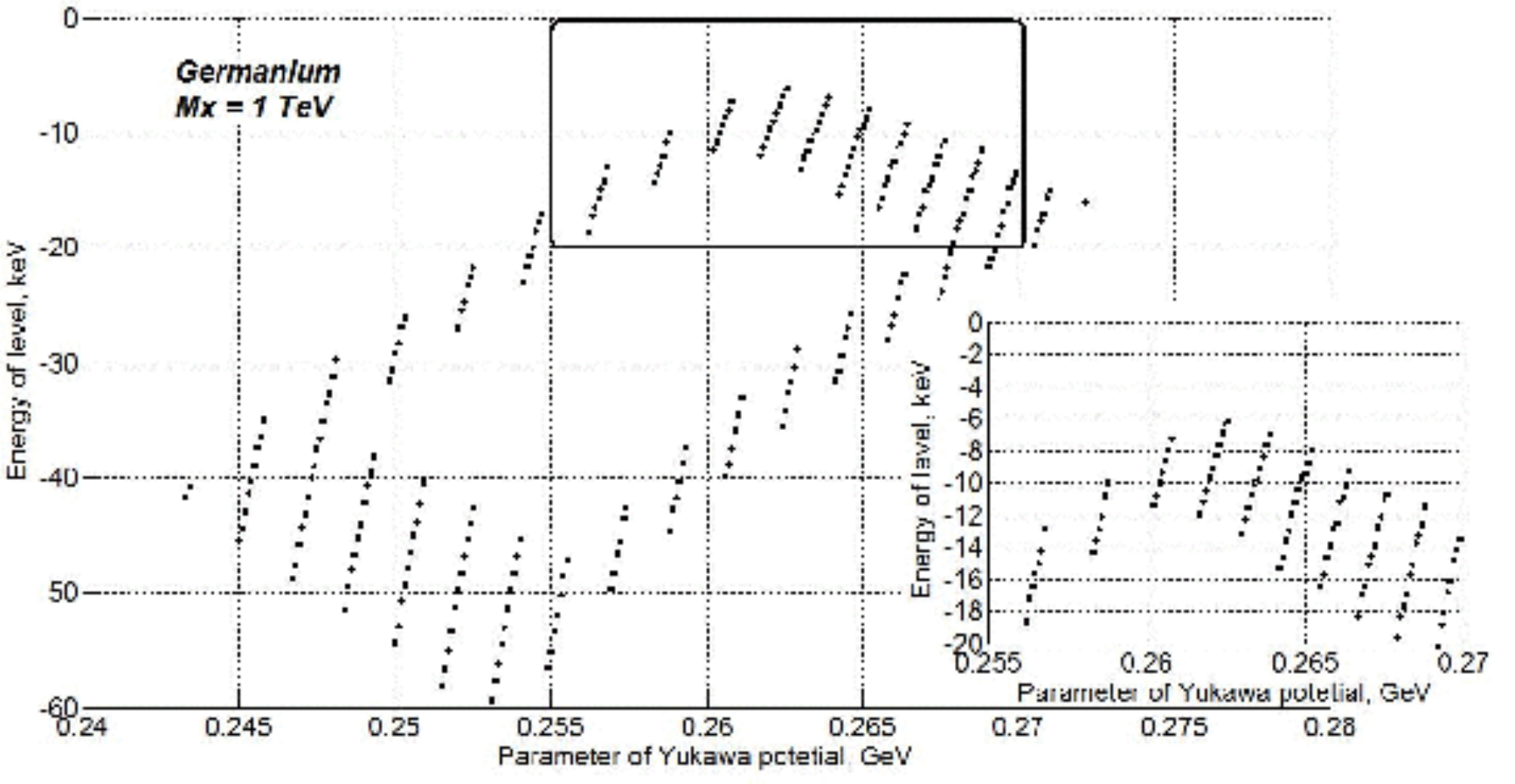}\\
        \caption{Energy levels in OHe bound system with germanium. The range of energies close to energy release in DAMA experiment is blown up to demonstrate that even in this range there is no formal intersection with DAMA results. }\label{mkpa2Ge}
    \end{center}
\end{figure}

For the chosen range of nuclear parameters, reproducing the results of
DAMA/ NaI and DAMA/LIBRA, we can calculate the binding energy of OHe-nucleus states in nuclei, corresponding to chemical composition of set-ups in other experiments. The results of such calculation for germanium, corresponding to the detector of CDMS experiment, are presented on Fig. \ref{mkpa2Ge}.
For all the parameters, reproducing results of DAMA experiment the predicted energy level of OHe-germanium bound state is beyond the range 2-6 keV, being dominantly in the range of tens - few-tens keV by absolute value. It makes elusive a possibility to test DAMA results by search for ionization signal in the same range 2-6 keV in other set-ups with content that differs from Na and I. In particular, our approach naturally predicts absence of ionization signal in the range 2-6 keV in accordance with the recent results of CDMS \cite{mkpa2Kamaev:2009gp}.

\begin{figure}
    \begin{center}
        \includegraphics[width=4in]{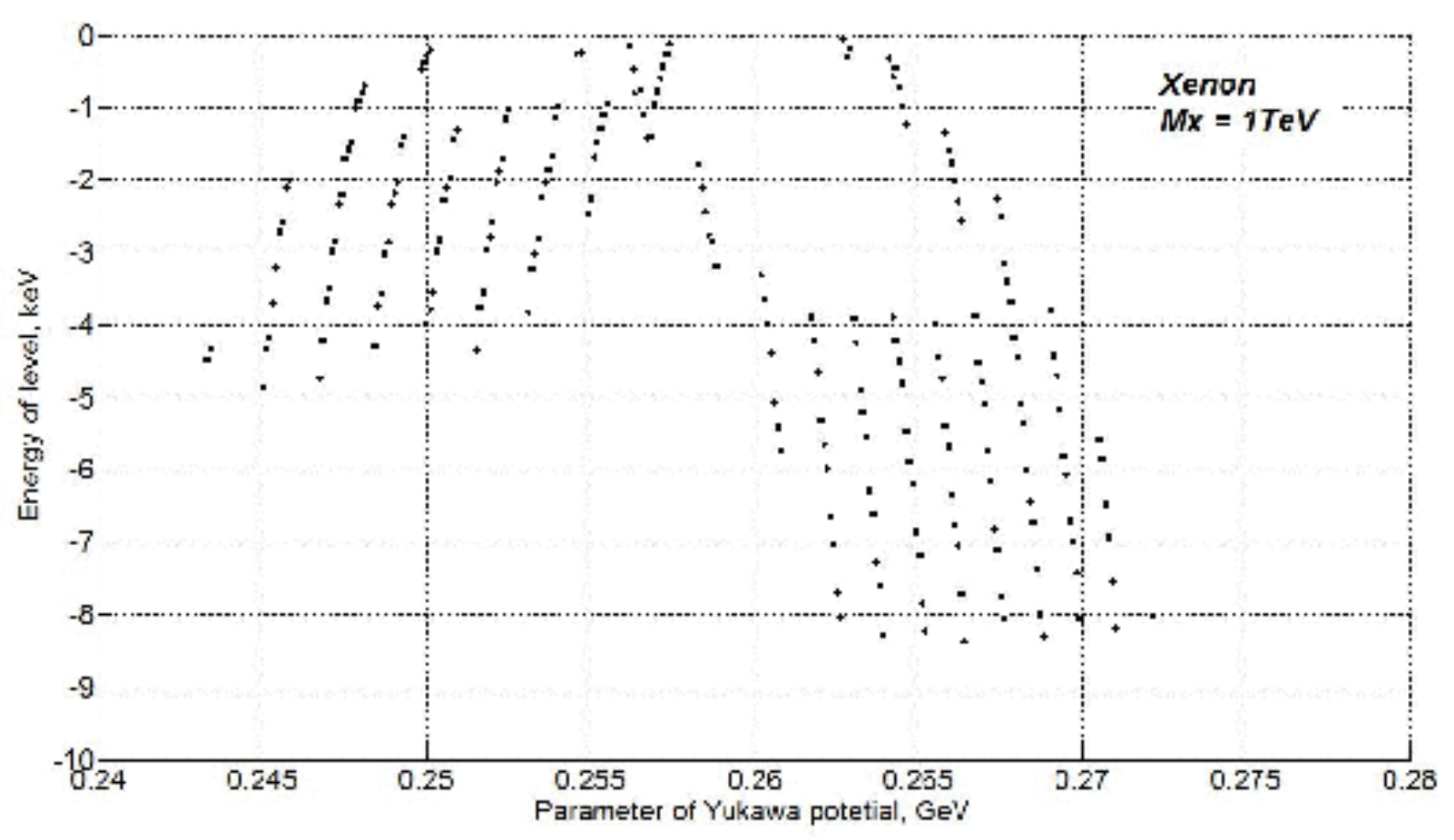}\\
        \caption{Energy levels in OHe bound system with xenon.}\label{mkpa2Xe}
    \end{center}
\end{figure}

\begin{figure}
    \begin{center}
        \includegraphics[width=4in]{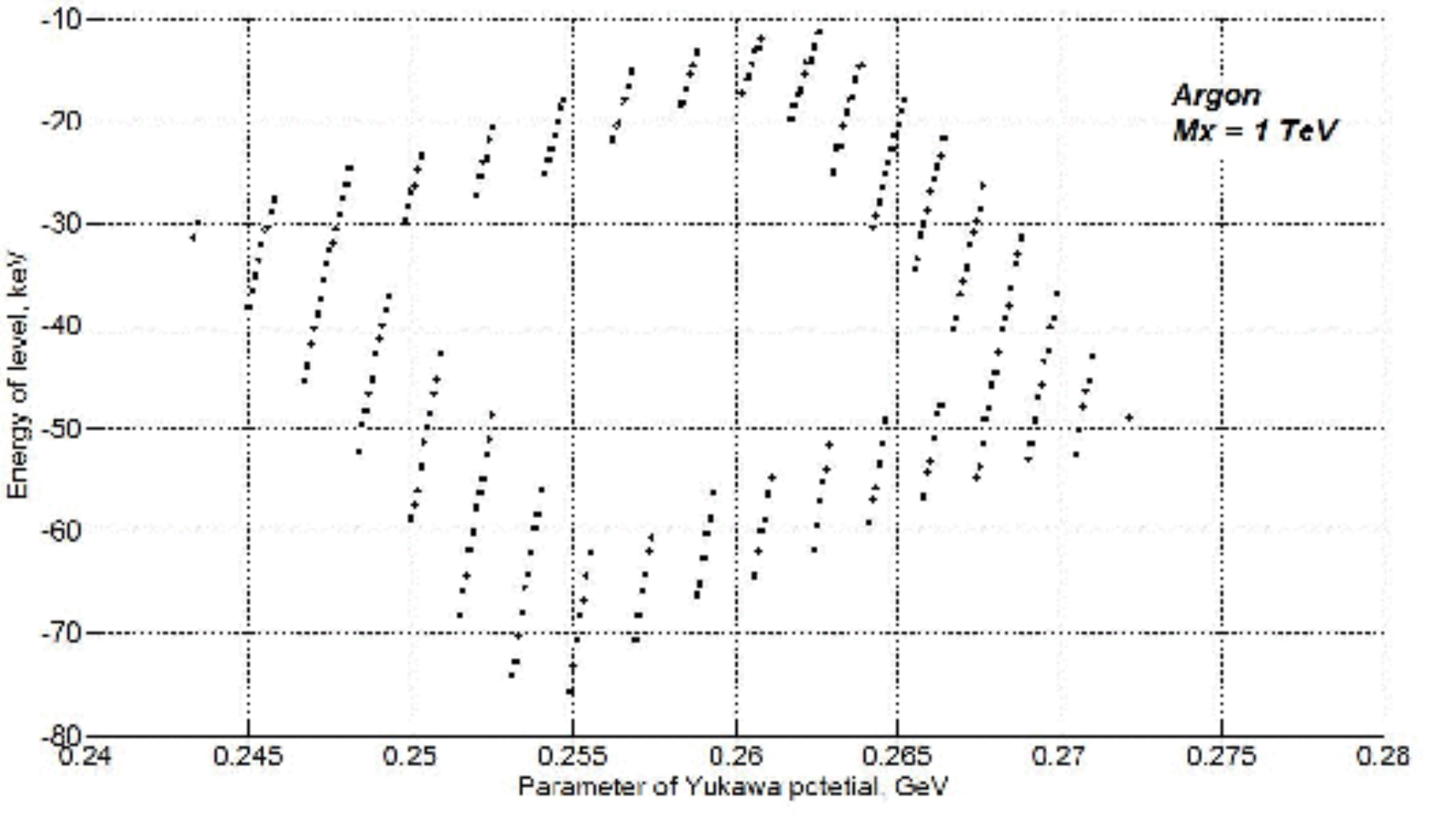}\\
        \caption{Energy levels in OHe bound system with argon.}\label{mkpa2Ar}
    \end{center}
\end{figure}

\begin{figure}
    \begin{center}
        \includegraphics[width=4in]{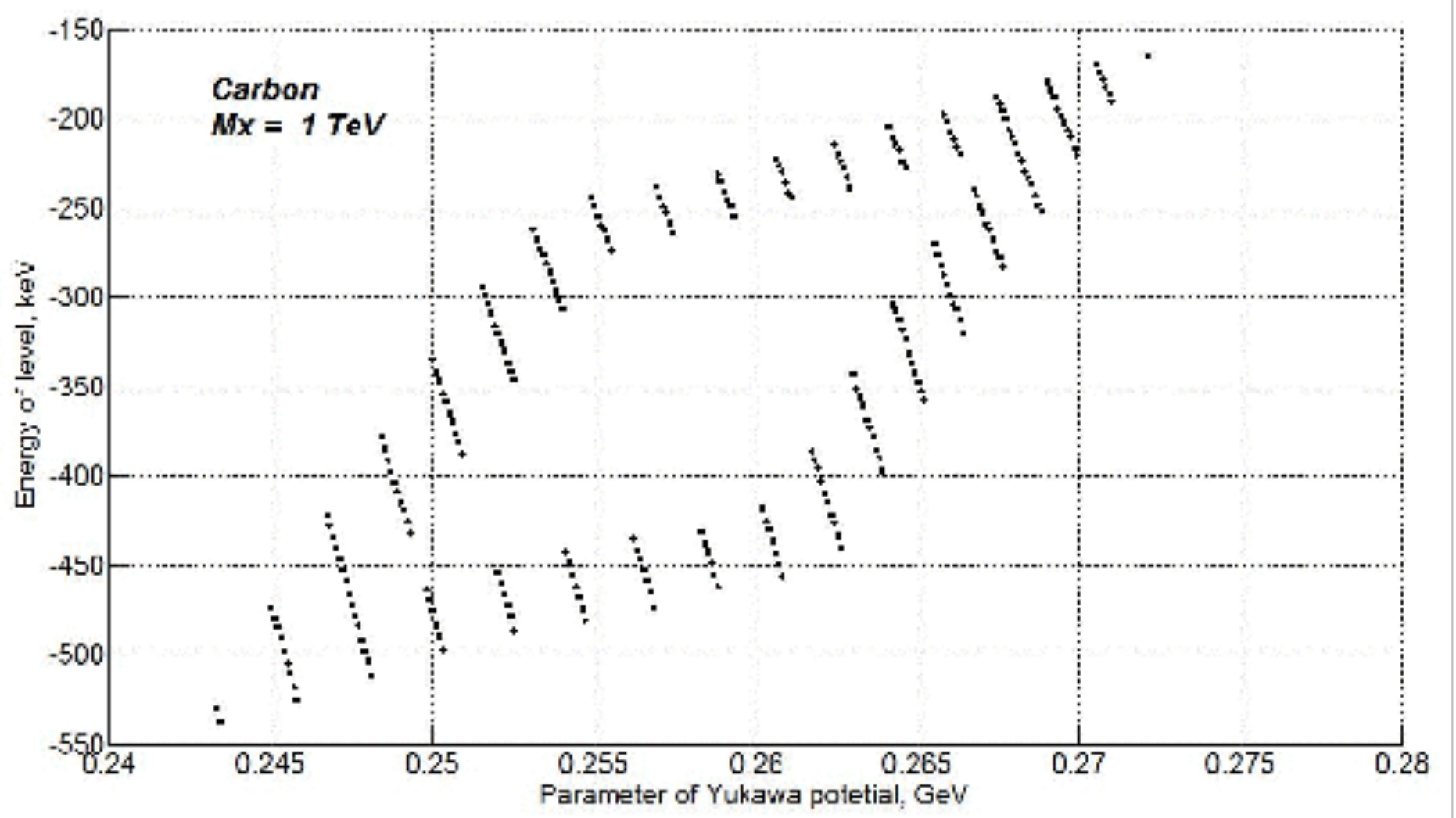}\\
        \caption{Energy levels in OHe bound system with carbon.}\label{mkpa2C}
    \end{center}
\end{figure}

\begin{figure}
    \begin{center}
        \includegraphics[width=4in]{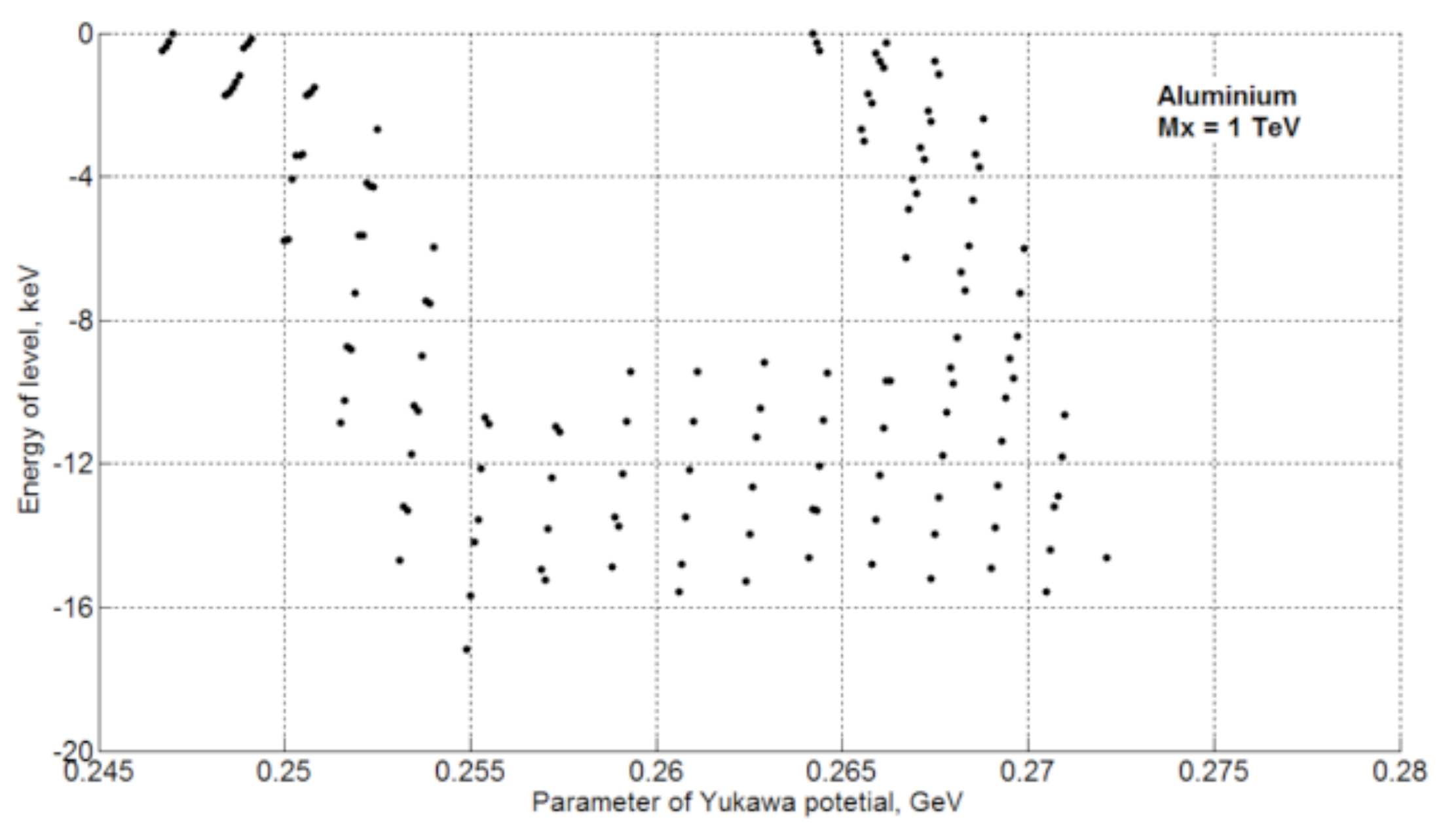}\\
        \caption{Energy levels in OHe bound system with aluminium.}\label{mkpa2Al}
    \end{center}
\end{figure}

\begin{figure}
    \begin{center}
        \includegraphics[width=4in]{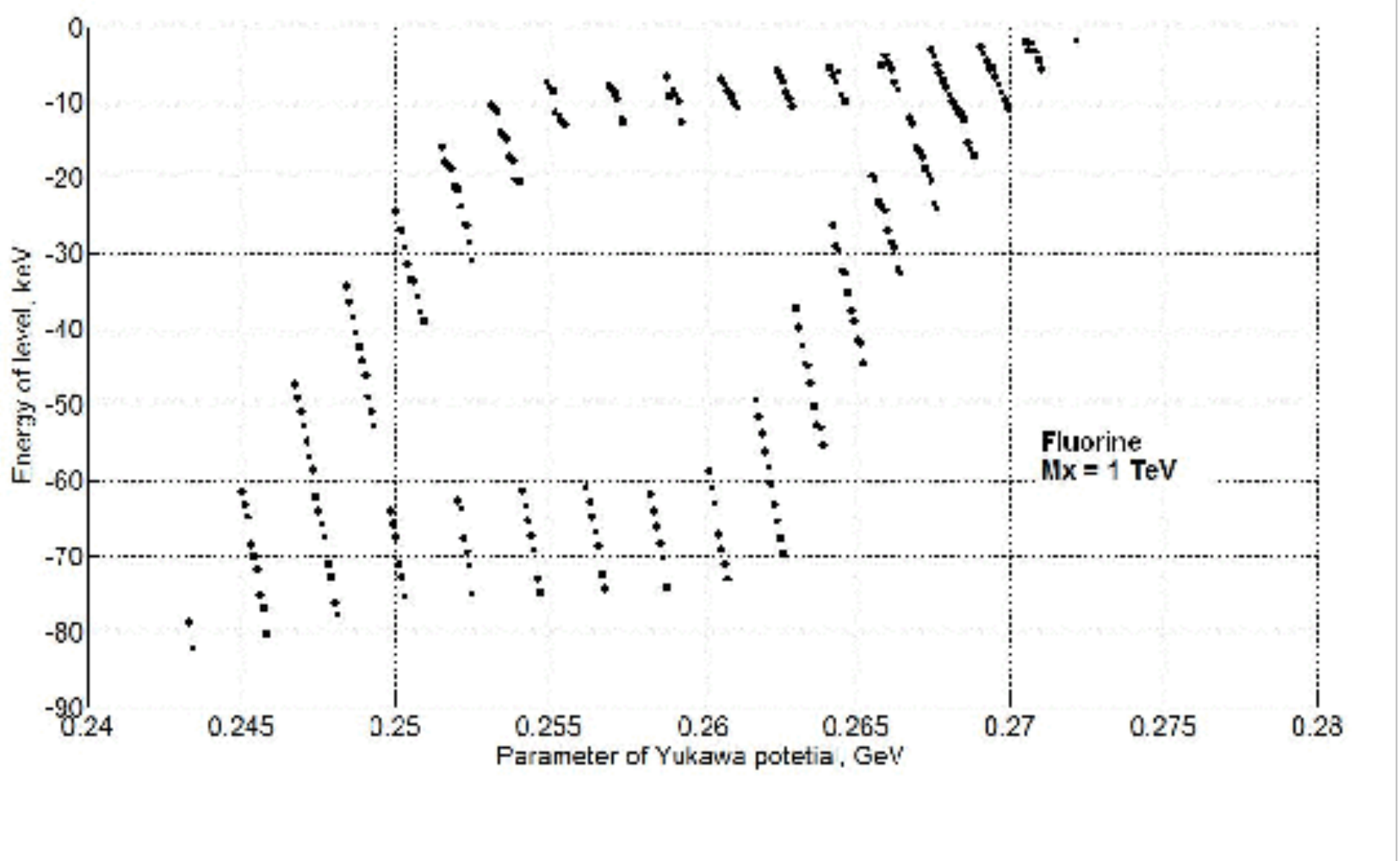}\\
        \caption{Energy levels in OHe bound system with fluorine.}\label{mkpa2F}
    \end{center}
\end{figure}

\begin{figure}
    \begin{center}
        \includegraphics[width=4in]{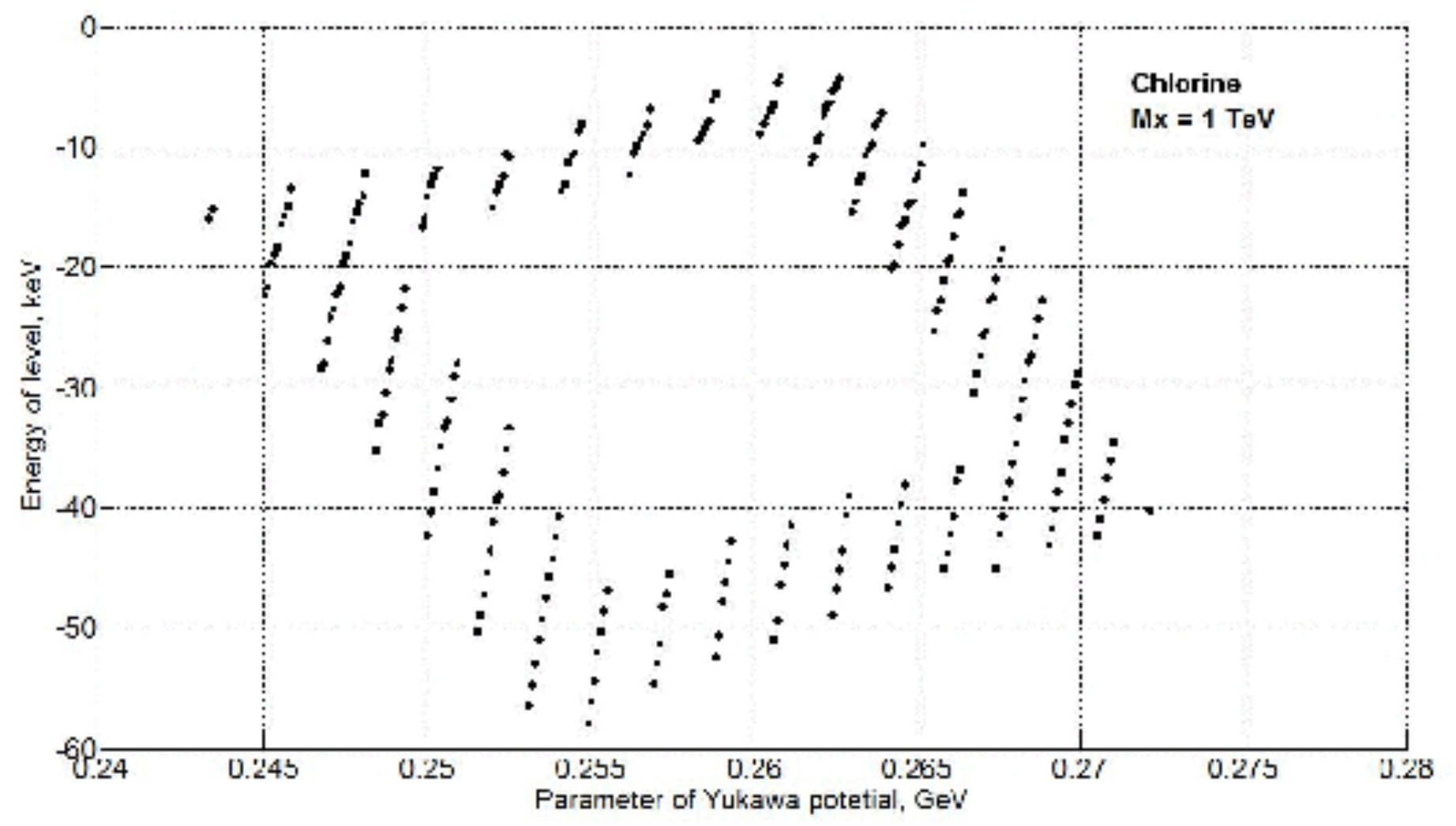}\\
        \caption{Energy levels in OHe bound system with chlorine.}\label{mkpa2Cl}
    \end{center}
\end{figure}

\begin{figure}
    \begin{center}
        \includegraphics[width=4in]{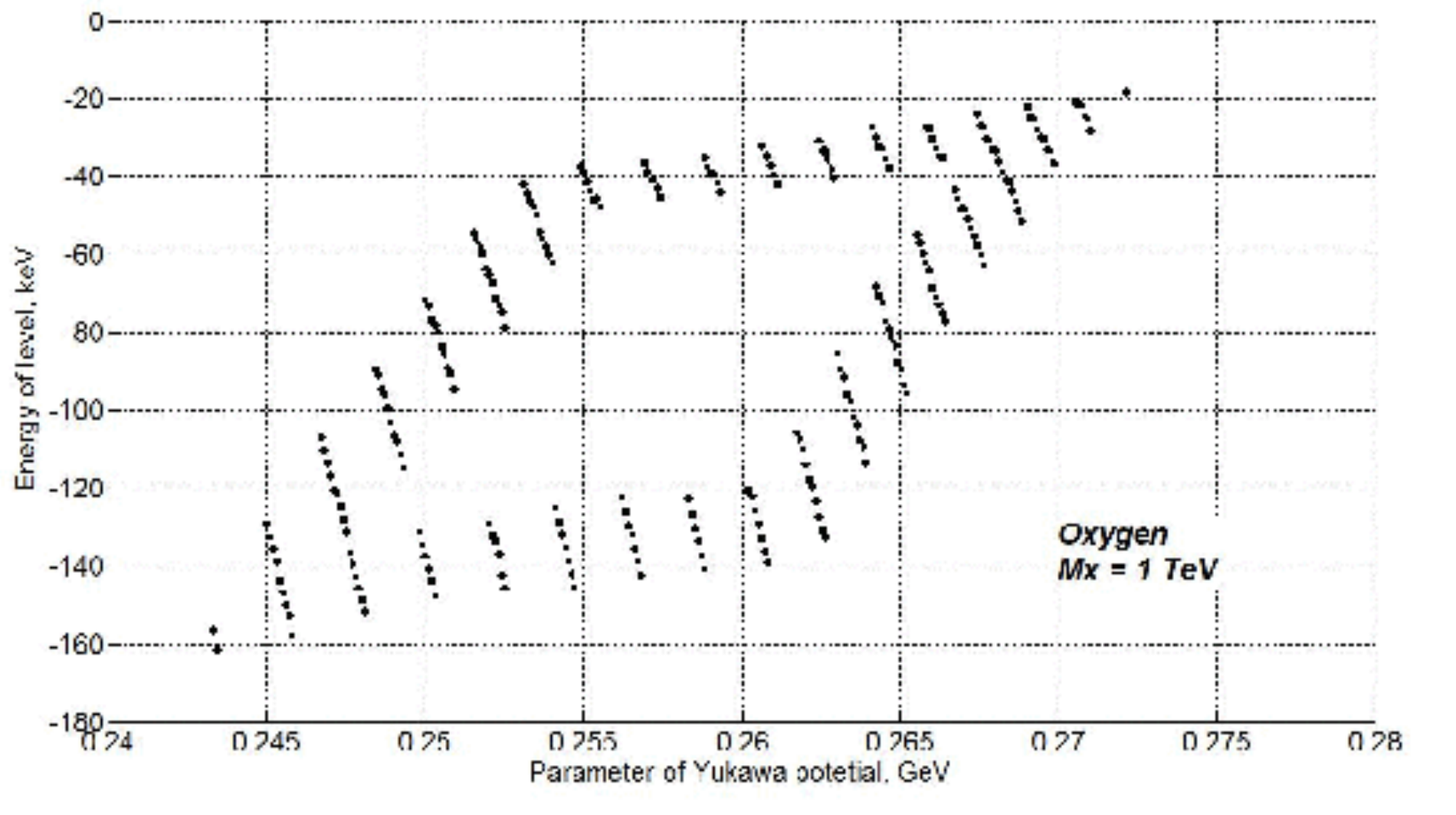}\\
        \caption{Energy levels in OHe bound system with oxygen.}\label{mkpa2O}
    \end{center}
\end{figure}

We have also calculated the energies of bound states of OHe with xenon (Fig. \ref{mkpa2Xe}), argon (Fig. \ref{mkpa2Ar}), carbon (Fig. \ref{mkpa2C}), aluminium (Fig. \ref{mkpa2Al}), fluorine (Fig. \ref{mkpa2F}), chlorine (Fig. \ref{mkpa2Cl}) and oxygen (Fig. \ref{mkpa2O}).

\subsection{Superheavy OHe}
\begin{figure}
    \begin{center}
        \includegraphics[width=4in]{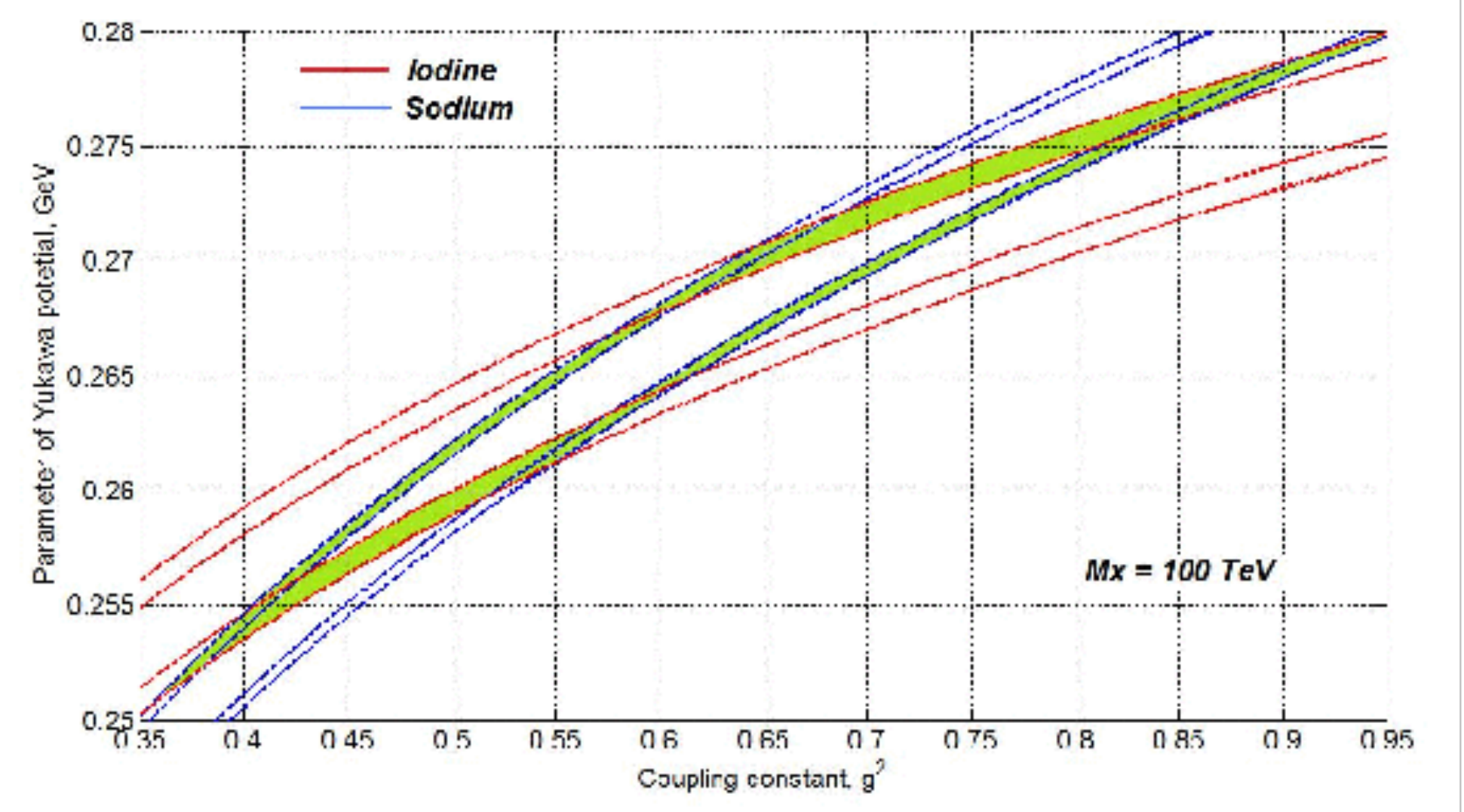}\\
        \caption{The range of parameters $\mu$ and $g^2$, for which Na and I have a level in the interval 2-6 keV for $S_3=100$. This range becomes a bit wider as compared with the case of $S_3=1$, presented on Fig. \ref{mkpa2NaI}. }\label{mkpa2NaI100}
    \end{center}
\end{figure}

\begin{figure}
    \begin{center}
        \includegraphics[width=4in]{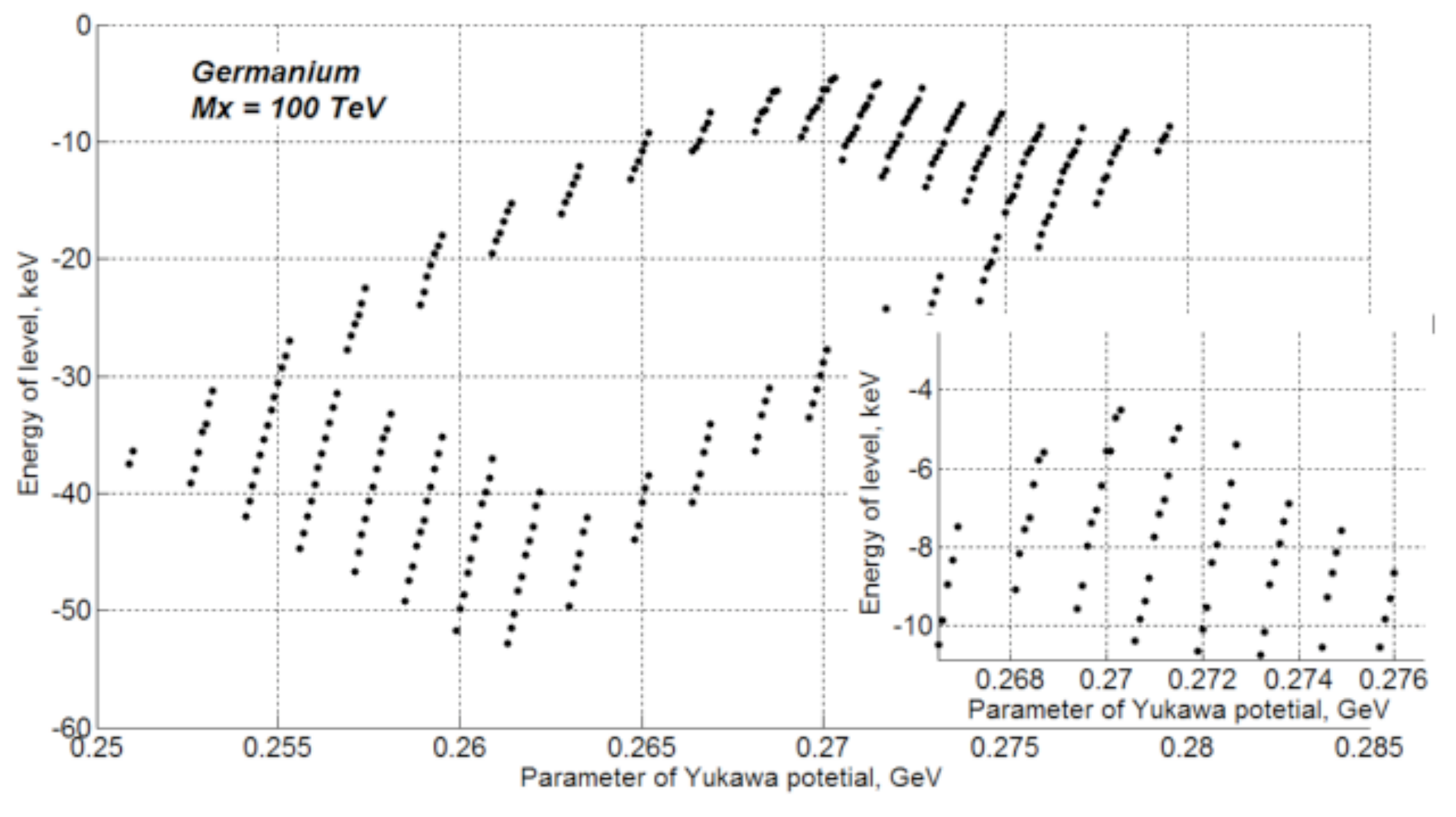}\\
        \caption{Energy levels in OHe bound system with germanium.}\label{mkpa2Ge100}
    \end{center}
\end{figure}

\begin{figure}
    \begin{center}
        \includegraphics[width=4in]{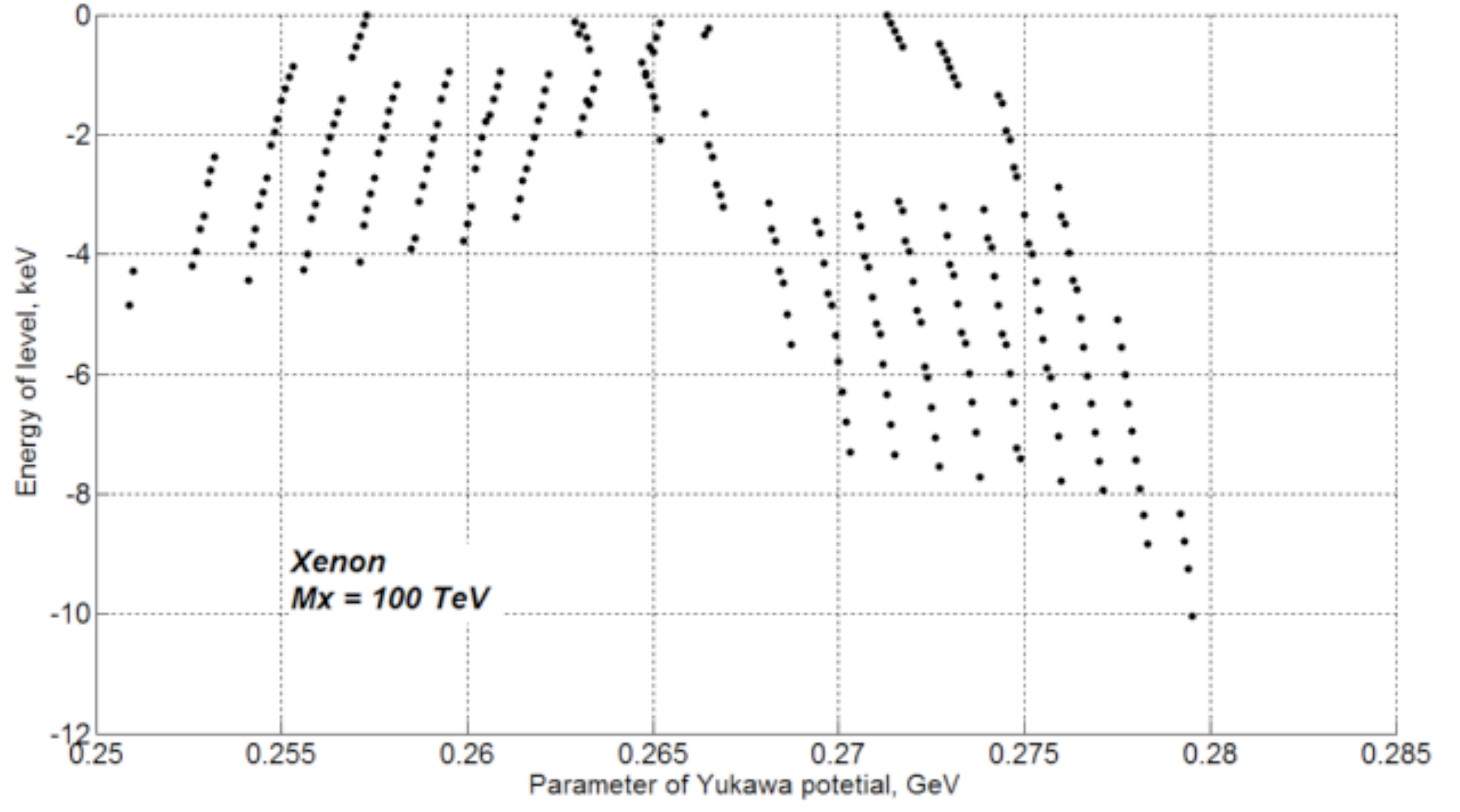}\\
        \caption{Energy levels in OHe bound system with xenon.}\label{mkpa2Xe100}
    \end{center}
\end{figure}

In view of possible applications for the approach, unifying spins and charges \cite{mkpa2Norma}, we consider here the case of superheavy OHe, since the candidate for $X^{--}$, coming from stable 5th generation ($\bar u_5 \bar u_5 \bar u_5$) is probably much heavier, than 1 TeV. With the growth of the mass of O-helium the reduced mass (\ref{mkpa2m}) slightly grows, approaching with higher accuracy the mass of nucleus. It extends a bit the range of nuclear parameters $\mu$ and $g^2$, at which the binding energy of OHe with sodium and/or iodine is within the range 2-6 keV (see Fig. \ref{mkpa2NaI100}).
At these parameters the binding energy of O-helium with germanium and xenon are presented on figures \ref{mkpa2Ge100} and \ref{mkpa2Xe100}, respectively. Qualitatively, these predictions are similar to the case of $S_3=1$. Though there appears a narrow window with OHe-Ge binding energy, below 6 keV for the dominant range of parameters energy release in CDMS is predicted to be of the order of few tens keV.
\section{Conclusions}

%\medskip

To conclude, the results of dark matter search in experiments
DAMA/NaI and DAMA/LIBRA can be explained in the framework of
composite dark matter scenario without contradiction with negative
results of other groups. This scenario can be realized in different
frameworks, in particular in Minimal Walking Technicolor model or in
the approach unifying spin and charges and contains distinct
features, by which the present explanation can be distinguished from
other recent approaches to this problem \cite{mkpa2Edward} (see also
review and more references in \cite{mkpa2Gelmini}).

Our explanation is based on the mechanism of low energy binding of OHe with nuclei.
We have found that within the uncertainty of nuclear physics parameters there exists a range at which OHe
binding energy with sodium and/or iodine is in the interval 2-6 keV. Radiative capture of OHe to this bound state leads to the corresponding energy release observed as an ionization signal
in DAMA detector.

OHe concentration in the matter of underground detectors is determined by the equilibrium between the incoming cosmic flux of OHe and diffusion towards the center of Earth. It is rapidly adjusted and follows the
change in this flux with the relaxation time of few
minutes. Therefore the rate of radiative capture of OHe should experience annual modulations reflected in annual modulations of the ionization signal from these reactions.

%The method to calculate the rate of OHe reactions was developed and
%the calculated total amount of such events is shown to be consistent
%with the results of DAMA/NaI and DAMA/LIBRA experiments for the mass
%of OHe around 1 TeV. This method can be applied to the analysis of
%the whole set of inelastic processes, induced by O-helium in matter.

An inevitable consequence of the proposed explanation is appearance
in the matter of DAMA/NaI or DAMA/LIBRA detector anomalous
superheavy isotopes of sodium and/or iodine,
having the mass roughly by $m_o$ larger, than ordinary isotopes of
these elements. If the atoms of these anomalous isotopes are not
completely ionized, their mobility is determined by atomic cross
sections and becomes about 9 orders of magnitude smaller, than for
O-helium. It provides their conservation in the matter of detector. Therefore mass-spectroscopic
analysis of this matter can provide additional test for the O-helium
nature of DAMA signal. Methods of such analysis should take into account
the fragile nature of OHe-Na (and/or OHe-I) bound states. Their binding energy is only few keV.

With the account for high sensitivity of our results to the values of uncertain nuclear parameters
and for the approximations, made in our calculations, the presented results can be considered
only as an illustration of the possibility to explain puzzles of dark matter search in
the framework of composite dark matter scenario. However, even at the present level of
our studies we can make a conclusion that the ionization signal expected in detectors
with the content, different from NaI, can be dominantly in the energy range beyond 2-6 keV.
Therefore test of results of DAMA/NaI and DAMA/LIBRA experiments by other experimental groups can become a very nontrivial task.

%\bigskip

%{\centering{ \large \textbf{Acknowledgments}} }

\section {Acknowledgments}

%\medskip

We would like to thank Jean Pierre Gazeau for discussions.

\section*{Appendix. Solution of Schrodinger equation for rectangular well}
In the 4 regions, indicated on Fig. \ref{mkpa2pic23}, Schrodinger
equation has the form
\begin{equation}
I: \frac{1}{r}\frac{d^2}{dr^2}(r\psi_1)+k_1(r)^2\psi_1=0,
k_1(r)=k_1=\sqrt{2m(U_1-|E|)};
\end{equation}

\begin{equation}
II: \frac{1}{r}\frac{d^2}{dr^2}(r\psi_2)+k_2(r)^2\psi_2=0,
k_2(r)=k_2=\sqrt{2m(U_2-|E|)};
\end{equation}

\begin{equation}
III: \frac{1}{r}\frac{d^2}{dr^2}(r\psi_3)+k_3(r)^2\psi_3=0,
k_3(r)=k_3=\sqrt{2m(U_3-|E|)};
\end{equation}

\begin{equation}
IV: \frac{1}{r}\frac{d^2}{dr^2}(r\psi_4)-k_4(r)^2\psi_4=0,
k_4(r)=k_4=\sqrt{2m|E|}.
\end{equation}

The wave functions in these regions with the account for the boundary conditions have the form \cite{mkpa2LL3}

\begin{equation}
I: \psi_1=A\frac{\sin (k_1 r)}{r};
\end{equation}

\begin{equation}
II: \psi_2=\frac{B_1\cdot exp(-k_2 r)+B_2\cdot exp(k_2 r)}{r};
\end{equation}

\begin{equation}
III: \psi_3=C\frac{\sin (k_3 r + \delta)}{r}
\end{equation}

\begin{equation}
IV: \psi_4=D\frac{\exp (-k_4 r)}{r}
\end{equation}

The conditions of continuity of a logarithmic derivative
$\frac{\psi_i'}{\psi_i}=\frac{\psi_{i+1}'}{\psi_{i+1}}$ îò
$\emph{r}\psi$ at the boundaries of these regions \emph{$r=R_A$},
\emph{$r=a$} and \emph{$r=b$} are given by

\begin{equation}
I - II: k_1\cdot ctg(k_1 R_A) = k_2\cdot \frac{exp(k_2 R_A)-F\cdot
exp(-k_2 R_A)}{exp(k_2 R_A)+F\cdot exp(-k_2 R_A)},
\label{mkpa2e12}
\end{equation}

\begin{equation}
II - III: k_3\cdot ctg(k_3 a + \delta) = k_2\cdot \frac{exp(k_2
a)-F\cdot exp(-k_2 a)}{exp(k_2 a)+F\cdot exp(-k_2 a)},
\label{mkpa2e13}
\end{equation}

\begin{equation}
III - IV: k_3\cdot ctg(k_3 b + \delta) = -k_4,
\label{mkpa2e14}
\end{equation}

where
\begin{equation}
F = B_1/B_2.
\end{equation}

Now we can solve this system of equations for 3 variables. It follows from Eq. (\ref{mkpa2e12}) that
\begin{equation}
F = exp(2 k_2 R_A) \cdot \frac{k_2-k_1\cdot ctg(k_1
R_A)}{k_2+k_1\cdot ctg(k_1 R_A)},
\end{equation}

and from Eq. (\ref{mkpa2e13})

\begin{equation}
\delta=-k_3 a + arcctg(\frac{k_2}{k_3}\cdot \frac{exp(k_2 a)-F\cdot
exp(-k_2 a)}{exp(k_2 a)+F\cdot exp(-k_2 a)}).
\end{equation}

Since

\begin{equation}
E=U(r)-\frac{k^2}{2m},
\label{mkpa2e18}
\end{equation}

one has

\begin{equation}
k_4=\sqrt{2mU-k_3^{2}},
\end{equation}

Then Eq.(\ref{mkpa2e14}) has the form
\begin{equation}
k_3^{2}[\frac{1}{sin^{2}(k_3 b + \delta)}-1]=2mU_3-k_3^{2},
\end{equation}

%\bigskip

%{\centering{ \large \textbf{References}} }

%\section*{References}

%\medskip

%% A. Kleppe, 02.11.2009
%\documentstyle[12pt,axodraw]{article}
%\textwidth 16cm
%\textheight 24cm
%\hoffset=-1cm
%\oddsidemargin 0.5cm
%\evensidemargin 0.5cm
%\topmargin -1cm
%
%%
%%************************* maketitle content *********************************
\title{On the Possibilities and Impossibilities of Random Dynamics}
\author{A. Kleppe}
\institute{%
SCAT, Oslo, Norway}
 
\titlerunning{On the Possibilities and Impossibilities of Random Dynamics}
\authorrunning{A. Kleppe}
\maketitle                                       

%****************************************************************************
\begin{abstract}
Random Dynamics is an anti-grand unification project, based on the assumption that at a fundamental scale Nature is not necessarily "simple", but probably enormously complicated and is most simply described in terms of randomness.  
The ambition is to "derive" all the known physical laws as an almost unavoidable consequence of a random fundamental "world machinery", which is a very general, random mathematical structure, which contains non-identical elements and some set-theoretical notions.

But how can one extract anything from something very general and random, which is not even well described in detail?

\end{abstract}
%%%%%%%%%%%%%%%%%%%%%%%%%%%%%%%%%%%%%%%%%%%%%%%%%%%%%%%%%%%%%%%%%%%%
\section{The notion of theory}

%%\begin{flushleft}
The ambition of the physicist's search for a 'theory of everything' is to formulate an ultimate, finite theory. Many, probably most physicists, favour the Grand Unified Theory (GUT)-scenario, based on the assumption that the symmetry increases with energy (in the sense of larger symmetry groups describing the dynamics). The idea is that there is a large group at $10^{15} GeV$, which spontaneously breaks down and eventually ends up as SU(3)xSU(2)xU(1) at the weak scale. The interactions that are observed as separate, different forces at our energy level, are thus believed to be unified at GUT level.     
Physics itself seems to point towards unification and "monocausality", the amazing success of the Standard Model and electroweak unification indeed seems to whisper: "Grand Unification".

But there are alternative approaches according to which physics does not become simpler and more symmetric at higher energies, and there is no unification at higher energy.

The Random Dynamics project \cite{ak1RD} is such an anti-grand-unification (AGUT) scheme, based on the assumption that the physical laws as we know them come about at low energy, from something un-describably complex that exists at high energy.  
The assumption is that at a fundamental scale, Nature is enormously complicated and most simply described in terms of {\it randomness}, and that the regular and fairly simple physics we observe comes about as one goes down from the high energy "fundamental" level to our lower energy level. The idea is that {\it any} sufficiently complex and general model for the fundamental physics at (or above) the Planck scale, will in the low energy limit (where we operate) yield the physics we know.
The reason is that as we go down the energy scale, the structure and complexity characteristic of the high energy level are shaved away. Only those features survive which are common for the long wavelength limit of any generic model of fundamental supra-Planck scale physics.
The ambition of Random Dynamics is to "derive" all the known physical laws as an almost unavoidable consequence of a random fundamental "world machinery".
 
When we call something {\it fundamental}, it's usually implied that it is {\it simple}, but
the 'simplicity' of Random Dynamics lies in simple formulations like "the fundamental world machinery is essentially random".
If one would be able to formulate the details of the "laws", they are probably exceedingly complicated!

The expectation that the fundamental level should be 'simple' or 'transparent' started with Euclid (300 B.C.), who found that he could bring back the theorems of geometry onto a small set of axioms.
The basic idea is that the information content of a theory is contained in a finite set of axioms/principles/elements which so to speak constitutes the truth content of the theory.

In the beginning of the 20th century, David Hilbert wanted to do the same thing for mathematics, by constructing a formal axiomatic system from which he was going to derive all of mathematics.

In 1931, Kurt G\"{o}del \cite{ak1Godel} however proved that Hilbert's program was impossible.
He showed that any finite formal system of axioms is either incomplete or inconsistent. If you assume that your formal axiomatic system only tells the truth, it will not tell the whole truth,
and if you assume that the axioms don't allow to prove false theorems, there will be true theorems that cannot be proven within your axiomatic system.
G\"{o}del's result concerns any formal axiomatic system. This is of course relevant for both mathematics and physics.

A physical theory is a mathematical description of some part of reality, allowing us to make accurate, verifiable predictions,
and a 'fundamental physical theory' is by definition a theory about the physics at very high energy $\sim$ Planck scale. But we of course don't know anything about what happens at Planck scale, albeit
we do know that physics looks very different at different energy levels. What is  elementary at one level is complex at another level. 
For example in chemistry the atom is fundamental, while in particle physics it is complex and non-fundamental in the sense that given the equations of elementary particle physics, it is not obvious that atoms should be constructed as they are. 
It is thus very optimistic to believe that we can guess what physics is like at Planck scale.

But let us nevertheless imagine that such a fundamental theory is formulated: an equation, an algorithm or some principle(s) constituting the ultimate theory.
{\it According to G\"{o}}{\it del, such a theory can however never encompass all of physics in a consistent way}.
There can exist excellent 'partial' theories, G\"{o}del doesn't imply that there is no scientific truth or insight. But it's not possible to formulate a theory that encompasses all of reality.
%\end{flushleft}

\section{The information content of a theory}

One way of dealing with the notion of theory, is to regard it as a (computer) program for predicting observations.
Occam's "the simplest theory is best" then reads "the most concise computer program is the best theory".

Now, it's very hard to make a theory about some part of reality - a "system" - unless this system has some regularity. The reason is that a theory is a formulation of the information content of the system, and in a programming context, to fish out the information content of a system corresponds to {\it compressing} \cite{ak1Chaitin} it. A very regular system obviously allows for more compression than a more chaotic system, so the computer program that describes a regular system is smaller than the program needed to describe a more chaotic system.

\noindent For example, a pattern like this:

$$
\includegraphics{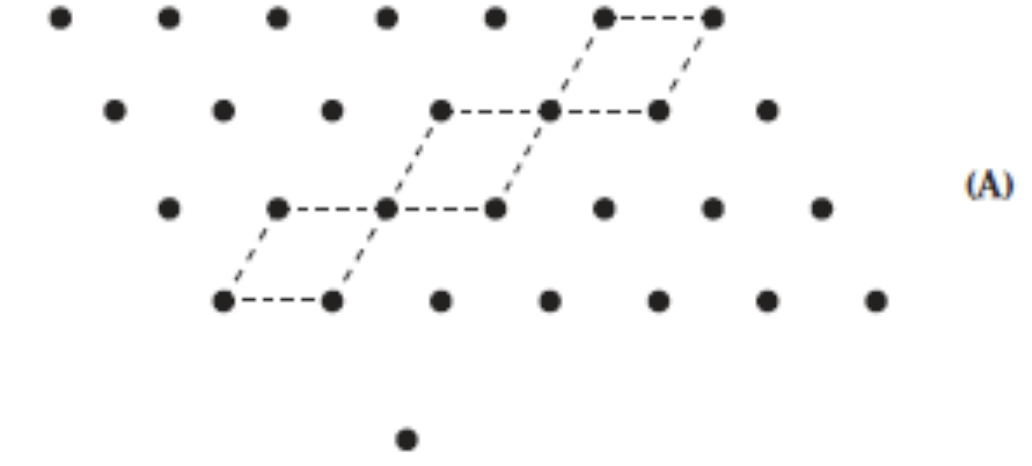}
$$
%%\begin{center}\begin{picture}(220,125)(0,0)
%%\SetScale{1.5}
%%\DashLine(65,60)(55,42){2}
%%\DashLine(85,60)(75,42){2}
%%\DashLine(55,42)(75,42){2}
%%\DashLine(65,60)(85,60){2}
%%
%%\DashLine(115,77)(105,60){2}
%%\DashLine(115,77)(95,77){2}
%%\DashLine(85,60)(95,77){2}
%%\DashLine(85,60)(105,60){2}
%%
%%\DashLine(35,42)(55,42){2}
%%\DashLine(45,25)(55,42){2}
%%\DashLine(35,42)(25,25){2}
%%\DashLine(45,25)(25,25){2}
%%
%%\Vertex(25,25){2}
%%\Vertex(145,25){2}
%%\Vertex(45,25){2}
%%\Vertex(65,25){2}
%%\Vertex(85,25){2}
%%\Vertex(105,25){2}
%%\Vertex(125,25){2}
%%\Vertex(35,42){2}
%%\Vertex(15,42){2}
%%\Vertex(55,42){2}
%%\Vertex(75,42){2}
%%\Vertex(95,42){2}
%%\Vertex(115,42){2}
%%\Vertex(135,42){2}
%%
%%\Vertex(5,60){2}
%%\Vertex(25,60){2}
%%\Vertex(45,60){2}
%%\Vertex(65,60){2}
%%\Vertex(85,60){2}
%%\Vertex(105,60){2}
%%\Vertex(125,60){2}
%%\Vertex(-5,77){2}
%%\Vertex(15,77){2}
%%\Vertex(115,77){2}
%%\Vertex(35,77){2}
%%\Vertex(55,77){2}
%%\Vertex(75,77){2}
%%\Vertex(95,77){2}
%%\Text(250,70)[]{{\bf{(A)}}}
%%
%%\end{picture}\end{center}

\noindent has less complexity than this one:
$$
\includegraphics{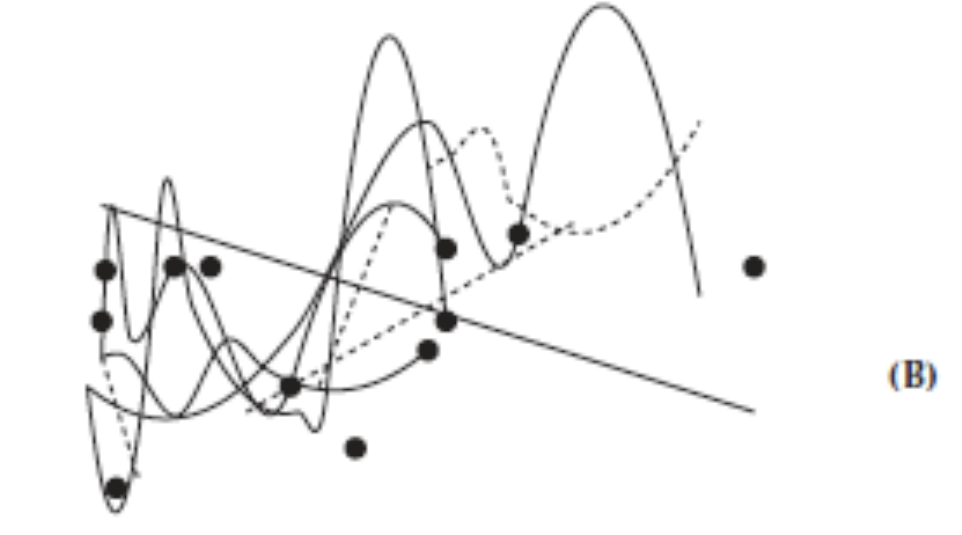}
$$
%%\begin{center}\begin{picture}(220,165)(0,0)
%%\Curve{(25,75)(30,76)(46,59)(60,80)(70,72)(115,77)}
%%\Curve{(21,67)(39,90)(41,115)(50,89)(79,60)(87,65)(90,96)(120,85)}
%%\DashLine(65,60)(155,112){2}
%%\Vertex(120,85){3}
%%\Vertex(25,85){3}
%%\Vertex(115,77){3}
%%\Vertex(140,109){3}
%%\Vertex(77,67){3}
%%\Vertex(26,99){3}
%%\Vertex(45,100){3}
%%\DashLine(85,67)(105,117){2}
%%\Vertex(55,100){3}
%%\Vertex(95,50){3}
%%\Vertex(88,200){3}
%%\Vertex(205,100){3}
%%\Vertex(29,39){3}
%%\DashLine(35,42)(25,75){2}
%%\DashCurve{(115,127)(120,130)(136,125)(137,120)(140,117)(190,140)}{2}
%%\Curve{(21,67)(60,64)(77,81)(115,140)(140,109)(145,133)(190,92)}
%%\Curve{(25,75)(26,99)(33,80)(45,100)(77,67)(80,77)(120,105)}
%%\Vertex(120,105){3}
%%\Line(25,117)(205,60)
%%\Text(250,70)[]{{\bf{(B)}}}
%%\end{picture}
%%\end{center}
%\begin{flushleft}

\noindent Therefore, while we easily can formulate an algorithm P(A) which describes/ge\-nerates the pattern ${\bf{A}}$, a description of ${\bf{B}}$ may require all the information in ${\bf{B}}$, unless we manage to find a way of compressing the pattern ${{\bf{B}}}$. 
We can conclude that most probably, $P(A) \leq P(B)$.
%\end{flushleft}
Now, unless the amount of information of the theory is smaller than the amount of information of the the described system, the theory strictly speaking is no theory.
In the pattern ${\bf{B}}$ above, there does not seem to be any clear regularity, thus no obvious compression is at hand, and there is no theory (or algorithm) for ${\bf{B}}$!
But there might come a theory! The day someone finds a regularity in ${\bf{B}}$ that allows a algorithm generating ${\bf{B}}$ to be formulated, then we would have a theory for ${\bf{B}}$.

Suppose that you have a theory: a program withe fewer bits than the system described by the theory.
It is good to have a theory, but you are very ambitious, and want to have a "fundamental theory", meaning the smallest theory that encompasses the (non-redundant) information of the described system.

In a computer context your "system" can for example be a string of tokens like 

$S_1$ = 1000110110101111100010... or

$S_2$ = ABCQQBCARQQAA.. or

$S_3$ = 1269789125482976502...

and your "theory" is an algorithm that gives rise to your string.
The fundamental theory for a string is the smallest program that generates this string, i.e. the program with the smallest complexity (i.e. the least information).

%\begin{flushleft}
So you have a system (a string of numbers) $S$, and
a small program $P_S$ which generates $S$. $P_S$ is rather small, so you may think that $P_S$ is your fundamental theory. But $P_S$ is perhaps still compressible - you just haven't realized it. And still worse: if $P_S$ were incompressible, you would not be able to prove it, precisely because you can never prove that nobody will never find a program $P_S'$ which is smaller than $P_S$, such that $P_S'$ in its turn generates $P_S$:   $P_S'$    $\rightarrow$ $P_S$.

And since you cannot know if there is such a program $P_S'$, you cannot decide on the complexity of $S$, because its complexity is defined as being equal to the complexity of the smallest program that generates $S$...

A "fundamental theory" which describes $S$ would thus have the same complexity as $S$. But since you cannot ever prove that a program $P_S$ that generates $S$ is the smallest possible, the complexity of $S$ is {\it undecidable}. You may have a small program $P_S$ that generates $S$, but you cannot know if $P_S$ is the smallest program generating $S$, and
therefore, you cannot know if a theory is fundamental or not.
This is true in any situation where we want to formulate a theory about some part of the world or the entire world, in terms of a finite set of axioms.

An axiom is a statement or a string that we simply have to {\it define} as fundamental (meaning that it cannot be defined in terms of something more fundamental). And a pattern is random if it has no (obvious) pattern, i. e. there is no obvious plan behind its structure (so it cannot be defined by something even more fundamental). As we saw above, we cannot prove that any string is fundamental (incompressible), so we cannot prove that an axiom is fundamental. Likewise, we cannot prove that a random pattern is random, but we can with cetrainty claim that an axiom should be (information-theoretically) random.

In sum: the search for a theory of everything can be regarded as a quest for an ultimate compressions of the world.
But since we cannot know when we have reached the limit of compressibility we can never know when or if our theory is a theory of everything, because {\em we cannot:} 

- prove incompressibility. 

- prove randomness.

- prove that some statement or string {\it A} is an axiom.
%\end{flushleft}

\section{Random Dynamics}

In Random Dynamics, the problems in formulating an ultimate theory are circumvened by not starting from a well-defined, finite set of formal axioms, but from "a random mathematical structure" ${\cal{M}}$, which is not described in great detail. All we know is that the fundamental "world machinery" ${\cal{M}}$ is a very general, random mathematical structure, which contains non-identical elements and some set-theoretical notions. There are also strong exchange forces present, but there is as yet no physics. At some stage ${\cal{M}}$ comes about, and then physics follows.

We have no detailed information about the nature of the elements of ${\cal{M}}$, but we nevertheless claim that the observed physics emerges from ${\cal{M}}$, defined as a generic set randomly chosen from a set of such sets, \{${\cal{M}}_1$, ${\cal{M}}_2$, ${\cal{M}}_3$..\} where every ${\cal{M}}_j$ gives rise to the known physics at low energy, and none of the ${\cal{M}}_j$ is known in elaborate detail.   

While Random Dynamics avoids the consistency and completeness problems of a formal ultimate theory by describing its basic assumptions in heuristic, non-formal terms, there is still something worrisome about this approach. The problem is the apparent paradox that we start with something which by definition is generic and not described in great detail. How can we ascribe properties - and such powerful properties - to something that not observable or even describable? 

Should we not demand that ${\cal{M}}$ be related it to something more substantial and well-known? The claim is after all that ${\cal{M}}$ underlies all of reality.

\section{An excursion into the real}

So let us consider something very well known: the real numbers, 
${\bf{\cal{R}}}$. 
A real number can be described as a length measured with arbitrary precision with up to an infinite number of digits, and a computable real number is a number for which there is a computer program that calculates its digits one by one.

Next consider the set of all possible computer programs, which is a {\em countable set}. 
We list the possible computer programs:
\begin{eqnarray}\label{programs}
{\hspace{50mm}}P_1\nonumber\\
{\hspace{50mm}}P_2\nonumber\\
{\hspace{50mm}}P_3\nonumber\\
{\hspace{50mm}}...\nonumber
\end{eqnarray}

\noindent Since each computable real corresponds to a  computer program, the set of computable reals is also countable.

But the set of all real numbers ${\cal{R}}$ has the power of the continuum, so ${\cal{R}}$ {\em is uncountable}, and thus 
the set of uncomputable reals, i. e. [${\cal{R}}$ $\backslash$ \{{\em computable reals}\}] is uncountable.
And there are many more uncomputable reals than computable reals, so if you randomly pick a real number from the number line, the number you pick will with probability 1 be an uncomputable number!

Proof: take all computable reals, take all computer programs that compute these reals, and make a list:
\\
\begin{eqnarray}
p_1 & \sim& first{\hspace{2mm}}computable{\hspace{2mm}} real{\hspace{2mm}}r_1\nonumber\\
p_2 & \sim& second{\hspace{2mm}}computable{\hspace{2mm}}real{\hspace{2mm}}r_2\nonumber\\
p_3 & \sim& third{\hspace{2mm}}computable{\hspace{2mm}} real{\hspace{2mm}}r_3\nonumber\\
    &....\nonumber\\                
p_k & \sim& kth{\hspace{2mm}}computable{\hspace{2mm}} real{\hspace{2mm}}r_k\nonumber\\
    &....\nonumber                
\end{eqnarray}
and cover each $r_j$ with the interval $\epsilon/2^j$.
The size of the sum of all the covering intervals is then  
$\epsilon/2+\epsilon/2^2+\epsilon/2^3+...\epsilon/2^k+..=\epsilon$, where $\epsilon$ can be made indefinitely small. The probability of randomly picking a computable real from the number line is thus indefinitely small - i.e. zero.
%\begin{flushleft}
Thus:

{\it The set of reals that can be individually named or specified or even defined or referred to within formal language, has probability zero.}   
%\end{flushleft}
%\begin{flushleft}
Thus: 

{\it Reals are un-nameable with probability 1.}
%\end{flushleft}

Even if we in practice talk about these un-nameable reals, only those reals that are defined by a finite amount of information can be spoken about on a formally secure ground. We can refer to the uncomputable reals, but we can by definition not specify the properties of an individual uncomputable real.

So, when we talk about 'the real numbers', we talk about entities that we mostly cannot label individually.
Only very few of them are tangible, computable, most of them are untouchable, un-nameable. Among these un-nameable reals we discern the {\it random reals}, which are maximally incompressible reals, meaning that few or none of the digits are computable by a program. 

So the few reals that are hands-on are those that are completely defined by a finite number of digits in the sense that there is a finite number of well-defined programs by which the digits can be calculated.

But as we know, that doesn't make the notion of real numbers useless, on the contrary.
The same is true for ${\cal{M}}$. We don't know its details, but we are nevertheless able to deduce a wealth of information from the assumption of its existence.

\section{Emergent phenomena}

The requirement that a fundamental theory should consist of a finite set of simple elements from which the observed physics can be deduced, means that the "fundamental" is supposed to be finite, transparent and handable.

According to the Random Dynamics approach there is however no such transparency at the fundamental level, which on the contrary is believed to be characterized by a lack of (visible) organized structure. Only the initial input can be perceived as simple and transparent, consisting of the set ${\cal{M}}$, some set theoretical notions and some exchange forces, from which the physics {\it emerges}.

%\begin{flushleft}
{\it Emergence} is a process reconstructing a system in such a way that some new - emergent - properties appear.

An example is deterministic chaos, where deterministic equations of motion lead to apparently unpredictable behaviour. 

The randomness originates from a situation where the effective dynamic which maps initial conditions to states at later times, becomes so complicated that an observer has no way to compute precisely enough to predict the future behaviour.

Opposite example: order arising from disorder, like in self-avoiding random walk in 2 dimensions, where the step-by-step behaviour of a particle is constrained by the demand that the next step is taken in a random direction, with the exclusion of the direction from which it comes. This results in a path tracing out a self-similar set of positions in the plane: a "fractal" structure emerges!

Deterministic chaos and self-avoiding random walk are examples of emergence of {\it pattern}. 
The emerging features are new and in direct opposition to the system's defining character.

Emergent phenomena also have to do with {\it scale}. The system as a whole may have properties that are not apparent at elementary levels of scale; the same goes for energy levels.
  
An emergent property which is well-defined at one level may be meaningless at another. For example: the M\"{o}bius strip. The strip may be cut up in smaller parts, each of which is orientable; while the entire M\"{o}bius strip is not. The non-orientability can be described as emerging while perceiving the strip, i. e. the sum of the parts, as a whole.
According to the philosophy of emergence, while there certainly exist certain global, fundamental principles, many of the notions we perceive as fundamental are only "locally fundamental".

In condensed matter physics we see many instances of emergence, and while investigating whether physics follows a GUT or an anti-GUT scheme, it can be useful to use {\it analogy}, by comparing high energy physics with the physics of quantum liquids, superconductors, superfluids, ferromagnets. Condensed matter systems display many properties reminiscent of high energy physics, both "GUT-features" and "AGUT-features" \cite{ak1Volovik}.

Superfluid $^3He-A$ is an example of this; at high temperatures, the $^3He$ gas, and at lower temperatures the $^3He$ liquid have all the symmetries of ordinary condensed matter: translational invariance, global U(1) group, etc.
When the temperature goes down, the liquid $^3He$ reaches the superfluid temperature $T_c \sim 1 mK$, and below $T_c$ all the symmetries disappear, except translational invariance: $^3He$ is still liquid. 
This low energy symmetry breaking resembles the one in particle physics - in accordance with the GUT scenario. But then, as $T_c \rightarrow 0$, the superfluid $^3He-A$ gradually acquires all the high energy symmetries.
From nothing it gets Lorenz invariance, local gauge invariance, etc, in a perfect AGUT spirit. 

Seemingly fundamental features may thus disappear or emerge with changing energy or scale, it is thus very hard to establish what is "fundamental" - hard to formulate an ultimate theory.
Some things that are scale invariant however remain:

- physical principles
(like the principle of least action, conservation of energy).

- mathematical rules
(integers, mathematical operations, logic).
%\end{flushleft}

\section{Conclusion}

The goal of the Random Dynamics project is to formulate the minimal set of assumptions needed for deriving the laws of nature, and thereafter derive the known physics. It may seem both too vague and to technical, but the notion of a set as general as the Random Dynamics fundamental "world machinery" ${\cal{M}}$ is not more empty or meaningless than the notion of real numbers.

The stumbling block of any ultimate theory, is its all-encompassing ambition. An ultimate theory is supposed to be finite, containing a finite amount of information (axioms, assumptions), but since a finite formal system of axioms is either incomplete or inconsistent, no finite theory can ever be "ultimate" in the sense of all-explanatory. 
The Random Dynamics random point of departure offers a way out of this dilemma.

The Random Dynamics philosphy that symmetries and seemingly fundamental laws of nature are in reality emergent phenomena, is moreover supported by data from condensed matter physics.

%% SNMB, 22.12.2009, first contribution (revised)
%%
\title{Spin Connection Makes Massless Spinor Chirally Coupled to 
Kaluza-Klein Gauge Field After Compactification 
of $M^{1+5}$ to $M^{1+3} \times$ Infinite Disc Curved on $S^2$}
\author{D. Lukman${}^1$, N. S. Manko\v c Bor\v stnik${}^1$ and H. B. Nielsen${}^2$}
\institute{%
${}^1$ Department of Physics, FMF, University of Ljubljana,\\
 Jadranska 19,Ljubljana, 1000\\
${}^2$ Department of Physics, Niels Bohr Institute,\\
Blegdamsvej 17,
Copenhagen, DK-2100}

\authorrunning{D. Lukman, N. S. Manko\v c Bor\v stnik and H. B. Nielsen}
\titlerunning{Spin Connection Makes Massless Spinor Chirally Coupled\ldots}
\maketitle
 
\begin{abstract} 
One step towards realistic Kaluza-Klein-like theories and a loop hole through the 
Witten's "no-go theorem" is presented for cases which we call "an effective two dimensionality" cases:  
We present the case of a spinor in $d=(1+5)$ compactified on an (formally) infinite disc with 
the zweibein which makes a disc curved on $S^2$ and with the spin connection 
field  which allows on such a sphere only one massless spinor state of a particular charge, 
which couples the spinor chirally to the corresponding Kaluza-Klein gauge field. 
In refs.~\cite{dhn1hnkk06,dhn1hn07} we achieved masslessness of spinors with  the appropriate 
choice of a boundary on a finite disc, in this paper the masslessness is achieved with 
the choice of a spin connection field on a curved infinite disc. In $d=2$, namely, the equations of motion 
following from the action with the linear curvature leave  spin connection and zweibein 
undetermined~\cite{dhn1dhnproc04}. 
\end{abstract}

\section{Introduction}
\label{dhn1introduction}
The idea of Kaluza and Klein of obtaining the electromagnetism 
- and under the influence of their idea in now a days also the  weak and colour fields - 
from purely gravitational degrees of freedom connected with having extra dimensions 
is very elegant, but were almost killed by a "no-go theorem" of E. Witten~\cite{dhn1witten}  telling that 
these kinds of theories have very severe difficulties with obtaining  massless fermions 
chirally coupled to the Kaluza-Klein-type gauge fields in $d=1+3$, 
as required by the standard model of the electroweak and colour interactions.
There may be  escapes from the "no-go theorem" by having torsion or by 
having an orbifold structure in the extra dimensional space. 
In refs.~\cite{dhn1hnkk06,dhn1hn07} we achieved masslessness of spinors with  the appropriate 
choice of a boundary on a finite disc. 

When we have no fermions present and only  the curvature in the Lagrange 
density the spin connections are determined from the vielbein fields and   
the torsion is zero.
A major point of the present article is that in some cases the 
spin connections, we  call  these cases "an effective two-dimensionality" is not
fully determined from the vielbeins. In such special cases  there is 
the possibility of having torsion in a gauge theory of gravity with spin connections and vielbeins 
and therefore for a possibility for a Kaluza-Klein-like model, which effectively in four dimensional 
space-time manifests the known gauge fields, while yet the Lagrange density  contains only the 
curvature. This opens a loop  hole through the 
Witten's "no-go theorem" even if there are no boundaries, which take care of maslesness~\cite{dhn1hnkk06,dhn1hn07}.

In the here proposed types of models is the chance for having chirally mass protected fermions 
in a theory in which the chirally 
protecting effective four dimensional gauge fields are true Kaluza-Klein-like fields 
the degrees of which inherit from the higher dimensional gravitational ones.
We are thus hoping for a revival of true Kaluza-Klein[like models as candidates for 
phenomenological viable models!

One of us has been trying for long to develop the  Approach unifying spins and charges so that 
spinors which carry in $d\ge 4$ nothing but two kinds of the spin (no charges), would manifest 
in $d=(1+3)$ all the properties assumed by the Standard model. The Approach  
proposes in $d=(1 + (d-1))$  a simple starting action for spinors  with the two kinds of the 
spin generators ($\gamma$ matrices): the Dirac one, which takes care of the spin and the charges, 
and the second one, 
anticommuting with the Dirac one, which generates families~\footnote{To understand the 
appearance of the two kinds of the spin generators we invite the reader to look at the 
refs.~\cite{dhn1pikanorma06,dhn1holgernorma2002,dhn1holgernorma2003}.}. A spinor couples in $d=1+13$ 
to only the vielbeins and (through two kinds of the spin generators to) the 
spin connection fields.  
Appropriate breaks of the starting symmetry lead to the 
left handed quarks and leptons in $d=(1+3)$, which carry the weak charge while the 
right handed ones are weak chargeless. The Approach might have the right answer to  
the questions about the origin of families of quarks and leptons, about the explicit 
values of their masses and mixing matrices  
as well as about the masses of the scalar  and the weak gauge fields, about the dark 
matter candidates, and about the breaking of the discrete symmetries\footnote{  
There are many possibilities in the Approach unifying spins and charges for breaking 
the starting symmetries to those of the Standard model. These problems were studied 
in some crude approximations in refs.~\cite{dhn1gmdn06,dhn1gmdn07}. It was also studied~\cite{dhn1hnm06}  
how does the Majorana  mass of spinors depend on the dimension of space-time 
if spinors carry only the spin and no charges. We have proven %in ref.~\cite{dhn1hnm06} 
that only in even dimensional spaces of  $d=2$ 
modulo $4$ dimensions (i.e.\ in $d=2(2n+1),$ $n=0,1,2,\cdots$) spinors  (they are allowed 
to be in families) of one handedness and with no conserved 
charges gain no  Majorana mass.}. 

Let us point out that in odd dimensional spaces and in even dimensional spaces devisible 
with four there is no mass protection in the Kaluza-Klein-like theories~\cite{dhn1hn06}. 
The spaces therefore, for which we can have a hope that the Kaluza-Klein-like theories lead to 
chirally protected fermions and accordingly to the effective theory of the standard model 
of the electroweak and colour interactions, have $2(2n+1)$ dimensions. 

Let us accordingly assume that we start with the $2(2n+1)$-dimensional space, with gravity only,  
described by the action 
  \begin{eqnarray}
         {\cal S} = \alpha \int \; d^d x \, E  {\cal\,  R}.
  \label{dhn1action}
  \end{eqnarray}
with the Riemann scalar ${\cal R} = {\cal R}_{abcd}\eta^{ac}\eta^{bd}$ determined by the Riemann tensor
\begin{eqnarray}
{\cal R}_{abcd} = f^{\alpha}{}_{[a} f^{\beta}{}_{b]}(\omega_{cd \beta, \alpha} 
- \omega_{ce \alpha} \omega^{e}{}_{d \beta} ), 
\label{dhn1RT}
\end{eqnarray}
with vielbeins $f^{\alpha}{\!}_{a}$ (the 
choice of the meaning of indices can be found in the footnote~ \footnote{$f^{\alpha}{}_{a}$ are inverted 
vielbeins to 
$e^{a}{}_{\alpha}$ with the properties $e^a{}_{\alpha} f^{\alpha}{\!}_b = \delta^a{}_b,\; 
e^a{}_{\alpha} f^{\beta}{}_a = \delta^{\beta}_{\alpha} $. 
Latin indices  
$a,b,..,m,n,..,s,t,..$ denote a tangent space (a flat index),
while Greek indices $\alpha, \beta,..,\mu, \nu,.. \sigma,\tau ..$ denote an Einstein 
index (a curved index). Letters  from the beginning of both the alphabets
indicate a general index ($a,b,c,..$   and $\alpha, \beta, \gamma,.. $ ), 
from the middle of both the alphabets   
the observed dimensions $0,1,2,3$ ($m,n,..$ and $\mu,\nu,..$), indices from 
the bottom of the alphabets
indicate the compactified dimensions ($s,t,..$ and $\sigma,\tau,..$). 
We assume the signature $\eta^{ab} =
diag\{1,-1,-1,\cdots,-1\}$.
}), the gauge fields of the infinitesimal generators of translation, and  spin connections $\omega_{ab\alpha}$
the gauge fields of the $S^{ab}= \frac{i}{4}(\gamma^a \gamma^b - \gamma^b \gamma^a)$. 
$[a\,\,b]$ means that the antisymmetrization must be performed over the two indices $a$ and $b$.

In the ref.~\cite{dhn1dhnproc04} we proved that in the absent of the fermion fields the action in Eq.(~\ref{dhn1action}
leads to the equations of motion (Eq. (6.10) in the ref.~\cite{dhn1dhnproc04})
\begin{equation}
\label{dhn1omegabcc}
(d-2)\,\omega_b{}^c{}_c = 
\frac{e^a{}_\alpha}{E} \partial_\beta\left(Ef^\alpha{}_{[a}f^\beta{}_{b]} \right),
%+ \frac{1}{2} \bar{\Psi}\gamma^a S_{ab}\Psi.
\end{equation}
which for $d=2$ clearly demonstrates that  any spin connection $\omega_b{}^c{}_c = 
f^{\alpha}{}_{d} \omega_b{}^d{}_c$ (which can in $d=2$ have  only two different indices) satisfies 
this equation. In the same ref.~\cite{dhn1dhnproc04}, Eq.(6.15), we also prove that for $d=2$ any 
zweibein fulfills the equations of motion 
\begin{eqnarray}
E f^{\alpha}{}_{[a} f^{\beta}{}_{b]} 
&=& \frac{1}{4} \varepsilon^{\alpha \beta} \varepsilon_{ab}. 
\label{dhn1varactd2}
\end{eqnarray}
It also can be shown~(\cite{dhn1dhnproc04}, Eq.(6.17)) that the variation of the action~(\ref{dhn1action})
  with respect to vielbeins leads to the equation
\begin{eqnarray}
-e^s{}_{\sigma} R + 4 f^{\tau t} \omega_{st \sigma,\tau} = 0, 
\label{dhn1fomegad2}
\end{eqnarray}
which is trivially zero for any $R$. This can be seen by multiplying the above equation by 
$ f^{\sigma}{}_{s}$ and summing over the two indices $\sigma$ and $s$. It follows then that $(d-2) R =0.$

\begin{figure}
\centering
\includegraphics{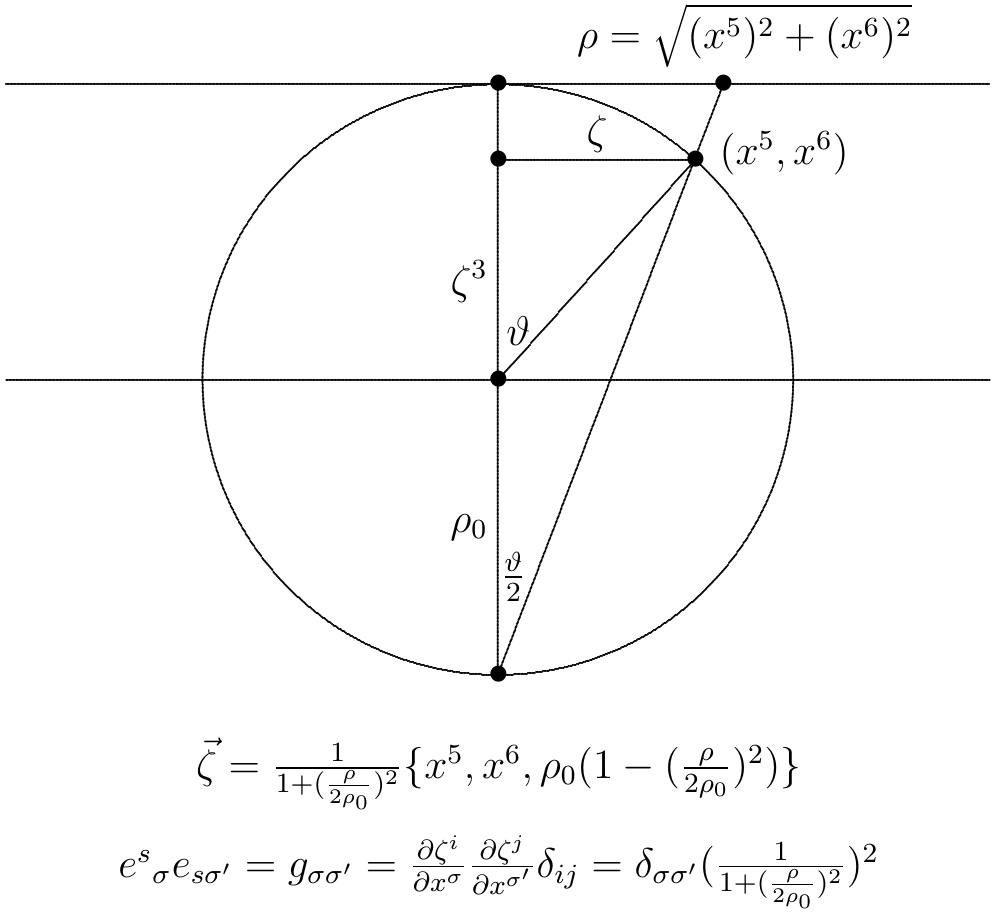}
\caption{The disc is curved on the sphere $S^2$. 
\label{dhn1discgrav}}
\end{figure}

We shall accordingly make a choice for $d=2$ of a zweibein, which curves an infinite disc 
(a two dimensional infinite plane with the rotational symmetry around the 
axes perpendicular to the plane) into 
a spehere $S^2$ with the radius $\rho_{0}$ 
\begin{eqnarray}
e^{s}{}_{\sigma} = f^{-1}
\begin{pmatrix}1  & 0 \\
 0 & 1 \end{pmatrix},
f^{\sigma}{}_{s} = f
\begin{pmatrix}1 & 0 \\
0 & 1 \end{pmatrix},
\label{dhn1fzwei}
\end{eqnarray}
with 
\begin{eqnarray}
\label{dhn1f}
f &=& 1+ (\frac{\rho}{2 \rho_0})^2= \frac{2}{1+\cos \vartheta},\nonumber\\ 
x^5 &=& \rho \,\cos \phi,\quad  x^6 = \rho \,\sin \phi, \quad E= f^{-2}.
\end{eqnarray}
The angle $\vartheta$ is the ordinary asimutal angle on a sphere. % as can be seen from Fig.~ref{Fig.I.}. 
The last relation follows  from $ds^2= 
e_{s \sigma}e^{s}{}_{\tau} dx^{\sigma} dx^{\tau}= f^{-2}(d\rho^{2} + \rho^2 d\phi^{2})$.
We use indices $s,t=5,6$ to describe the flat index in the space of an infinite plane, and 
$\sigma, \tau = (5), (6), $ to describe the Einstein index.  
$\phi$ determines the angle of rotations around  the axis through the two poles of a sphere, 
while $\rho = 2 \rho_0 \, \sqrt{\frac{1- \cos \vartheta}{1+ \cos \vartheta}}$, where 
$\tan \frac{\vartheta}{2} = \frac{\rho}{2\rho_0}$.
Fig.(~\ref{dhn1discgrav}) shows the (well known) relation between $\rho$ and $\vartheta$.

We make a 
choice  of the spin connection field
\begin{align}
  f^{\sigma}{}_{s'}\, \omega_{st \sigma} &= i F\, f \, \varepsilon_{st}\; 
  \frac{e_{s' \sigma} x^{\sigma}}{(\rho_0)^2}
  = i \varepsilon_{st}\, \frac{F\,\sin \vartheta}{\rho_0}\, (\cos \phi,\sin \phi)
  ,\quad s=5,6,\,\,\; \sigma=(5),(6), 
\label{dhn1omegas}
\end{align}
which for  the choice $0 <2F \le 1$ allows only one massless spinor of a particular charge on such 
$S^2$, as we shall see in sect.~\ref{dhn1equations}.

Accordingly, if we have a Weyl spinor in $d=(1+5)$, which breaks into $M^{1+3}$ cross an infinite 
disc, which by a zweibein is curved on $S^2$,
then at least for this case we know the solutions for the gauge 
fields fulfilling the equations of motion for the action linear in the curvature, 
where the vielbein and  
spin connection guarantee masslessness of spinors in the space $d=1+3$.

{\it Let us point out that the two dimensionality can be simulated in any dimension larger than two,
if vielbeins and spin connections are completely flat in all but two dimensions.}

We take (in this paper we do not study the appearance of families, as 
we also did not in the refs.~\cite{dhn1hnkk06,dhn1hn07})  for the covariant momentum of a 
spinor 
\begin{eqnarray}
p_{0 a} &=& f^{\alpha}{\!}_{a}p_{0 \alpha}, \quad p_{0 \alpha} \psi = 
p_{ \alpha} - \frac{1}{2} S^{cd} 
\omega_{cd \alpha}\nonumber\\
a &=& 0,1,2,3,5,\cdots d,\quad \alpha=(0),(1),(2),(3),(5),\cdots (d), . 
\label{dhn1covp}
\end{eqnarray}
A spinor carries  in $d\ge 4$ nothing but 
a spin and interacts accordingly with only the gauge fields of the corresponding 
generators of the infinitesimal transformations (of translations and  the 
Lorentz transformations in the space of spinors), that is 
with vielbeins $f^{\alpha}{\!}_{a}$ and  spin connections $\omega_{ab\alpha}$.

The corresponding Lagrange density 
for   a Weyl spinor has the form
${\cal L}_{W} = \frac{1}{2} [(\psi^{\dagger} E \gamma^0 \gamma^a p_{0a} \psi) + 
(\psi^{\dagger} E \gamma^0\gamma^a p_{0 a}
\psi)^{\dagger}]$, leading to the equation of motion
\begin{eqnarray}
{\cal L}_{W}&=& \psi\, \gamma^0 \gamma^a E \{f^{\alpha}{}_a p_{\alpha} +
\frac{1}{2E} \{p_{\alpha},f^{\alpha}{}_a E\}_-   -\frac{1}{2} S^{cd}  \omega_{cda} 
 \}\psi =0,
\label{dhn1weylL}
\end{eqnarray}
with $ E = \det(e^a{\!}_{\alpha}), $ where  
\begin{eqnarray}
  \omega_{cda} &=& \Re e \;\omega_{cda}, \;\; {\rm if \;\; c,d,a\;\; all\;\; different} \nonumber\\
               &=& i\,\Im m\; \omega_{cda},\;\; \rm{otherwise}.
              % -i F  x^{s} \frac{f-1}{\rho^2},\quad s=5,6, 
\label{dhn1p0s}
\end{eqnarray}
Let us have no gravity in $d=(1+3)$ ($f^{\mu}{}_m = \delta^{\mu}_m$ and  
$\omega_{mn\mu}=0$ for $ m,n=0,1,2,3, \mu =0,1,2,3 $) and let us make  a choice of  a 
zweibein and spin connection on our disc as written in Eqs.~(\ref{dhn1f},\ref{dhn1omegas}). 
($S^2$ does not break the rotational symmetry on the disc, it breaks the translational symmetry 
after making a choice of the nothern ans southern pole.)

Although for any $0 <2F \le 1$ only one massless spinor on $S^2$ is allowed, it will be demonstrated 
that in the  particular case that $2F=1$ the 
spin connection term $- S^{56} \omega_{56 \sigma} $ 
compensates the term  $\frac{1}{2Ef} \{ p_{\sigma},E f \}_-$ for the left handed spinor  
with respect to $d=1+3$, while for the spinor 
of the opposite handedness  the 
spin connection term doubles the term $\frac{1}{2Ef} \{p_{\sigma},E f \}_-$.

The vielbeins and spin connection fields of Eqs.~(\ref{dhn1f},\ref{dhn1omegas}) are invariant 
under the rotation around the north pole to south pole axis of the $S^2$ sphere. 
The infinitesimal coordinate 
transformations manifesting this symmetry are: $x^{'\mu}= x^{\mu}, $ $x^{'\sigma}= x^{\sigma} + 
\phi_{A} \,K^{A \sigma}$, with $\phi_{A}$ the parameter of rotations 
 around the axis which goes through both poles and with the 
infinitesimal generators of rotations around this axis $M^{(5)(6)}(= x^{(5)} p^{(6)}- 
x^{(6)} p^{(5)} + S^{(5)(6)})$
\begin{eqnarray}
 K^{A \sigma}= K^{(56) \sigma} = -i M^{(5)(6)} x^{\sigma} =
 \varepsilon^{\sigma}{}_{\tau} x^{\tau}, 
\label{dhn1killings}
\end{eqnarray}
with $\varepsilon^{\sigma}{}_{\tau}= -1 = - \varepsilon_{\tau}{}^{\sigma}, \, 
\varepsilon^{(5) (6)}=1. $ The operators 
$K^{A}_{ \sigma}=f^{-2} 
\varepsilon_{\sigma \tau} x^{\tau}$ fulfil the Killing relation 
$$K^{A}_{ \sigma, \tau} + 
\Gamma^{\sigma'}{}_{\sigma \tau} K^{A}_{ \sigma'} + K^{A}_{ \tau, \sigma} + 
\Gamma^{\sigma'}{}_{\tau \sigma} K^{A}_{ \sigma'}=0,$$ 
(with $\Gamma^{\sigma'}{}_{\sigma \tau}= - 
\frac{1}{2} \, g^{\rho \sigma'} (g_{\tau \rho,\sigma} + g_{\sigma \rho,\tau} - 
g_{\sigma \tau,\rho})$).

The equations of motion for spinors (the Weyl equations) which follow from the Lagrange density 
(Eq.~\ref{dhn1weylL}) are then
\begin{eqnarray}
&&\{E\gamma^0 \gamma^m p_m + E f \gamma^0 \gamma^s \delta^{\sigma}_s  ( p_{0\sigma} 
+  \frac{1}{2 E f}
\{p_{\sigma}, E f\}_- )\}\psi=0,\quad {\rm with} \nonumber\\
&& p_{0\sigma} = p_{\sigma}- \frac{1}{2} S^{st}\omega_{st \sigma},
\label{dhn1weylE1}
\end{eqnarray}
with $f$ from  Eq.~(\ref{dhn1f})
and with $ \omega_{st \sigma}$ from Eq.~(\ref{dhn1omegas}).
From $\gamma^a p_{0a}\gamma^b p_{0b}= p_{0a} p_{0}{}^a - 
i S^{ab} S^{cd}\, {\cal R}_{abcd} + S^{ab}\,{\cal T}^{\beta}{}_{ab} \,p_{0 \beta}$ 
we find for the Riemann tensor of Eq.~(\ref{dhn1RT}) and the torsion 
\begin{eqnarray}
 {\cal T}^{\beta}{}_{ab}&= &f^{\alpha}{}_{[a} (f^{\beta}{}_{b]})_{, \alpha} + \omega_{[a}{}^{c}{}_{b]} 
f^{\beta}_{c}. 
\label{dhn1T}
\end{eqnarray}

 From Eq.~(\ref{dhn1RT}) we read that to the torsion on $S^2$ both, the zweibein 
 $f^{\sigma}_{\tau}$  and the spin connection $\omega_{st \sigma}$,  
 contribute. While we have on $S^2$ for ${\cal R}_{\sigma \tau} = 
 f^{-2} \eta_{\sigma \tau} \frac{1}{\rho^2} $ and correspondingly for the curvature 
 ${\cal R}= \frac{-2}{(\rho_0)^2}$, we find for the torsion 
 % Eq.~(\ref{dhn1RT}) 
 ${\cal T}^{s}{}_{t s'} = {\cal T}^{s}{}_{t \sigma} f^{\sigma}_{s'}$ with 
 %
 %\begin{eqnarray}
 %\label{dhn1TF}
 ${\cal T}^{5}{}_{ss} =  
 0 = {\cal T}^{6}{}_{ss},\quad s=5,6, $ 
 ${\cal T}^{5}{}_{65} = - {\cal T}^{5}{}_{56} =   -(f_{,6} + \frac{4iF (f-1)}{\rho^2 } x_5), $
 ${\cal T}^{6}{}_{56} = -  
  {\cal T}^{6}{}_{65} = - f_{,5} + \frac{4iF (f-1)}{\rho^2 } x_6.$
  %\end{eqnarray}
  %
  The torsion ${\cal T}^2 = {\cal T}^{s}{}_{t s'} {\cal T}_{s}{}^{t s'} $ 
  is for our particular choice of the zweibein and spin connection fields 
  from Eqs.~(\ref{dhn1f},\ref{dhn1omegas}) correspondingly   equal to  
  $-\frac{2 \rho^2}{ (\rho_0)^4} (1-  (2F)^2)$. If we take the model~\cite{dhn1mil} with 
  ${\cal T}^2 = {\cal T}^{s}{}_{t s'} {\cal T}_{s}{}^{t s'} + 
  2 {\cal T}^{s}{}_{t s'} {\cal T}^{t}{}_{s}{}^{s'} - 
  4 {\cal T}^{s}{}_{t s} {\cal T}_{s'}{}^{ts'}$, we obtain for the choice of fields from 
  Eqs.~(\ref{dhn1f},\ref{dhn1omegas}) $\;\;{\cal T}^2=0$.

\section{Equations of motion  for spinors and the solutions}
\label{dhn1equations}

Let the spinor  "feel" the zweibein $f^{\sigma}{\!}_{s}= 
\delta^{\sigma}{\!}_{s} f(\rho), \,f(\rho)= 1+ (\frac{\rho}{2 \rho_0})^2 =
\frac{2}{1+ \cos \vartheta}$   and 
the spin connection $\omega_{st \sigma}= iF  \varepsilon_{st} \frac{x_{\sigma}}{f \, \rho^{2}_{0}}\, 
= iF \frac{\sin \vartheta }{\rho_0}\, (\cos \phi, \sin \phi)$.
The  solution  of the equations of motion (\ref{dhn1weylE1}) for a spinor   
in $(1+5)$-dimensional space, which breaks into  
$M^{(1+3)}$ $\times S^2$, should be written as a superposition
of all  four ($2^{6/2 -1}$) states of a single Weyl representation. (We kindly ask the 
reader to see the technical details  about how to write 
a Weyl representation 
in terms of the Clifford algebra objects after making a choice of the Cartan subalgebra,  
for which we take: $S^{03}, S^{12}, S^{56}$ in the refs.~\cite{dhn1holgernorma2002,dhn1hn07}.)
In our technique~\cite{dhn1holgernorma2002} one spinor representation---the four 
states, which all are the eigenstates of the chosen Cartan subalgebra with the eigenvalues $\frac{k}{2}$, 
correspondingly---are  
the following four products of projections $\stackrel{ab}{[k]}$ and nilpotents 
$\stackrel{ab}{(k)}$: 
\begin{eqnarray}
\varphi^{1}_{1} &=& \stackrel{56}{(+)} \stackrel{03}{(+i)} \stackrel{12}{(+)}\psi_0,\nonumber\\
\varphi^{1}_{2} &=&\stackrel{56}{(+)}  \stackrel{03}{[-i]} \stackrel{12}{[-]}\psi_0,\nonumber\\
\varphi^{2}_{1} &=&\stackrel{56}{[-]}  \stackrel{03}{[-i]} \stackrel{12}{(+)}\psi_0,\nonumber\\
\varphi^{2}_{2} &=&\stackrel{56}{[-]} \stackrel{03}{(+i)} \stackrel{12}{[-]}\psi_0,
\label{dhn1weylrep}
\end{eqnarray}
where  $\psi_0$ is a vacuum state for the spinor state.
If we write the operators of handedness in $d=(1+5)$ as $\Gamma^{(1+5)} = \gamma^0 \gamma^1 
\gamma^2 \gamma^3 \gamma^5 \gamma^6$ ($= 2^3 i S^{03} S^{12} S^{56}$), in $d=(1+3)$ 
as $\Gamma^{(1+3)}= -i\gamma^0\gamma^1\gamma^2\gamma^3 $ ($= 2^2 i S^{03} S^{12}$) 
and in the two dimensional space as $\Gamma^{(2)} = i\gamma^5 \gamma^6$ 
($= 2 S^{56}$), we find that all four states are left handed with respect to 
$\Gamma^{(1+5)}$, with the eigenvalue $-1$, the first two states are right handed and the second two 
 states are left handed with respect to 
$\Gamma^{(2)}$, with  the eigenvalues $1$ and $-1$, respectively, while the first two are 
left handed 
and the second two right handed with respect to $\Gamma^{(1+3)}$ with the eigenvalues $-1$ and $1$, 
respectively. 
Taking into account Eq.~(\ref{dhn1weylrep}) we may write~\cite{dhn1hn07} the most general wave function  
$\psi^{(6)}$ obeying Eq.~(\ref{dhn1weylE1}) in $d=(1+5)$ as
\begin{eqnarray}
\psi^{(6)} = {\cal A} \,{\stackrel{56}{(+)}}\,\psi^{(4)}_{(+)} + 
{\cal B} \,{\stackrel{56}{[-]}}\, \psi^{(4)}_{(-)}, 
\label{dhn1psi6}
\end{eqnarray}
where ${\cal A}$ and ${\cal B}$ depend on $x^{\sigma}$, while $\psi^{(4)}_{(+)}$ 
and $\psi^{(4)}_{(-)}$  determine the spin 
and the coordinate dependent parts of the wave function $\psi^{(6)}$ in $d=(1+3)$ 
\begin{eqnarray}
\psi^{(4)}_{(+)} &=& \alpha_+ \; {\stackrel{03}{(+i)}}\, {\stackrel{12}{(+)}} + 
\beta_+ \; {\stackrel{03}{[-i]}}\, {\stackrel{12}{[-]}}, \nonumber\\ 
\psi^{(4)}_{(-)}&=& \alpha_- \; {\stackrel{03}{[-i]}}\, {\stackrel{12}{(+)}} + 
\beta_- \; {\stackrel{03}{(+i)}}\, {\stackrel{12}{[-]}}. 
\label{dhn1psi4}
\end{eqnarray}
Using $\psi^{(6)}$ in Eq.~(\ref{dhn1weylE1}) and separating dynamics in $1+3$ and on $S^2$, 
the following relations follow, from which we recognize the mass term $m$:  
$\frac{\alpha_+}{\alpha_-} (p^0-p^3) - \frac{\beta_+}{\alpha_-} (p^1-ip^2)= m,$ 
$\frac{\beta_+}{\beta_-} (p^0+p^3) - \frac{\alpha_+}{\beta_-} (p^1+ip^2)= m,$ 
$\frac{\alpha_-}{\alpha_+} (p^0+p^3) + \frac{\beta_-}{\alpha_+} (p^1-ip^2)= m,$
$\frac{\beta_-}{\beta_+} (p^0-p^3) + \frac{\alpha_-}{\beta_+} (p^1-ip^2)= m.$ 
(One notices that for massless solutions  ($m=0$)  $\psi^{(4)}_{(+)}$ 
and $\psi^{(4)}_{(-)}$ 
decouple.) 
Taking into account that $S^{56} \stackrel{56}{(+)}= \frac{1}{2} \stackrel{56}{(+)}$, while 
$S^{56} \stackrel{56}{[-]}= -\frac{1}{2} \stackrel{56}{[-]}$, 
we end up with the equations of motion
 for ${\cal A}$ and ${\cal B}$ as follows  
\begin{eqnarray}
-2i\,f\, (\frac{\partial}{\partial z} + \frac{\partial \ln \sqrt{Ef}}{\partial z} -    
\frac{e^{-i\phi}}{\rho} G) \, {\cal B}  
+ m \;{\cal A} =0,\nonumber\\
-2i\,f\, (\frac{\partial}{\partial \bar{z}} + \frac{\partial\ln \sqrt{Ef}}{\partial \bar{z}}
+ \frac{e^{i\phi}}{\rho} G)\, {\cal A}  
+ m \;{\cal B} =0,  
\label{dhn1equationm56gen}
\end{eqnarray}
where $z: = x^5 + i x^6 = \rho \,e^{i\phi}$, $\bar{z}: = x^5 - i x^6 = \rho \, e^{-i\phi}$ 
and $\frac{\partial}{\partial z}   = \frac{1}{2}\,(\frac{\partial}{\partial x^5} - i 
\frac{\partial}{\partial x^6}) = \frac{e^{-i\phi}}{2} \;
(\frac{\partial}{\partial \rho} - \frac{i}{\rho}\,\frac{\partial}{\partial \phi} ) $, 
$\frac{\partial}{\partial \bar{z}} = \frac{1}{2}\, (\frac{\partial}{\partial x^5} + i 
\frac{\partial}{\partial x^6}) = \frac{e^{ i\phi}}{2} \;
(\frac{\partial}{\partial \rho} + \frac{i}{\rho}\,\frac{\partial}{\partial \phi} ) $. 
Eq.~(\ref{dhn1equationm56gen}) can be rewritten as follows
\begin{eqnarray}
&&-if \, e^{-i \phi}\, (\frac{\partial}{\partial \rho} - \frac{i}{\rho}\,(\frac{\partial}{\partial \phi} -  
 i 2G ) + \frac{\partial}{\partial \rho} \ln {\sqrt{Ef}}) {\cal B} + m {\cal A} = 0,  
\nonumber\\
&&-if \, e^{i \phi}\, (\frac{\partial}{\partial \rho} + \frac{i}{\rho}\,(\frac{\partial}{\partial \phi} - 
 i 2G ) + \frac{\partial}{\partial \rho} \ln {\sqrt{Ef}}) {\cal A} + m {\cal B}= 0,
\label{dhn1equationm56gen1}
\end{eqnarray}
with $G=F \frac{f-1}{f}(=\frac{1}{2}\,F(1-\cos \vartheta))$. 
Having the rotational symmetry around the axis perpendicular to the plane of the fifth and the sixth 
dimension we require that $\psi^{(6)}$ is the eigenfunction of the total angular momentum
operator $M^{56}$
\begin{eqnarray}
M^{56}\psi^{(6)}= (n+\frac{1}{2})\,\psi^{(6)}, \quad M^{56} = x^5 p^6-x^6 p^5  + 
S^{56}.
\label{dhn1mabx}
\end{eqnarray}
Let ${\cal A}={\cal A}_n(\rho) \,\rho^n \, e^{i n \phi}  $ and ${\cal B}= {\cal B}_n(\rho)\,\rho^{-n}\,
e^{i n \phi}$. 
%The orbifolding condition Eq.~(\ref{dhn1orbifolding}) means that only {\em even} $n $ are allowed. 

Let us treat first the massless case ($m=0$). Taking into account that $\frac{G}{\rho} = 
\frac{\partial}{\partial \rho} \ln f^{\frac{F}{2}}$ and that $E=f^{-2}$ it  follows  
\begin{eqnarray}
\frac{\partial \, \ln ({\cal B} \,f^{-F -1/2})}{\partial \rho}=0, \qquad%\nonumber\\
\frac{\partial \, \ln ({\cal A} \,f^{F -1/2})}{\partial \rho}=0.
\label{dhn1masslesseq}
\end{eqnarray}
We get correspondingly the solutions
\begin{eqnarray}
{\cal B}_n = {\cal B}_0 \, e^{in \phi}\, \rho^{-n} f^{F+1/2}, \qquad%\nonumber\\
{\cal A}_n = {\cal A}_0 \, e^{in \phi}\, \rho^{n} f^{-F+1/2}. 
\label{dhn1masslesseqsol}
\end{eqnarray}
Requiring that only normalizable (square integrable) solutions are acceptable 
\begin{eqnarray}
2\pi \, \int^{\infty}_{0} \,E\, \rho d\rho {\cal A}^{\star}_{n} {\cal A}_{n}  < \infty, \qquad%\nonumber\\
2\pi \, \int^{\infty}_{0} \,E\, \rho d\rho {\cal B}^{\star}_{n} {\cal B}_{n}  < \infty, 
\label{dhn1masslesseqsolf}
\end{eqnarray}
it follows 
\begin{eqnarray}
&&{\rm for}\; {\cal A}_{n}: -1 < n < 2F, \nonumber\\
&&{\rm for}\; {\cal B}_{n}: 2F < n < 1, \quad n \;\; {\rm is \;\; an \;\;integer}.
\label{dhn1masslesseqsolf1}
\end{eqnarray}
Eq.~(\ref{dhn1masslesseqsolf1}) tells us that the strength $F$ of the spin connection field 
$\omega_{56 \sigma}$ can make a choice between the two massless solutions ${\cal A}_n$ and ${\cal B}_n$: 
For $0< 2F \le 1$ the only massless solution is the left handed spinor with respect to 
$(1+3)$
\begin{eqnarray}
\psi^{(6)m=0}_{\frac{1}{2}} ={\cal N}_0  f^{-F+1/2} 
\stackrel{56}{(+)}\psi^{(4)}_{(+)}.
\label{dhn1Massless}
\end{eqnarray} 
It is the eigenfunction  of $M^{56}$ with the eigenvalue $1/2$. 
No right handed massless 
solution is allowed for $0< 2F \le 1$.
For the  particular choice  $2F=1$ the spin connection field $-S^{56} \omega_{56\sigma}$ 
compensates the term $\frac{1}{2Ef} \{p_{\sigma}, Ef \}_- $ and the  left handed spinor
with respect to $d=1+3$ becomes a constant with respect to $\rho $ and $\phi$.  

For $2F=1$ it is easy to find also all the massive solutions of Eq.~(\ref{dhn1equationm56gen1}).
To see this let us rewrite Eq.~(\ref{dhn1equationm56gen1}) in terms of the parameter $\vartheta$. 
Taking into account that 
$f= \frac{2}{1+\cos \vartheta}$, $\omega_{56 \sigma}= 
-iF \frac{\sin \vartheta}{\rho_0} \;(\cos \phi, \sin \phi)$
and assuming that ${\cal A}={\cal A}_n(\rho) \, e^{i n \phi}  $ and 
${\cal B}= {\cal B}_{n+1}(\rho)\, e^{i (n+1) \phi}$, which guarantees that the states will be  
the eigenstates of $M^{56}$, it follows
\begin{eqnarray}
&&(\frac{\partial}{\partial \vartheta} +  \frac{n+1 -(F+1/2)(1-\cos \vartheta)}{\sin \vartheta} ) 
{\cal B}_{n+1} + i \tilde{m} {\cal A}_n = 0,  
\nonumber\\
&&(\frac{\partial}{\partial \vartheta} +  \frac{-n +(F-1/2)(1-\cos \vartheta)}{\sin \vartheta} ) 
{\cal A}_{n} + i \tilde{m} {\cal B}_{n+1} = 0,
\label{dhn1equationm56theta}
\end{eqnarray}
with $\tilde{m}=\rho_0 m$. For the particular choice of $2F= 1$ the equations simplify to
\begin{eqnarray}
&&(\frac{\partial}{\partial \vartheta} +  \frac{n +\cos \vartheta}{\sin \vartheta} ) 
{\cal B}_{n+1} + i \tilde{m} {\cal A}_n = 0,  
\nonumber\\
&&(\frac{\partial}{\partial \vartheta} - \frac{n }{\sin \vartheta} ) 
{\cal A}_{n} + i \tilde{m} {\cal B}_{n+1} = 0,
\label{dhn1eqthetaF}
\end{eqnarray}
from where we obtain 
\begin{eqnarray}
&&\{\frac{1}{\sin{\vartheta}} \frac{\partial}{\partial \vartheta}(\sin{\vartheta} 
\frac{\partial}{\partial \vartheta} ) + [\tilde{m}^2 + \frac{(-n^2-1-2n 
\cos \vartheta)}{\sin^2\vartheta}]\} {\cal B}_{n+1} =0, 
\nonumber\\
&&\{\frac{1}{\sin \vartheta} \frac{\partial}{\partial \vartheta}(\sin \vartheta 
\frac{\partial}{\partial \vartheta} ) + [\tilde{m}^2 - \frac{n^2}{\sin^2\vartheta}]\} {\cal A}_{n} =0.
\label{dhn1sphtheta}
\end{eqnarray}
From above  equations 
we see that for $\tilde{m}=0$, that is for the massless 
case, the only solution 
with $n=0$ exists, which is $Y^{0}{}_{0}$, the spherical harmonics, which is a constant (in  
agreement with our discussions  above). 
All the massive solutions have $\tilde{m}^2= l(l+1), \, l=1,2,3,..$ and $-l\le n \le l$.  
Legendre polynomials are the solutions for ${\cal A}_{n}= P^{l}_n$, 
as it can be read from the second of the equations Eq.~(\ref{dhn1sphtheta}), while 
we read from the second equation of Eq.~(\ref{dhn1eqthetaF}) that ${\cal B}_{n+1}= 
\frac{i}{\sqrt {l(l+1}}\, (\frac{\partial}{\partial \vartheta} - \frac{n }{\sin \vartheta} ) 
P^{l}_{n}$.

Accordingly  the massive solution  
with the mass equal to $m  = l (l+1)/\rho_0$ (we use the units in which $c=1=\hbar$) 
and the eigenvalues of $M^{56}$ 
((Eq.~\ref{dhn1mabx}))---which is 
the charge as we shall see later---equal to $(\frac{1}{2}+n)$, with $-l \le n \le l$, $l=1,2,..$, 
are
\begin{align}
\psi^{(6)\tilde{m}^2=l(l+1)}_{n+1/2} &= {\cal N}^{l}_{n+1/2} \{ 
\stackrel{56}{(+)} \psi^{(4)}_{(+)} + 
\frac{i}{\sqrt{l(l+1)}}\, \stackrel{56}{[-]} \psi^{(4)}_{(-)}\, e^{i\phi}
(\frac{\partial}{\partial \vartheta} \, -\frac{n}{\sin \vartheta})\}
Y^{l}_{n}). 
\label{dhn1knsol}
\end{align}
with $Y^{l}_{n}$, which are the spherical harmonics. 
Rewriting the mass operator $\hat{m}= \gamma^0 \gamma^s f^{\sigma}{}_{s} (p_{\sigma} - 
S^{56} \omega_{56 \sigma} + \frac{1}{2Ef} \{p_{\sigma}, Ef\}_-)$ as a function of 
$\vartheta $ and $\phi$  
\begin{align}
\label{dhn1m}
\rho_0 \hat{m} &= i \gamma^0\, \{\stackrel{56}{(+)} e^{-i\phi} 
(\frac{\partial}{\partial \vartheta} \, -\frac{i}{\sin \vartheta}
\frac{\partial}{\partial \phi } \, - \frac{1-\cos \vartheta}{\sin \vartheta}) +
\stackrel{56}{(-)} e^{i\phi} 
(\frac{\partial}{\partial \vartheta} \, + \frac{i}{\sin \vartheta}
\frac{\partial}{\partial \phi } ) \}, 
\end{align}
one can easily show that when  applying $\rho_0 \hat{m}$ and $M^{56}$ 
on $\psi^{(6)\tilde{m}^2=k(k+1)}_{n+1/2}$, for $-k \le n \le k$,  one obtains
from Eq.~(\ref{dhn1knsol})
\begin{eqnarray}
\label{dhn1eigenmass}
\rho_0 \hat{m}\, \psi^{(6)\tilde{m}^2=k(k+1)}_{n+1/2} &=&
k(k+1) \psi^{(6)\tilde{m}^2=k(k+1)}_{n+1/2}, \nonumber\\ 
M^{56}\, \psi^{(6)\tilde{m}^2= (n+1/2)k(k+1)}_{n+1/2} &=&
(n+1/2) \psi^{(6)\tilde{m}^2=k(k+1)}_{n+1/2}.
\end{eqnarray}
A  wave packet,  which is the eigen function of $M^{56}$ with the eigenvalue 
$1/2$,  for example,  can be written as
\begin{align}
\label{dhn1gaussian}
\psi^{(6)}_{1/2} &=\sum_{k=0,  \infty} C_{1/2}^k \;\,
{\cal N}_{1/2} \{ 
\stackrel{56}{(+)} \psi^{(4)}_{(+)} + (1 - \delta^{k}_0)
\frac{i}{\sqrt{k(k+1)}}\, \stackrel{56}{[-]} \psi^{(4)}_{(-)}\,e^{i\phi}
\frac{\partial}{\partial \vartheta} \}
Y^{k}_{0}. 
\end{align}
The expectation value of the mass operator $ \hat{m}$ on such a wave packet is 
$$\sum_{k=0,  \infty} C_{1/2}^{k*} C_{1/2}^{k} \sqrt{k(k+1)}/\rho_0.$$ 

Let us start from the southern pole by  rewriting  Eq.~(\ref{dhn1eqthetaF}) 
and the second equation of Eq.~(\ref{dhn1sphtheta}) so that $\vartheta$ is replaced by 
$(\pi - \vartheta)$ 

\begin{eqnarray}
&&(\frac{\partial}{\partial (\pi-\vartheta)} +  
\frac{- n +\cos (\pi - \vartheta)}{\sin(\pi - \vartheta)} ) 
(-){\cal B}_{-n+1} + i \tilde{m} {\cal A}_{-n} = 0,  
\nonumber\\
&&(\frac{\partial}{\partial(\pi- \vartheta)} - \frac{-n }{\sin \vartheta} ) 
{\cal A}_{-n} + i \tilde{m} (-){\cal B}_{-n+1} = 0,
\label{dhn1equationm56thetaFissouth}
\end{eqnarray}
and
\begin{align}
&\{\frac{1}{\sin(\pi- \vartheta)} \frac{\partial}{\partial (\pi- \vartheta)}
(\sin (\pi- \vartheta )
\frac{\partial}{\partial (\pi - \vartheta)} ) + [\tilde{m}^2 - 
\frac{(-n)^2}{\sin^2 (\pi- \vartheta)}]\} {\cal A}_{-n} =0.
\label{dhn1sphsouth}
\end{align}
Since ${\cal A}_{-n} (\pi- \vartheta)= P^{l}_{-n} (\pi-\vartheta) = (-1)^{l+n} 
P^{l}_{n} (\vartheta)$  
are the solutions of Eq.~(\ref{dhn1sphsouth}) and since $P^{l}_{-n} (\pi-\vartheta)= (-)^{l+2n} 
P^{l}_{n}(\theta)$, the solutions of Eq.~(\ref{dhn1sphsouth}) coincide with the solutions 
of Eq.~(\ref{dhn1sphtheta}). Correspondingly also the solutions 
for $(-){\cal B}_{-n+1} (\pi - \vartheta)= \frac{i}{\tilde{m}} 
(\frac{\partial}{\partial(\pi- \vartheta)} - \frac{-n }{\sin \vartheta} ) 
{\cal A}_{-n} (\pi - \vartheta)$ coincide with the solutions of ${\cal B}_{n+1} (\vartheta)$, 
which proves %after patching 
that the one missing point on $S^2$ makes no harm.
%%%%%%%%%%%%%%G
\section{Gauge transformations from the northern to the southern pole}
\label{dhn1gaugetrans}

\begin{figure}
\centering
\includegraphics{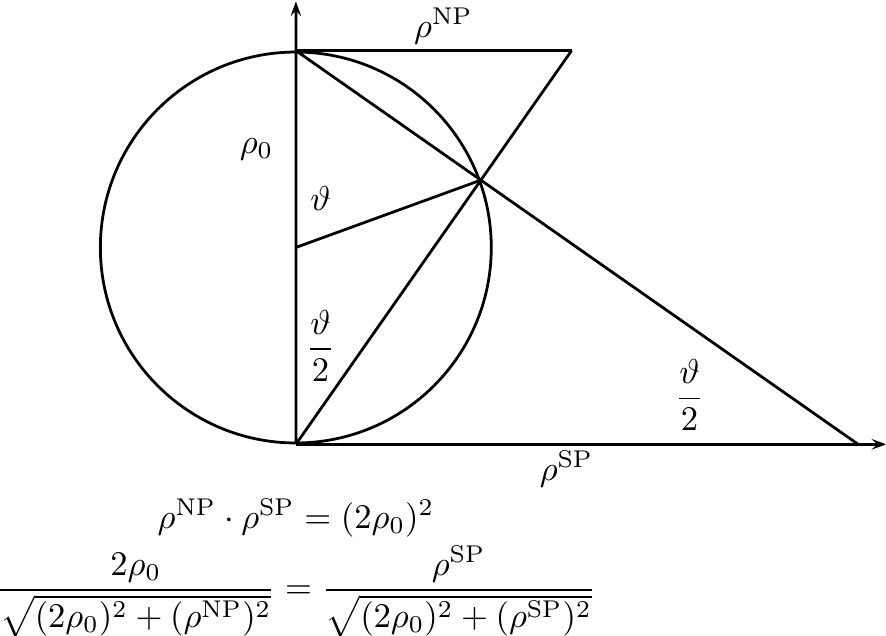}
\caption{Transforming coordinates from the north to the south pole on $S^2$. 
\label{dhn1northsouthpole}}
\end{figure}

Let us transform the coordinate system from the northern to the southern 
pole of the sphere $S^2$ and look at how do the equations of motion and the wave functions 
transform correspondingly. 
From Fig.~\ref{dhn1northsouthpole} we read
\begin{eqnarray}
\label{dhn1xsp}
x^{NP(5)}&=& (\frac{2\rho_0}{\rho^{SP}})^2\,x^{SP(5)},\quad
x^{NP(6)} = -(\frac{2\rho_0}{\rho^{SP}})^2\,x^{SP(6)},
\end{eqnarray}
and
\begin{eqnarray}
\rho^{SP}\rho^{NP}&=& (2\rho_0)^2,\quad E^{NP} \, d^2 x^{NP} =
E^{SP} \, d^2 x^{SP}, 
\end{eqnarray}
where $x^{NP\sigma }, \sigma=(5),(6)$ stay for up to now used  $x^{\sigma}, \sigma=(5),(6)$, 
while $x^{SP \sigma }, \sigma=(5),(6)$ stay for coordinates when we put our coordinate system 
on the southern pole and $\rho_0$ is the radius of $S^2$ as before. 
We have $E^{SP}=(1+ (\frac{\rho^{SP}}{2 \rho_0})^2)$  
and $E^{NP}=(1+ (\frac{\rho^{NP}}{2 \rho_0})^2)= (\frac{\rho^{SP}}{2 \rho_0})^4 \,E^{SP}$.   
We also can write 
$x^{NP\sigma}= - (\frac{2\rho_0}{\rho^{SP}})^2\, \varepsilon^{\sigma}{}_{\tau} x^{SP \tau}$, with
the antisymmetric tensor $\varepsilon^{(5)(6)}=1= -\varepsilon^{(5)}{}_{(6)}$.

We ought to transform the Lagrange density %on the $S^2$ 
expressed with respect to the coordinates on the northern pole (Eq.(\ref{dhn1weylL}))
\begin{eqnarray}
{\cal L}^{NP}_{W}&=&\psi^{NP \dagger}E^{NP}  \gamma^0 \gamma^s \, ( f^{NP \sigma}{}_s \, p^{NP}_{0 \sigma} 
+  \frac{1}{2 E^{NP} } \, \{p^{NP}_{\sigma}, E^{NP} \, f^{NP \sigma}{}_{s}\}_- )\,\psi^{NP}, \nonumber\\
p^{NP}_{0 \sigma} &=& p^{NP}_{\sigma}- \frac{1}{2} S^{st}\, \omega^{NP}_{st \sigma},\nonumber\\
f^{NP \sigma}{}_{s}\, \omega^{NP}_{s't' \sigma} &=& 
\frac{iF \delta^{\sigma}_{s}\, \varepsilon_{s' t'} x^{NP}_{\sigma}}{\rho^{2}_{0}}  
\label{dhn1weylE}
\end{eqnarray}
to the corresponding Lagrange density ${\cal L}^{SP}_{W}$
expressed with respect to the coordinates on the southern pole by assuming 
\begin{eqnarray}
\psi^{NP} &=& S \, \psi^{SP}. %,\quad \psi^{NP \dagger} }=  \psi^{SP \dagger}\, U^{-1}.
\label{dhn1psinpsp}
\end{eqnarray}
We recognize that 
\begin{eqnarray}
\label{dhn1unpsp}
f^{SP \sigma}{}_{s}&=&
f^{NP \sigma'}{}_{t} \;\frac{\partial x^{SP \sigma}}{\partial x^{NP \sigma'}}
\; O^{t}{}_{s} = f^{SP} \, \delta^{\sigma}_{s}, \quad f^{SP}= (1+(\frac{\rho^{SP}}{2 \rho_0})^2).
\end{eqnarray}
The matrix $O$ takes care that the zweibein  expressed with 
respect to the coordinate system in the southern pole 
is diagonal: $f^{SP \sigma}{}_{s} = f^{SP}\, \delta^{\sigma}_{s}$ 
\begin{eqnarray}
O = 
\begin{pmatrix}- \cos(2 \phi +\pi)  & - \sin (2 \phi +\pi) \\
          \;\;\; \sin(2 \phi +\pi)  & - \cos(2 \phi +\pi)  \end{pmatrix}. 
\label{dhn1o}
\end{eqnarray}
Requiring that $S^{-1}  \gamma^0 \gamma^s S \, O^{t}{}_{s}= \gamma^0 \gamma^t$,
from where it follows that 
$S^{-1}  S^{st} S O_{s}{}^{s'} O_{t}{}^{t'}$  $= S^{s't'}$, 
and recognizing that $p^{NP}_{\sigma}= 
\frac{\partial x^{SP \sigma'}}{\partial x^{NP \sigma}} \, p^{SP}_{\sigma'}$,  
with $p^{SP}_{\sigma}= i\frac{\partial}{\partial x^{SP}_\sigma}$, 
we find that 
$\gamma^s \, f^{NP \sigma}{}_{s}\, p^{NP}_{0 \sigma}
(=\gamma^s \, f^{NP \sigma}{}_{s}\,  (p^{NP}_{ \sigma}  - \frac{1}{2}   
S^{st}\; \omega^{NP}_{s t  \sigma}))$  
transforms into 
$\gamma^{s} \,f^{SP \sigma}{}_{s} \, p^{SP}_{0 \sigma}$
\begin{align} 
\gamma^{s} \,f^{SP \sigma}{}_{s} \, p^{SP}_{0 \sigma}   
&=\gamma^{s} \,f^{SP \sigma}{}_{s} \, \{p^{SP}_{ \sigma} - \frac{1}{2} \, S^{s't'}\,
i \varepsilon_{s' t'} ( \frac{F\, x^{SP}_{\sigma}(-)^{\sigma}
}{f^{SP}\,(f^{SP}-1) \rho^{2}_0} \nonumber\\
 &\qquad + 
2i \frac{ \varepsilon_{\sigma}{}^{\tau}\,x^{SP}_{\tau}(-)^{\tau +1}}{ (2\,\rho_{0})^{2} (f^{SP}-1)} )\}.
\label{dhn1gammafposigma}
\end{align}
In the above equation we took into account that $\omega^{NP}_{s' t'  \sigma }$ transforms into 
\begin{eqnarray}
O^{s"}{}_{s'}\,O^{t"}{}_{t'}\,O^{\sigma"}{}_{\sigma}\;
\omega^{SP}_{s" t"  \sigma" }. %+  i \frac{ \varepsilon_{s' t'}
%\varepsilon_{\sigma}{}^{\tau}\,x^{SP}_{\tau}}{ \rho^{2}_{0}} ),
\label{dhn1omegasp}
\end{eqnarray}
%
%with $\omega^{SP}_{s" t"  \sigma" }= i \frac{F \varepsilon_{s" t"}
%\varepsilon_{\sigma"}{}^{\tau}\,x^{SP}_{\tau}}{f^{SP} \rho^{2}_{0}}$. 
%

Similarly we  transform the term 
$\gamma^s \, \frac{1}{2 E^{NP} } \, \{p^{NP}_{\sigma}, E^{NP}\, 
f^{NP \sigma}{}_{s}\}_- $ into 
\begin{eqnarray}
 \gamma^s ( \frac{1}{2 E^{SP} } \,  \{p^{SP}_{\sigma}, E^{SP}\, 
f^{SP \sigma}{}_{s}\}_-  + \frac{1}{2} f^{SP \sigma}{}_{s} 
\{p^{SP}_{\sigma}, \ln(\frac{\rho^{SP}}{2 \rho_0})^2\}_{-}\,).
\end{eqnarray}
The Lagrange density from Eq.(\ref{dhn1weylL}) reads, when the coordinate system 
is put in the southern pole, as 
\begin{align}
{\cal L}^{SP}_{W}&=\psi^{SP \dagger}E^{SP}  \gamma^0 \gamma^s
\,\nonumber\\
 &  ( f^{SP \sigma}{}_s \, p^{SP}_{0 \sigma} 
+  \frac{1}{2 E^{SP} } \, \{p^{SP}_{\sigma}, E^{SP} \, f^{SP \sigma}{}_{s}\}_- + \frac{1}{2}
 f^{SP \sigma}{}_{s} 
\{p^{SP}_{\sigma}, \ln(\frac{\rho^{SP}}{2 \rho_0})^2\}_{-}\,)\,\psi^{SP}.
%p^{SP}_{0 \sigma} &=& p^{SP}_{\sigma}- \frac{1}{2} S^{st}\, \omega^{SP}_{st \sigma} (1 + \frac{f^{SP}}{F}).
\label{dhn1weylESP}
\end{align}
The requirement that $S^{-1} \gamma^0 \gamma^a \,S \,  O_{a}{}^{b} = \gamma^0 \gamma^b$
is fulfilled by the operator 
$$S^{-1} \gamma^0 \gamma^a \,S \,  O_{a}{}^{b} = \gamma^0 \gamma^b,$$ 
with 
$S= e^{-i S^{56} \omega_{56}}$, and $\omega_{56}= 2 \phi +\pi$, so that in the space 
of the two vectors $(\stackrel{56}{(+)}\psi^{(4)}_{(+)}, \stackrel{56}{([-]}\psi^{(4)}_{(-)})$ 
\begin{eqnarray}
S = 
\begin{pmatrix}e^{i (\phi^{NP}+ \frac{ \pi}{2})}  & 0 \\
          0  & e^{-i (\phi^{NP}+ \frac{ \pi}{2})}  \end{pmatrix}, 
\label{dhn1s}
\end{eqnarray}

with $\phi^{NP} = - \phi^{SP}$, while we have
\begin{eqnarray}
\gamma^0 \gamma^5 = 
\begin{pmatrix} 0  & -1 \\
          -1 &  0 \end{pmatrix},
\gamma^0 \gamma^6 = 
\begin{pmatrix} 0  & \; i \\
          -i &  0 \end{pmatrix}.          
\label{dhn1gamma56}
\end{eqnarray}

Let us look  how does an eigenstate of $M^{ab}$ from Eq.~(\ref{dhn1mabx}) expressed with respect to 
the coordinate on the northern pole 
\begin{eqnarray}
\psi^{NP(6)}_{n+\frac{1}{2}}= 
(\alpha_{n}(\rho^{NP}) \stackrel{56}{(+)} \psi^{(4)}_{(+)} + i \beta_{n}(\rho^{NP}) \stackrel{56}{[-]}
\psi^{(4)}_{(-)} \; e^{i \phi^{NP}})\, e^{i n \phi^{NP}}, 
\label{dhn1psi6np}
\end{eqnarray}
with the property $M^{NP 56} \psi^{NP(6)}_{n+\frac{1}{2}}= (n+\frac{1}{2})\,\psi^{NP(6)}_{n+\frac{1}{2}}$, 
where 
$ M^{NP 56} = S^{56} -i \frac{\partial}{\partial \phi^{NP}}$ looks like  
when we put the coordinate system on the southern pole. 
Taking into account Eqs.~(\ref{dhn1s}, \ref{dhn1o}) we obtain
\begin{eqnarray}
&&\psi^{SP (6)}_{n'+\frac{1}{2}} (x^{NP \tau})= 
S \;\psi^{NP(6)}_{n+\frac{1}{2}}(x^{NP \tau}(x^{SP \tau})) \nonumber\\
&& =
(i \alpha_{n}(\frac{(2 \rho_0)^2}{\rho^{SP}})\, e^{-i \phi^{SP}}\, \stackrel{56}{(+)} \psi^{(4)}_{(+)} + 
 \beta_{n}(\frac{(2 \rho_0)^2}{\rho^{SP}})\, \stackrel{56}{[-]} \psi^{(4)}_{(-)} )\, e^{-i n \phi^{SP}}. 
\label{dhn1spsi6sp}
\end{eqnarray}
When evaluating  
\begin{align}
(S^{56}- i \frac{\partial}{\partial \phi^{SP}})\; S 
\;\psi^{NP(6)}_{n+\frac{1}{2}}(x^{NP \tau}(x^{SP \tau})) &=
-(n+\frac{1}{2}) \; S \;\psi^{NP(6)}_{n+\frac{1}{2}} = -(n+\frac{1}{2}) \; \psi^{SP(6)}_{-(n+\frac{1}{2})}
\label{dhn1spsi6spA}
\end{align}
we recognize that the eigenvalue $(n'+ \frac{1}{2})$ of $M^{SP 56}$ on the state on the southern pole 
$\psi^{SP(6)}_{n'+ \frac{1}{2}}
= S \, \psi^{NP(6)}_{n+ \frac{1}{2}}$ 
is related to the eigenvalue $(n+ \frac{1}{2})$ of $M^{NP 56}$ on  the state $  \psi^{NP(6)}_{n+ \frac{1}{2}}$      
as follows: $(n'+ \frac{1}{2})= -(n+ \frac{1}{2})$, from where it follows $n' = -(n+1)$.

Accordingly the massless state $\psi^{NP(6)m=0}_{\frac{1}{2}} = {\cal N}^{NP}_0 \, f^{NP (-F +\frac{1}{2})}\, 
\stackrel{56}{(+)}\, \psi^{(4)}_{(+)}$ from Eq.~(\ref{dhn1Massless})  looks, when transforming  the 
coordinate system from the northern to the southern pole, as 
\begin{eqnarray}
\psi^{SP(6)m=0}_{-\frac{1}{2}} &=& {\cal N}^{SP}_0\, 
(f^{SP} \,(\frac{2 \rho_0}{\rho^{SP}})^2)^{(-F +\frac{1}{2})}\, \stackrel{56}{(+)}\, \psi^{(4)}_{(+)} \, e^{-i \phi^{SP}}.
\label{dhn1psisp}
\end{eqnarray}

\section{Spinors and the gauge fields in $d=(1+3)$}
\label{dhn1properties1+3}
To study how do spinors couple to the Kaluza-Klein gauge fields in the case of 
$M^{(1+5)}$, ``broken'' to 
$M^{(1+3)} \times S^2$ with the radius of $S^2$ equal to  $\rho_0$ and with 
the spin connection field 
$\omega_{st \sigma} = i4F \varepsilon_{st} \frac{x_{\sigma}}{\rho}\frac{f-1}{\rho f}$
we first look for (background) gauge gravitational fields, which preserve the rotational symmetry 
around the axis through the northern and southern pole.
Requiring that the symmetry determined by the Killing vectors of Eq.~(\ref{dhn1killings}) 
(following ref.~\cite{dhn1hnkk06}) with $f^{\sigma}{}_{s} = f \delta^{\sigma}_{s}, f^{\mu}{}_s=0, 
e^{s}{}_{\sigma}= f^{-1} \delta^{s}_{\sigma}, e^{m}{}_{\sigma}=0,$ is preserved, we find 
for the background vielbein field  
\begin{eqnarray}
e^a{}_{\alpha} = 
\begin{pmatrix}\delta^{m}{}_{\mu}  & e^{m}{}_{\sigma}=0 \\
 e^{s}{}_{\mu} & e^s{}_{\sigma} \end{pmatrix},
f^{\alpha}{}_{a} =
\begin{pmatrix}\delta^{\mu}{}_{m}  & f^{\sigma}{}_{m} \\
0= f^{\mu}{}_{s} & f^{\sigma}{}_{s} \end{pmatrix},
\label{dhn1f6}
\end{eqnarray}
with $f^{\sigma}{}_{m} = K^{(56)\sigma} B^{(5)(6)}_{\mu} f^{\mu}{}_{m} = 
\varepsilon^{\sigma}{}_{\tau} x^{\tau} A_{\mu} \delta^{\mu}_{m}$, 
$e^{s}{}_{\mu} = - \varepsilon^{\sigma}{}_{\tau} x^{\tau} A_{\mu} e^{s}{}_{\sigma}$, 
$ s=5,6; \sigma = (5),(6)$.  
Requiring that correspondingly the only nonzero torsion fields are those from 
Eq.~(\ref{dhn1RT}) 
we find for the spin connection fields 
\begin{eqnarray}
\omega_{st \mu} =  \varepsilon_{st}  A_{\mu},
\quad \omega_{sm \mu} = 
\frac{1}{2}f^{-1}\varepsilon_{s \sigma } x^{\sigma} \delta^{\nu}{}_{m} F_{\mu \nu},
\label{dhn1omega6}
\end{eqnarray}
$F_{\mu \nu}= A_{[\nu,\mu]}$. 
 The $U(1)$ gauge field $A_{\mu}$ depends only on $x^{\mu}$.
All the other components of the spin connection fields, except (by the 
Killing symmetry preserved)  $\omega_{st\sigma}$ from Eq.~(\ref{dhn1weylE}), are zero, 
since for simplicity we allow no gravity in
$(1+3)$ dimensional space. 
The corresponding nonzero torsion fields ${\cal T}^{a}{}_{bc}$ are presented in 
Eq.~(\ref{dhn1RT}), all the other components are zero.

To determine the current, which couples the spinor to the Kaluza-Klein gauge fields 
$A_{\mu}$, we
analyze (as in the refs.~\cite{dhn1hnkk06,dhn1hn07}) the spinor action (Eq.~\ref{dhn1weylL})
\begin{eqnarray}
{\cal S} &=& \int \; d^dx  \bar{\psi}^{(6)} E \gamma^a p_{0a} \psi^{(6)} =\nonumber\\
&& \int \, d^dx  \bar{\psi}^{(6)} \gamma^s  p_{s} \psi^{(6)}+ \nonumber \\  
&& \int \, d^dx  \bar{\psi}^{(6)} \gamma^m \delta^{\mu}{}_{m} p_{\mu} \psi^{(6)} + \nonumber\\
% && \int \; d^dx   \bar{\psi}^{(6)} \gamma^m (-)S^{sm} \omega_{sm \mu} \psi^{(6)}  + 
%\int \, d^dx  \bar{\psi}^{(6)} \gamma^s \delta^{\sigma}{}_{s} p_{\sigma} \psi^{(6)} +\nonumber\\
&& \int \, d^dx   \bar{\psi}^{(6)} \gamma^m  \delta^{\mu}{}_{m} A_{\mu} 
(\varepsilon^{\sigma}{}_{\tau} x^{\tau}
 p_{\sigma} + S^{56}) \psi^{(6)} + \nonumber\\
&& {\rm \; terms } \propto  x^{\sigma} \,{\rm or } \propto   x^{5}  x^{6}.
 %- \gamma^m S^{sn}\omega_{snm}- \gamma^m f^{\sigma}{}_{m} \frac{1}{2}S^{st}\omega_{st \sigma}
\label{dhn1spinoractioncurrent}
\end{eqnarray}
 Here $\psi^{(6)}$ is a spinor state  in $d=(1+5)$ after the break of $M^{1+5}$ 
 into $M^{1+3} \times $ $S^2$.
 $E$ is for $f^{\alpha}{}_{a}$ from Eq.~(\ref{dhn1f6}) equal to $f^{-2}$. 
The  term in the second row in Eq.~(\ref{dhn1spinoractioncurrent}) is the mass term  
(equal to zero for the massless spinor), the term in the third row is the kinetic term, 
together with the term in the fourth row  defines  
the  covariant derivative $p_{0 \mu}$ in $d=(1+3)$.  
The terms in the last row  contribute nothing when the integration over 
the disk (curved into a sphere $S^2$) is performed, since they all 
are proportional to $x^{\sigma}$ or to $ \varepsilon_{\sigma \tau} x^{\sigma} x^{\tau}\;$ 
($-\gamma^{m} \,\frac{1}{2}S^{sm} \omega_{s m n} = -\gamma^{m}\,\frac{1}{2}\,f^{-1}
F_{m n}  \varepsilon_{s \sigma} x^{\sigma}$ and $-\gamma^m \,f^{\sigma}{}_{m}\frac{1}{2}
\,S^{st} \omega_{st \sigma}= 
\gamma^m A_m   x^{5}x^{6} S^{st} \varepsilon_{s t} \frac{4iF(f-1)}{f \rho^2}$).

We end up with the current in $(1+3)$
\begin{eqnarray}
j^{\mu} = \int \;E  d^2x \bar{\psi}^{(6)} \gamma^m \delta^{\mu}{}_{m} M^{56}  \psi^{(6)}.
\label{dhn1currentdisk}
\end{eqnarray}
 The charge in $d=(1+3)$ is  proportional to the total 
angular momentum  $M^{56} =L^{56} + S^{56}$ around the axis from the southern to the 
northern  pole of $S^2$, but since for the choice of 
$  2 F =1$ (and for any $0 < 2F \le 1 $) in Eq.~(\ref{dhn1masslesseqsolf1}) only a left 
handed massless spinor exists,  
with the angular momentum zero, the charge of a massless 
spinor in $d=(1+3)$ is equal to  $1/2$.
The Riemann scalar is for 
the vielbein of Eq.~(\ref{dhn1f6}) equal to 
${\cal R}= -\frac{1}{2} \rho^2 f^{-2} F^{mn}F_{mn}$. 
If we integrate the Riemann scalar 
over the fifth and the sixth dimension, we get $-\frac{8\pi}{3} (\rho_0)^4 F^{mn}F_{mn}$.

\section{Conclusions}
\label{dhn1conclusion}

We prove in this paper that  one can  escape from the "no-go theorem" of Witten~\cite{dhn1witten}, 
that is one can guarantee the masslessness of spinors and their chiral coupling  to the 
Kaluza-Klein-like gauge fields when breaking the symmetry from the $d$-dimensional one to 
$(1+3)\times M^{d-4}$ space, in cases which we call the "effective two dimensionality" 
even without boundaries, as we proposed in the 
references~\cite{dhn1hnm06,dhn1hn07}. Namely, we can guarantee above properties of spinors, 
when $M^{d-4}$, $d-4> 2$ breaks in a way that vielbeins and 
spin connections are completely flat in all but two dimensions. Taking in the absent of fermions the 
action with the linear curvature for $d=2$ we proved that any zweibein and any spin connection fulfills the 
corresponding equations of motion. We make a choice of the zweibein, which curves the flat disc on $S^2$ 
and the spin connection, which then allows spinors of only one handedness to be a normalizable state
on such $S^2$. This leads to nonzero torsion .

The possibility  (besides the particular choice of boundaries) on a flat two dimensional manifold  
of a special choice of the spin connection and the zweibein, which curves a two dimensional infinite 
manifold on $S^2$, opens, to our understanding, a new hope to the Kaluza-Klein-theories and will help to revival 
the Kaluza-Klein-like theories, to which also the ''approach unifying spins and char\-ges'' proposed by of one of 
the authors of this paper (S.N.M.B.) belongs (and which offers also the explanation for the appearance 
of families).

 We study in this paper the case, when  a left handed spinor  
carrying in $d=1+5$ nothing but a spin, with the symmetry of 
$M^{(1+5)}$, which breaks to $M^{(1+3)} \times$ the infinite disc with the zweibein, which curves 
the disc on $S^2$ ($f= 1+ (\frac{\rho}{2 \rho_0})^2$, with $\rho_0$ the radius of $S^2$), 
and with the spin connection field on the disc equal to $\omega_{st \sigma} = i \,\varepsilon_{st}\,
4F \frac{f-1}{\rho f} \frac{x_{\sigma}}{\rho}, \sigma=(5),(6); s,t=5,6,$ which allows for 
$0 < 2F \le 1 $ one massless spinor of the charge $1/2$ and of the left handedness with respect to 
$d=(1+3)$. This spinor state couples chirally to the 
corresponding Kaluza-Klein gauge field. 
There are infinitely many massive states, which are at the same time 
the eigen states of $M^{56}= x^{5} p^{6}- x^{6} p^{5} + S^{56}$, with the eigen values 
$(n+1/2)$, carrying the Kaluza-Klein charge $(n+1/2)$. 
For the choice of $2F=1$ the massive states have the mass equal to $k(k+1)/\rho_0, k=1,2,3,..$,
 with $-k \le n \le k$. 
We found the expression for the massless eigen state  and for 
the particular choice of $2F=1$ also for all the massive states.  

We therefore found an example, in which the internal gauge fields---spin connections 
and zweibeins---allow only one massless state, that is the spinor of 
one handedness and of one charge with respect to 
$d=1+3$ space. Since for the zweibein  curving the infinite disc on $S^2$,    
the spin connection  field  $\omega_{st \sigma} = 
i 4F \frac{f-1}{\rho f} \frac{x_{\sigma}}{\rho}, $ with any $2F$ 
fulfilling the condition 
$\;0< 2F \le 1$  ensures that a massless spinor state of only one handedness and one 
charge in $d=(1+3)$ exists (only one massless state is normalizable),  it is not 
a fine tuning what we propose. To find simple solutions for the massive states, we made 
a choice  of  $2F= 1$. The massless state is in this case a constant with respect to 
the two angles on $S^2$, while  the angular dependence of the massive states, 
with the masses equal to $l(l+1)/\rho_0$, 
are expressible with the spherical harmonics $Y^{l}_{n},\;\; -l \le n \le l$, and 
with the 
$e^{i\phi} \, \frac{i}{\sqrt{l(l+1)}}\; (\frac{\partial}{\partial \vartheta} - 
\frac{i}{\sin \vartheta} ) Y^{l}_{n}$ (Eq.~(\ref{dhn1knsol})).

The zweibein and the spin connection fulfills the equations of motion following from the  action 
with the linear curvature and produce the nonzero torsion. 
We study the gauge transformation which transforms the coordinates and correspondingly the 
zweibein and spin connection when the coordinate system is put on the north pole to the 
case, when the coordinate system is put on the south pole. We look the transformation properties 
of any state under the above transformations, recognizing that the massless  (left handed) state, which 
carry the momentum $M^{56}$, and accordingly the Kaluza-Klein charge equal to $\frac{1}{2}$ if we use 
the coordinate system put on the north pole, transforms 
to a state of the same handedness but with the charge equal to 
$-\frac{1}{2}$ if  the coordinate system is put on the south pole.

Let us conclude the paper by pointing out that while in the two papers~\cite{dhn1hnkk06,dhn1hn07} 
we  achieved the masslessness 
of a spinor, its mass protection and the chiral coupling to the 
corresponding Kaluza-Klein gauge field  after a break of a symmetry from 
$d=1+5$ to $d=(1+3)$, with the choice of the boundary condition on a flat (finite)
disk 
in this paper the massless spinor and its chiral coupling to the 
corresponding Kaluza-Klein gauge field is  achieved by the choice of the 
appropriate spin connection and zweibein fields 
which fulfill the equations of motion following from the action with the linear curvature, which 
in the two dimensional case allow any zweibein and any spin connection.

Although we do not discuss the problem of the families in this paper (we kindly ask the 
reader to take a look on the 
refs.~\cite{dhn1norma92,dhn1norma93,dhn1norma94,dhn1norma95,dhn1Portoroz03,dhn1pikanorma06,dhn1gmdn06,dhn1gmdn07}
where the proposal for solving the problem of families is presented) we believe that the 
present paper is opening a new hope to the wonderful idea of the Kaluza-Klein-like theories 
through "an effective two dimensional" cases, when in all the higher dimensions but two 
the vielbeins and spin connections are flat.

\section{Acknowledgement } One of the authors (N.S.M.B.) 
would like to warmly thank Jo\v ze Vrabec 
for  fruitful discussions.

%% SNMB, second contribution, 13.11.2009, Proceedings BLED
%%
%%
\title{Offering the Mechanism for Generating Families --- the Approach Unifying Spins and Charges
Predicts New Families}
\author{N.S. Manko\v c Bor\v stnik\thanks{E-mail: norma.mankoc@fmf.uni-lj.si}}
\institute{%
Department of Physics, FMF, University of
Ljubljana,\\
 Jadranska 19, 1000 Ljubljana, Slovenia\\
http://www.fmf.uni-lj.si}
 
\authorrunning{N.S. Manko\v c Bor\v stnik}
\titlerunning{Offering the Mechanism for Generating Families\ldots}
\maketitle

\begin{abstract}
The "approach unifying spin and charges"~\cite{sn2snmb:n92,sn2snmb:pn06,sn2snmb:gmdn07,sn2snmb:n07} offers 
the explanation for all the internal degrees of freedom---the 
spin, all the charges and the family quantum number---by introducing two kinds of the 
spin, the Dirac kind and the second kind anticommuting with the Dirac one. 
It offers  a new way of 
understanding the properties of quarks and leptons: their charges and their 
connection to the corresponding gauge fields and their appearance in families and 
their Yukawa couplings. 
In this talk I present the way from a simple starting Lagrange density for a spinor---carrying 
in $d=1+13$ only two kinds of the spin, no charges, and interacting with 
the vielbeins and the two kinds of the spin connection fields---to 
the effective Lagrangean, postulated by the "standard model of the electroweak and colour interactions".
The way of breaking the starting symmetries determines the observed properties of 
the families of spinors and of the gauge fields, predicting that there are four 
families at low energies and that a much heavier fifth family with zero Yukawa 
couplings to the lower four families, might, by forming baryons in the evolution 
of the universe, contribute a major part to the dark matter. 
I comment on properties of the Yukawa couplings following from the simple starting Lagrangean, as well 
as on the possibility that the starting Lagrangean for spin connection and vielbeins fields linear in the 
curvature might lead to the observable properties of the gauge fields and their couplings to  
almost massless observed fermions. 
\end{abstract}

%%\keywords{Keywords: Origin of families; New stable family; Dark matter candidate; 
%%Higher dimensional spaces; Unifying theories; Dark matter candidates; Kaluza-Klein-like theories.}  

%\bodymatter
%\maketitle 

%
\section{Introduction}
\label{sn2snmb:introduction}

The standard model of the electroweak and colour interactions (extended by the right handed  
neutrinos) fits with around 30 parameters and constraints all the existing experimental 
data. It leaves, however,  unanswered many open questions, among which are also the questions   
about the origin of charges ($U(1), SU(2), SU(3)$), of families, and correspondingly 
of the Yukawa couplings 
of quarks and leptons and the Higgs mechanism. Answering the question about 
the origin of families and their masses is the most promising way leading beyond the 
today knowledge about the elementary fermionic and bosonic fields.

A simple Lagrange density for spinors, which carry  in $d=1+13$ two  kinds 
of the spin, represented by  two kinds of the 
Clifford algebra objects\cite{sn2snmb:hn02hn03} $S^{ab}= \frac{i}{4} (\gamma^a \gamma^b - \gamma^b \gamma^a)$ and 
$\tilde{S}^{ab}= \frac{i}{4} (\tilde{\gamma}^a \tilde{\gamma}^b - \tilde{\gamma}^b \tilde{\gamma}^a)$, 
with $\{\gamma^a, \gamma^b \}_{+}= 2 \eta^{ab} = \{\tilde{\gamma}^a, \tilde{\gamma}^b \}_{+}, 
\{\gamma^a, \tilde{\gamma}^b \}_{+}=0 $,  and no charges, 
and interact correspondingly only with the vielbeins and the two kinds of spin connection fields,
of the "approach unifying spins and charges"~\cite{sn2snmb:n92,sn2snmb:pn06,sn2snmb:gmdn07,sn2snmb:n07} offers the 
possibility to lead at observable 
energies to the observed families of quarks and leptons coupled through the charges to the 
known gauge fields in the way assumed by the standard model, and carrying masses, 
determined by a part of a simple starting action~\footnote{This is the only theory in the literature  
to my knowledge, which does not 
explain the appearance of families by just postulating their numbers on one or another way, 
but by offering the mechanism for generating families.}. 
The approach predicts an even number of families, among which is the candidate 
for forming the dark matter clusters. 

The  approach confronts several problems (some of them are the problems common to all 
the Kaluza-Klein-like theories), which we~\footnote{I started the project named the approach 
unifying spins and charges fifteen years ago, proving alone or together with collaborators %and 
%students, 
step by step, that such a theory has the chance to answer the open questions 
of the Standard model. The names of the collaborators and students can be found on the 
cited papers. } are studying step by step when  
searching for possible 
ways of spontaneous breaking of the starting symmetries and conditions, which might lead to  
the observed properties of families of fermions and of gauge and scalar 
fields, and looking for predictions the approach might make.

In what follows I briefly present in the first part of the talk the starting action 
of the approach %unifying spins and charges 
for fermions and the corresponding gauge fields. 
The representation of one Weyl spinor %(carrying in $d=1+13$ only spins) 
of the group $SO(1,13)$ in $d=1+13$, 
analysed with respect to the properties of the subgroups $SO(1,7) \times SU(3) \times U(1)$ of this 
group and further with respect to $SO(1,3)\times SU(2) \times U(1) \times SU(3)$ %and the second U(1) 
manifests the left handed weak charged quarks and leptons and the right handed weak chargeless quarks and 
leptons.

The way of braking symmetries leads first to eight families at low energy region and 
then to twice four families. It is a part of the starting Lagrange density 
for a spinor in $d=1+13$ which manifests as Yukawa couplings in $d=1+3$. 
The lowest three of the lower 
four families are the observed families of quarks and leptons, with all the known 
properties assumed by the Standard model. Our rough estimations predict that there is 
the fourth family with possibly low enough masses that it might be seen at LHC.

The fifth family, which decouples in the Yukawa couplings from the lower four families,   
has a chance in the evolution of our universe %due to our rough estimations  
to form baryons and is accordingly  the candidate to form the dark matter. 

I comment on the way of breaking symmetries, including the effects beyond the tree level 
and possible phase transitions. I also comment on the possibility that the Kaluza-Klein-like 
theories, to which  the "approach unifying spin and charges"  also belongs, make 
a loop hole through the Witten's "no-go theorem" through "an effective two dimensionality" 
cases~\cite{sn2dhnBled08,sn2dhnBled09} or with the boundaries~\cite{sn2snmb:hn05hn07}.
In the second part of the talk I present properties of the stable fifth family, 
as required by the approach 
and as limited by the cosmological evidences and the direct 
measurements~\cite{sn2snmb:gnBled08}. 

Although a lot of work is already done on this topic, all estimates are still very 
approximate and need serious additional studies. Yet  
these  rough estimations give a hope that the approach is the right way beyond the 
standard model of the electroweak and colour interaction and also good guide to further studies. 

\section{The approach unifying spin and charges}
\label{sn2snmb:approach}

The approach~\cite{sn2snmb:n92,sn2snmb:pn06,sn2snmb:gmdn07,sn2snmb:n07} assumes that in $d\ge(1+13)$-dimensional space a 
Weyl spinor carries nothing but two kinds of the spin (no charges): The Dirac spin described by 
$\gamma^a$'s defines the ordinary spinor representation,  the second kind of 
spin~\cite{sn2snmb:hn02hn03},  described by $\tilde{\gamma}^a$'s and anticommuting  
with the Dirac one, defines the families 
of spinors~\footnote{There is no  third kind of the Clifford algebra objects: If the Dirac one 
corresponds to the multiplication of any object (any product of the Dirac $\gamma^a$'s) 
from the left hand side, the second kind of the Clifford object is  understood (up to a factor) 
as the multiplication of any object from the right hand side.}.  
\begin{eqnarray}
&& \{ \gamma^a, \gamma^b\}_{+} = 2\eta^{ab} =  
\{ \tilde{\gamma}^a, \tilde{\gamma}^b\}_{+},\quad
\{ \gamma^a, \tilde{\gamma}^b\}_{+} = 0,\nonumber\\
%\quad \quad \tilde{\gamma}^a B : &=& i(-)^{n_B} B \gamma^a, \nonumber
&&S^{ab}: = (i/4) (\gamma^a \gamma^b - \gamma^b \gamma^a), \quad
\tilde{S}^{ab}: = (i/4) (\tilde{\gamma}^a \tilde{\gamma}^b 
- \tilde{\gamma}^b \tilde{\gamma}^a).
\label{sn2snmb:tildegclifford}
\end{eqnarray}
Defining the vectors (the nilpotents and projector)~\cite{sn2snmb:hn02hn03}
\begin{eqnarray}
\stackrel{ab}{(\pm i)}: &=& \frac{1}{2}(\gamma^a \mp  \gamma^b),  \; 
\stackrel{ab}{[\pm i]}: = \frac{1}{2}(1 \pm \gamma^a \gamma^b), \quad
{\rm for} \,\; \eta^{aa} \eta^{bb} = -1, \nonumber\\
\stackrel{ab}{(\pm )}: &= &\frac{1}{2}(\gamma^a \pm i \gamma^b),  \; 
\stackrel{ab}{[\pm ]}: = \frac{1}{2}(1 \pm i\gamma^a \gamma^b), \quad
{\rm for} \,\; \eta^{aa} \eta^{bb} =1,
\label{sn2snmb:eigensab}
\end{eqnarray} 
and noticing that the above vectors are eigen vectors of $S^{ab}$ as well as of $\tilde{S}^{ab}$
\begin{eqnarray}
S^{ab} \stackrel{ab}{(k)} =  \frac{k}{2} \stackrel{ab}{(k)}, \quad 
S^{ab} \stackrel{ab}{[k]} =  \frac{k}{2} \stackrel{ab}{[k]}, %\nonumber\\
\tilde{S}^{ab} \stackrel{ab}{(k)}  = \frac{k}{2} \stackrel{ab}{(k)},  \quad 
\tilde{S}^{ab} \stackrel{ab}{[k]}  =   - \frac{k}{2} \stackrel{ab}{[k]},\;\;
\label{sn2snmb:eigensabev}
\end{eqnarray}
and recognizing that $\gamma^a$ transform   
$\stackrel{ab}{(k)}$ into  $\stackrel{ab}{[-k]}$, while 
$\tilde{\gamma}^a$ transform  $\stackrel{ab}{(k)}$ 
into $\stackrel{ab}{[k]}$ 
\begin{eqnarray}
\gamma^a \stackrel{ab}{(k)}= \eta^{aa}\stackrel{ab}{[-k]},\; 
\gamma^b \stackrel{ab}{(k)}= -ik \stackrel{ab}{[-k]}, \; 
\gamma^a \stackrel{ab}{[k]}= \stackrel{ab}{(-k)},\; 
\gamma^b \stackrel{ab}{[k]}= -ik \eta^{aa} \stackrel{ab}{(-k)},\;\;\;
\label{sn2snmb:graphgammaaction}
\end{eqnarray}
\begin{eqnarray}
\tilde{\gamma^a} \stackrel{ab}{(k)} = - i\eta^{aa}\stackrel{ab}{[k]},\;
\tilde{\gamma^b} \stackrel{ab}{(k)} =  - k \stackrel{ab}{[k]}, \;
\tilde{\gamma^a} \stackrel{ab}{[k]} =  \;\;i\stackrel{ab}{(k)},\; 
\tilde{\gamma^b} \stackrel{ab}{[k]} =  -k \eta^{aa} \stackrel{ab}{(k)},\; 
\label{sn2snmb:gammatilde}
\end{eqnarray}
one  sees that $\tilde{S}^{ab}$ form the equivalent representations with respect 
to $S^{ab}$ and the families of quarks and leptons certainly do 
(before the break of the electroweak symmetry in the standard model of the 
electroweak and colour interactions) 
manifest the equivalent representations.

Let us make a choice of the Cartan subalgebra   set of the algebra $S^{ab}$ as follows:
$ S^{03}, S^{12}, S^{56}, S^{78}, S^{9 \;10}, S^{11\;12}, S^{13\; 14}$. 
Then we can write as a starting basic vector  of one left handed ($\Gamma^{(1,13)} =-1$)  
Weyl representation of the group $SO(1,13)$, the quark $ u_{R}^{c1}$. It is the  eigen 
state of all the members of the 
Cartan  subalgebra and it is the right handed (with respect to $\Gamma^{(1+3)}$), and has  
the properties: $Y \;u_{R}^{c1} = 2/3 \; \,u_{R}^{c1}$, $\tau^{2i}\;u_{R}^{c1}=0$ and  
$(\tau^{33}, \tau^{38})\;u_{R}^{c1}= (\frac{1}{2}, \frac{1}{2 \sqrt{3}})\;\, u_{R}^{c1}$. 
Written in terms of nilpotents and projectors it looks like:
\begin{equation}
\begin{gathered}
\stackrel{03}{(+i)}\stackrel{12}{(+)}|\stackrel{56}{(+)}\stackrel{78}{(+)}
||\stackrel{9 \;10}{(+)}\stackrel{11\;12}{(-)}\stackrel{13\;14}{(-)}
|\psi \rangle =\\
 \frac{1}{2^7} 
(\gamma^0 -\gamma^3)(\gamma^1 +i \gamma^2)| (\gamma^5 +
i\gamma^6)(\gamma^7 +i \gamma^8)||
\\
(\gamma^9 +i\gamma^{10})(\gamma^{11} -i \gamma^{12})(\gamma^{13}-i\gamma^{14})
|\psi \rangle 
\end{gathered}
\label{sn2urc1}
\end{equation}
The eightplet (the representation of $SO(1,7)$ with the fixed colour charge, $\tau^{33}=1/2$, 
$\tau^{38}=1/(2\sqrt{3})$),  of one of the 
eight families (equivalent representations),  looks like in Table~\ref{sn2snmb:Table I}. 

\begin{table}
%\tbl
\centering
{\begin{tabular}{|r|c||c||c|c|c|c||}
\hline
i&$$&$|^a\psi_i>$&$\Gamma^{(1,3)}$&$ S^{12}$&
$\tau^{23}$&$Y$\\
\hline\hline
&& ${\rm Octet}\; {\rm of \; quarks}$&&&&\\
\hline\hline
1&$u_{R}^{c1}$&$
\stackrel{03}{(+i)} \stackrel{12}{(+)}|\stackrel{56}{(+)}\stackrel{78}{(+)}
||\stackrel{9 \;10}{(+)}\stackrel{11\;12}{(-)}\stackrel{13\;14}{(-)}$
& 1&$ \frac{1}{2}$&           0&$ \frac{2}{3}$\\
\hline 
2&$u_{R}^{c1}$&$\stackrel{03}{[-i]}\stackrel{12}{[-]}|\stackrel{56}{(+)}\stackrel{78}{(+)}
||\stackrel{9 \;10}{(+)}\stackrel{11\;12}{(-)}\stackrel{13\;14}{(-)}$
& 1&$-\frac{1}{2}$&           0&$ \frac{2}{3}$\\
\hline
3&$d_{R}^{c1}$&$\stackrel{03}{(+i)}\stackrel{12}{(+)}|\stackrel{56}{[-]}\stackrel{78}{[-]}
||\stackrel{9 \;10}{(+)}\stackrel{11\;12}{(-)}\stackrel{13\;14}{(-)}$
& 1&$ \frac{1}{2}$&           0&$-\frac{1}{3}$\\
\hline 
4&$d_{R}^{c1}$&$\stackrel{03}{[-i]}\stackrel{12}{[-]}|\stackrel{56}{[-]}\stackrel{78}{[-]}
||\stackrel{9 \;10}{(+)}\stackrel{11\;12}{(-)}\stackrel{13\;14}{(-)}$
& 1&$-\frac{1}{2}$&           0&$-\frac{1}{3}$\\
\hline
5&$d_{L}^{c1}$&$\stackrel{03}{[-i]}\stackrel{12}{(+)}|\stackrel{56}{[-]}\stackrel{78}{(+)}
||\stackrel{9 \;10}{(+)}\stackrel{11\;12}{(-)}\stackrel{13\;14}{(-)}$
&-1&$ \frac{1}{2}$&$-\frac{1}{2}$&$\frac{1}{6}$\\
\hline
6&$d_{L}^{c1}$&$\stackrel{03}{(+i)}\stackrel{12}{[-]}|\stackrel{56}{[-]}\stackrel{78}{(+)}
||\stackrel{9 \;10}{(+)}\stackrel{11\;12}{(-)}\stackrel{13\;14}{(-)}$
&-1&$-\frac{1}{2}$&$-\frac{1}{2}$&$\frac{1}{6}$\\
\hline
7&$u_{L}^{c1}$&$ \stackrel{03}{[-i]}\stackrel{12}{(+)}|\stackrel{56}{(+)}\stackrel{78}{[-]}
||\stackrel{9 \;10}{(+)}\stackrel{11\;12}{(-)}\stackrel{13\;14}{(-)}$
&-1&$ \frac{1}{2}$&$ \frac{1}{2}$&$\frac{1}{6}$\\
\hline
8&$u_{L}^{c1}$&$\stackrel{03}{(+i)}\stackrel{12}{[-]}|\stackrel{56}{(+)}\stackrel{78}{[-]}
||\stackrel{9 \;10}{(+)}\stackrel{11\;12}{(-)}\stackrel{13\;14}{(-)}$
&-1&$-\frac{1}{2}$&$ \frac{1}{2}$&$\frac{1}{6}$\\
\hline\hline
\end{tabular}}
%\end{center}
\caption{\label{sn2snmb:Table I} %
The 8-plet of quarks - the members of $SO(1,7)$ subgroup, belonging to one Weyl left 
handed ($\Gamma^{(1,13)} = -1 = \Gamma^{(1,7)} \times \Gamma^{(6)}$) spinor representation of 
$SO(1,13)$. It contains the left handed weak charged quarks and the right handed weak chargeless quarks 
of a particular colour ($(1/2,1/(2\sqrt{3}))$). Here  $\Gamma^{(1,3)}$ defines the handedness in $(1+3)$ space, 
$ S^{12}$  the ordinary spin (which can also be read directly from the basic vector), 
$\tau^{23}$  the weak charge and $Y$ defines the hyper charge. Let the reader notice 
(by taking into account the relations $\gamma^a \stackrel{ab}{(k)}=\eta^{aa}\stackrel{ab}{[-k]}, \, 
\stackrel{ab}{(-k)}\stackrel{ab}{(k)} = \eta^{aa}  \stackrel{ab}{[-k]}$) that 
$\gamma^0 \stackrel{78}{(-)}$ (appearing 
in $- {\mathcal L}_{Y} = \psi^{\dagger} \; \gamma^0 \{
 \stackrel{78}{(+)} \; p_{0+}\; + \; \stackrel{78}{(-)} \;p_{0-} \}\psi$) transforms  
%$\gamma^0 \stackrel{78}{(-)}$ transforms 
$u_{R}^{c1}$ of the $1^{st}$ row 
into $ u_{L}^{c1}$ of the $7^{th}$ row, while $ \gamma^0 \stackrel{78}{(+)}$ transforms 
$d_{R}^{c1}$ of the $3^{rd}$ row into $d_{L}^{c1}$ of the $5^{th}$ row, doing what the 
Higgs and $\gamma^0$ do in the standard model. }
\end{table}
One can notice (when using Eq.(\ref{sn2snmb:graphgammaaction})) that $\gamma^0 \gamma^7$ and 
$\gamma^0 \gamma^8$ rotate the right handed weak 
chargeless quark into the left handed weak charged quark of the same colour charge and the same spin. 

The generators $\tilde{S}^{ab}$ transform one vector of the representation of $S^{ab}$ 
into the vector with the same properties with respect to $S^{ab}$, in particular
both vectors bellow describe a right handed $u_R$-quark of the same colour and the same spin and the same 
hyper charge
\begin{eqnarray}
2i \tilde{S}^{01}\; 
\stackrel{03}{(+i)}\stackrel{12}{(+)}| \stackrel{56}{(+)} \stackrel{78}{(+)}||
\stackrel{9 10}{(+)} \stackrel{11 12}{(-)} \stackrel{13 14}{(-)}  = 
\stackrel{03}{[\,+i]} \stackrel{12}{[\,+\,]}| \stackrel{56}{(+)} \stackrel{78}{(+)}||
\stackrel{9 10}{(+)} \stackrel{11 12}{(-)} \stackrel{13 14}{(-)}. \;\;
\end{eqnarray}
Since the term $- \frac{1}{2}\,\tilde{S}^{ab} \tilde{\omega}_{abc}$ transforms in general 
one equivalent representation into all the others, we expect that it generates, together with
the corresponding gauge fields,  the Yukawa couplings.
%
%\vspace{1mm}
%
Let us~\cite{sn2snmb:n92,sn2snmb:pn06,sn2snmb:gmdn07,sn2snmb:n07} now make a choice of a simple action for  a spinor 
which carries in $d=(1+13)$ only two kinds of the   
spin (no charges) %as follows
\begin{eqnarray}
S &=& \int \; d^dx \; {\mathcal L}_{f}, \quad 
{\mathcal L}_{f} = \frac{1}{2} (E \bar{\psi}\gamma^a p_{0a} \psi) + h.c. \nonumber\\
p_{0a }&=& f^{\alpha}{}_a p_{0\alpha} + \frac{1}{2 E} \{p_{\alpha}, E f^{\alpha}{}_a \}_-, \quad  
p_{0\alpha} = p_{\alpha} - \frac{1}{2}S^{ab} \omega_{ab\alpha} - 
\frac{1}{2}\tilde{S}^{ab} \tilde{\omega}_{ab\alpha}.\;\;\;\;
\label{sn2snmb:lagrange}
\end{eqnarray}
The above action can further be rewritten as
\begin{eqnarray}
{\mathcal L}_{f} =  \bar{\psi} \,\gamma^{m} 
(p_{m}- \sum_{A,i}\; g^{A}\tau^{Ai} A^{Ai}_{m}) \psi + 
\{ \sum_{s=7,8} \; \bar{\psi} \gamma^{s} p_{0s} \; \psi \} + {\rm the \;rest}, 
\label{sn2snmb:lagrangeafter}
\end{eqnarray}
with the meaning  
\begin{eqnarray}
\tau^{Ai} = \sum_{a,b} \;c^{Ai}{ }_{ab} \; S^{ab},
\quad
\{\tau^{Ai}, \tau^{Bj}\}_- = i \delta^{AB} f^{Aijk} \tau^{Ak},
\label{sn2snmb:tau}
\end{eqnarray}
where $A=1$ stays for $ U(1)$, $i=\{1\}$, which is the hyper charge $Y$ in the standard model 
notation, $A=2$ stays for the $SU(2)$ weak charge, $i=\{1,2,3\}$, 
$A=3$ stays for the colour $SU(3)$ charge, $i=\{1,\cdots,8\}$. All the spinors, 
which appear in $2^{8/2-1}$ families before the break of the $SO(1,7)$ symmetry, are massless
\footnote{In the references~\cite{sn2snmb:hn05hn07} we present for the toy model the proof that 
the break of symmetry can preserve masslessness.}, 
while the term $\sum_{s=7,8} \; \bar{\psi} \gamma^{s} p_{0s} \; \psi $ 
in Eq.(\ref{sn2snmb:lagrangeafter}) form what the standard model 
postulates as the Yukawa couplings. Let us rewrite it, naming it  ${\mathcal L}_{Y}$
\begin{eqnarray}
- {\mathcal L}_{Y} &=& \psi^{\dagger} \gamma^0 \gamma^s p_{0 s}  \psi \quad
=  \psi^{\dagger} \; \gamma^0 \{
 \stackrel{78}{(+)} \; p_{0+}\; + \; \stackrel{78}{(-)} \;p_{0-} \}\psi,
\label{sn2snmb:yukawatildeyukawa}
\end{eqnarray}
with
\begin{eqnarray}
p_{0\pm} &=& (p_{7} \mp i \; p_{8}) - \frac{1}{2}  S^{ab}  \omega_{ab\pm} 
- \frac{1}{2} \tilde{S}^{ab} \tilde{\omega}_{ab\pm}; \nonumber\\
\omega_{ab\pm} &=&  \omega_{ab7} \mp i \;  \omega_{ab8}, \quad
\tilde{\omega}_{ab\pm} = \tilde{\omega}_{ab 7} 
\mp i \;  \tilde{\omega}_{ab8}. 
\label{sn2snmb:yukawatildeyukawadet}
\end{eqnarray}
%
%When studying the masses and mixing matrices we put $p_{7}=0= p_{8}$. 
One can see  in ref.~\cite{sn2snmb:pn06,sn2snmb:gmdn07,sn2snmb:n07} how does this term behave after 
particular breaks of symmetries and 
what predictions for the masses and the mixing matrices does it make. 

The action for the gauge fields is  the Einstein one~\cite{sn2snmb:hn05hn07}: linear in the curvature 
\begin{eqnarray}
S &=& \int \; d^d{} x \; E \; (R + \tilde{R}), \nonumber\\  
R &=& \frac{1}{2}\, [f^{\alpha [a} f^{\beta b]} \;(\omega_{a b \alpha,\beta} - \omega_{c a \alpha}
\omega^{c}{}_{b \beta})] + h.c., \nonumber\\
\tilde{R} &=& \frac{1}{2}\, [f^{\alpha [a} f^{\beta b]} \;(\tilde{\omega}_{a b \alpha,\beta} - 
\tilde{\omega}_{c a \alpha} \tilde{\omega}^{c}{}_{b \beta})] + h.c.. \nonumber
\label{sn2snmb:Riemannaction}
\end{eqnarray}
Here~\footnote{$f^{\alpha}{}_{a}$ are inverted 
vielbeins to 
$e^{a}{}_{\alpha}$ with the properties $e^a{}_{\alpha} f^{\alpha}{\!}_b = \delta^a{\!}_b,\; 
e^a{\!}_{\alpha} f^{\beta}{\!}_a = \delta^{\beta}_{\alpha} $. 
Latin indices  
$a,b,..,m,n,..,s,t,..$ denote a tangent space (a flat index),
while Greek indices $\alpha, \beta,..,\mu, \nu,.. \sigma,\tau ..$ denote an Einstein 
index (a curved index). Letters  from the beginning of both the alphabets
indicate a general index ($a,b,c,..$   and $\alpha, \beta, \gamma,.. $ ), 
from the middle of both the alphabets   
the observed dimensions $0,1,2,3$ ($m,n,..$ and $\mu,\nu,..$), indices from 
the bottom of the alphabets
indicate the compactified dimensions ($s,t,..$ and $\sigma,\tau,..$). 
We assume the signature $\eta^{ab} =
diag\{1,-1,-1,\cdots,-1\}$.} 
$f^{\alpha [a} f^{\beta b]}= f^{\alpha a} f^{\beta b} - f^{\alpha b} f^{\beta a}$. 
The action~(Eq.(\ref{sn2snmb:Riemannaction})) manifests after the break of symmetries all the known 
gauge fields and the Higgs fields~\footnote{I am studying how does the break of symmetries 
of $SO(1,7) \times SU(3) \times U(1)$ to $SO(1,3) \times U(1)\times U(1) \times SU(3)$ 
influence the gauge fields, leading to not only all the gauge fields, but also to  
(since the symmetry breaks twice to two kinds of) scalar (that is Higgs) fields.}. 

The question arises, whether one can at all with the action linear in the curvature  
(without any torsion) "force" spinors that after the break from say $SO(1,13)$ to 
$SO(1,7)\times SU(3)\times U(1)$ stay massless and chirally coupled to the 
$SU(3)$ and $U(1)$ gauge fields. We shall later comment on this problem.

\section{The Yukawa couplings, the masses of families and the mixing matrices}
\label{sn2snmb:Yukawa}

Let us analyze the Yukawa couplings 
$$-{\cal L}_Y =  \psi \quad
=  \psi^{\dagger} \; \gamma^0 \{
 \stackrel{78}{(+)} \; p_{0+}\; + \; \stackrel{78}{(-)} \;p_{0-} \}\psi$$ 
 (Eq.\ref{sn2snmb:yukawatildeyukawa},  \ref{sn2snmb:yukawatildeyukawadet}) and see what 
 predictions we can make.  The break of symmetries from the starting one  of 
 $SO(1,13)$ to the symmetries assumed by the standard model 
 occur spontaneusly, under the influence of the break of symmetries of the part of spin connection 
 and vielbein fields which  in $d=1+3$ manifest as scalar fields. 
 Since these breakings can be highly nonperturbative, 
 it is hard to know the way of breaking the starting symmetry $SO(1,13)$, but it  
 should be the way, which leads to all the starting assumptions of the standard model 
 of the electroweak and colour interaction. 
 Since the handedness in $d=1+3$, which obviously concerns the spin,  and the weak charge are assumed to be 
 related in the standard model, the breaking must go through $SO(1,7)$, where 
 the spin and the handedness are manifestly correlated as seen in TABLE~\ref{sn2snmb:Table I}. We assume 
 accordingly~\cite{sn2snmb:n92,sn2snmb:pn06,sn2snmb:gmdn07,sn2snmb:n07} the following way of breaking: First 
 $SO(1,13) \rightarrow  SO(1,7) \times  SU(3) \times U(1), $ then  $ 
 SO(1,7) \times  SU(3) \times U(1) $ $\rightarrow SO(1,3) \times SU(2)\times U(1) \times SU(3),$
 and finally  $\rightarrow SO(1,3) \times U(1) \times SU(3)$, which is just the 
 observed symmetry. These breaking must appear in both sectors: 
 $\omega_{ab \alpha}$ and $\tilde{\omega}_{ab \alpha}$, 
 not necessarily with the same parameters. 
After the first break $SO(1,7) \times  SU(3) \times U(1) $ 
$\rightarrow SO(1,3) \times SU(2)\times U(1) \times SU(3),$ 
the superpositions of fields in both sectors apear and new quantum numbers manifest. In $S^{ab}$
sector we expect
\begin{eqnarray}
\label{sn2newfieldssab} 
A^{23}_{a}  &=& A^{Y}_{a} \sin \theta_2 + A^{Y'}_{a} \cos \theta_2, \nonumber\\
A^{4}_{a} &=& A^{Y}_{a} \cos \theta_2 - A^{Y'}_{a} \sin \theta_2, \nonumber\\
A^{2\pm}_a &=& \frac{1}{\sqrt{2}}(A^{21}_a \mp  i A^{22}_a),
\end{eqnarray}
for $a=m,s$. The  corresponding new  operators are then 
\begin{eqnarray}
\label{sn2newoperatorssab}
Y&=& \tau^{4}+ \tau^{23}, \quad Y'= \tau^{23} - \tau^{4} \tan^{2} \theta_{2}, 
\quad \tau^{2\pm} = \tau^{21}\pm i \tau^{22}. 
\end{eqnarray}
Correspondingly we find in  the $\tilde{S}^{ab}$ sector 
\begin{eqnarray}
\label{sn2newfieldstildesab} 
\tilde{A}^{23}_{s} &=& \tilde{A}^{Y}_{s} \sin \tilde{\theta}_2 + 
\tilde{A}^{Y'}_{s} \cos \tilde{\theta}_2, \nonumber\\
\tilde{A}^{4}_{s} &=& \tilde{A}^{Y}_{s} \cos \tilde{\theta}_2 -  
\tilde{A}^{Y'}_{s} \sin \tilde{\theta}_2,\nonumber\\
\tilde{A}^{2\pm}_s &=& \frac{1}{\sqrt{2}}(\tilde{A}^{21}_s \mp i \tilde{A}^{22}_s)
\end{eqnarray}
with   
\begin{eqnarray}
\label{sn2newoperatorstildesab}
\tilde{Y}&=& \tilde{\tau}^{4}+ \tilde{\tau}^{2}, \quad 
\tilde{Y'}= \tilde{\tau}^{23} - \tilde{\tau}^{4} \tan^{2} \tilde{\theta}_2,
\quad \tilde{\tau}^{2\pm} = \tilde{\tau}^{21}\pm i \tilde{\tau}^{22}, 
\end{eqnarray}
and 
$\tilde{\tau}^{23} = \frac{1}{2}(\tilde{S}^{56}+ \tilde{S}^{78}),
\quad \tilde{\tau}^{4}= -\frac{1}{3}(\tilde{S}^{9\,10} + \tilde{S}^{11\,12} + 
\tilde{S}^{13\,14}).$ 

 The way of the above suggesting breaking leads in the sector 
 $-\frac{1}{2}\,S^{ab}\, \omega_{ab \alpha}$ 
 to the charges and gauge fields as assumed in Eq.(\ref{sn2snmb:lagrangeafter}), while it 
 leads in the sector  $-\frac{1}{2}\,\tilde{S}^{ab}\, \tilde{\omega}_{ab \alpha}$ to 
 two times four decoupled families. For $\tilde{\theta}_2=0$ the lower four of the 
 two decoupled four families are massless. 
\begin{table}
\centering
 %\begin{center}
%\end{center}
%\tiny{
%\renewcommand{\arraystretch}{1.5}
\begin{tabular}{|r||c|c|c|c|c|c|c|c||}
\hline
 &$ I $&$ II $&$ III $&$ IV $&$ V $&$ VI $
 &$ VII $&$ VIII$\\
\hline\hline
$I  $ & $ 0 $ & $ 0 $ & $ 0 $ & $ 0 $
          & $ 0 $ & $ 0 $ & $ 0 $ & $ 0 $\\
\hline
$II $ & $ 0 $ & $ 0 $ & $ 0 $ & $ 0 $
          & $ 0 $ & $ 0 $ & $ 0 $ & $ 0 $\\
\hline
$III$ & $ 0 $ & $ 0 $ & $ 0 $ & $ 0 $
          & $ 0 $ & $ 0 $ & $ 0 $ & $ 0 $\\ 
\hline
$IV $ & $ 0 $ & $ 0 $ & $ 0 $ & $ 0 $
          & $ 0 $ & $ 0 $ & $ 0 $ & $ 0 $\\
\hline\hline
$ V  $ & $ 0 $ & $ 0 $ & $ 0 $ & $ 0 $ & 
$ -\frac{\tilde{g}}{c}\, \tilde{A}^{23}_{-}$ & 
$  \frac{\tilde{g}}{\sqrt{2} \,c}\, \tilde{A}^{2-}_{-}$ & 
$ 0 $ & $ 0 $\\
\hline
$ VI $ & $ 0 $ & $ 0 $ & $ 0 $ & $ 0 $ &
$ -\frac{\tilde{g}}{\sqrt{2} \,c}\, \tilde{A}^{2+}_{-}$ & 
$ -\frac{\tilde{g}}{c}\, \tilde{A}^{23}_{-}$ & 
$ 0 $ & $ 0 $\\ 
\hline
$VII $ & $ 0 $ & $ 0 $ & $ 0 $ & $ 0 $ & $ 0 $ & $ 0 $ &
$  \frac{\tilde{g}}{c}\, \tilde{A}^{23}_{-}$ & 
$  \frac{\tilde{g}}{\sqrt{2} \,c}\, \tilde{A}^{2-}_{-}$ \\
\hline
$VIII$ & $ 0 $ & $ 0 $ & $ 0 $ & $ 0 $ & $ 0 $ & $ 0 $ &  
$ -\frac{\tilde{g}}{\sqrt{2} \,c}\, \tilde{A}^{2+}_{-}$ & 
$  \frac{\tilde{g}}{c}\, \tilde{A}^{23}_{-}$ \\
\hline\hline
\end{tabular}
%}
%\end{center}
% 
\caption{\label{sn2Table2}%
The {\bf Yukawa couplings} for $u-$quarks after the  {\bf break of $SO(1,7)\times U(1)$ into
$SO(1,3)\times SU(2) \times U(1)$}.} 
\end{table}
 
 The starting Lagrange density for fermions transforms into
 \begin{eqnarray}
 {\mathcal L}_{f} &=& \bar{\psi}\,\{ \gamma^{m}\; [ \,p_{m}- g^{3}\,\sum_{i}\; \tau^{3i} A^{3i}_{m} 
 - g^{Y}\;\tau^{Y} A^{Y}_{m}  -   g^{Y'}\;Y' A^{Y'}_{m} - 
 g^{1}\, \sum_{i=1,2,3}\;  \tau^{1i} A^{1i}_{m} -  \, \nonumber \\
 &&\frac{g^2}{\sqrt{2}} ( \tau^{2+} A^{2+}_m  + \tau^{2-}  A^{2-}_m ) ] + \nonumber\\
 && \gamma^{s}\; [ \, p_{s}  - g^{Y}\;Y A^{Y}_{s} -  g^{Y'}\;Y' A^{Y'}_{s}  - \nonumber\\
 && \tilde{g}^{Y} \tilde{Y} \tilde{A}^{Y}_{s} - 
   \tilde{g}^{Y'} \tilde{Y'} \tilde{A}^{Y'}_{s} - 
   \frac{\tilde{g}^2}{\sqrt{2}} (\tilde{\tau}^{2+} \tilde{A}^{2+}_s  + 
   \tilde{\tau}^{2-} \tilde{A}^{2-}_s ) - \nonumber\\ 
  && \tilde{g}^{1} \, \sum_{i=1,2,3}\; \tilde{\tau}^{1i} \tilde{A}^{1i}_{s} -  
 \frac{\tilde{g}^{(1+3)}}{2} \tilde{S}^{mm'}\tilde{\omega}_{mm's} \,] 
 \,\} \psi, \nonumber\\
 && m,m'\in\{0,1,2,3\},\; s,s',t \in\{5,6,7,8\}.
 %+ \nonumber\\
 %& &  \sum_{s=7,8}\; 
 %\bar{\psi} \gamma^{s} p_{0s} \; \psi + {\rm the \;rest}.
 \label{sn2lagrange3}
 \end{eqnarray}
 For $\theta_{2}=0$ and $\tilde{\theta}_2=0$ the new fields are 
 $ A^{Y}_{a}  =  A^{4}_{a},  \,
 A^{Y'}_{a} = A^{23}_{a} $, $a=m,s$; 
 $ \tilde{A}^{Y}_{s}  =  \tilde{A}^{4}_{s},
 \tilde{A}^{Y'}_{s} = \tilde{A}^{23}_{s}$, with the  
 coupling constants expressible with the previous ones. % $g^Y= g^{2} \,\sin \theta_2$, 
 %$g^{Y'}= g^{2} \, \cos \theta_2 $, $ \tan \theta_2 = \frac{g^4}{g^{2}}$ 
 %$\tilde{g}^Y= \tilde{g}^{2} \, \sin \tilde{\theta}_2$, 
 %$\tilde{g}^{Y'}= \tilde{g}^{2} \cos \tilde{\theta}_2 $, 
 %$ \tan \tilde{\theta}_2 = \frac{\tilde{g}^4}{\tilde{g}^{2}}$.  
 %
 %
 %Since the operators $\tau^{2\pm}$ transform right handed quarks into right handed 
 %quarks, changing the charge $Y$ and accordingly the $Q$ charge, and equivalently 
 %for the leptons, and since the mass terms must  conserve the $Q$ charge, we  put the 
 %
  %fourth row of Eq.(\ref{sn2lagrange3}) equal to zero. 
%%%%%

 The last step to the massive observable fields follows after the break of $SU(2)\times U(1)$ 
 to $U(1)$ at the weak scale. New fields in the $S^{ab}$ sector  
 \begin{eqnarray}
 \label{sn2newfieldsweaksab}
 A^{13}_{a} &=& A_{a} \sin \theta_1 + Z_{a} \cos \theta_1,\nonumber\\ 
 A^{Y}_{a}  &=& A_{a} \cos \theta_1 -  Z_{a} \sin \theta_1,\nonumber\\ 
 W^{1\pm}_a &=& \frac{1}{\sqrt{2}}(A^{11}_a \mp i  A^{12}_a),
 \end{eqnarray}
 with $a=m,s$ appear as the gauge fields of  new operators 
 \begin{eqnarray}
 \label{sn2newoperatorsweaksab}
 Q  &=&  \tau^{13}+ Y = S^{56} +  \tau^{4},\nonumber\\
 Q' &=& -Y \tan^2 \theta_1 + \tau^{13}, \nonumber\\
 \tau^{1\pm}&=& \tau^{11} \pm i\tau^{12}
 \end{eqnarray}
 and with new coupling constants $e = 
 g^{Y} \cos \theta_1, g' = g^{1}\cos \theta_1$  and $\tan \theta_1 = 
 \frac{g^{Y}}{g^1} $. 
 
 Similarly also new fields in the $\tilde{S}^{ab}$ sector appear
 \begin{eqnarray}
 \label{sn2newfieldsweaktildesab}
 \tilde{A}^{13}_{s} &=& \tilde{A}_{s} \sin \tilde{\theta}_1 + 
 \tilde{Z}_{s} \cos \tilde{\theta}_1,\nonumber\\ 
 \tilde{A}^{Y}_{s} &=& \tilde{A}_{s} \cos \tilde{\theta}_1 -  
 \tilde{Z}_{s} \sin \tilde{\theta}_1, \nonumber\\
 \tilde{W}^{\pm}_a &=& \frac{1}{\sqrt{2}}(\tilde{A}^{11}_a \mp i  \tilde{A}^{12}_a),
 \end{eqnarray}
 and new operators 
 \begin{eqnarray}
 \label{sn2newoperatorsweaktildesab}
 \tilde{Q}  &=&  \tilde{\tau}^{13}+ \tilde{Y} = \tilde{S}^{56} +  \tilde{\tau}^{4},\nonumber\\
 \tilde{Q'} &=& -\tilde{Y} \tan^2 \tilde{\theta}_1 + \tilde{\tau}^{13},\nonumber\\
 \tilde{\tau}^{1\pm}&=& \tilde{\tau}^{11} \pm i\tilde{\tau}^{12}
 \end{eqnarray}
 with new coupling constants $\tilde{e} = 
 \tilde{g}^{Y}\cos \tilde{\theta}_1, \tilde{g'} = 
 \tilde{g}^{1}\cos \tilde{\theta}_1$  and $\tan \tilde{\theta}_1 = 
\frac{\tilde{g}^{Y}}{\tilde{g}^1} $.

The Yukawa coupling 
\begin{eqnarray}
- {\mathcal L}_{Y} &=& \psi^{\dagger} \gamma^0 \gamma^s p_{0 s} \psi\; \nonumber\\
&=& \psi^{\dagger} \; \gamma^0 \{
 \stackrel{78}{(+)} \; p_{0+}\; + \; \stackrel{78}{(-)} \;p_{0-} \}\psi,\nonumber
\label{sn2yukawatildeyukawa}
\end{eqnarray} 
can be rewritten as follows
\begin{eqnarray}
{\mathcal L}_{Y} &=& \psi^{\dagger} \gamma^0 %\nonumber\\  
\{ \stackrel{78}{(+)}  (\sum_{y=Y,Y'}\; y A^{y}_{+} + 
\frac{-1}{2}\sum_{(ab)}\; \tilde{S}^{ab} \tilde{\omega}_{ab+}))
%\sum_{\tilde{y}=\tilde{N}^{+}_{3},\tilde{N}^{-}_{3},\tilde{\tau}^{13},\tilde{Y},\tilde{Y'}} 
%\tilde{y} \tilde{A}^{\tilde{y}}_{+}\;)\; +
+  \nonumber\\
& &  \stackrel{78}{(-)}  (\sum_{y=Y,Y'}\;y  A^{y}_{-} + 
  \frac{-1}{2}\sum_{(ab)}\; \tilde{S}^{ab} \tilde{\omega}_{ab-})
%\sum_{\tilde{y}= \tilde{N}^{+}_{3},\tilde{N}^{-}_{3},\tilde{\tau}^{13},\tilde{Y},\tilde{Y'}} 
%\tilde{y} \tilde{A}^{\tilde{y}}_{-}\;) + 
\nonumber\\
 & & \stackrel{78}{(+)} \sum_{\{(ac)(bd) \},k,l} 
 \; \stackrel{ac}{\tilde{(k)}} \stackrel{bd}{\tilde{(l)}}
\tilde{{A}}^{kl}_{+}((ac),(bd)) \;\;+  \nonumber\\
 & & \stackrel{78}{(-)} \sum_{\{(ac)(bd) \},k,l} \; \stackrel{ac}{\tilde{(k)}}\stackrel{bd}{\tilde{(l)}}
\tilde{{A}}^{kl}_{-}((ac),(bd))\}\psi,\nonumber
\label{sn2yukawa4tilde0}
\end{eqnarray}
with $k,l=\pm 1,$ if $\eta^{aa}\eta^{bb}=1$ and 
$ \pm i,$ if $\eta^{aa}\eta^{bb}=-1$, while $Y=\tau^{21} +\tau^{41}$ and $Y'= -\tau^{21} +\tau^{41}$,
$(ab),(cd),\cdots$  {\bf Cartan only}. 
%%%%%%%%%%%%%5
 
 In references~\cite{sn2snmb:pn06,sn2snmb:gmdn07,sn2snmb:n07} this 
 decoupling is analyzed and  the predictions made. 
The way of breaking and correspondingly the symmetries imposed on the 
 fields $\omega_{ab \sigma}$ and $\tilde{\omega}_{ab \sigma}$, $\sigma = \{5,6,7,8\}$ influence 
 properties estimated for quarks and leptons of the first four families. Our rough estimations 
 did not go beyond  the  tree level, when predicting properties of the masses of the fourth 
 family and the mixing matrices of the first four families. 
 This rough estimation~\cite{sn2snmb:pn06,sn2snmb:gmdn07,sn2snmb:n07} predicts  
 the masses of the fourth family quarks to lie at around $250$ GeV or higher, the fourth family neutrino
 mass at around $80$ GeV or higher and the fourth family electron mass at around $200$ GeV or higher. 
 We predict  the mixing matrices for quarks and leptons. The fourth family quarks have possibly a 
 chance to be seen at LHC.
 
 The lower of the upper four families, which is stable (has zero Yukawa couplings to the lower four families),  
 must have accordingly the masses  above $1$ TeV. Being stable the neutral (with respect to the weak and 
 colour charge) clusters of the fifth family members are candidates for forming the dark matter.  
 
 These rough estimations, although to my understanding a good guide to the properties of families, need much 
 more sophisticated calculations to be really trustful.

 The numerical results for the Yukawa couplings of the lower four families and correspondingly for  
 their masses and mixing matrices can be found in the referece~\cite{sn2snmb:gmdn07}. We took   
  the symmetries of the Yukawa couplings as discussed above (determined by the way of breaking 
  symmetries) and assumed that the calculations  
 beyond the tree level  would bring the expected differences in the nondiagonal (in the basis of the four 
 family members) Yukawa couplings (that is among  the 
members of families), so that the experimental data for the known three families can be fitted. We were able to 
predict that the quark masses of the four family lie at around or above $250$ GeV, while the fifth family electron 
has a mass above $100$ GeV and the corresponding neutrino mass is above $50$ GeV.

 \section{Yukawa couplings beyond the tree level}
 \label{sn2snmb:beyondtree}

 To understand how does the Yukawa couplings change when going beyond the tree level one must see 
 how do the scalar fields occur spontaneously, manifesting in the effective Lagrangean in $d=1+3$ 
 the Higgs field of the standard model. It is the vielbein in $d>(1+3)$, in interaction with
  the spin connection fields of both sectors, those with the indices $\sigma=(5),(6)\cdots$, 
  which manifests  properties of  scalar fields, while those with indices 
  $\mu= 0,1,2,3$ manifest as  gauge fields of the corresponding charges 
 \begin{eqnarray}
 \label{sn2higgsgauge}
  e^{a}{}_{\alpha} = 
 \begin{pmatrix} \delta^{m}{}_{\mu}  & e^{m}{}_{\sigma}=0  \\
  e^{s}{}_{\mu}= e^{s}{}_{\sigma} E^{\sigma}{}_{Ai} A^{Ai}_{\mu} & e^s{}_{\sigma} \end{pmatrix}.
    \end{eqnarray}
 We started with the analyse of the scalar fields dynamics and their influence on the 
 Yukawa couplings, treating Yukawa couplings beyond the tree level in this Bled workshop, hoping 
 that the Yukawa couplings beyond the tree level,  arranged as the four times four matrices at the operators 
 $Q$, $Q'$ and the powers of these operators, will manifest the measured  differences of the 
 properties of the members of one family. Although we have during the Bled workshop started 
 with these studies, we have not succeeded to come to the point to publish the results in this 
 proceedings.

  \section{Kaluza-Klein-like theories and massless fermions}
  \label{sn2snmb:kk} 
 The approach unifying spins and charges shares with the Kaluza-Klein-like theories the 
 difficulties with forcing  massless spinors to stay massless also 
 after the breaking of the starting symmetry 
 (determined in d-dimensional space). Let in our case speak about the breaking of  $SO(1,13)$ to 
 $SO(1,7)\times SU(3)\times U(1)$. The {\it no-go} theorem of witten~\cite{sn2witten} suggest that there 
 is no hope for the Kaluza-Klein-like theories to lead to the observed masses of the three families of 
 quarks as long as the break occurs at high energy scale as it is  $10^{17}$ GeV or even higher, since then 
 the masses of the families would be of this order (divided by $c^2$) or higher. The offer of 
 a possible  solution of this problem can be found in our papers~\cite{sn2hn05,sn2hn06,sn2hn08,sn2dhnBled09,sn2dhnBled08}, 
 one of them included also in this proceedings. We have solved this problem either with a choice of 
 a particular boundary conditions, or with the choice of decoupled vielbeins and spin connections, which 
 in all the cases allow only one massless spinor of one handedness to chorally couple to the Kaluza-Klein 
 gauge field of a particular charge as manifested in the vielbein of Eq.(\ref{sn2higgsgauge}). 
 We speak in~\cite{sn2dhnBled09} about the "effective two dimensionality", since in two-dimensional manifolds   
 the action for  free vielbein and spin connection fields which is linear in the 
 curvature leads to the equations of motion which any vielbein and any spin connection fulfils.    
 Making a choice of the zweibein, which curves the infinite disc on $S^2$, we were able to find the 
 spin connection field, which allows only one massless spinor. The reader can find further explanation in this 
 paper.

 \section{The fifth family as the candidate for forming the dark matter clusters}
  \label{sn2snmb:darkmatter}
This section is meant as a short overview of the work, which is in details presented in this 
proceedings~\cite{sn2snmb:gnBled08proc}, 
(it will also be  published in Phys. Rev. D~\cite{sn2snmb:gnBled08}) and concerns the study of  
the properties of the fifth family members, predicted by the approach unifying spins and charges. 
This family, having zero Yukawa couplings to the lower four family members, 
is the candidate to constitute the dark matter. I study in the ref.~\cite{sn2snmb:gnBled08proc}, together 
with Gregor Bregar, the behaviour of this family members during the evolution of the expanding universe. 
Although we have not yet studied the properties of the five 
family members in details, it is clear from what it is presented and discussed in the section~\ref{sn2snmb:Yukawa}
that the masses of the fifth family members are expected to be above $1$ TeV/$c^2$, since already the 
fourth family quark masses are close to  $300$ GeV or even above. 
Accordingly we follow the fifth family quarks and leptons through the 
expansion of the universe, starting when the temperature  of the plasma is above the 
fifth family members' masses (times $c^2/k_{b}$), under the assumption that 
the fifth family masses are all above $1$ TeV/$c^2$. At this temperature all the fifth family members,  
as do also the members of all the lower mass families and all the gauge fields, contribute 
in the thermal equilibrium 
to the plasma. When the plasma's temperature falls bellow the fifth family quarks' masses, the 
quarks start to decouple from the plasma, since the formation of quark-antiquark's pairs 
out of the plasma start to be less and less possible. When the temperature of the plasma falls  
bellow the binding energy
of the two and correspondingly three quarks clusters, quarks start to form colourless clusters, since  
scattering of fermions and bosons in the plasma on these fifth family clusters results in 
destroying the clusters with  less and less probabilities. 
For large enough fifth family masses the colourless fifth family baryons as well as the  
neutrinos (if there are lighter than the fifth family electrons) start to decouple from the plasma far before 
the colourless phase transition (which starts at approximately  $T= 1 $ GeV/$k_{b}$). 

We make in this study the assumption that the lightest fifth family baryons are neutrons and that the 
neutrino is the lightest lepton.  Other possibilities are under considerations. 
For known masses of quarks and leptons all the other properties should follow. 
Although at high enough temperatures of the cosmic plasma quarks predominantly interact with one gluon 
exchange, while the weak and $U(1)$ (before the break of the electroweak symmetry this U(1) gauge field 
carries the hyper $Y$ charge as it follows from the section~\ref{sn2snmb:Yukawa}) interactions are, due to 
the much weaker couplings constants, negligible, yet the calculations are not simple. There is the 
$SU(2)\times U(1)$ breaking into $U(1)$, which is very probably nonperturbative and needs to be studied 
in details causing a possible phase transition). It is also the phase transition of the lowest four 
massless families and the massless weak fields into massive four families and weak massive bosons, 
caused by the vielbeins and the spin connections of two kinds, which manifest as scalar fields, 
which should be studied seriously. And it is also the colour phase transition which starts bellow 
$1$ GeV, and which might or not force all the fifth family quarks and anti-quarks to annihilate (or to form 
the colourless fifth family clusters if it is the fifth family baryon-antibaryon asymmetry) 
above the temperature, when the first family quarks and antiquarks start to form 
hadrons.

We evaluated properties of the fifth family hadrons if masses are larger than $1$ TeV/$c^2 $, while we 
estimated  that the fifth family neutrinos with masses above TeV/$c^2$ and bellow $200$ TeV/$c^2$ 
contribute to the dark matter and to the direct measurements 
less than the fifth family neutrons.

We estimated that the {\it nuclear force} of the fifth family baryons  
manifests for the quark mass, let say, in the region 
$(1--500)$ TeV/$c^2 $ the scattering cross section $(10^{-5}-- 10^{-12})$ fm$^2$, respectively, 
while the binding energies are in the region $((-.02)-- (-2))$ TeV. 

%%%%%%%%%%%%%%%%%
%12.11.2009

To solve the coupled Boltzmann equations for the numbers of the fifth family quarks and the colourless 
clusters of the quarks in the plasma of all the other fermions and bosons in the thermal equilibrium in the 
expanding universe, we aught to estimate the cross sections for the annihilation 
of quarks with antiquarks  and for forming the clusters. We did this within some 
uncertainty intervals, which we took into account by parameters. 
We solved the Boltzmann equations for several values 
of quark masses and several values of the parameters correcting the roughness of the estimated cross 
sections and  following these decoupling of the fifth family quarks and the fifth family neutrons out 
of the plasma down to the temperature
$1$ GeV/$c^2$ when the colour phase transition of the plasma starts. 

The fifth family neutrons, packed into very tinny 
clusters so that they are totally decoupled of the plasma, 
do not feel the colour phase transition of the plasma, while the fifth family quarks 
and coloured clusters of quarks do. Their scattering cross section  grew due to the nonperturbative 
behaviour of gluons as did the scattering cross section of all the other quarks. The quarks 
"dressed" into the constituent mass. While the three of the lowest four families decayed into the 
first family quarks, due to the corresponding Yukawa couplings, the fifth family quarks can not. 
Although the "dressing" do not influence the scattering of the very heavy fifth family quarks  
the very much enlarged scattering cross section does. 
 Having the binding energy a few orders of magnitude larger the $1$ GeV and moving in the rest of  
plasma of the first family quarks and antiquarks and gluons as a very heavy objects with a very 
large scattering cross section the fifth family coloured objects annihilated with their partners 
or formed the colourless objects (which results in the decoupling from the plasma)
long before the temperature fell bellow a few MeV/$k_{b}$, when the first family quarks 
could start to form bound states. 

Following further the fifth family neutrons in the expanding universe up to today and 
equating the today's dark matter density with the calculated one, we estimated  the mass 
interval of the fifth family quarks to be 
\begin{eqnarray}
\label{sn2massinterval}
10 \;\; {\rm TeV} < m_{q_5}\, c^2 < {\rm a\, few} \cdot 10^2 {\rm TeV}.
\end{eqnarray}
The detailed calculations with all the needed explanations can be found in the 
paper~\cite{sn2snmb:gnBled08proc,sn2snmb:gnBled08}.

\section{Dynamics of a heavy family baryons in our galaxy and the direct measurements}
\label{sn2snmb:dynamics}

Although the evarage properties of the dark matter in the Milky way are pretty well known 
(the everage dark matter density, which is approximately  spherically symmetrically distributed around 
the center of the galaxy and is dropping with the distance from the galaxy center with the 
second power of the distance keeping the velocities of the suns arround the center of the Milky way constant,
is at the position of the Sun expected to be $\rho_0 \approx 0.3 \,{\rm GeV} /(c^2 \,{\rm cm}^3)$,  
and the everage velocity of the dark matter consituents around the center of our 
galaksy is expected to be approximately velocity of our  Sun), their real local properties are 
known much less accurate, its density may be within the fastor of $10$ and its velocity may be 
a little better. 
 
 When evaluating the number of events which our fifth family members  triggered in the direct 
 measurements of DAMA~\cite{sn2snmb:rita0708} and CDMS~\cite{sn2snmb:cdms} experiments, we took all these  
 uncertainties into account. Let the dark matter member hitts the Earth with the  velocity 
 $\vec{v}_{dm \,i}$. The velocity of the Earth around the center of the galaxy is equal to:  
$\vec{v}_{E}= \vec{v}_{S} + \vec{v}_{ES} $, with $v_{ES}= 30$ km/s and 
$\frac{\vec{v}_{S}\cdot \vec{v}_{ES}}{v_S v_{ES}}\approx \cos \theta, \theta = 60^0$. 
The dark matter cluster of the $i$- th  velocity class  hits the Earth with the velocity:  
$\vec{v}_{dmE\,i}= \vec{v}_{dm\,i} - \vec{v}_{E}$. 
Then the flux of our dark matter clusters hitting the Earth is:    
$\Phi_{dm} = \sum_i \,\frac{\rho_{dm i}}{m_{c_5}}  \,
|\vec{v}_{dm \,i} - \vec{v}_{E}|  $, which  (for $\frac{v_{ES}}{|\vec{v}_{dm i}- \vec{v}_S|}$ 
 small) equals to  
$\Phi_{dm}\approx \sum_i \,\frac{\rho_{dm i}}{m_{c_5}}  \,
\{|\vec{v}_{dm \,i} - \vec{v}_{S}| - \vec{v}_{ES} \cdot \frac{\vec{v}_{dm i}- \vec{v}_S}{
|\vec{v}_{dm i}- \vec{v}_S|} \}.$ One can take approximately  that
$\sum_i \, |\vec{v_{dm i}}- \vec{v_S}| \,\rho_{dm i} = \varepsilon_{v_{dmS}} 
\, \varepsilon_{\rho}\,  v_S\, \rho_0 $, 
and further %correspondingly 
$ \sum_i \, \vec{v}_{ES}  \cdot \frac{\vec{v}_{dm i}- \vec{v}_S}{
|\vec{v}_{dm i}- \vec{v}_S|} = v_{ES} \varepsilon_{v_{dmES}}
\cos \theta \, \sin \omega t $.  % with $v_{ES} = 30$ km/s, 
%$\theta = 60^0$, where $\omega $ is the  rotation velocity of our Earth around the Sun.   
We estimate (due to experimental data and our theoretical evaluations) that 
$\frac{1}{3} < \varepsilon_{v_{dmS}} < 
3$ and $\frac{1}{3} < \frac{\varepsilon_{v_{dmES}}}{\varepsilon_{v_{dmS}}} < 3$. 
This last term determines the annual modulations observed by DAMA~\cite{sn2snmb:rita0708}.

The cross section for our fifth family baryon to elastically
(the excited states of nuclei,  
which we shall treat, I and Ge, are at $\approx 50$ keV 
or higher and are very narrow, while the average recoil energy of Iodine is expected to be 
$30$ keV) scatter  on an ordinary nucleus with $A$ nucleons is 
$\sigma_{A} = 
\frac{1}{\pi \hbar^2} <|M_{c_5 A}|>^2 \, m_{A}^2$. %(****PREVIDNOST V $A^2$ IN KOHERENCI****)
For our fifth family neutrons  %with a small cross section $\sigma_{c_{5}}$  
is  $m_{A}  $  approximately the mass of the ordinary nucleus. In the case of a 
coherent scattering (if recognizing that $\lambda= \frac{h}{p_A}$ is for a nucleus large enough 
to make scattering coherent, when the mass of  the cluster is 
 $1$ TeV or more and its velocity 
$\approx v_{S}$), the cross section is  almost independent of the recoil 
velocity of the nucleus. 
For the case that the fifth family ''nuclear force'' as manifesting  
in the cross section discussed above (which is proportional $10^{-6}$ fm$^2$ for the masses 
of the fifth family quarks let say $1$ TeV/$c^2$ or to $10^{-12}$ fm$^2$ for the masses of quarks $500$ 
TeV/$c^2$)    
%in Eq.(\ref{sn2snmb:bohr}) 
brings the main contribution, the cross section  is  proportional to $(3A)^2$ 
(due to the square of the matrix element) times $(A)^2$ (due to the mass of the nuclei 
$m_A\approx 3 A \,m_{q_1}$, with $m_{q_1}\, c^2 \approx \frac{1 {\rm GeV}}{3}$).  
When $m_{q_5}$ is  heavier than $10^4 \, {\rm TeV}/c^2$, 
the weak interaction dominates and $\sigma_{A}$ is proportional to $(A-Z)^2 \, A^2$, 
since to $Z^0$ boson exchange only neutron gives an appreciable contribution. 
Accordingly we have that $\sigma(A) \approx \sigma_{0} \, A^4 \, \varepsilon_{\sigma}$, with 
$\sigma_{0}\, \varepsilon_{\sigma}$, which is $9\,\pi r_{c_5}^2 \, 
\varepsilon_{\sigma_{nucl}} $, with 
 $\frac{1}{30} < \varepsilon_{\sigma_{nucl}} < 30$ (taking into account the roughness 
with which we treat our  heavy baryon's properties and the scattering procedure) 
when the ''nuclear force'' dominates, while $ \sigma_{0}\,\varepsilon_{\sigma}$ is $  
 (\frac{m_{n_1} G_F}{\sqrt{2 \pi}} 
\frac{A-Z}{A})^2 \,\varepsilon_{\sigma_{weak}}  $
($=( 10^{-6} \,\frac{A-Z}{ A} \, {\rm fm} )^2 \,\varepsilon_{\sigma_{weak}} $), 
$ \frac{1}{10}\, <\,  \varepsilon_{\sigma_{weak}} \,< 1$ (the weak force is pretty accurately 
 evaluated, but the way of averaging is not), when the weak interaction dominates.

 Let $N_A$ be the number of nuclei of a type $A$ in the %measurement 
 apparatus  
 (of either DAMA~\cite{sn2snmb:rita0708}, which has $4\cdot 10^{24}$ nuclei per kg of $I$, 
 with $A_I=127$,  
  and  $Na$, with $A_{Na}= 23$ (we shall neglect $Na$), 
 or of CDMS~\cite{sn2snmb:cdms}, which has $8.3 \cdot 10^{24}$ of $Ge$ nuclei 
 per kg,  with $A_{Ge}\approx 73$). 
 At velocities  of a dark matter cluster  $v_{dmE}$ $\approx$ $200$ km/s  
 are the $3A$ scatterers strongly bound in the nucleus,    
 so that the whole nucleus with $A$ nucleons elastically scatters on a 
 heavy dark matter cluster.  
Then the number of events per second  ($R_A$) taking place 
in $N_A$ nuclei   is  due to the flux $\Phi_{dm}$ and the recognition that the cross section 
is at these energies almost independent 
of the velocity %(and depends accordingly only  on $A$ of the nucleus),  
equal to
\begin{eqnarray}
\label{sn2snmb:ra}
R_A = \, N_A \,  \frac{\rho_{0}}{m_{c_5}} \;
\sigma(A) \, v_S \, \varepsilon_{v_{dmS}}\, \varepsilon_{\rho} \, ( 1 + 
\frac{\varepsilon_{v_{dmES}}}{\varepsilon_{v_{dmS}}} \, \frac{v_{ES}}{v_S}\, \cos \theta
\, \sin \omega t).
\end{eqnarray}

Let $\Delta R_A$ mean the amplitude of the annual modulation of $R_A$, 
$R_A(\omega t = \frac{\pi}{2}) - R_A(\omega t = 0) =$
$N_A \, R_0 \, A^4\, 
\frac{\varepsilon_{v_{dmES}}}{\varepsilon_{v_{dmS}}}\, \frac{v_{ES}}{v_S}\, \cos \theta$, where 
$ R_0 = \sigma_{0} \, \frac{\rho_0}{
m_{q_{c_5}}} \,  v_S\, \varepsilon$, and  %$R_0$ is for the case that the ''nuclear force'' 
%dominates $R_0 =  \pi\, (\frac{\hbar\, c}{\alpha_c \, m_{q_5}\, c^2})^2\, 
%\frac{\rho_0}{m_{q_5}} \, v_S\, \varepsilon$, with 
$\varepsilon = 
\varepsilon_{\rho} \, \varepsilon_{v_{dmES}} \varepsilon_{\sigma} $.  %$R_0$ is therefore 
%proportional to $m_{q_5}^{-3}$. 
Let  $\frac{1}{300} < \varepsilon < 300$     
demonstrates  the uncertainties in the knowledge about the dark matter dynamics 
in our galaxy and our approximate treating of the dark matter properties.  
%When for $m_{q_5} \, c^2 > 10^4$ TeV the weak interaction determines the cross section  
%$R_0 $ is in this case proportional to $m_{q_5}^{-1}$. 
%
An experiment with $N_A$ scatterers  should  measure %the amplitude
$R_A \varepsilon_{cut\, A}$, with $\varepsilon_{cut \, A}$ determining  the efficiency  of 
a particular experiment to detect a dark matter cluster collision. 
For small enough $\frac{\varepsilon_{v_{dmES}}}{\varepsilon_{v_{dmS}}}\, 
\frac{v_{ES}}{v_S}\, \cos \theta$ it follows:  
\begin{eqnarray}
R_A \, \varepsilon_{cut \, A}  \approx  N_{A}\, R_0\, A^4\, 
 \varepsilon_{cut\, A} = \Delta R_A \varepsilon_{cut\, A} \,
 \frac{\varepsilon_{v_{dmS}}}{\varepsilon_{v_{dmES}}} \, \frac{v_{S}}{v_{ES}\, \cos \theta}. 
\label{sn2snmb:measure}
\end{eqnarray}
If DAMA~\cite{sn2snmb:rita0708}   is measuring 
our  heavy  family baryons %with weak enough scattering cross section 
then 
$$R_{I} \varepsilon_{cut\, dama} \approx  \Delta R_{dama} % \; \varepsilon_{cut\, dama}\,
\frac{\varepsilon_{v_{dmS}}}{\varepsilon_{v_{dmES}}}\,
\frac{v_{S}  }{v_{SE}\, \cos 60^0 } $$, with $\Delta R_{dama}\approx 
\Delta R_{I}  \, \varepsilon_{cut\, dama}$.  
Most of unknowns about the dark matter properties, except the local velocity of our Sun,  
the cut off procedure ($\varepsilon_{cut\, dama}$) and 
$\frac{\varepsilon_{v_{dmS}}}{\varepsilon_{v_{dmES}}}$,
%(estiameted to be $\frac{1}{3} < \frac{\varepsilon_{v_{dmS}}}{\varepsilon_{v_{dmES}}} < 3$), 
 are hidden in $\Delta R_{dama}$. Taking for the Sun's velocity  
$v_{S}=100, 170, 220, 270$ km/s,  we find   $\frac{v_S}{v_{SE} \cos \theta}= 7,10,14,18, $ 
respectively. %(The recoil energy of the nucleus $A=I$ changes correspondingly %in the average 
%with the square of   $v_S $.)
%
DAMA/NaI, DAMA/ LIBRA~\cite{sn2snmb:rita0708} publishes %with $4. 10^{24}$ scatterers  per kg 
%$\varepsilon_{cut \,I}$ \, 
$ \Delta R_{dama}= 0,052  $ counts per day and per kg of NaI. 
Correspondingly  is $R_I \, \varepsilon_{cut\, dama}  = 
 0,052 \, \frac{\varepsilon_{v_{dmS}}}{\varepsilon_{v_{dmES}}}\, \frac{v_S}{v_{SE} \cos \theta} $ 
counts per day and per kg. 
CDMS should then in $121$ days with 1 kg of Ge ($A=73$) detect   
$$R_{Ge}\, \varepsilon_{cut\, cdms}
\approx \frac{8.3}{4.0} \, 
 (\frac{73}{127})^4 \; \frac{\varepsilon_{cut\,cdms}}{\varepsilon_{cut \,dama}}\, 
 \frac{\varepsilon_{v_{dmS}}}{\varepsilon_{v_{dmES}}}\;
 \frac{v_S}{v_{SE} \cos \theta} \;  0.052 \cdot 
 121 $$ 
events, 
which is for the above measured velocities equal to 
$$(10,16,21,25)
\, \frac{\varepsilon_{cut\, cdms}}{\varepsilon_{cut\,dama}}\;
\frac{\varepsilon_{v_{dmS}}}{\varepsilon_{v_{dmES}}}.$$
 CDMS~\cite{sn2snmb:cdms} 
has found no event.

The approximations we made might cause that the expected  numbers 
$(10$, $16$, $21$, $25)$ multiplied by $\frac{\varepsilon_{cut\,Ge}}{\varepsilon_{cut\,I}}\;
\frac{\varepsilon_{v_{dmS}}}{\varepsilon_{v_{dmES}}}$  
are too high (or too low!!) for a factor let us say $4$ or $10$. 
%%(But they also might be too low for the same fastor!)  
If in the near future  
CDMS (or some other experiment) 
will measure the above predicted events, then there might be  heavy 
family clusters which form the dark matter. In this case the DAMA experiment   
puts the limit on our heavy family masses: We  evaluate 
the lower limit for the mass $m_{q_5}\, c^2> 200$ TeV.

\section{Concluding remarks}
\label{sn2snmb:conclusions}
I presented in my talk very briefly the approach unifying spins and 
charges~\cite{sn2snmb:n92,sn2snmb:pn06,sn2snmb:gmdn07}, 
which is offering the way to explain the assumptions of the standard model 
of the electroweak and colour interactions, 
with the appearance of family included by proposing  the mechanism (the only one in the literature so far) 
for generating families. 
It is a  simple starting Lagrange density with spinors which carry only two kinds of the spin,
no charges, and interact with vielbeins and the two kinds of spin connection fields, which manifests  
at observed energies all the observed properties of fermions and bosons. 

Rough estimations, made up to now on a tree level under the assumption that 
the calculations beyond the tree level will manifest the differences in the off diagonal matrix 
elements in the Yukawa couplings of different members of one family, predict the fourth 
family to be possibly seen at LHC and
the stable fifth family neutrons and neutrinos to form the dark matter clusters. 

Predictions depend on the way of breaking the starting symmetries, and on the perturbative 
and nonperturbative effects, which follow the breaking. Accordingly future %(certainly needed) 
more sophisticated calculations will be  
very demanding. And we have just start some steps.

With the simple Bohr-like model we evaluated the 
properties of  the fifth family baryons and the "nuclear" interaction among these baryons as well as
with the ordinary  nuclei, recognizing that %in the latter case 
the weak interaction dominates over the 
''nuclear interaction'' for massive enough clusters ($m_{q_5}> 10^4$ TeV), while  nonrelativistic 
clusters interact among themselves with the weak force only.

Following the evolution of the number  density of the fifth family quarks and neutrinos in the 
plasma of the expanding universe, and assuming that their is our fifth family, which form the dark matter, 
we estimated that the masses of quarks and neutrinos lie in the interval 
of a few TeV$/c^2$ to a few $100$ TeV$/c^2$. 

Assuming that the DAMA and CDMS experiments  measure our fifth family baryons and neutrinos, 
we find the limit on the fifth family quark mass:  $ 200 \,\rm{TeV} < m_{q_{5}}c^2 < 10^5\, \rm{TeV}$.  
If the weak 
interaction determines the $n_5$ cross section we find:     
$10 {\rm TeV} < m_{q_5} \, c^2< 10^5$ TeV, which is as well the limit for $m_{\nu_{5}}$.   
In this case is the estimated cross section for the dark matter cluster to 
(elastically, coherently and nonrelativisically) scatter on the nucleus 
determined on the lower mass limit by the ''nuclear force'' %and is equal to 
%($ (3\cdot 10^{-5}\,A^2\, {\rm fm} )^2$) % while 
and on the higher mass limit by the weak force.  %determines %the cross 
%section, which is equal to  
%($ ( A (A-Z)\, 10^{-6} \, {\rm fm} )^2 $). 
%

Our rough estimations predict that, if the DAMA experiments~\cite{sn2snmb:rita0708} 
observes the events due to our (any) 
heavy family members, %(or any heavy enough family cluster with 
%small enough cross section),  
the CDMS experiments~\cite{sn2snmb:cdms} will observe  
a few events as well in the near future. 

 The fact that the fifth family baryons might form the dark matter does not contradict  
 the measured (first family) baryon number and its ratio to the photon 
 energy density as long as the fifth family members are 
 heavy enough ($>$ few TeV). Then they form neutral clusters far before 
 the colour phase transition at around $1$ GeV, while the coloured fifth family clusters either 
 annihilate or contribute (if it is fifth quark-antiquark asymmetry) to the dark matter. 
 Also the stable fifth family neutrino % (with masses higher than a few TeV) 
 does not in this case contradict observations, either electroweak or cosmological or the direct 
 measurements. 

 Our  fifth family baryons are not the objects---WIMPS---which 
  would interact with only the weak interaction, 
 since their decoupling from the rest of the 
 plasma in the expanding universe is determined by the colour force and  
 their interaction with the ordinary matter is determined with the  
 fifth family "nuclear force" (the force among the fifth family nucleons, 
 manifesting much smaller cross section than does the ordinary 
 "nuclear force") as long as their mass is not higher than $10^{4} $ TeV, when the weak interaction starts to 
 dominate and they interact in the today dark matter among themselves with the weak force.

Let me conclude this talk saying:   
 If the approach unifying spins and charges is the right way beyond the 
 standard model of the electroweak and colour interactions,  
 then more than three 
 families of quarks and leptons do exist. The fourth family will sooner or latter be measured, 
 while  we already see through the gravitational force  the stable 
 (with respect to the age of the universe) fifth family, whose neutrons and neutrinos form the dark matter. 

\section*{ Acknowledgments}

The author acknowledge very useful collaboration of the participants 
of the annual  workshops entitled "What comes beyond the Standard models", 
starting in 1998, and in particular of H. B. Nielsen. 

%snmb:mdnbled06 snmb:n92,snmb:pn06,snmb:gdmn07%snmb:gmdn07,snmb:mdnBled06,snmb:okun,n07bled,hn02hn03
%snmb:gmdn07,snmb:mdnBled06,snmb:okun

%% R. Mirman, 08.10.2009
%%\documentclass[11pt]{article}
%%\begin{document}
\author{R. Mirman\thanks{sssbbg@gmail.com}}
\title{Confusions That are so Unnecessary}
\institute{%
14U\\
155 E 34 Street\\
New York, NY  10016
}

\titlerunning{Confusions That are so Unnecessary}
\authorrunning{R. Mirman}
\maketitle

\begin{abstract}
There are too many aspects of science, particularly quantum mechanics, that should be obvious but are quite unclear to too many people (especially physicists and journalists who seem to enjoy flaunting their confusions). We summarize and analyze these here; detailed discussion and proofs are well known. 
\end{abstract}

%%\section{}\label{rm1con-sgv}

There has been much misunderstanding, usually quite unnecessary, about physics, especially quantum mechanics. Here we consider some of the misunderstandings trying to see why they arise and to clarify what physics, mathematics and logic actually require. This we do by raising questions that have puzzled, often unconsciously, too many people, providing answers and explanations. Many of these considerations, but far from all, are discussed in greater depth, often with proofs, elsewhere~(\cite{rm1con-nmb}; \cite{rm1con-nm2}; \cite{rm1con-nm3}; \cite{rm1con-nm4}; \cite{rm1con-ia}; \cite{rm1con-gf}; \cite{rm1con-ml}; \cite{rm1con-pt}; \cite{rm1con-qm}; \cite{rm1con-cnfr}; \cite{rm1con-bna}; \cite{rm1con-bnb}; \cite{rm1con-bnc}; \cite{rm1con-op}; \cite{rm1con-cg}; \cite{rm1con-imp}; \cite{rm1con-msi}; \cite{rm1con-obn}; \cite{rm1con-rn}).

1. Why is there so much difficulty interpreting quantum mechanics?
A. Because physicists try to interpret it as if it were classical physics with classical objects. But of course it cannot be. Thus there is no wave-particle duality since there are no waves and no particles. These are classical concepts which do not apply. If physicists assumed that electrons were people they would also have immense difficulty with interpretation. That is just what physicists are doing with quantum mechanics. It is like saying that sometimes an electron is hungry, sometimes sleepy. Quite unlikely. 

2. Are electrons, protons, and so on point particles?
A. Of course not. Where in the formalism is there even the slightest hint of particles, let alone point particles? If anyone disagrees they can show where in the formalism these appear. What objects are is considered elsewhere. 

3. If there is nothing in the formalism to indicate objects are particles, let alone point particles, why do physicists keep thinking of them as such?
A. Because in kindergarten science was explained in pictures and the electron was drawn as a point. Physicists seem to have learned nothing since kindergarten, and can't believe that what they learned as 5-year olds is misleading. That is a reason the mathematically impossible string theory is so popular. It ``solves'' the nonexistent problems caused by physicists inability to forget what they learned in kindergarten and learn the correct facts about nature.

4. Why are physicists trying so hard to ``quantize'' gravity even though it is so clear that is the quantum theory of gravity, the only possible one?      
A. Because when they first started studying quantum mechanics they ``quantized'' some specific cases and that is all they know how to do. Also general relativity was discovered before quantum mechanics. If it were to have been discovered fifteen years later it might have been realized what it really is: the quantum theory of gravity. Physicists believe that what determines the nature of a theory is when it was discovered. Also they do not understand quantum mechanics. They think it has something to do some uncertainty, without knowing where that comes from or what that means. And they believe that something must fluctuate, but do not know what. (It is of course results of repeated experiments, not space). Physicists are having much difficulty ``quantizing'' gravity. They are discovering that it is very difficult to solve a nonexistent problem. But being physicists they will keep trying.

5. What is quantitization?
A. When quantum mechanics was first developed the correct formalism was found by taking the classical formalism and guessing the quantum formalism from it (usually by substituting operators for variables. But such guessing is now unnecessary since quantum mechanics is understood. However since it is traditional physicists still love to do it, often producing nonsense.

6. What does quantum mechanics require to vary? 
A. It is of course results of repeated experiments. Consider a box of decaying nuclei. The number decaying in each unit time, fluctuates, that is varies from one instant to the next (slightly). But the nuclei don't fluctuate.

6. Why do physicists believe so strongly in the Higgs boson?
A. Physicists really, really like gauge transformations. And they feel that if they like these so much they must be universal. Of course these are a trivial property of massless objects, and are possible for these only~\cite{rm1con-imp}. But since they are so enthusiastic about them they feel they must hold for all systems, even though they obviously do not and the other objects are not massless. Physicists also like to generalize from one example, here masslessness. And physicists believe that if their theories disagree with nature, then nature must be changed (as shown also by their attempts to change the dimension to agree with their theories). Thus all objects must be massless which they are not. However since their theories must hold they change nature by introducing the Higgs boson to make these massless objects massive, thereby saving gauge transformations, which it does not do. But it does cover up the fact that these transformations are not universal, allowing them to continue their love affair with them. Isn't that the whole reason for working in ``physics''?    

7. Are there problems in physics that string theory is supposed to solve?
A. In perturbation theory in some intermediate steps there occur integrals with lower limits of 0. That makes it looks like they diverge so there are infinities that cause problems (but only for physicists who get very upset by these, not for the theory). The calculational method provides algorithms for calculating quantities and when these are carried to completion the results are finite and agree with experiment to a large number of significant figures (in quantum electrodynamics). There are  no infinities or problems (to be solved). The lesson is not that nature has to revised but that it makes no sense to stop in the middle of an algorithm. Also this ``problem'' occurs for a particular approximation scheme; if other schemes were used it would not appear. Physicists believe, as we see again and again, that what determines the laws of nature is their choice of approximation method. String theory is an unmotivated, mathematically impossible theory, violently disagreeing with experiment, that is absolutely certain to be wrong, that cannot possibly have anything to do with reality. Perhaps that is why physicists are so enthusiastic about it.  

8. Does quantum mechanics show that that there are particles popping in and out of the vacuum (!), that the vacuum is full of energy? 
A. The fundamental equations of quantum mechanics say that the vacuum is empty. There are diagrams in a particular approximation scheme, which again physicists believe determine the laws of nature, given names (which always confuse physicists) like ``vacuum expectation values''. If these had been given different names, like class A diagrams, or a different approximation scheme were used, then this nonsense about the vacuum would never have arisen. Despite physicists' strong opinion to the contrary, their choice of approximation scheme, or names, does not determine the laws of nature. 

9. Do scientists really believe that particles pop out of the vacuum to change the solutions of equations, or that the vacuum  has energy? 
A. No. People who say such things are not scientists but crackpots. These violate the fundamental laws and equations of quantum mechanics. In a particular approximation scheme (and physicists strongly believe, as we see again, that their choice of approximation scheme determines the laws of nature) there are diagrams that have been given names like vacuum expectation value. Physicists of course are very confused by names. If a different approximation scheme were used, or these diagrams given different names, all this nonsense about the vacuum would never have occurred. 

10. Do particles pop out of the vacuum to change the solutions of equations? 
A. The electron statefunction does not obey the free-particle Dirac equation, but one with interactions, obviously otherwise we would never know of it. Different equations have different solution, something that physicists never realized. The actual equation cannot be solved so the solution must be approximated. To keep the bookkeeping straight there are diagrams (pictures) including ones that have been given names like vacuum expectation values. These picture, but only that, particles doing things like ``poping out of the vacuum''. But these (pictures) do not cause the solutions to be different. That is the result of the equations being different. 

11. If the potential, rather than the electromagnetic field, is the physical object, how could that be since it is not gauge invariant?
A. The electromagnetic filed is not a physical object, not gauge invariant and not measurable, the potential is. The potential is not gauge invariant but the system, the potential plus the charged object, is. If we consider an object in a gravitational field then space seems not translationally invariant. An object moves to a particular point when dropped. But to study invariance we must consider the entire system, here the object plus the earth. That --- ignoring other objects in the universe --- is invariant. It is the same with gauge invariance. The system, the field plus the charges, is gauge invariant. Each part is not by itself. 

12. Doesn't Maxwell's equations have a hole that the magnetic monopole is needed to fill?
A. Maxwell's equations are irrelevant. They are classical and nature is quantum mechanical (as we should know by now). There is no hole.The magnetic monopole is not possible because it cannot be coupled to charges so cannot be observed so does not exist.

13. Is there a cosmological constant? 
A. No because with it in Einstein's equation the two parts of the equation do not transform the same so making it inconsistent, among other problems like an object reacting to a gravitational field an infinitely long time before it is emitted.

14. Why do physicists believe that the proton decays even though it has been proved that it cannot?
A. They think that they are putting the proton in a multiplet with other objects that do decay so that it also does. However using the symbol p for a proton does make it one. It has to have the properties of the proton including the strong interactions. The letter p does not. If they wanted to do that they could have put George Bush in the same multiplet and then he would have decayed. Unfortunately that does not happen either, at least not in that sense.

15. Do physicists really believe that 1 had a different value early in the history of the universe?
A. There are ``theories'' in which the (unfortunately named) speed of light was greater then. However this speed is not a property of a physical object, but rather of geometry. If the units of space and time were taking the same, as is often done, this value would be 1. A space with dimension 3+1 is divided into parts. One in which distances and masses are real (in which we live), another in which they are imaginary, in which we definitely cannot live. Thus there must be a boundary, cones, forward and back, in which they are both 0. Light and gravitation being massless both travel on these cones, the boundary cones, so regrettably called the light cones. If light had a greater speed it would be outside the cone, so with imaginary mass. Such objects are indeed imaginary. The numeral value of this speed has no physical significance, but is purely a conversion factor between two units (like feet and meters). It is very common in physics to take these units, of space and time, the same so this speed is then 1. To say that this speed is different then means that the value of 1 is different at different times. Quite unlikely, but still quite popular.

16. Is a pilot wave theory, in which a deterministic wave tells the particle wave where to go, possible? 
A. No because a wave and particle cannot interact so the wave cannot tell the particle anything. That is why classical physics is not possible and quantum mechanics necessary.

17. Why do physicists believe so strongly in the ``standard model'' claiming that it explains everything, everything (except of course the experimental results)?
A. This model consists of two parts, that relating to the weak interaction, which works (at least) fairly well, and quantum chromodynamics, which is too complicated to give experimental results. However because one part works physicists take that as success of both parts. It is also part of the standard model that the US government has three parts. But that is known to be true. That proves that quantum chromodynamics must be correct. 

18. Does quantum mechanics require action-at-a-distance, so if the spin of one of a pair of particles is measured that determines the spin direction of the other?

19. No, quantum mechanics forbids it. That violates an uncertainty principle (number-phase). What that argument shows is the spooky action-at-a-distance of classical physics. 

20. Does anti-matter result from field theory?
A. No, it occurs because quantum wavefunctions are complex. 

21. Can there be alternate theories of gravity? 
A. It would be very difficult, at best. General relativity is a property of the basic foundations of geometry~\cite{rm1con-ml}. If it were wrong there would have to be very major revisions in our views of nature. It is difficult to see how that can be, and is very unlikely. 

22. Are the ``symmetries'' of theoretical physics the result of symmetries?
A. Not necessarily. Some cases are spectrum-generating algebras but in other cases are properties of geometry, with no need of invariance. For example no matter how badly rotational symmetry is broken, comparing the expansion of a statefunction in spherical harmonics in two different systems we find that states of one angular momentum representation go into states of the same representation with no mixing of representations. This is a property of geometry; that coordinates are real. Also because of geometry, not invariance, there can be no particle with spin ${1 \over 3}$ or $\pi$.

\section*{Acknowledgements}

This discussion could not have existed without Norma Manko\v c Bor\v stnik.

\cleardoublepage
\chapter*{\Huge DISCUSSION SECTION}
\addcontentsline{toc}{chapter}{Discussion Section}
\vspace{1 cm}
\textit{Authors:} The participants of the workshop, either actually
present at Bled, besides in 
an enjoyable working atmosphere also in long walks in a beautiful country and in 
mountaineering, or virtually on talks and discussions through the video conferences.

\vspace{1cm}
In the discussion sections we present  discussions on those topics, 
which  have hopefully a chance to 
end up as  articles up to the next thirteenth Bled workshop  or will be continued 
to be discussed again in the next year  Bled workshop, as well as the discussions 
which have taken place through the video conferences, organized by 
the Virtual Institute for 
Astrophysics (www.cosmovia.org). These discussions can be followed, 
together with talks, also on
 
\vspace{0.5 cm}
http://viavca.in2p3.fr/bled\_09.html
\vspace{0.5 cm}

\noindent We namely were learning for the second time (we started with 
video conferences 
 on the eleventh Bled workshop "What comes beyond the standard models") 
how to enable participants, present only virtually  at least in a part of our workshop, 
to discuss with comments and questions and to present talks. 

\newpage

\cleardoublepage

%% Introduction to the discussion sessions
%% M. Khlopov
\title{A Short Overview of Videoconferences at Bled}
\author{by the Discussion Participants}
\institute{%
(http://viavca.in2p3.fr/bled\_09.html)}

\authorrunning{by the Discussion Participants}
\titlerunning{A Short Overview of Videoconferences at Bled}
\maketitle

\begin{abstract} 
A short report on talks and discussions taken place at Bled 
through the video conferences, organized by the Virtual Institute for 
Astrophysics (www.cosmovia.org) is presented. Talks and discussions can be found 
on http://viavca.in2p3.fr/bled\_09.html. The list of open questions proposed for wide 
discussions with the use of VIA facility is added. 
\end{abstract}

\section{The list of open questions, 
proposed for wide discussions with the use of VIA facility}

VIA discussion sessions have developed the earlier experience of such discussion 
at XI Bled Workshop, at which the puzzles of dark matter searches were discussed 
\cite{dsOBregar:2008zz}. These sessions took place during the second working week 
of XII Bled Workshop from $19^{th}$  to $24^{th}$  of July 2009 and lasted each day from one to several hours.

In the course of Bled Workshop meeting The list of open questions 
proposed for wide discussions with the use of VIA facility. 

\begin{enumerate}

\item [1.] Where do families of quarks and leptons come from? 
(The Standard model of the electroweak and colour interaction postulates the 
existence of families and so do in one or another way almost all the proposals 
up to now. Answering this question is one of the most promising way beyond 
the Standard model. The Approach unifying spin and charges, 
do offer the answer to this open question predicting the number of families 
and soon also the Yukawa couplings.)

\item [2.] Where do the Yukawa couplings come from? (In the Standard model the Yukawa 
couplings are just put by hand. Can we answer this question?)

\item [3.] What does determine the strength of the Yukawa couplings and 
accordingly the weak scale? (In the Standard model the scale is put by hand. Can we say more?)

\item [4.] Why do only the left handed spinors carry the weak charge, while the 
right handed are weak chargeless? (This assures the mass protection mechanism in the 
Standard model until the Higgs - by ''dressing'' the right handed fermions with the weak charge - 
destroys this protection.)

\item [5.] How many families appear at (soon) observable energies? 
What are the properties of the heavy families, if they are stable?

\item [6.] Are among the members of the families the candidates for the dark matter 
clusters? What are properties of such clusters?

\item [7.] Where do charges come from?

\item [8.] What makes the supersymmetry appearing at 
observable energy scale?

\item [9.] What are physical grounds for inflation, baryosynthesis, 
dark matter and dark energy?

\item [10.] Looking at the list of observed elementary fields – fermions and bosons – 
we can conclude that all the observed elementary particles which are fermions have 
charges in the fundamental representation while the observed bosons (all of them are 
the gauge fields) have charges in the adjoint representation. Standard model assumes 
Higgs particles, which are scalars and have the weak charge in the fundamental 
representation of the weak group. Assuming that the supersymmetry does show up at 
the measurable low energy scale, we shall be able to see bosons in the fundamental 
representation with respect to the charge groups and fermions in the adjoint 
representation with respect to the charge groups. But if Higgs is the elementary 
field one would say that we already have one supersymmetric partner-namely the 
Higgs field. Its ‘’ordinary’’ partner then lies higher in the mass scale. Why MSSM 
does not admit Higgs already as a possible supersymmetric particle?
\end{enumerate}

The list of these questions was put on the VIA site and all the participants of 
VIA sessions were invited to address them during VIA discussions.

\subsection{VIA  talks and discussions}

The sessions started on $19^{th}$ of July with the Introduction into the via conference of the 
twelfth Bled workshop What comes beyond the standard models
by N.S. Manko\v c Bor\v stnik and M. Khlopov.

Next day, on $20^{th}$ of July N.S. Manko\v c Bor\v stnik~\cite{dsOnormatalk,dsOpn06,dsOgmdn07},  
presented her "Approach unifying spins and charges" and the answers to the 
above open questions, which her approach is offering. The Approach predicts, 
having the mechanism for generating families, more than the so far measured three families.  
The fourth family could possibly be seen at the LHC, while a  stable family is 
a severe candidate to form the dark matter.

 On $21^{st}$ of July C. Balazs took part from Australia in discussion of some questions of 
 SUSY physics. 
 
 In the framework of the program of Bled Workshop John Ellis, from  CERN, gave on $22^{nd}$ 
 July his talk "Beyond the Standard Model and the LHC" and took part in the 
 successive discussion. VIA sessions were finished on $23^{rd}$
 July by the discussion 
 of puzzles of dark matter searches. N.S. Manko\v c Bor\v stnik presented 
 possible dark matter candidates that follow from the Approach unifying spin and charges", 
 and M. Khlopov presented composite dark matter scenario, mentioning that it 
 can offer the solution for the puzzles of direct dark matter searches as well as 
 that it can find physical basis in the Norma's approach. Their arguments are 
 presented in these proceedings~\cite{dsOgnbled}.

VIA sessions provided participation at distance in Bled discussions for John Ellis 
and A.S.Sakharov (CERN, Switzerland), K.Belotsky  A.Mayorov and E. Soldatov 
(MEPhI, Moscow), J.-R. Cudell (Liege, Belgium), R.Weiner (Marburg, Germany) and 
many others. For C. Balasz, who attended Bled Workshop the first week, VIA 
videoconferencing gave the opportunity to continue discussions during the 
second week, when he returned to Australia.

\begin{figure}
    \begin{center}
        \includegraphics[scale=0.35]{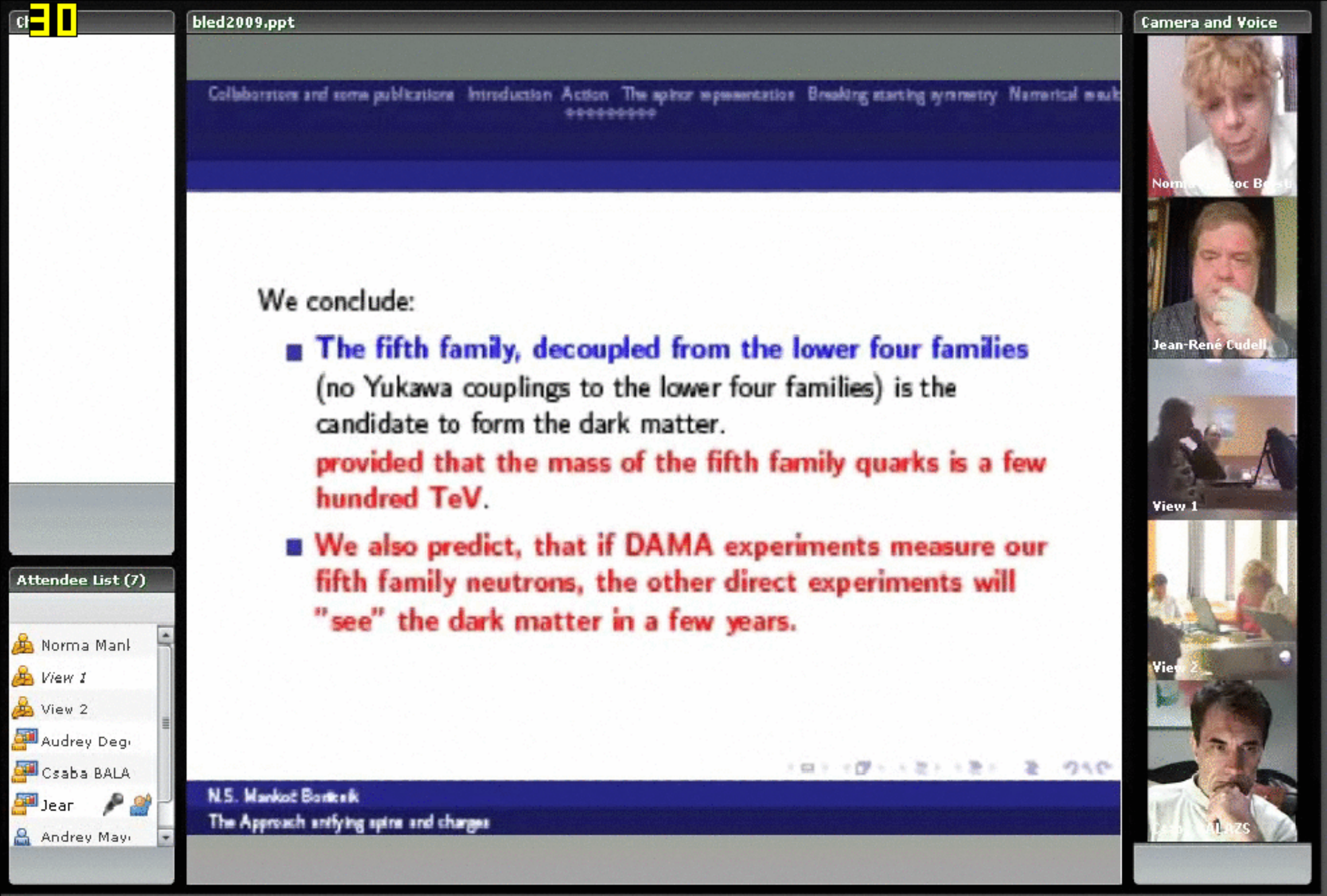}
        \caption{Bled Conference Discussion Bled-Moscow-CERN-Australia-Marburg-Liege}
        \label{dsOc}
    \end{center}
\end{figure}

\section{Conclusions}

Staring to learn how one can use very efficiently the via conference facilities for 
discussing at least on very well defined questions the organizers and participants 
enjoyed very much this possibility. One can learn more about the possibilities of the 
video conferences in the explanation of M. Khlopov.

%% Discussion section1, 16.11.2009
%%
\title{Discussion Section On the Witten's No Go Theorem for the Kaluza-Klein-like Theories} 
\author{D. Lukman${}^1$, N. S. Manko\v c Bor\v stnik${}^1$ and H. B. Nielsen${}^2$}
\institute{%
${}^1$ Department of Physics, FMF, University of Ljubljana,\\
 Jadranska 19,Ljubljana, 1000\\
${}^2$ Department of Physics, Niels Bohr Institute,%\\
Blegdamsvej 17,
Copenhagen, DK-2100}

\authorrunning{D. Lukman, N.S. Manko\v c Borstnik and H.B. Nielsen}
\titlerunning{Discussion section On the Witten's No Go Theorem\ldots}
\maketitle

\vspace{0.5 cm}
%\begin{abstract} 
In this proceedings there is the talk of the same three authors, in which 
one step further towards realistic Kaluza-Klein-like theories is presented. 
The idea of Kaluza and Klein that the charges follow from the properties 
(dynamics) in higher dimensions  is such a beautiful idea,  that the authors 
can hardly accept the possibility that it has no  application in nature. 
Particularly one of the authors, with her proposed "approach unifying spin 
and charges",  offering also the mechanism for generating families, which 
is a kind of the Kaluza-Klein-like theory, is trying hard to find the 
way out of the "no-go" theorem of Witten. 
The problem with Kaluza-Klein-like theories  is 
 that any break of symmetries seems to cause, if there are no special protections,  
that a massless fermion of one handedness gain after the break the mass of 
the scale of breaking. 

The authors discuss in several papers~\cite{hnkk06,hn07} and published talks 
possibilities that  breaks of the symmetries  might not always end up with 
massive fermions. The proposed loop hole through the 
Witten's "no-go theorem"  was the appearance of the appropriate boundaries 
for a toy model with $(1+5)$-dimensional space. 

In the work presented in this proceedings (and also in the previous one, except that 
a further step was made) the hope for cases when the manifold in all the higher 
dimensions, except two, is flat, and which the authors  call  "an effective two 
dimensionality" cases is found. Namely, in the case of a spinor in $d=(1+(d-1))$ 
compactified on an (formally) infinite disc with 
the zweibein which makes a disc curved on $S^2$ and with the spin connection 
field  which allows on such a sphere only one massless spinor state of a particular charge, 
the massless spinor does chirally couple to the corresponding Kaluza-Klein gauge field. 
 In $d=2$, namely, the equations of motion 
following from the action with the linear curvature leave  spin connection and zweibein 
undetermined

%The discussions were very intense and also fruitful.

%% discussion section II: 
\title{Discussions Section On the Fifth Family Proposed by the "Approach 
Unifying Spin and Charges" and the Dark Matter Content}
\author{G. Bregar and N.S. Manko\v c Bor\v stnik}
\institute{%
Department of Physics, FMF, University of Ljubljana,\\
 Jadranska 19,Ljubljana, 1000}

\authorrunning{G. Bregar and N.S. Manko\v c Bor\v stnik}
\titlerunning{Discussions Section On the Fifth Family\ldots}
\maketitle

\begin{abstract} 
The "approach unifying spin and charges", proposed by Norma Susana Manko\v c 
Bor\v stnik~\cite{gn2n09,gn2pn06,gn2gmdn08} , 
predicts four families, which are connected with the (non zero) Yukawa couplings. The masses 
of the fourth family quarks lie above a few $100 $ GeV/$c^2$, the masses of the fourth family 
leptons are at around $100$ GeV/$c^2$ or above. 
The masses of quarks might be low enough to be possibly measured at the LHC~\cite{gn2gmdn08} .  
The approach predicts 
also the stable fifth family (with no Yukawa couplings to the lower four families), which 
is the candidate  to form the main part of the dark matter. 
The work done by Gregor Bregar and Norma Susana Manko\v c Bor\v stnik~\cite{gn2gn}  
assumes  that the {\it neutron 
is  the lightest fifth family baryon and the neutrino the lightest fifth family lepton}.
Following the evolution of the fifth family members in the expanding universe, and analysing 
carefully the interaction of the fifth family neutrons and neutrinos with the ordinary matter 
in the direct measurements of the DAMA and the CDMS experiments and in  other published 
measurements which could concern our fifth family members as the dark matter constituents and 
accordingly their properties,  the authors of the paper~Phys. Rev. D 80, 083534 (2009)
predict that the fifth family quarks with the masses of a few $100 $ TeV/$c^2$ and the 
fifth family neutrinos with the mass of a few  TeV/$c^2$ are the  candidates for 
forming the dark matter. This is true also for not too large interval of matter-antimatter 
asymmetry of the fifth family baryons (which could contradict the measured dark matterdensity) 
Possible weak points pf the evaluations in the work~\cite{gn2gn} are discussed bellow by Gregor and Norma. 
 \end{abstract}

\section{What speaks for the conclusion that the fifth family members with the 
quark masses of a few hundred  TeV/$c^2$ and the neutrino mass of a few  TeV/$c^2$ 
are the candidates to form the dark matter, and what might speak against it? What 
speaks for the antibaryon $\bar{u}_5 \bar{u}_5 \bar{u}_5 $ to be the stable particle?}
\label{gn2gn}
\subsection{A short review of the "approach unifying spin and charges" from the point of view of the 
dark matter candidates.}
\label{gn2review}

Let us first point out those details  of the "approach unifying spin and charges",  which seem to be 
connected with possible answers  to the question put in the title of this section~\ref{gn2gn}:\\
What speaks for the conclusion that the fifth family members with the 
quark masses of a few hundred  TeV/$c^2$ and the neutrino mass of a few  TeV/$c^2$ 
are the candidates to form the dark matter? 

The reader can
find more about the "approach" in the talk of Norma and the references therein, as well as in the 
other two talks, whose coauthor is Norma and in~\cite{gn2pn06}.

First let us point out that the "approach" is offering the  mechanism for 
the appearance of families by introducing the  second kind  
of the Clifford algebra objects,  which generates families as the  
 equivalent representations to the Dirac spinor representation. Accordingly the number of families is 
 determined by the "approach" and there is no freedom to make a choice of the number of families, let say,  
 by the choice of an appropriate group, which would allow a chosen  number of stable or unstable families. 
 Let us say that this is not the case for other models where the number of families are put in 
 by hand, at least by a choice of an appropriate group.
 
The "approach" predicts from the simple starting action 
in the energy region bellow the unification scale 
of the three observed charges two times four families. The upper four families 
are decoupled in the Yukawa couplings from the lower four families.

Due to the "approach"  particular  spontaneous break of the starting symmetries of 
the spinor and the gauge fields (vielbeins and the two kinds of the spin connection fields), 
leads to massive upper four families and massless lower four families. The lower four 
families have all the properties assumed by the  ''standard model of the electroweak 
and colour interactions" before the electroweak break. The electroweak break influences the properties 
of  the lower and upper four families.  The quarks of the fourth of the lower four 
families are predicted (the references are 
in the talk of Norma) to have 
masses at around $250$ GeV/$c^2$ or above and the lepton masses are predicted to be at around $100$ GeV/$c^2$ or 
above. 
The fifth family, with no  Yukawa couplings to the lower four families, is  accordingly stable 
and therefore the candidate for forming  the dark matter constituents.

The accurate prediction of the fifth family masses is at this stage of the development of the "approach"  not yet 
possible. Too many problems have to be solved first, like: \\
i. We must treat in a trustful way the nonperturbative breaking of a starting 
symmetry, explaining how does the break occur and why. \\
ii. We must derive the Yukawa couplings beyond the tree level and show that this calculations   explain 
drastic differences in the properties of u-quark, d-quark, neutrino and electron.  \\
iii. We must  understand  all the discrete symmetries following from the "approach". \\
iv. We must study possible phase transitions connected with the groups, which  symmetries break.\\
v. And several others.

Following the evolution of the fifth family members in the expanding universe up to present dark matter 
density for different choices of the fifth family masses and evaluating 
the properties of the fifth family members when scattering on the ordinary 
(mainly formed of the first family members) matter can help to better understand the problems 
presented above and to easier find the way of solving them. 

We start to follow the number density of our fifth family members when the temperature was 
high enough that the fifth family quarks and antiquarks, leptons and antileptons were in thermal 
equilibrium with the plasma to which all the massless gauge and scalar 
fields and  all the massless families contribute. The expansion causes that the plasma cools down 
and makes less and less possible the generation of massive quarks and antiquarks out of the plasma. 
Massive fifth family members start to decouple since they have less and less occasion to meet their 
%corresponding 
antiparticles. Scattering of the plasma constituents on  clusters of the fifth family quarks 
destroys the clusters unless the temperature falls appreciably 
bellow the binding energy of the clusters. Then the clusters start to form and decouple out of plasma.

If the  fifth family quark masses are of the order $100 $ TeV/$c^2$, is the binding energy of the 
order of $1$ TeV/$c^2$ and our calculations show that the number density of the colourless 
fifth family neutrons is, when the temperature lowers to the colour phase transition temperature 
(to $1$ GeV/$k_b$), of 
the same order of magnitude as the number density of the fifth family quarks and antiquarks. 
The colour phase transition causes huge enlargement of the scattering cross sections of all the 
 quarks and antiquarks and of  coloured objects and dresses the quarks with $\approx 300$ MeV/$c^2$, 
 while the colourless neutrons are too strongly 
 bound to feel the phase transition at all. 
Due to huge enlargement of the scattering cross sections the fifth family 
quarks and antiquarks either annihilate or form colourless objects and deplete out of the rest of plasma 
long before the temperature of the 
plasma falls bellow $1$ MeV/$k_b$ when the first family quarks can start to form  bound states either 
among themselves or with the fifth family members.

%%%%%%%%%%%%%%%%%

If the  fifth family quark masses are of the order $300 $ MeV/$c^2$, as Maxim is assuming,
then their binding energy is of the 
order of  a few  MeV/$c^2$ and the number density of the colourless baryons is at the colour 
phase transition (T=$1$ GeV/$k_b$) negligible. 
The colour phase transition,  enlarging very much the scattering cross section, causes  
annihilation  of a large amount of the fifth family quarks and antiquarks, the formation of the 
colourless objects and, if the 
fifth family antibaryon-baryon asymmetry is assumed, as Maxim and his group does, also the 
colourless object (for a particular choice of the baryon-antibaryon asymmetry) of $\bar{u}_5
\bar{u}_5 \bar{u}_5$. 
Those that succeed to survive as an  coloured object at $1$ MeV/$k_b$ 
start to form the colourless  objects with the first family 
quarks. Maxim  and his group claim that there are $\bar{u}_5
\bar{u}_5 \bar{u}_5$ that mostly survive forming with He nuclei the electric chargeless objects.
%%%%%%%%%%%%%%%%%%%%%

For masses of the fifth family quarks and antiquarks above few hundred TeV/$c^2$ the fifth family 
baryon-antibaryon asymmetry makes no difference, as long as the approximations we  made  
when evaluating properties 
of the fifth family members in the evolution of the universe are meaningful. 

%%%%%%%%%%
For masses close to one TeV/$c^2$ or bellow the baryon-antibaryon asymmetry starts to be  essential  
(as it is for the first family members).
%%%%%%%%%%%%5

In the next subsection we discuss the evaluations we made to estimate properties of the fifth family 
members by studying their behaviour in the evolution of the universe and when scattering on the 
ordinary matter. We shall point out those approximations, which need to be treated more 
accurately, although for most of  these points more accurate treatment appears as a very 
demanding project. 

Let us point out that we started to follow the behaviour of the fifth family members for 
the masses far above $1$ TeV/$c^2$ after 
all the breaks except the electroweak break took place. %We should look at  the history 
%before the last break  to be sure that the starting point of ours is the  acceptable one. 

%
\subsection{The stable fifth family members in the expanding universe if the fifth family neutron
and the fifth family neutrino is the lightest baryon and lepton, respectively.}
\label{gn2GNSC}

If the mass of the stable fifth family neutrino is a few TeV/$c^2$ or above and 
of the stable fifth family neutron a few hundred  TeV/$c^2$ or above,  these neutrons and neutrinos 
are the candidates to be the constituents of the dark matter, fulfilling 
all the requirements for the dark matter, from either   
the cosmological observations or direct measurements of any kind. However, in the paper of Gregor and 
Norma~\cite{gn2gn}, we were not yet able to determine the relative contribution of these two components 
of the dark matter.

Let us point out the approximations we have done when treating the fifth family members 
as the candidates to form the dark matter:

\begin{itemize}

\item We assumed that in the interval region of temperatures from $\frac{m_{q_5} c^2}{k_b}$ to 
a GeV/$k_b$ (which is the temperature of the $SU(3)$ phase transition), in which we calculated the 
number density of the fifth family quarks and antiquarks, 
and the fifth family neutrons, the one gluon exchange is the dominant contribution to the 
interaction among quarks.  This assumption is the meaningful one.

\item We evaluated in the Bohr like model the binding energy and the potential 
among the fifth family baryons with the assumption that the fifth family quarks interact 
dominantly with the one gluon exchange. Also this assumption does not seem questionable.

\item  When solving the coupled Boltzmann equations for the number density of quarks and 
the number density of coloured and colourless clusters, we needed the scattering amplitudes, presented 
in Eq.(2) of the paper~\cite{gn2gn}. 
The expressions for the scattering cross sections of Eq.(2) are very approximate, and 
we corrected their accuracy with the parameters $\eta_{c_5}$,  which takes into account 
that the clusters of two quarks bind into the clusters of three quarks,  and  $\eta_{(q\bar{q})_{b}}$, 
which takes into account the roughness of the estimation. These two cross sections should be 
calculated  more precisely, 
so that we could limit the interval, within which both $\eta$'s lie. These calculations might influence 
considerably the conclusions.

\item The colourless fifth family baryons, tied strongly into very small clusters, do not feel 
colour phase transition when it occurs at around $1$ GeV, while the coloured quarks and antiquarks 
or the coloured clusters of two quarks or antiquarks do. At the colour phase transition all the quarks 
of any family start to enlarge very much  the scattering cross section. But while 
the fifth family quarks with masses  of 
several hundred TeV/$c^2$ and accordingly of the binding energy into the corresponding clusters of several 
TeV  start to bind at $1$ GeV, the first family quarks can not until  the temperature 
falls below the binding energy of  "dressed" first family quarks, that is  bellow a few  MeV. 
We evaluated that the fifth family quarks and antiquarks either annihilate or form the 
colourless objects and deplete soon after $1$ GeV, so that there is a negligible amount available below 
a few MeV to form clusters with the first family members. 
It is a hard project to treat the colour phase transition, although we should do this.

\item The references treating  neutrinos with masses above few TeV as candidates for the dark matter 
constituents~\cite{gn2dolgov,gn2maxim} report that the large scattering amplitudes for such neutrinos 
cause strong annihilation of neutrinos lowering accordingly their possible contribution to the dark matter. 
For the mass region of the neutrino  from $10$ TeV/$c^2$ to $100$ TeV/$c^2$ the scattering 
amplitudes were only roughly estimated so far. The estimation in the mass interval of a few TeV/$c^2$ 
up to $100$TeV/$c^2$ seems accordingly  not to contradict 
the measured dark matter density, 
which means that the fifth family neutrinos with masses in this region do not contribute more 
than our fifth family neutrons. 
 In the case of the fifth family quark masses  of a few $100$ TeV/$c^2$ and the fifth family 
 neutrino mass of a few TeV/$c^2$ it seems reasonable to conclude that 
 the dark matter  consists either mostly 
 of the fifth family neutrons, or of the fifth family neutrinos or of both. 
 However, more in-depth studies are needed 
 to make final conclusions. 
 
 \item The evaluations of the interaction of the fifth family neutrons with the 
 ordinary matter, although very approximate, brought the conclusion that the fifth family 
 baryons are the acceptable dark matter constituents. Also these estimations are quite rough.  
  Taking into account the uncertainties in knowing the local properties of the dark matter,  we 
  can conclude 
 that the  fifth family members are the right candidates to form the dark matter, provided that 
 the neutron is the lightest baryon and the neutrino the lightest lepton.

\end{itemize}

Can it be that not the neutron but, let say, proton or some other fifth family baryon is the stable 
fifth family baryon? Would this change 
the conclusion that the fifth family  is an acceptable candidate to form 
the dark matter?  $n_5$ is the lightest baryon when the masses 
of $u_5$ and $d_5$ are in relative separation of the order of magnitude $10^{-4}$. The  
 possibilities that $u_5 u_5 u_5$ or $d_5 d_5 d_5$ or $u_5 u_5 d_5$ are
the lightest baryons are under considerations now. 

Let us point out that our evaluations with the stable $n_5$ and $\nu_5$ predict that the CDMS or 
other experiments will measure the
dark matter signals which will not contradict the DAMA results.

 Can even  very light fifth family members with masses of quarks of a few hundred MeV$/c^2$, 
 as assumed by Maxim and his group, be a possible solution, if one assumes in addition 
 a very particular fifth family antibaryon-baryon asymmetry? Maxim claims that it does. 
 Norma has severe doubts that such assumptions can be fulfilled at all, not only within the 
 "approach unifying spins and charges" but within any model, which "wants to be 
 elegant".
But to say anything about the assumptions of the Maxim's  group in the context of the "approach 
unifying spins and charges"  one should study first the  discrete symmetries, as well as the 
matter-antimatter asymmetry within this approach.

%% Maxim's part of Discussion II
%%
\title{Discussions section On the Fifth Family Proposed by the "Approach
Unifying Spin and Charges" and the Dark Matter Content}
\author{M.Yu. Khlopov}
\institute{%
Centre for Cosmoparticle Physics  Cosmion, 115409 Moscow, Russia;\\ 
Moscow Engineering Physics Institute (National Nuclear Research University), 115409 Moscow, Russia \\
APC laboratory 10, rue Alice Domon et L\'eonie Duquet \\75205
Paris Cedex 13, France}

\authorrunning{M.Yu. Khlopov}
\titlerunning{Discussions Section On the Fifth Family\ldots}
\maketitle

\begin{abstract}
The "approach unifying spin and charges", proposed by Norma Susana Manko\v c
Bor\v stnik~\cite{mkd2n09,mkd2pn06,mkd2gmdn08,mkd2gn} predicts
the stable fifth family (with no Yukawa couplings to the lower four families). The conclusion on stability
of this family is strongly motivated in this approach and the extensive study of possible candidates for 
the dark matter is challenging.

In view of the uncertainty of fifth family masses all the possible variants for the lightest stable particle 
can be considered following the methods, developed in \cite{mkd2N,mkd2Q,mkd2FK,mkd2FKS}.
 The possibility of stable charged leptons and quarks is generally in serious 
 trouble, related with inevitable presence of stable positively charged species 
 that behave as anomalous isotopes of hydrogen. However there is one exception. 
 It is the solution of composite dark matter, which assumed an excess of -2 charged species, 
 bound in atom like systems with He nuclei that formed in primordial nucleosynthesis. 
 This O-helium (OHe) nuclear interacting form of dark matter was shown to avoid any 
 direct contradiction with experimental constraints \cite{mkd2I,mkd2Q,mkd2FKS,mkd2KK}. 
 It provides Warmer than Cold Dark Matter scenario, can explain the excess 
 of positron annihilation line observed by INTEGRAL and can resolve the puzzles of 
 direct and indirect dark matter searches. It was shown that electroweak $SU(2)_{ew}$ 
 sphaleron transitions in very early Universe can provide relationship between the 
 observed baryon asymmetry and excess of -2 charged species over their antiparticles, 
 if these species have nontrivial $SU(2)_{ew}$ charges. If sphaleron transitions are 
 possible for the fifth family members, predicted by the  "approach unifying spin and charges" 
 of N.S.M.B. and having nontrivial $SU(2)_{ew}$ charges,  and if their masses assure 
 that $\bar{u}_5 \bar{u}_5 \bar{u}_5 $ is the lightest stable fifth family antibaryon, 
 the excess of $\bar{u}_5$ over $u_5$ can be generated in the early Universe and 
 OHe composite dark matter scenario with $\bar{u}_5 \bar{u}_5 \bar{u}_5 $ constituent 
 can be realized. For highly improbable masses of the fifth family quarks at around 
 $300$ GeV/$c^2$, such scenario can reproduce all the features of composite dark matter scenario. 
 For case of quarks with the masses of a few $100 $ TeV/$c^2$ that are assumed more realistic 
 for the "approach"  some of these features still hold true, with the lack of explanation for 
 the excess of positron annihilation line and of anomalies in spectra of cosmic high energy 
 electrons and positrons. These astrophysical data may not, however, require dark matter 
 solution and can be explained by natural astrophysical sources.

The problems of composite dark matter solution for the puzzles of direct dark 
matter searches and of realization of this scenario with the use of stable fifth family are discussed.
\end{abstract}

\section{Composite dark matter with $\bar{u}_5 \bar{u}_5 \bar{u}_5$ constituent.}
\label{mkd2true}

The "approach unifying spin and charges"  predicts that the lightest particles of fifth family are  stable
and are therefore the candidates for the dark matter. The "approach"  assumes that they are 
very heavy and can be hardly produced and studied at accelerators. One has to use analysis 
of cosmological evolution for different mass ratios and make a conclusion on the consistency 
of its results with observations.

In view of unknown mass ratio of the fifth family quarks the possibilities can be considered 
that charged  $u_5 u_5 u_5$ or $d_5 d_5 d_5$ or $u_5 u_5 d_5$ are
the lightest baryons. If they are treated as dark matter candidates in charge symmetric case,
the results of the analysis \cite{mkd2Q} leave practically no room for consistency of such 
possibilities with cosmology.
The only possibility that does not meet immediate troubles is to use composite dark matter 
scenario with excessive $\bar{u}_5 \bar{u}_5 \bar{u}_5 $ as its constituent.

\subsection{Brief review of composite dark matter models}
\label{mkd2Comp}

It was shown in \cite{mkd2I,mkd2Q,mkd2FKS,mkd2KK} that the existence of heavy stable -2 
charged particles, being in excess over their antiparticles and forming atom-like 
neutral O-helium bound state with primordial helium, is compatible with all the 
experimental constraints. In this case composite dark matter scenario of nuclear 
interacting Warmer than Cold Dark Matter. Such scenario can be realized for a wide 
range of -2 charged particles masses, including 100 TeV range. For the masses of 
OHe $ m_o \sim 1 $TeV this new form of dark matter can provide explanation of excess 
of positron annihilation line radiation, observed by INTEGRAL in the galactic bulge. 
Such explanation \cite{mkd2KK} is based on the calculation of the rate of E0 transitions in 
O-helium atoms, excited in collisions in the central part of Galaxy. The rate of such 
collisions decreases as $\propto m_o^{-2}$ and cannot explain INTEGRAL data for masses 
about 100 TeV. The search for stable -2 charge component of cosmic rays is challenging 
for PAMELA and AMS02 experiments. However such fraction decreases inversely proportional 
$m_o$ and can also be out reach of cosmic ray experiments for the mass around 100 TeV. 
Decays of heavy charged constituents of composite dark matter can provide explanation 
for anomalies in spectra of cosmic high energy positrons and electrons, observed by PAMELA, 
FERMI and ATIC. For the "approach unifying spins and charges" this possibility needs special 
study, but seems hardly possible. In the context of the approach \cite{mkd2I,mkd2Q,mkd2FKS,mkd2KK} search for 
heavy stable charged quarks and leptons at LHC acquires the significance of experimental probe 
for components of cosmological composite dark matter. Such search is restricted by masses 
$ m_o \le 1 $TeV and is impossible for 100 TeV quarks of fifth family.

The results of dark matter search in experiments
DAMA/NaI and DAMA/ LIBRA can be explained in the framework of
composite dark matter scenario without contradiction with negative
results of other groups. This scenario can be realized in different
frameworks, in particular, in the extensions of Standard Model, 
based on the approach of almost commutative geometry \cite{mkd2FKS}, 
in the model of stable quarks of 4th generation \cite{mkd2I,mkd2Q} that 
can be naturally embedded in the heterotic superstring phenomenology, 
in the models of stable technileptons and/or techniquarks \cite{mkd2KK}, 
following from Minimal Walking Technicolor model. It might be also possible in
the approach unifying spin and charges.

The proposed explanation of the puzzles of direct dark matter searches 
is based on the mechanism of low energy binding of OHe with nuclei. The 
following picture is assumed: at the distances larger, than its size,
OHe is neutral and it feels only Yukawa exponential tail of nuclear attraction,
due to scalar-isoscalar nuclear potential. It should be noted that scalar-isoscalar
nature of He nucleus excludes its nuclear interaction due to $\pi$ or $\rho$ meson exchange,
so that the main role in its nuclear interaction outside the nucleus plays $\sigma$ meson exchange,
on which nuclear physics data are not very definite. When the distance from the 
surface of nucleus becomes
smaller than the size of OHe, the mutual attraction of nucleus and OHe is changed by 
dipole Coulomb repulsion. Inside the nucleus strong nuclear attraction takes place. 
In the result a specific spherically symmetric potential appears and the solution of 
Schrodinger equation with such potential for the OHe- nucleus suystem can be found.
Within the uncertainty of nuclear physics parameters there exists a range at which OHe
binding energy with sodium and/or iodine is in the interval 2-6 keV. Radiative capture 
of OHe to this bound state leads to the corresponding energy release observed as an 
ionization signal
in DAMA detector.

OHe concentration in the matter of underground detectors is determined by the equilibrium
between the incoming cosmic flux of OHe and diffusion towards the center of Earth. It is 
rapidly adjusted and follows the
change in this flux with the relaxation time of few
minutes. Therefore the rate of radiative capture of OHe should experience annual 
modulations reflected in annual modulations of the ionization signal from these reactions.

%The method to calculate the rate of OHe reactions was developed and
%the calculated total amount of such events is shown to be consistent
%with the results of DAMA/NaI and DAMA/LIBRA experiments for the mass
%of OHe around 1 TeV. This method can be applied to the analysis of
%the whole set of inelastic processes, induced by O-helium in matter.

An inevitable consequence of the proposed explanation is appearance
in the matter of DAMA/NaI or DAMA/LIBRA detector anomalous
superheavy isotopes of sodium and/or iodine,
having the mass roughly by $m_o$ larger, than ordinary isotopes of
these elements. If the atoms of these anomalous isotopes are not
completely ionized, their mobility is determined by atomic cross
sections and becomes about 9 orders of magnitude smaller, than for
O-helium. It provides their conservation in the matter of detector. Therefore mass-spectroscopic
analysis of this matter can provide additional test for the O-helium
nature of DAMA signal. Methods of such analysis should take into account
the fragile nature of OHe-Na bound states, since their binding energy is only few keV.

With the account for high sensitivity of the numerical results to the values of nuclear parameters
and for the approximations, made in the calculations, the presented results \cite{mkd2Bled09} can be considered
only as an illustration of the possibility to explain puzzles of dark matter search in
the framework of composite dark matter scenario. An interesting feature of this explanation 
is a conclusion that the ionization signal expected in detectors
with the content, different from NaI, can be dominantly in the energy range beyond 2-6 keV.
Therefore test of results of DAMA/NaI and DAMA/LIBRA experiments by other experimental 
groups can become a very nontrivial task. In particular, energy release in reaction of 
OHe binding with germanium in CDMS detector is beyond the range 2-6 keV and this 
conclusion becomes stronger with the growth of $m_o$ as show the results of our 
calculations presented in these Proceedings. This feature corresponds to the recent 
analysis of CDMS data \cite{mkd2Kamaev:2009gp}, claiming that ionization energy release 
in the range of DAMA signal (2-6 keV) is excluded with the significance of 6 standard deviations.

To prove to be an explanation for DAMA results, the composite dark matter scenario 
should reproduce the detected signal. A straightforward calculation of the rate of 
radiative capture of nuclei by OHe is now under way. The number of events is determined 
by the product of this rate and the equilibrium concentration of OHe in detector, which 
in turn is adjusted to the incoming flux. The latter is inversely proportional to the 
mass of OHe. Therefore the results of this calculations will provide information on the 
preferable mass of OHe, determined by its -2 charged constituent.

\subsection{Can composite dark matter scenario take place in the approach, unifying spins and charges?}
\label{mkd2Comp1}

In the case of approach, unifying spins and charges, composite dark matter scenario 
\cite{mkd2I,mkd2Q,mkd2FKS,mkd2KK} can be realized completely for masses of $u_5$ about few hundred GeV. 
This realization assumes the necessary excess of $\bar{u}_5$ and can provide 
explanation for DAMA/CDMS controversy and positron excess observed by INTEGRAL. 
Decays of $\bar{u}_5$ can explain excess of high energy electrons observed by 
FERMI and ATIC, but in the absence of subdominant +2 charged component positron 
anomalies can not be explained. This scenario with low mass quarks is, however, 
very implausible in the framework of "approach", since the masses of the stable 
family are assumed to be very close to the
third family masses.

For the case of 100 TeV mass quarks, the possibility to explain INTEGRAL data and 
high energy cosmic electron anomaly is lost. However, these phenomena can find 
explanation with the use of natural astrophysical sources and may not imply effects 
of dark matter. Then only DAMA/CDMS controversy should be explained, and such 
explanation is shown to be possible  \cite{mkd2Bled09}.

Let us stipulate some necessary steps in further development of this scenario:

\begin{itemize}

\item The mechanism of baryosynthesis should be developed in the "approach". 
This mechanism can be directly applied also to the fifth family. If not, and only 
first family baryon asymmetry is initially formed, sphaleron transitions would 
redistribute the excess of particles and create the excess of fifth quarks. 
For composite dark matter scenario the excess of antiquarks is needed and the 
conditions under which the asymmetry in baryons of fifth family has opposite sign 
relative to the first family baryon asymmetry should be studied.

\item Self-consistent analysis should also clarify the role of fifth neutrino in this scenario.

 \item If OHe hypothesis is correct, it should give the amount of events in NaI, 
 corresponding to the detected signal. It implies quantum- mechanical calculation 
 of the rate of OHe-nucleus radiative capture.

 \item The calculated rate of OHe-nucleus radiative capture should be used for 
 reproduction of DAMA signal with account for all the physical and astrophysical uncertainties.
 \end{itemize}

\section{Some conclusions for future work}
It is mutually agreed that stability of the fifth family is very well motivated in 
the approach, unifying spins and charges. It gives rise to various possible candidates 
for stable lightest particles (heavy quark clusters) and correspondingly different 
dark matter scenarios. Tests of these scenarios along the lines of the present 
discussion are challenging for our future joint work.

%% Discussion Section III
\title{Discussions Section On the Yukawa Couplings Proposed by the "Approach 
Unifying Spin and Charges" Beyond the Tree Level}
\author{A. Hern\'andez-Galeana${}^1$ and N.S. Manko\v c Bor\v stnik${}^2$}
\institute{%
${}^1$Departamento de F\'{\i}sica,   Escuela Superior de
F\'{\i}sica y Matem\'aticas, I.P.N., \\
U. P. "Adolfo L\'opez Mateos". C. P. 07738, M\'exico, D.F.,
M\'exico\\
${}^2$Department of Physics, FMF, University of Ljubljana,\\
 Jadranska 19,Ljubljana, 1000}

\authorrunning{A. Hern\'andez-Galeana and N.S. Manko\v c Bor\v stnik}
\titlerunning{Discussions Section On the Yukawa Couplings\ldots}
\maketitle

\begin{abstract}
The "approach unifying spin and charges", proposed by Norma Susana Manko\v c Bor\v stnik, 
predicts two kinds of the Yukawa couplings. One kind distinguishes on the tree level only 
among the members of one family (among the u-quark, d-quark, neutrino and electron), while 
the other kind distinguishes only 
among the families. 
Long discussions at the present workshop between Norma and Albino lead 
to the first step of collaboration presented in this contribution: to a toy model 
with evaluated contributions bellow the tree level, done by Albino. 
\end{abstract}

\section{A short introduction I, written by Norma} 
The "approach unifying spin and charges", proposed by Norma Susana Manko\v c Bor\v stnik, 
predicts two kinds of the Yukawa couplings. One kind distinguishes on the tree level only 
among the members of one family (among the u-quark, d-quark, neutrino and electron), while 
the other kind distinguishes only 
among the families. Beyond the tree level both kinds of the Yukawa couplings start to 
contribute coherently and a detailed  study should   
manifest the drastic differences in properties of quarks and leptons: in their masses and 
mixing matrices. The reader can find the explanation for this statement in the contribution 
presented in this proceedings on page 119 by Norma and in the references therein). 
This is a very demanding project. To understand how does this occur we start this 
study first on a toy model. 
This work is the introduction into first steps towards understanding the 
properties of the lower four families of quarks and leptons as predicted by the "approach", 
by using a toy model. Albino has made first step which could help to do calculations beyond the 
tree level for the "approach unifying spin and charges". 

\section{The introduction II and all the rest, written by Albino}
We make the first step which could help to do calculations beyond the 
tree level for the "approach unifying spin and charges". We propose a tentative hierarchical mass
generation mechanism for one sector; u, d, e or $\nu$, where the
mass of the heaviest ordinary family is generated from a See-saw
mechanism, meanwhile light fermions obtain masses from radiative
corrections, at one and two loops, respectively. These radiative
corrections could be mediated either by scalar or gauge boson
fields within the "approach unifying spin and charges".

\section{Tree level mass matrix }

For a given sector; u (up quarks), d (down quarks), e
(charged leptons) or $\nu$ (neutrinos), $f_1\;,f_2\;,f_3$ denote
ordinary families, and F corresponds to the fourth very heavy
family. Let us start by {\bf assuming} the tree level mass terms
\begin{equation}
m_{34}\:\bar{f_{3L}^o}\:F_R^o+m_{43}\:\bar{F_L^o}\:f_{3R}^o+M\:\bar{F_L^o}\:F_R^o
+ h.c.=\bar{\psi_L^o}\:{\cal M}_o\:\psi_R^o +h.c.\label{ah1eq1}\end{equation}
where
\begin{equation} \psi_{oL}^T=(f_1^o,f_2^o,f_3^o,F^o)_L \qquad , \qquad
\psi_{oR}^T=(f_1^o,f_2^o,f_3^o,F^o)_R \label{ah1eq2} \end{equation}
are weak or interaction eigenfields. For the
sake of simplicity, let us assume that it is possible to set
$m_{34}=m_{43}\equiv p$, such that, we may write ${\cal M}_o$ in Eq.
(1) as the real and symmetric "See-saw" type mass matrix
\begin{equation} {\cal M}_o=
\begin{pmatrix} 0 & 0 & 0 & 0\\ 0 & 0 & 0 & 0\\ 0 & 0 & 0 &
p \\ 0 & 0 & p & M \end{pmatrix} \; ;\qquad  M>p>0 \:. \label{ah1eq3}\end{equation}
Using an orthogonal matrix $V^o$ to
diagonalize ${\cal M}_o$; $ \Psi^{o}_L = V^o \:\chi_L^o \; ,\;
\Psi^{o}_R = V^o \:\chi_R^o \;,$
\begin{equation} \bar{\psi_L^o}\:{\cal M}_o\: \psi_R^o=
\bar{\chi_{L}^o}\:{V^o}^T\:{\cal M}_o\:V^o \:\chi_{R}^o \:,\label{ah1eq4}\end{equation}
where $T$ means transpose, and we write
$V^o$ as
\begin{equation} V^o = \left( \begin{array}{ccrr}
1 & 0 & 0& 0\\0 & 1& 0 &0\\ 0 & 0& \cos{\alpha} & \sin{\alpha}\\
0 & 0 & -\sin{\alpha} & \cos{\alpha}
\end{array} \right) \:,\label{ah1eq5}\end{equation}
where the two nonzero eigenvalues $\lambda_3$
and $\lambda_4$ of ${\cal M}_o$ satisfy
\begin{equation} \lambda^2 - M \lambda - p^2=0 , \quad M=\lambda_3+\lambda_4 ,
\quad - p^2=\lambda_3 \:\lambda_4  \label{ah1eq6}\end{equation}

\begin{equation} \lambda_3=\frac{1}{2} \left(M-\sqrt{M^2+4 p^2}\right)<0 \qquad
, \qquad \lambda_4=\frac{1}{2} \left(M+\sqrt{M^2+4 p^2}\right)>0 \label{ah1eq7}
\end{equation}

\begin{equation} \cos{\alpha} = \sqrt{\frac{\lambda_4}{\lambda_4 - \lambda_3}}
\quad , \quad  \sin{\alpha} = \sqrt{\frac{- \lambda_3}{\lambda_4 -
\lambda_3}} \quad , \quad
\cos{\alpha}\sin{\alpha}=\frac{p}{\lambda_4 - \lambda_3} \label{ah1eq8} \end{equation}

\begin{equation} -
\lambda_3\:\cos^2{\alpha}=\lambda_4\:\sin^2{\alpha}=\frac{p^2}{\lambda_4
- \lambda_3} \equiv m_o  \label{ah1eq9}\end{equation}

\begin{equation} {V^o}^T\:{\cal M}_o\:V^o = Diag(0,0,\lambda_3,\lambda_4) \label{ah1eq10}\end{equation}
Eqs. (\ref{ah1eq5}-\ref{ah1eq10}) are exact analytic results from
the diagonalization of ${\cal M}_o$.

Note that if we impose the hierarchy
$\frac{|\lambda_3|}{\lambda_4} \ll 1$, then\footnote{From now on
we are going to assume this hierarchy.}
\begin{equation} \frac{p^2}{M^2} = \frac{|- p^2|}{M^2}= \frac{|\lambda_3
\lambda_4|}{(\lambda_3+\lambda_4)^2}=
\frac{|\lambda_3|}{\lambda_4}\frac{1}{(1+\frac{\lambda_3}{\lambda_4})^2}
\ll 1 \;. \label{ah1eq11}\end{equation}

In this limit, we may approach
\begin{eqnarray} \lambda_3 \thickapprox - \frac{p^2}{M} \qquad , \qquad \lambda_4
\thickapprox M + \frac{p^2}{M}\thickapprox M \;, \nonumber\\
                                                \nonumber\\
\sin{\alpha} = \sqrt{\frac{- \lambda_3}{\lambda_4 -
\lambda_3}}\approx \sqrt{\frac{p^2}{M^2}}=\frac{p}{M} \ll 1 \;,\label{ah1eq12}\\
                      \nonumber\\
\cos{\alpha}\sin{\alpha}=\frac{p}{\lambda_4 - \lambda_3} \approx
\frac{p}{M} \ll 1  \;,\nonumber\end{eqnarray}
in agreement with the well known results
from See-saw mass matrix. $- \lambda_3$ may be associated, in good
approximation, with the mass for the heaviest ordinary fermion
$m_t$, $m_b$, $m_\tau$ or $m_3$, and $\lambda_4$ with the mass of
the heavy fourth fermion in a given sector.

\section{One loop corrections}

\begin{figure}[htp]
\begin{center}
\includegraphics[width=0.9\columnwidth]{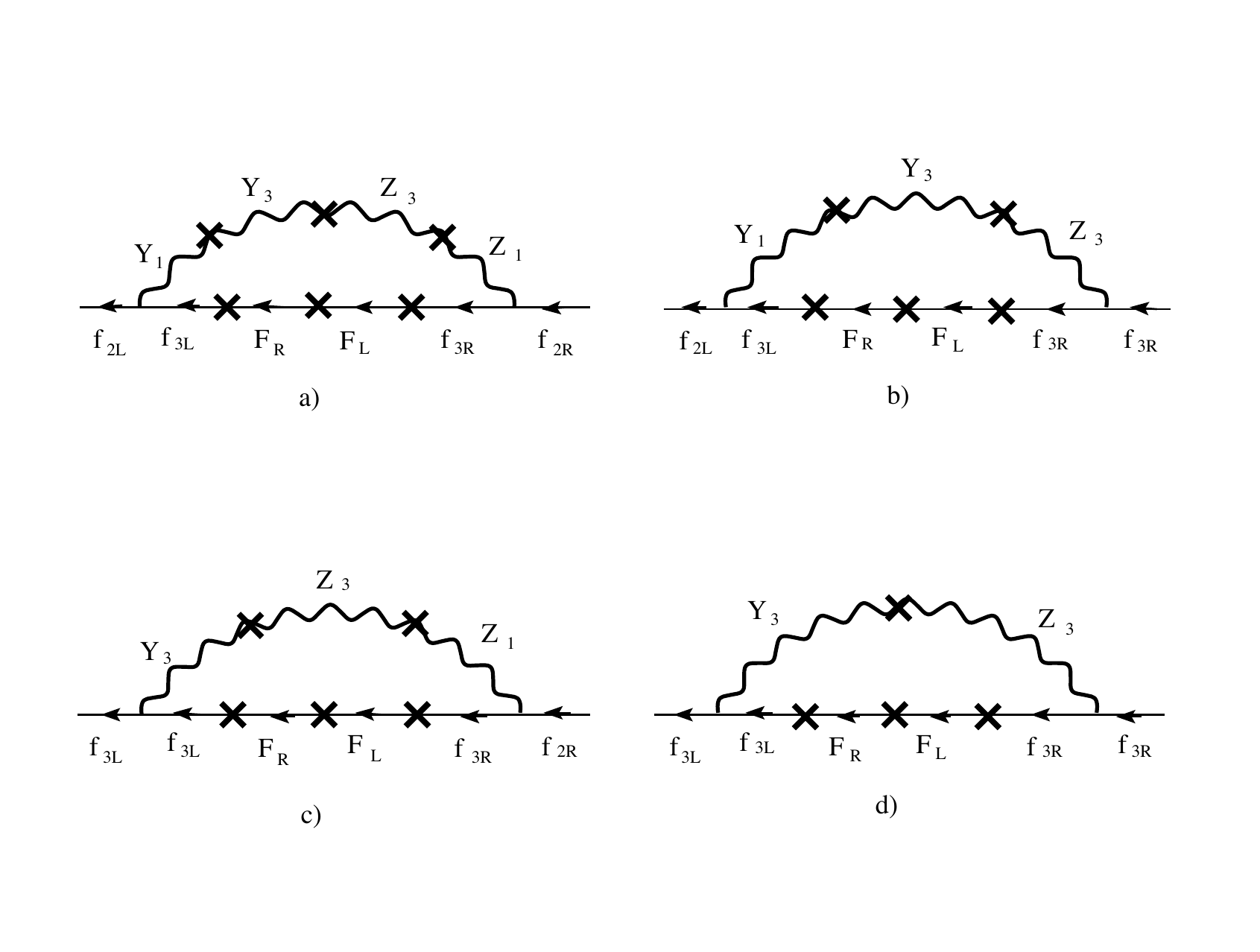}
\caption{\label{ah1fig1}One loop contributions to: a)
$m_{22}^{(1)}\:\bar{f_{2L}^o}\:f_{2R}^o$, b)
$m_{23}^{(1)}\:\bar{f_{2L}^o}\:f_{3R}^o$, c)
$m_{32}^{(1)}\:\bar{f_{3L}^o}\:f_{2R}^o$, d)
$m_{33}^{(1)}\:\bar{f_{3L}^o}\:f_{3R}^o$}
\end{center}
\end{figure}

Subsequently, the masses for the light fermions would arise through
one and two loops radiative corrections, respectively. To achieve
this goal, let us {\bf introduce} the gauge bosons
$Y_1\;,Y_3\;,Z_1\;,Z_3$, with the gauge couplings to fermions in the
interaction basis as\footnote{Analogous Yukawa couplings could be
introduced if radiative corrections were mediated by scalar fields;
See for example Ref.\cite{ah1u1}}

\begin{multline} \left(h_{12} \:\bar{f_{1L}^o} \gamma_\mu \:f_{2L}^o +h_{23}
\:\bar{f_{2L}^o} \gamma_\mu \:f_{3L}^o \right) Y_1^\mu +
h_{33}\:\bar{f_{3L}^o} \gamma_\mu \:f_{3L}^o \:Y_3^\mu  \\
                                                       \\
+ \left(H_{12} \:\bar{f_{1R}^o} \gamma_\mu \:f_{2R}^o + H_{23}
\:\bar{f_{2R}^o} \gamma_\mu \:f_{3R}^o \right) Z_1^\mu +
H_{33}\:\bar{f_{3R}^o} \gamma_\mu \:f_{3R}^o \:Z_3^\mu + h.c.
\label{ah1eq13}
\end{multline}
where $h_{12}$, $h_{23}$, $h_{33}$, $H_{12}$, $H_{23}$,
and $H_{33}$ are gauge coupling constants\footnote{of same order of
magnitude}. We also {\bf assume} the gauge boson mass matrix:
\begin{equation} X^T \:M_B^2\:X \qquad , \qquad X^T=(Y_1,Y_3,Z_3,Z_1) \;,\label{ah1eq14}\end{equation}
with
\begin{equation} M_B^2 =\left( \begin{array}{cccc} a_1 & b
& 0 & 0\\ b & a_2 & c & 0\\ 0 & c & a_3 & d\\
0 &  0 & d & a_4
\end{array} \right). \label{ah1eq15}
\end{equation}
These mass terms for gauge(or scalar)
fields in Eq.(\ref{ah1eq15}) should be generated at some stage of symmetry
breaking at the scale $\Lambda \:> (\mbox{or may be}\:\gg$) than
the electroweak scale $v\approx 246\:GeV$. The diagonalization of
$M_B^2$ is performed in the Appendix B. Using the gauge boson
couplings, Eq.(\ref{ah1eq13}), and the tree level mass terms (See-saw
mechanism), Eq.(\ref{ah1eq1}), we can construct the one loop mass diagrams of
Fig.~\ref{ah1fig1}. 
The evaluation of these diagrams yields the one loop mass
terms contributions
\begin{equation}
m_{22}^{(1)}\:\bar{f_{2L}^o}\:f_{2R}^o+m_{23}^{(1)}\:\bar{f_{2L}^o}\:f_{3R}^o+
m_{32}^{(1)}\:\bar{f_{3L}^o}\:f_{2R}^o+m_{33}^{(1)}\:\bar{f_{3L}^o}\:f_{3R}^o
\;,\label{ah1eq16}\end{equation}
with
\begin{equation}
m_{22}^{(1)} =\frac{h_{23} H_{23} }{16\pi^{2}}
\sum_{k=1,2,3,4;\,i=3,4} m_i^{(o)}\:(V^o)_{3i}^2 \:U_{1k} U_{4k}\:
f(M_k, m_i^{(o)}),\label{ah1eq17}
\end{equation}

\begin{equation}
m_{23}^{(1)} =\frac{h_{23} H_{33} }{16\pi^{2}}
\sum_{k=1,2,3,4;\,i=3,4} m_i^{(o)}\:(V^o)_{3i}^2 \:U_{1k} U_{3k}\:
f(M_k, m_i^{(o)}),\label{ah1eq18}
\end{equation}

\begin{equation}
m_{32}^{(1)} =\frac{h_{33} H_{23} }{16\pi^{2}}
\sum_{k=1,2,3,4;\,i=3,4} m_i^{(o)}\:(V^o)_{3i}^2 \:U_{2k} U_{4k}\:
f(M_k, m_i^{(o)}),\label{ah1eq19}
\end{equation}

\begin{equation}
m_{33}^{(1)} =\frac{h_{33} H_{33} }{16\pi^{2}}
\sum_{k=1,2,3,4;\,i=3,4} m_i^{(o)}\:(V^o)_{3i}^2 \:U_{2k} U_{3k}\:
f(M_k, m_i^{(o)}),\label{ah1eq20}
\end{equation}
where $m_3^{(o)}=\lambda_3$, $m_4^{(o)}=\lambda_4$,
Eqs.(\ref{ah1eq7},\ref {ah1eq11}), $U$ is the orthogonal matrix which diagonalizes
$M_B^2$,Eqs.(\ref{ah1eq71},\ref{ah1eq78}), with
\begin{equation}
B_i=U_{ij}\omega_j \quad , \quad B_1 = Y_1 \quad , \quad B_2 = Y_3
\quad , \quad B_3 = Z_3 \quad , \quad B_4 = Z_1 \label{ah1eq21}
\end{equation}
being the relation between interaction and mass boson
eigenfields $\omega_i$, $i,j=1,2,3,4$, $M_k^2$ are the eigenvalues
of $M_B^2$, and
\begin{equation}
f(a,b) \equiv \frac{a^2}{a^2-b^2}\ln{\frac{a^2}{b^2}} \;. \label{ah1eq22}
\end{equation}
Performing the summation over the index
$i=3,4$ in Eqs.(\ref{ah1eq17}-\ref{ah1eq20}), and using the relations in Eqs.(\ref{ah1eq8},\ref{ah1eq9}), we
may write
\begin{equation} m_{22}^{(1)} =\frac{h_{23} H_{23} }{16\pi^{2}} m_o \sum_{k}
U_{1k} U_{4k} F(M_k) \quad , \quad m_{23}^{(1)} =\frac{h_{23}
H_{33} }{16\pi^{2}} m_o \sum_{k} U_{1k} U_{3k} F(M_k) \:,\label{ah1eq23}\end{equation}

\begin{equation} m_{32}^{(1)} =\frac{h_{33} H_{23} }{16\pi^{2}} m_o \sum_{k}
U_{2k} U_{4k} F(M_k) \quad , \quad m_{33}^{(1)} =\frac{h_{33}
H_{33} }{16\pi^{2}} m_o \sum_{k} U_{2k} U_{3k} F(M_k) \:,\label{ah1eq24}\end{equation}
where the mass parameter $m_o$ is defined
in Eq.(\ref{ah1eq9}), and
\begin{equation} F(M_k) \equiv \frac{M_k^2}{M_k^2-\lambda_4^2}
\ln{\frac{M_k^2}{\lambda_4^2}} - \frac{M_k^2}{M_k^2-\lambda_3^2}
\ln{\frac{M_k^2}{\lambda_3^2}} \:. \label{ah1eq25}\end{equation}
So, the one loop contribution in the
interaction basis reads
\begin{equation} \bar{\psi_{L}^o}\:{\cal M}_1^o\: \psi_{R}^o=
\bar{\chi_{L}^o}\:{V^o}^T\:{\cal M}_1^o\:V^o \:\chi_{R}^o \qquad ,
\qquad {\cal M}_1^o= \left( \begin{array}{ccrr}
0 & 0 & 0& 0\\0 & m_{22}^{(1)} & m_{23}^{(1)} &0\\ 0 &  m_{32}^{(1)}& m_{33}^{(1)} & 0 \\
0 & 0 & 0 & 0
\end{array} \right)  \:,\label{ah1eq26}\end{equation}
and thus, up to one loop corrections, we
get the mass terms
\begin{equation} \bar{\chi_{L}^o} \left[ {V^o}^T\:{\cal M}_1^o\:V^o
+Diag(0,0,\lambda_3,\lambda_4) \right] \:\chi_{R}^o \equiv
\bar{\chi_{L}^o}\:{\cal M}_1\:\chi_{R}^o \:,\label{ah1eq27}\end{equation}
with
\begin{equation} {\cal{M}}_1= \left( \begin{array}{clll}
0 & 0 & 0& 0\\0 & m_{22}^{(1)} & m_{23}^{(1)} \cos\alpha & m_{23}^{(1)} \sin\alpha \\
0 &  m_{32}^{(1)} \cos\alpha& \lambda_3 + m_{33}^{(1)} \cos^2\alpha& m_{33}^{(1)} \cos\alpha \sin\alpha \\
0 & m_{32}^{(1)} \sin\alpha & m_{33}^{(1)} \cos\alpha \sin\alpha
&\lambda_4 + m_{33}^{(1)} \sin^2\alpha
\end{array} \right)  \:.\label{ah1eq28}\end{equation}
Remember now from Eq.(\ref{ah1eq12}) the value $\sin\alpha \ll 1$. In
consistency with this tiny mixing angle, we assume that mixing
between ordinary fermions with the fourth family, in each sector,
is defined to leading order at the tree level by the see-saw mass
matrix ${\cal M}_o$ in Eq.(\ref{ah1eq3}), and so, in this approach, we may
neglect one loop corrections in what concern the mass and mixing
of the fourth family with the ordinary ones, and then we may set
$\sin\alpha =0$ in the mass matrix ${\cal M}_1$. Hence, we may
approximate
\begin{equation} {\cal{M}}_1\approx \left( \begin{array}{clll}
0 & 0 & 0& 0\\0 & m_{22}^{(1)} & m_{23}^{(1)} & 0 \\
0 &  m_{32}^{(1)} & \lambda_3 + m_{33}^{(1)} & 0 \\
0 & 0 & 0 &\lambda_4
\end{array} \right)  \:.\label{ah1eq29}\end{equation}
Thus, the diagonalization of ${\cal M}_1$ in
this approach reduces to the diagonalization of a $2\times 2$ mass
matrix as is done in the Appendix A. In terms of a biunitary
transformation $\chi_L^o= V_L^{(1)} \;\chi_L^1$ and $\chi_R^o=
V_R^{(1)} \;\chi_R^1$,
\begin{equation} \bar{\chi_L^o} \: {\cal{M}}_1 \:\chi_R^o =\bar{\chi_L^1}
\;{V_L^{(1)}}^T {\cal{M}}_1 \:V_R^{(1)} \;\chi_R^1   \:. \label{ah1eq30}\end{equation}
With $c_{L,R}=\cos\theta_{L,R}$,
$s_{L,R}=\sin\theta_{L,R}$ we may write
\begin{equation}
V_L^{(1)}=\left( \begin{array}{ccrc} 1&0&0& 0\\
0&c_L & s_L  & 0\\ 0& - s_L & c_L &0\\
0& 0& 0&1
\end{array} \right) \quad \text{and} \quad
V_R^{(1)}=\left( \begin{array}{ccrc} 1&0&0& 0\\
0&c_R & s_R  & 0\\ 0& - s_R & c_R &0\\
0& 0& 0&1
\end{array} \right) \;,\label{ah1eq31}\end{equation}
where mixing angles $\sin\theta_L$ and
$\sin\theta_R$ are defined in the Appendix A in terms of the
eigenvalues $\sigma_2$ and $\sigma_3$, Eq.(\ref{ah1eq56}), and the parameters
of ${\cal M}_1 \:{\cal{M}}_1^T$ and ${{\cal M}_1}^T {\cal{M}}_1$,
respectively. From $V^{(1)}_L$ and $V^{(1)}_R$ one computes
\begin{equation} {V^{(1)}_L}^T {\cal{M}}_1 \:V^{(1)}_R
=Diag(0,\sqrt{\sigma_2},- \sqrt{\sigma_3},\sqrt{\lambda_+})\label{ah1eq32}
\end{equation}
\begin{equation} {V^{(1)}_L}^T {\cal{M}}_1 \:{\cal{M}}_1^T \;V^{(1)}_L =
{V^{(1)}_R}^T {\cal{M}}_1^T \:{\cal{M}}_1 \;V^{(1)}_R =
Diag(0,\sigma_2,\sigma_3,\lambda_+)    \:.\label{ah1eq33}\end{equation}
Here $- \sqrt{\sigma_3}$ is a tiny correction to
$\lambda_3$ in Eqs.(\ref{ah1eq7},\ref{ah1eq12}).

\subsection{Two loop contributions}

\begin{figure}[htp]
\begin{center}
\includegraphics[width=0.9\columnwidth]{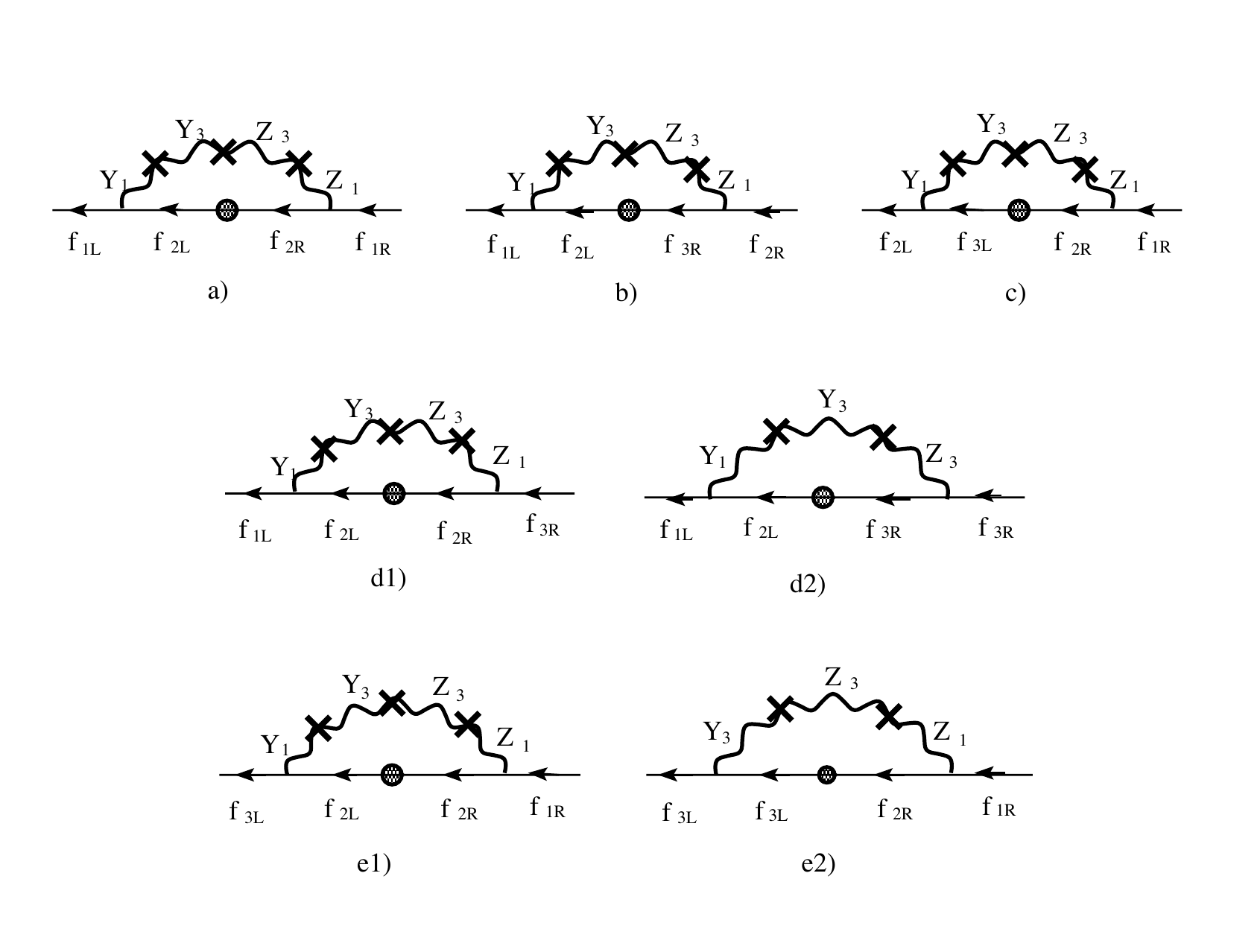}
\caption{\label{ah1fig2}Two loops contributions.}
\end{center}
\end{figure}

We see from Eqs.(\ref{ah1eq32},\ref{ah1eq33}) that up to one loop corrections the first
family of ordinary fermions, $m_u$, $m_d$, $m_e$ or $m_{\nu 1}$
remain massless, and so we need to consider two loop
contributions. We consider the two loop diagrams\footnote{We are
neglecting tiny two loop contributions to the entries
$m_{22}^{(2)}$, $m_{23}^{(2)}$,$m_{32}^{(2)}$ and $m_{33}^{(2)}$.}
given in the Fig.~\ref{ah1fig2}.
Let us recall that the transformations from
massless (interaction) to mass eigenfields up to one loop are
given by $\Psi^{o}_L = V^{o} \:\chi_L^o=V^{o} \:V^{(1)}_L
\:\chi_L^1$ and $\Psi^{o}_R = V^{o} \:\chi_R^o=V^{o} \:V^{(1)}_R
\:\chi_R^1 $, where, explicitly
\begin{align}
V^o\:V_L^{(1)} &=\left( \begin{array}{crrc} 1&0&0& 0\\
0&c_L & s_L  & 0\\ 0& - \cos\alpha\:s_L & \cos\alpha \:c_L &
\sin\alpha \\ 0& \sin\alpha \:s_L &- \sin\alpha \:c_L&\cos\alpha
\end{array} \right) \; , \;\nonumber\\
V^o\:V_R^{(1)} &= \left( \begin{array}{crrc} 1&0&0& 0\\
0&c_R & s_R  & 0\\ 0& - \cos\alpha\:s_R & \cos\alpha \:c_R &
\sin\alpha \\ 0& \sin\alpha \:s_R &- \sin\alpha \:c_R&\cos\alpha
\end{array} \right)   \:.\label{ah1eq34}\end{align}
Using these field transformations to write
the internal fermion lines in Fig.~\ref{ah1fig2} in terms of the one loop mass
eigenfieds, and performing a similar analysis as before, the two
loop diagrams yields the contributions
\begin{equation}
m_{11}^{(2)} =\frac{h_{12} H_{12}}{16\pi^{2}} \sum_{k=1,2,3,4 ;
\;i=2,3} m_i^{(1)}\:(V^o\:V_L^{(1)})_{2i}\:(V^o\:V_R^{(1)})_{2i}
\:U_{1k} U_{4k} \:f(M_k, m_i^{(1)}),\label{ah1eq35}
\end{equation}

\begin{equation}
m_{21}^{(2)} =\frac{h_{23} H_{12}}{16\pi^{2}} \sum_{k=1,2,3,4 ;
\;i=2,3} m_i^{(1)}\:(V^o\:V_L^{(1)})_{3i}\:(V^o\:V_R^{(1)})_{2i}
\:U_{1k} U_{4k} \:f(M_k, m_i^{(1)}),\label{ah1eq36}
\end{equation}

\begin{equation}
m_{12}^{(2)} =\frac{h_{12} H_{23}}{16\pi^{2}} \sum_{k=1,2,3,4 ;
\;i=2,3} m_i^{(1)}\:(V^o\:V_L^{(1)})_{2i}\:(V^o\:V_R^{(1)})_{3i}
\:U_{1k} U_{4k} \:f(M_k, m_i^{(1)}),\label{ah1eq37}
\end{equation}

\begin{multline}
m_{31}^{(2)} =\frac{h_{23} H_{12}}{16\pi^{2}} \sum_{k=1,2,3,4 ;
\;i=2,3} m_i^{(1)}\:(V^o\:V_L^{(1)})_{2i}\:(V^o\:V_R^{(1)})_{2i}
\:U_{1k} U_{4k} \:f(M_k, m_i^{(1)}) \\ + \frac{h_{33}
H_{12}}{16\pi^{2}} \sum_{k=1,2,3,4 ; \;i=2,3}
m_i^{(1)}\:(V^o\:V_L^{(1)})_{3i}\:(V^o\:V_R^{(1)})_{2i} \:U_{2k}
U_{4k} \:f(M_k, m_i^{(1)}),\label{ah1eq38}
\end{multline}

\begin{multline}
m_{13}^{(2)} =\frac{h_{12} H_{23}}{16\pi^{2}} \sum_{k=1,2,3,4 ;
\;i=2,3} m_i^{(1)}\:(V^o\:V_L^{(1)})_{2i}\:(V^o\:V_R^{(1)})_{2i}
\:U_{1k} U_{4k} \:f(M_k, m_i^{(1)}) \\ + \frac{h_{12}
H_{33}}{16\pi^{2}} \sum_{k=1,2,3,4 ; \;i=2,3}
m_i^{(1)}\:(V^o\:V_L^{(1)})_{2i}\:(V^o:V_R^{(1)})_{3i} \:U_{1k}
U_{3k} \:f(M_k, m_i^{(1)}) \:.\label{ah1eq39}
\end{multline}
Note that in the limit $M_k \gg
m_i^{(1)}\;,i=2,3$, the function $f(a,b)$ behaves as
$\ln{\frac{a^2}{b^2}}$. In this limit, using the one loop mass
eigenvalues, $m_2^{(1)}= \sqrt{\sigma_2}$, $m_3^{(1)}= -
\sqrt{\sigma_3}$, Eqs.(\ref{ah1eq32},\ref{ah1eq33}), $V^o\:V_L^{(1)}$ and
$V^o\:V_R^{(1)}$,Eq.(\ref{ah1eq34}), the relationships in Eqs.(\ref{ah1eq63},\ref{ah1eq64}) and
using the orthogonality of $U$, one gets
\begin{equation}
m_{11}^{(2)} =\frac{h_{12} H_{12}}{16\pi^{2}} m_{22}^{(1)} \:G_{14},\label{ah1eq40}
\end{equation}

\begin{equation}
m_{21}^{(2)} =\frac{h_{23} H_{12}}{16\pi^{2}} \cos\alpha
\:m_{32}^{(1)} \:G_{14},\label{ah1eq41}
\end{equation}

\begin{equation} m_{12}^{(2)} =\frac{h_{12} H_{23}}{16\pi^{2}} \cos\alpha
\:m_{23}^{(1)} \:G_{14}, \label{ah1eq42}\end{equation}

\begin{equation} m_{31}^{(2)} =\frac{h_{23} H_{12}}{16\pi^{2}}  \:m_{22}^{(1)}
\:G_{14}+ \frac{h_{33} H_{12}}{16\pi^{2}} \cos\alpha \:m_{32}^{(1)}
\:G_{24}, \label{ah1eq43}\end{equation}

\begin{equation} m_{13}^{(2)} =\frac{h_{12} H_{23}}{16\pi^{2}} \:m_{22}^{(1)}
\:G_{14}+ \frac{h_{12} H_{33}}{16\pi^{2}} \cos\alpha \:m_{23}^{(1)}
\:G_{13} \label{ah1eq44}\end{equation}
where the parameters $G_{14}$, $G_{24}$ and
$G_{13}$ are defined as
\begin{align} G_{14}  &\equiv \sum_{k} U_{1k} U_{4k} \ln{\frac{M_k^2}{m_o^2}}
\quad , \quad\nonumber \\ G_{24} &\equiv \sum_{k} U_{2k} U_{4k}
\ln{\frac{M_k^2}{m_o^2}} \quad , \quad \label{ah1eq45}\\ G_{13} &\equiv \sum_{k}
U_{1k} U_{3k} \ln{\frac{M_k^2}{m_o^2}}\nonumber
 \end{align}
Hence, the leading order two loop
contributions in the weak basis is written as $\bar{\psi_L^o}
\:{\cal M}_2^o \:\psi_R^o$,
\begin{equation} {\cal M}_2^o \approx \left( \begin{array}{clll}
m_{11}^{(2)} & m_{12}^{(2)} & m_{13}^{(2)} & 0\\ m_{21}^{(2)}& 0& 0 & 0 \\
m_{31}^{(2)}& 0& 0 & 0  \\ 0 & 0 & 0 &0 \end{array} \right)
\:.\label{ah1eq46}\end{equation}
So, up to two loops we obtain the mass
terms
\begin{equation} \bar{\chi_L^1} [ \:(V_o V_L^{(1)})^T \:{\cal M}_2^o \: V_o
V_R^{(1)} + Diag(0,\sqrt{\sigma_2},-
\sqrt{\sigma_3},\lambda_4)\:]\;\chi_R^1 \equiv \bar{\chi_L^1} {\cal
M}_2 \;\chi_R^1 \:,\label{ah1eq47}\end{equation}
where ${\cal{M}}_2 \thickapprox$
\begin{equation} 
%{\small
%\\ 
\left(
%\begin{array}{cccc}
\begin{smallmatrix}
m_{11}^{(2)}& m_{12}^{(2)} \:c_R - m_{13}^{(2)} \:\cos\alpha \:s_R &m_{13}^{(2)} \:\cos\alpha \:c_R + m_{12}^{(2)} \:s_R  &
\sin\alpha \:m_{13}^{(2)}\\
m_{21}^{(2)} \:c_L - m_{31}^{(2)} \:\cos\alpha \:s_L & \sqrt{\sigma_2}&0& 0 \\
m_{31}^{(2)} \:\cos\alpha \:c_L + m_{21}^{(2)} \:s_L &0
& - \sqrt{\sigma_3}&0\\ \sin\alpha \:m_{31}^{(2)}& 0& 0&\lambda_4
%\end{array} \right) .
\end{smallmatrix}\right).
%} %%% end tiny\
\label{ah1eq48}\end{equation}
The diagonalization of ${\cal M}_2$ yields
the physical masses for fermions in each sector u, d, e or $\nu$.
Using the same arguments as before, we can perform this
diagonalization in good approximation in the limit $\sin\alpha=0$.
In this approach, the diagonalization of ${\cal{M}}_2$ reduces to
diagonalize a $3\times 3$ mass matrix, and the results and/or method
of diagonalization introduced in Ref.\cite{ah1u1} may be applied.
Defining a new biunitary transformation $\chi_L^1=V_L^{(2)} \psi_L
\; \mbox{and} \; \chi_R^1=V_R^{(2)} \psi_R $, then
\begin{equation} \bar{\chi_L^1} \;{\cal{M}}_2\;\chi_R^1= \bar{\psi_L}
\:{V_L^{(2)}}^T {\cal{M}}_2\: V_R^{(2)} \:\psi_R\label{ah1eq49}\end{equation}
where now
\begin{equation} {\Psi_L}^T = ( f_{1L} , f_{2L} , f_{3L} , F_L ) \quad , \quad
{\Psi_R}^T = ( f_{1R} , f_{2R} , f_{3R} , F_R ) \label{ah1eq50}\end{equation}
are the mass eigenfields, that is
\begin{equation} {V_L^{(2)}}^T {\cal M}_2\; V_R^{(2)}=Diag(m_1, m_2, - m_3,
M_F) \;.\label{ah1eq51}\end{equation}

\begin{equation} {V^{(2)}_L}^T {\cal{M}}_2 \:{\cal M}_2^T \;V^{(2)}_L =
{V^{(2)}_R}^T {\cal M}_2^T \:{\cal{M}}_2 \;V^{(2)}_R =
Diag(m_1^2,m_2^2,m_3^2,M_F^2)    \:.\label{ah1eq52}\end{equation}
For example: $m_1=m_e \; , \; m_2=m_\mu \; ,
\; m_3= - m_\tau \; , \; M_F=M_E\;$ for charged leptons.

Thus, the final transformations from
massless (interaction) to mass fermion eigenfields are
\begin{equation} \Psi_L^o = V^{o} \:V^{(1)}_L \: V^{(2)}_L \:\Psi_L \quad
\mbox{and} \quad  \Psi_R^o = V^{o} \:V^{(1)}_R \: V^{(2)}_R
\:\Psi_R \label{ah1eq53}\end{equation}

\section{Discussion}

The {\bf basic assumptions} in this {Toy Radiative Corrections}
are: the tree level mass terms assumed in Eq.(\ref{ah1eq1}), the introduction
of the gauge bosons fields $Y_1$, $Y_3$, $Z_1$, $Z_3$ and their
couplings to fermions introduced in Eq.(\ref{ah1eq13}), as well as the
structure of their mass matrix assumed in Eqs.(\ref{ah1eq14},\ref{ah1eq15}). I have
already taken a look to the ``approach unifying spin and
charges''; Proceedings, Portoroz, Slovenia 2003 and arXiv:0708.2846.

From this reading, it is not clear yet to me whether it is
possible or not to accomplish (or give some arguments to justify)
these {\bf basic assumptions} for at least one of the sectors; u,
d, e or $\nu$. In order to go further in this task, I would like
to understand more details about the implementation of symmetry
breaking within your approach. In any case, we can take this
manuscript as one "Toy specific example" which points out the way
radiative corrections could arise, and the role they can play to
implement a hierarchical spectrum of fermion masses.

\section{Appendix A: \\ Diagonalization of a 2x2 mass matrix}

Let us consider the diagonalization of the mass matrix
\begin{equation} m =\left(\begin{array}{cc}
m_{22}^{(1)}&m_{23}^{(1)}\\m_{32}^{(1)}& \lambda_3+m_{33}^{(1)}
\end{array}\right) \equiv
\left(\begin{array}{cc}q_{2}&q_{23}\\q_{32}& q_3
\end{array}\right) \;.\label{ah1eq54}\end{equation}
We assume the signs:  $q_2>0$, $q_{23}<0$, $q_{32}<0$, 
$q_3=\lambda_3+m_{33}^{(1)}\approx -\frac{p^2}{M} + m_{33}^{(1)} <0$
\begin{equation}\begin{aligned} a_L&=q_2^2+q_{23}^2 , &  b_L&=q_3^2+q_{32}^2,
  & c_L&=q_2 q_{32}+q_3 q_{23}\\
a_R&=q_2^2+q_{32}^2,  & b_R&=q_3^2+q_{23}^2,
 & c_R&=q_2 q_{23}+q_3 q_{32}\end{aligned}  \label{ah1eq55}  \end{equation}

\begin{equation}\begin{aligned} \sigma_2 &= \frac{1}{2} \left( P - \sqrt{P^2 -4Q}
  \right), \\
  \sigma_3 &= \frac{1}{2} \left( P + \sqrt{P^2 -4Q}
\right)\end{aligned} \label{ah1eq56}\end{equation}

\begin{equation} \begin{aligned}
P=a_L+b_L=a_R+b_R=q_2^2+q_3^2+q_{23}^2+q_{32}^2=\sigma_2+\sigma_3 \\
Q=a_L b_L - c_L^2 = a_R b_R - c_R^2 = (- q_2 q_3 + q_{23}
q_{32})^2=\sigma_2 \sigma_3 \end{aligned} \label{ah1eq57}
\end{equation}

\begin{equation} \begin{aligned} (\sqrt{\sigma_3} - \sqrt{\sigma_2})^2&=(- q_3 - q_2)^2+(q_{23}
- q_{32})^2, \\ \sqrt{\sigma_2} \sqrt{\sigma_3} &= - q_2
q_3 + q_{23} q_{32} > 0 \end{aligned} \label{ah1eq58} \end{equation}

\begin{equation} \begin{aligned} \cos\theta_L&=\sqrt{\frac{\sigma_3 - a_L}{\sigma_3
- \sigma_2}},  & \sin\theta_L&=\sqrt{\frac{\sigma_3 -
b_L}{\sigma_3 - \sigma_2}}    \\
\cos\theta_R&=\sqrt{\frac{\sigma_3 - a_R}{\sigma_3 - \sigma_2}}, 
& \sin\theta_R&= \sqrt{\frac{\sigma_3 - b_R}{\sigma_3 -
\sigma_2}} \end{aligned}\label{ah1eq59}\end{equation}

\begin{equation}\begin{aligned} \cos\theta_L \sin\theta_L &= \frac{c_L}{\sigma_3 -
    \sigma_2}, \\
 \cos\theta_R \sin\theta_R &= \frac{c_R}{\sigma_3 - \sigma_2} \end{aligned}\label{ah1eq60}
\end{equation}

Assuming $q_2 \sqrt{\sigma_3} + q_3 \sqrt{\sigma_2} <
0$ one gets the useful relationships
\begin{equation} \begin{aligned} \cos\theta_L \:\cos\theta_R &= \frac{- q_3 \sqrt{\sigma_3} -
q_2 \sqrt{\sigma_2}}{\sigma_3 - \sigma_2} ,\\
\sin\theta_L \:\sin\theta_R &= \frac{- q_2 \sqrt{\sigma_3} -  q_3
\sqrt{\sigma_2}}{\sigma_3 - \sigma_2} \end{aligned}\label{ah1eq61}\end{equation}

\begin{equation} \begin{aligned}\cos\theta_L \:\sin\theta_R &= \frac{- q_{32} \sqrt{\sigma_3} +
q_{23} \sqrt{\sigma_2}}{\sigma_3 - \sigma_2} ,\\
\sin\theta_L \:\cos\theta_R &= \frac{- q_{23} \sqrt{\sigma_3} +
q_{32} \sqrt{\sigma_2}}{\sigma_3 - \sigma_2} \end{aligned}\label{ah1eq62}\end{equation}

\begin{equation} \begin{aligned}\sqrt{\sigma_2} \:\cos\theta_L \:\cos\theta_R  -
\sqrt{\sigma_3} \:\sin\theta_L \:\sin\theta_R &= q_2 ,\\
- \sqrt{\sigma_2} \:\sin\theta_L \:\sin\theta_R + \sqrt{\sigma_3}
\:\cos\theta_L \:\cos\theta_R &= - q_3 \end{aligned}\label{ah1eq63} \end{equation}

\begin{equation} \begin{aligned}\sqrt{\sigma_2} \:\cos\theta_L \:\sin\theta_R  +
\sqrt{\sigma_3} \:\sin\theta_L \:\cos\theta_R &= - q_{23} ,\\
\sqrt{\sigma_2} \:\sin\theta_L \:\cos\theta_R +
\sqrt{\sigma_3} \:\cos\theta_L \:\sin\theta_R &= - q_{32}\end{aligned} \label{ah1eq64}\end{equation}

\begin{equation} \begin{aligned}\cos\theta_L \:\cos\theta_R  + \sin\theta_L \:\sin\theta_R &=
\frac{- q_3 - q_2}{\sqrt{\sigma_3} - \sqrt{\sigma_2}} ,\\
\cos\theta_L \:\sin\theta_R  - \sin\theta_L \:\cos\theta_R &=
\frac{q_{23} - q_{32}}{\sqrt{\sigma_3} - \sqrt{\sigma_2}} \end{aligned}\label{ah1eq65}\end{equation}

\section{Appendix B}

{\bf  Diagonalization of the gauge boson mass matrix}

\begin{equation} M_B^2 =\left( \begin{array}{cccc} a_1 & b
& 0 & 0\\ b & a_2 & c & 0\\ 0 & c & a_3 & d\\
0 &  0 & d & a_4
\end{array} \right).\label{ah1eq66}
\end{equation}
This matrix may be diagonalize through the orthogonal
matrix $U$ as
\begin{equation} U^T M_B^2 \:U = Diag(\eta_1, \eta_2, \eta_3, \eta_4) \;,\label{ah1eq67}\end{equation}
$\eta_i \equiv M_i^2 \;, i = 1, 2, 3, 4$
being the eigenvalues of $M_B^2$. The determinant equation
\begin{equation} det|M_B^2 - \eta|= 0 \label{ah1eq68}\end{equation}
yields
\begin{equation} \begin{aligned}(a_1 - \eta)(a_2 - \eta)(a_3 -
    \eta)(a_4 - \eta)  &\quad\\ -(a_1 -
\eta)(a_2 - \eta) \:d^2 - (a_1 - \eta)(a_4 - \eta) \:c^2 &\quad \\ - (a_3 - 
\eta)(a_4 - \eta) \:b^2 + b^2 d^2 &= 0 \;,\end{aligned}\label{ah1eq69}\end{equation}
and then imposes the relationships
\begin{equation} \begin{aligned} \eta_1 + \eta_2 + \eta_3 + \eta_4 &=  a_1 + a_2 + a_3
  + a_4 \\
 \eta_1\eta_2 + (\eta_1 + \eta_2)(\eta_3 + \eta_4) + \eta_3\eta_4 
 &=  a_1 a_2 + (a_1 + a_2)(a_3 + a_4) \\ &\quad{}+ a_3 a_4 - b^2 - c^2 - d^2\\
(\eta_1 + \eta_2) \eta_3 \eta_4 + \eta_1 \eta_2(\eta_3 + \eta_4) &=
(a_1 + a_2)a_3 a_4 + a_1 a_2 (a_3 + a_4)  \\ &\quad{} - (a_1 + a_2)d^2  -
(a_1 + a_4)c^2   - (a_3 + a_4)b^2 \\
\eta_1\eta_2\eta_3\eta_4  &=   a_1a_2a_3a_4 - a_1a_2d^2 \\
 &\quad{} -a_1a_4c^2 - a_3a_4 b^2 + b^2 d^2 \end{aligned} \label{ah1eq70}
\end{equation}
on the eigenvalues $\eta_i \;, i=1,2,3,4$
and the parameters of $M_B^2$.

Computing the eigenvectors, the orthogonal matrix $U$ may be writing
as
\begin{equation} U = \left( \begin{array}{cccc} x & y\:
\frac{f_2(\eta_2)}{\Delta_2(\eta_2)} & z\:
\frac{f_3(\eta_3)}{\Delta_3(\eta_3)} & r\:
\frac{f_4(\eta_4)}{\Delta_4(\eta_4)} \\
    &              &                      \\
x\: \frac{f_2(\eta_1)}{\Delta_1(\eta_1)} & y & z\:
\frac{g_3(\eta_3)}{\Delta_3(\eta_3)} & r\:
\frac{g_4(\eta_4)}{\Delta_4(\eta_4)}  \\
    &              &                      \\
x\: \frac{f_3(\eta_1)}{\Delta_1(\eta_1)} & y\:
\frac{g_3(\eta_2)}{\Delta_2(\eta_2)} & z & r\:
\frac{h_4(\eta_4)}{\Delta_4(\eta_4)}  \\
    &              &                 \\
x\: \frac{f_4(\eta_1)}{\Delta_1(\eta_1)} & y\:
\frac{g_4(\eta_2)}{\Delta_2(\eta_2)} & z\:
\frac{h_4(\eta_3)}{\Delta_3(\eta_3)} & r\:
\end{array} \right) \:, \label{ah1eq71}\end{equation}
where $x$, $y$, $z$ and $r$ are normalization constants,
and the functions involved  are defined as
\begin{equation} \begin{array}{ccl} \Delta_1(\eta) & \equiv &
(a_2-\eta)(a_3-\eta)(a_4-\eta) -
(a_2-\eta)d^2 - (a_4-\eta)c^2          \;,            \\
                                                 \\
\Delta_2(\eta) & \equiv & (a_1-\eta)\left[(a_3-\eta)(a_4-\eta) - d^2
\right]                                  \;,        \\
                                                 \\
\Delta_3(\eta) & \equiv & (a_4-\eta)\left[(a_1-\eta)(a_2-\eta) - b^2
\right]                                     \;,     \\
                                                 \\
\Delta_4(\eta) & \equiv & (a_1-\eta)(a_2-\eta)(a_3-\eta) -
(a_1-\eta)c^2 - (a_3-\eta)b^2  \;, \end{array} \label{ah1eq72}\end{equation}
and
\begin{equation} \begin{array}{ccl} f_2(\eta) & \equiv & - b
\left[(a_3 - \eta)(a_4 - \eta) - d^2 \right]          \;,         \\
f_3(\eta) & \equiv & bc(a_4 - \eta)              \;, \\
f_4(\eta) & \equiv & - bcd                       \;, \\
g_3(\eta) & \equiv & - c(a_1 - \eta)(a_4 - \eta)  \;, \\
g_4(\eta) & \equiv & cd(a_1 - \eta)              \;,  \\
h_4(\eta) & \equiv & - d \left[(a_1 - \eta)(a_2 - \eta) - b^2
\right] \;. \end{array} \label{ah1eq73}\end{equation}
The above defined functions satisfy the relationships
\begin{equation} \begin{array}{ccc} f_2^2(\eta)=\Delta_1(\eta)\Delta_2(\eta)
& , & g_3^2(\eta)=\Delta_2(\eta)\Delta_3(\eta) \;, \\
f_3^2(\eta)=\Delta_1(\eta)\Delta_3(\eta) & , &
g_4^2(\eta)=\Delta_2(\eta)\Delta_4(\eta)      \;, \\
f_4^2(\eta)=\Delta_1(\eta)\Delta_4(\eta) & , &
h_4^2(\eta)=\Delta_3(\eta)\Delta_4(\eta)  \;,
\end{array}  \label{ah1eq74}\end{equation}

\begin{equation} \begin{array}{ccc} f_2(\eta) f_3(\eta)=\Delta_1(\eta)
g_3(\eta)
& , & f_2(\eta) g_3(\eta)=\Delta_2(\eta)f_3(\eta) \;, \\
f_2(\eta) f_4(\eta)=\Delta_1(\eta) g_4(\eta) & , &
f_2(\eta) g_4(\eta)=\Delta_2(\eta) f_4(\eta)      \;, \\
f_3(\eta) f_4(\eta)=\Delta_1(\eta) h_4(\eta) & , & g_3(\eta)
g_4(\eta)=\Delta_2(\eta) h_4(\eta)  \;,
\end{array} \label{ah1eq75} \end{equation}

\begin{equation} \begin{array}{ccc} f_3(\eta) g_3(\eta)=\Delta_3(\eta)
f_2(\eta)
& , & f_4(\eta) g_4(\eta)=\Delta_4(\eta) f_2(\eta) \;, \\
f_3(\eta) h_4(\eta)=\Delta_3(\eta) f_4(\eta) & , &
f_4(\eta) h_4(\eta)=\Delta_4(\eta) f_3(\eta)      \;, \\
g_3(\eta) h_4(\eta)=\Delta_3(\eta) g_4(\eta) & , & g_4(\eta)
h_4(\eta)=\Delta_4(\eta) g_3(\eta)      \;.
\end{array}  \label{ah1eq76}\end{equation}
Using now these equations, one obtains the normalization
constants
\begin{equation} \begin{aligned} x&=\sqrt{\frac{\Delta_1(\eta_1)}{h(\eta_1)}}
> 0 , & y&=\sqrt{\frac{\Delta_2(\eta_2)}{h(\eta_2)}} > 0  ,\\
z&=\sqrt{\frac{\Delta_3(\eta_3)}{h(\eta_3)}} > 0 , 
 & r&=\sqrt{\frac{\Delta_4(\eta_4)}{h(\eta_4)}} > 0  ,\end{aligned}\label{ah1eq77}\end{equation}
and in general
\begin{equation} U_{ij}^2 = \frac{\Delta_i(\eta_j)}{h(\eta_j)} \geqq 0
\qquad,\qquad \text{and hence} \qquad |U_{ij}|=
\sqrt{\frac{\Delta_i(\eta_j)}{h(\eta_j)}} \;, \label{ah1eq78}\end{equation}
where
\begin{eqnarray}  h(\eta) & \equiv & \Delta_1(\eta) +
\Delta_2(\eta) + \Delta_3(\eta) + \Delta_4(\eta)  \nonumber         \\
                & = & - 4 \eta^3 + 3(\eta_1 + \eta_2 + \eta_3 +
                 \eta_4) \eta^2 \nonumber\\
              & &  - 2[\eta_1 \eta_2 + (\eta_1 +
                 \eta_2)(\eta_3 + \eta_4) + \eta_3 \eta_4] \eta \nonumber\\ 
               & &  + \:(\eta_1 + \eta_2) \eta_3 \eta_4 +
                 \eta_1 \eta_2 (\eta_3 + \eta_4) \;.  \label{ah1eq79}\end{eqnarray}
Explicitly
\begin{equation} \begin{array}{ccc} h(\eta_1) & = & (\eta_2 - \eta_1)(\eta_3 -
\eta_1)(\eta_4 - \eta_1)          \;,    \\
                                     \\
h(\eta_2) & = & (\eta_1 - \eta_2)(\eta_3 -
\eta_2)(\eta_4 - \eta_2)            \;, \\
                                     \\
h(\eta_3) & = & (\eta_1 - \eta_3)(\eta_2 -
\eta_3)(\eta_4 - \eta_3)           \;,   \\
                                     \\
h(\eta_4) & = & (\eta_1 - \eta_4)(\eta_2 -
\eta_4)(\eta_3 - \eta_4) \;.
\end{array} \label{ah1eq80}\end{equation}
The above relationships among the defined functions,Eqs.(\ref{ah1eq74}-\ref{ah1eq76}), allows one
to check the orthogonality of $U$, $U^T U = 1$, the Eq. (\ref{ah1eq67}), as well
as the useful equalities:
\begin{equation} \begin{aligned} U_{1k} U_{4k}&=\frac{f_4(\eta_k)}{h(\eta_k)}, &  U_{2k} U_{4k}&=\frac{g_4(\eta_k)}{h(\eta_k)},\\
 U_{1k} U_{3k}&=\frac{f_3(\eta_k)}{h(\eta_k)}, 
 & U_{2k} U_{3k}&=\frac{g_3(\eta_k)}{h(\eta_k)} .\end{aligned}\label{ah1eq81}
\end{equation}
This procedure to diagonalize $M_B^2$ is the
generalization of the method introduced in the Appendix A in
Ref.\cite{ah1u1}, and it may be extended to diagonalize a generic real
and symmetric 4x4 mass matrix.

\subsection{Simplified Parameter Space: $a_3=a_1$ and $a_4=a_2$}

Note that for this particular case, the Eq.(\ref{ah1eq69}) reduces to
\begin{equation} \left[(a_1 - \eta)(a_2 - \eta)\right]^2 - (b^2+c^2+d^2)\:(a_1 -
\eta)(a_2 - \eta) + b^2 d^2=0 \:,\label{ah1eq82}\end{equation}
and hence, this simplified parameter space
allows one to compute the eigenvalues $\eta_i=M_i^2$ of $M_B^2$ in
exact analytical form in terms of the parameters $a_1$, $a_2$, $b$,
$c$, $d$.

%% R. Mirman, discussion questions, 08.10.2009
\author{R. Mirman\thanks{sssbbg@gmail.com}}
\title{Additional Open Questions}
\institute{%
14U\\
155 E 34 Street\\
New York, NY  10016
}

\titlerunning{Additional Open Questions}
\authorrunning{R. Mirman}
\maketitle

\vspace{0.5 cm}
1. There is a mass level formula for the elementary particles,
(Quantum Field Theory, Conformal Group Theory, Conformal Field Theory)
$m = n (1/\alpha + a)\times (m_e)$, where $n$ is an integer, or half-integer, $m_e$, the electron mass, $\alpha$ the fine structure constant, and $a = 1$, $0$ or $-1$, with no relationship (apparently) among these values. Most charged particles lie close to these values, but neutral ones agree poorly. Thus the proton differs by $0.1724m_e$, while the neutron by $1.1209m_e$. The charged pion is off by $0.1462m_e$, the neutral one by $5.0742m_e$. This is true in general although other discrepancies may be larger (or in some cases smaller). Can a model be created that gives results like these?
 
2. It is clear that group theory, especially for groups related to geometry, provides much information and constraints on the laws of nature. Can the groups be reasonably generalized, especially with geometrical motivation, to provide further information about, and requirements on, physics?

3. There have been attempts to generalize quantum mechanics. Considering what quantum mechanics is, can it possible be generalized? How?

4. Is it possible to have a theory of gravitation (which is a massless helicity-2 Poincar group representation) besides general relativity aside from some questions about the coupling? Or does the group (required by geometry) impose such strong conditions to make any other theory impossible?

%%%
\cleardoublepage
\chapter*{\Huge PRESENTATION OF \\
VIRTUAL INSTITUTE OF ASTROPARTICLE PHYSICS \\
AND\\
 BLED 2009 WORKSHOP VIDEOCONFERENCES}
\addcontentsline{toc}{chapter}{Presentation of Virtual Institute of
  Astroparticle Physics and Bled 2009 Workshop Videoconferences}
\newpage

\cleardoublepage
%%%
%% presentation of VIA and discussions on VIA
%%
%%\begin{document}
\title{Virtual Institute of Astroparticle Physics at Bled Workshop}
\author{M.Yu. Khlopov$^{1,2,3}$}
\institute{%
$^{1}$Moscow Engineering Physics Institute (National Nuclear Research University), 115409 Moscow, Russia \\
 $^{2}$ Centre for Cosmoparticle Physics "Cosmion" 125047 Moscow, Russia \\
$^{3}$ APC laboratory 10, rue Alice Domon et L\'eonie Duquet \\75205
Paris Cedex 13, France}

\authorrunning{M.Yu. Khlopov}
\titlerunning{Virtual Institute of Astroparticle Physics at Bled Workshop}
\maketitle

\begin{abstract}

Virtual Institute of Astroparticle Physics (VIA) has evolved in a
unique
 multi-functional complex, combining various forms of collaborative
 scientific work with programs of education on distance.
The activity on VIA website includes regular videoconferences with
systematic basic courses and lectures on various issues of
astroparticle physics, participation at distance in various scientific meetings and conferences,
library of their records and presentations, a
multilingual forum. VIA virtual rooms are open for meetings of
scientific groups and for individual work of supervisors with their
students. The format of a VIA videoconferences was effectively used in the program
of Bled Workshop to discuss the open questions of physics beyond the standard model.

\end{abstract}

\section{Introduction}
Studies in astroparticle physics link astrophysics, cosmology and
particle physics and involve hundreds of scientific groups linked by
regional networks (like ASPERA/ApPEC \cite{vd5aspera}) and national
centers. The exciting progress in these studies will have impact on the fundamental knowledge on the
structure of microworld and Universe and on the basic, still
unknown, physical laws of Nature (see e.g. \cite{vd5book} for review).

In the proposal \cite{vd5Khlopov:2008vd} it was suggested to organize a
Virtual Institute of Astroparticle Physics (VIA), which can play the
role of an unifying and coordinating structure for astroparticle
physics. Starting from the January of 2008 the activity of the
Institute takes place on its website \cite{vd5VIA} in a form of regular
weekly videoconferences with VIA lectures, covering all the
theoretical and experimental activities in astroparticle physics and
related topics. In 2008 VIA complex was effectively used for participation on distance in XI Bled Workshop and
Gran Sasso Summer Institute on Astroparticle physics \cite{vd5archiVIA}. The library of records of these lectures, talks and their presentations is now accomplished by multi-lingual forum. Here the
general structure of VIA complex and the format of its
videoconferences are stipulated to clarify the way in which VIA
discussion of open questions beyond the standard model took place in the
framework of Bled Workshop.
\section{The structure of VIA complex}
\begin{figure}
    \begin{center}
        \includegraphics[scale=0.3]{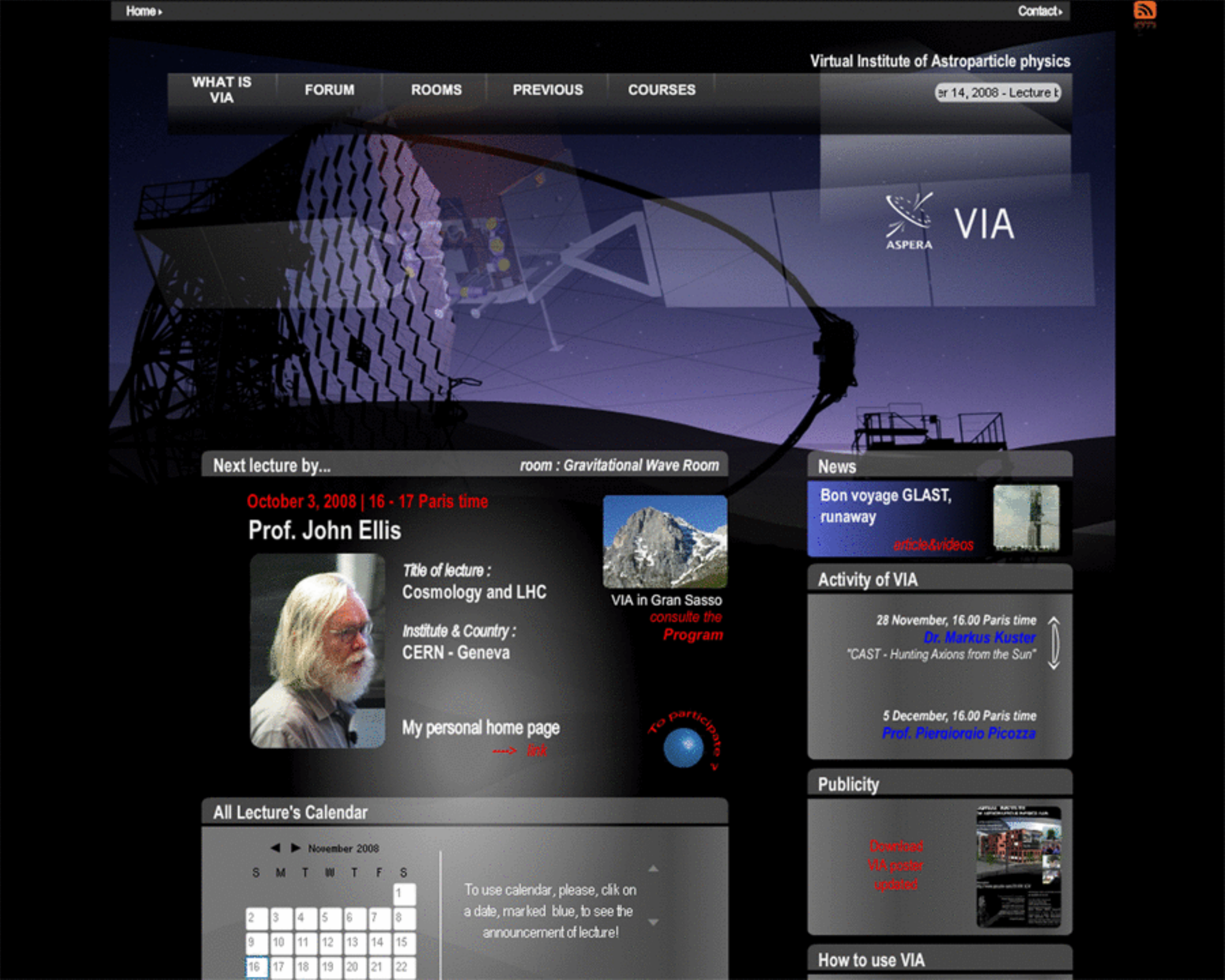}
        \caption{The home page of VIA site}
        \label{vd5a}
    \end{center}
\end{figure}

The structure of VIA complex is illustrated on Fig.~\ref{vd5a}.
The home page, presented on this figure, contains the information on
VIA activity and menu, linking to directories (along the upper line
from left to right): with general information on VIA (What is VIA),
to Forum, to VIA virtual lecture hall and meeting rooms (Rooms), to
the library of records and presentations of VIA lectures and courses
(Previous) and to contact information (Contacts). The announcement
of the next Virtual meeting, the calender with the program of future
lectures and courses together with the links to VIA news and posters
as well as the instructions How to use VIA are also present on the
home page. The VIA forum is intended to cover the
topics: beyond the standard model, astroparticle physics, cosmology,
gravitational wave experiments, astrophysics, neutrinos. Presently
activated in English, French and Russian with trivial extension
to other languages, the Forum represents a first
step on the way to multi-lingual character of VIA complex and its activity.
One of the interesting forms of forum activity is work on small thesis,
which students of Moscow Engineering Physics Institute should prepare to pass
their exam on course "Introduction to Cosmoparticle physics".
The record of videoconference with their oral exam is also put in the corresponding directory of forum.

\section{VIA lectures and virtual meetings}
\begin{figure}
    \begin{center}
        \includegraphics[scale=0.3]{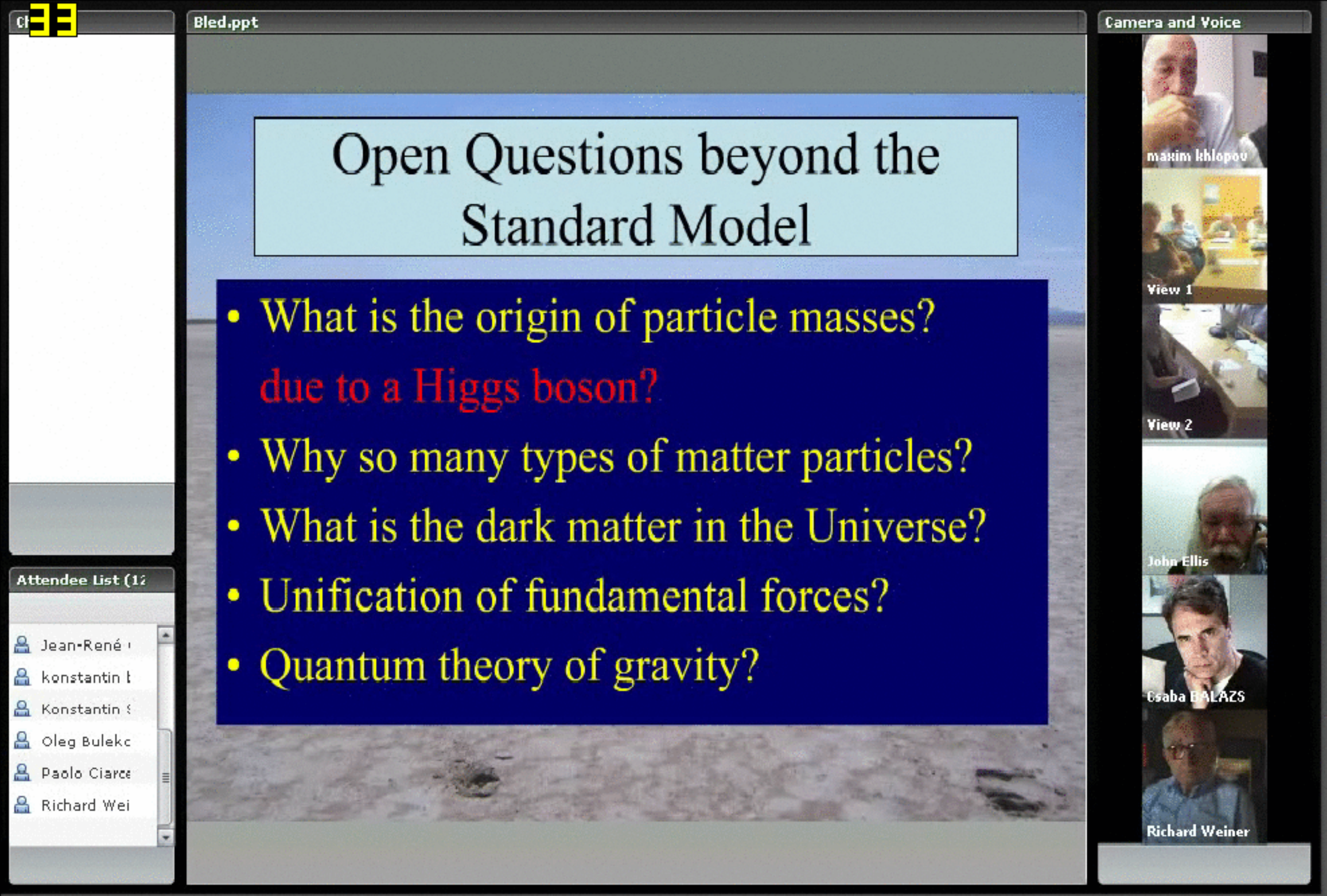}
        \caption{Videoconference with lecture by John Ellis, which he gave from his office in CERN,
        Switzerland, became a part of the program of
           XII Bled Workshop.}
        \label{vd5b}
    \end{center}
\end{figure}

First tests of VIA system, described in \cite{vd5Khlopov:2008vd,vd5archiVIA}, involved various
systems of videoconferencing. They included skype, VRVS, EVO, WEBEX, marratech
and adobe Connect. In the result of these tests the adobe Connect system
was chosen and properly acquired. Its advantages are: relatively easy use for participants,
a possibility to make presentation in a video contact between presenter and audience,
a possibility to make high quality records and edit them, removing from records occasional
and rather rare disturbances of sound or connection, to use a whiteboard facility for discussions,
the option to open desktop and to work online with texts in any format.
The regular form of VIA meetings assumes that their time and Virtual room are announced in advance.
Since the access to the Virtual room is strictly controlled by administration,
the invited participants should enter the Room as Guests, typing their names,
and their entrance and successive ability to use video and audio system is authorized by the Host
of the meeting.
The format of VIA lectures and discussions is shown on Fig.~\ref{vd5b}, illustrating the talk given by John Ellis from CERN in the framework of XII Workshop. The complete record of this talk and other VIA discussions are available on VIA website \cite{vd5VIAbled}

%\begin {figure} \vspace{7cm}\caption{Adobe videoconference Paris-Lyon-Geneve-Hamburg-Moscow-Amsterdam} \label{vd5adobe} \end {figure}
The ppt file of presentation is uploaded in the system in advance and then demonstrated in the central window.
Video images of presenter and participants appear in the right window, while in the lower left window the list of all the attendees
is given. To protect the quality of sound and record, the participants are required to switch out their audio system during presentation
and to use upper left Chat window for immediate comments and urgent questions. The Chat window can be also used by participants, having no microphone,
 for questions and comments during Discussion. In the end of presentation the central window can be used for a whiteboard utility
 as well as the whole structure of windows can be changed, e.g. by making full screen the window with the images of participants of discussion.

\section{Conclusions}
The exciting experiment of VIA Discussions at Bled Workshop, the three days of permanent online transmissions
and distant participation in the Gran Sasso Summer Institute on Astroparticle physics \cite{vd5VIAGS}, four days of VIA interactive online transmission of series of seminars by M.Khlopov in Liege \cite{vd5VIAliege}, online transmission from International Workshop on Astronomy and Relativistic Astrophysics (IWARA09, Maresias, Brazil), the stable
regular weekly videoconferences with VIA lectures and the solid library of their records and presentations,
creation of multi-lingual VIA Internet forum, regular basic courses and individual work on distance with students of MEPhI prove that the Scientific-Educational complex of Virtual Institute of Astroparticle physics can provide regular
communications between different groups and scientists, working in
different scientific fields and parts of the world, get the first-hand information on the newest scientific results,
as well as to support various educational programs on distance. This activity
would easily allow finding mutual interest and organizing task
forces for different scientific topics of astroparticle physics and related topics. It
can help in the elaboration of strategy of experimental particle,
nuclear, astrophysical and cosmological studies as well as in proper
analysis of experimental data. It can provide young talented people from all over the world to get the highest level education,
come in direct interactive contact with the
world known scientists and to find their place in the fundamental research. To conclude the VIA complex is in operation and
ready for a wide use and extension of its applications.
\section*{Acknowledgements}
 The initial step of creation of VIA was
 supported by ASPERA. I am grateful to S.Katsanevas for permanent stimulating support, to P.Binetruy, J.Ellis, F.Fidecaro, M.Pohl, Y. Giraud-Heraud for help in development of the project, to K.Belotsky, A.Kirillov and K.Shibaev for assistance in educational
 VIA program and to C. Kouvaris, A.Mayorov and E.Soldatov for cooperation in its applications, to Z.Be\-rezhiani and all Organizers of XIII Summer School on Astroparticle Physics in Gran Sasso, to J.-R.Cudell and all organizers of series of seminars in Liege, to Cesar Augusto Zen Vasconcellos, Rafael Martins Villas B\^oas and Mario Luiz Lopes da Silva and all organizers of IWARA09 Workshop
 for cooperation in the use of VIA for online distant participation and to D.Rouable for help in technical realization and support of VIA complex.
 I express my gratitude to N.S. Manko\v c Bor\v stnik, G.Bregar, D. Lukman and all
 Organizers of Bled Workshop for cooperation in the exciting experiment of
 VIA Discussion Session at XII Bled Workshop.

%%%%%%%%%%%%%%%%%%%%%%%%%%%%%%%%%%%%%%%%%%%%%%%%%%%%%%%%%%%%%%%%%%%%%%

\backmatter

\thispagestyle{empty}
\parindent=0pt
\begin{flushleft}
\mbox{}
\vfill
\vrule height 1pt width \textwidth depth 0pt
{\parskip 6pt

{\sc Blejske Delavnice Iz Fizike, \ \ Letnik~10, \v{s}t. 2,} 
\ \ \ \ ISSN 1580-4992

{\sc Bled Workshops in Physics, \ \  Vol.~10, No.~2}

\bigskip

Zbornik 12. delavnice `What Comes Beyond the Standard Models', 
Bled, 14.~-- 24.~julij 2009

Proceedings to the 12th workshop 'What Comes Beyond the Standard Models', 
Bled, July 14.--24.,  2009

\bigskip

Uredili Norma Susana Manko\v c Bor\v stnik, Holger Bech Nielsen in Dragan Lukman 

Publikacijo sofinancira  Javna agencija za knjigo Republike Slovenije 

Brezpla\v cni izvod za udele\v zence 

Tehni\v{c}ni urednik Vladimir Bensa

\bigskip

Zalo\v{z}ilo: DMFA -- zalo\v{z}ni\v{s}tvo, Jadranska 19,
1000 Ljubljana, Slovenija

Natisnila ALFAGRAF TRADE v nakladi 150 izvodov

\bigskip

Publikacija DMFA \v{s}tevilka 1762

\vrule height 1pt width \textwidth depth 0pt}
\end{flushleft}

%%%%%%%%%%%%%%%%%%%%%%%%%%%%%%%%%%%%%%%%%%%%%%%%%%%%%%%%%%%%%%%%%%%%%

\end{document}